\documentclass[12pt,english,american]{article}
\usepackage[T1]{fontenc}
\usepackage[latin9]{inputenc}
\usepackage{geometry}
\usepackage{color}
\usepackage{yfonts}
\usepackage{slashed}
\usepackage{babel}
\usepackage{amsthm}
\usepackage{mathrsfs}
\usepackage{amstext}
\usepackage{amssymb}
\usepackage[unicode=true,pdfusetitle,
 bookmarks=true,bookmarksnumbered=false,bookmarksopen=false,
 breaklinks=true,pdfborder={0 0 0},backref=false,colorlinks=true]
 {hyperref}
\hypersetup{
 linkcolor=blue,citecolor=blue, urlcolor=blue}

\makeatletter

\def\qed{$\Box$\medskip}

\newtheorem{theoreme}{Theorem } [section]
\newtheorem{proposition}[theoreme]{Proposition}
\newtheorem{lemma}[theoreme]{Lemma}
\newtheorem{definition}[theoreme]{Definition}
\newtheorem{corollary}[theoreme]{Corollary}
\newtheorem{remark}[theoreme]{Remark}
\newtheorem{example}[theoreme]{Example}
\newtheorem{criterion}[theoreme]{Criterion}

\newcommand{\beq}{\begin{equation}}
\newcommand{\eeq}{\end{equation}}
\newcommand{\beqa}{\begin{eqnarray}}
\newcommand{\eeqa}{\end{eqnarray}}
\newcommand{\ben}{\begin{arabicenumerate}}
\newcommand{\een}{\end{arabicenumerate}}
\newcommand{\bex}{\begin{example}}
\newcommand{\eex}{\end{example}}
\newcommand{\ber}{\begin{remark}}
\newcommand{\eer}{\end{remark}}
\newcommand{\bec}{\begin{corollary}}
\newcommand{\eec}{\end{corollary}}
\newcommand{\bep}{\begin{proposition}}
\newcommand{\eep}{\end{proposition}}
\newcommand{\becr}{\begin{criterion}}
\newcommand{\eecr}{\end{criterion}}

\def\bel{\begin{lem} } 
\def\eel{\end{lem} }
\def\bet{\begin{theoreme}}
\def\eet{\end{theoreme}}
\def\bed{\begin{defn}}
\def\eed{\end{defn} }

\usepackage{enumitem}		% customizable list environments
\theoremstyle{plain}
\newtheorem{thm}{\protect\theoremname}[section]
\theoremstyle{definition}
\newtheorem{defn}[thm]{\protect\definitionname}
\theoremstyle{plain}
\newtheorem{prop}[thm]{\protect\propositionname}
\theoremstyle{remark}

\theoremstyle{plain}
\newtheorem{lem}[thm]{\protect\lemmaname}
\theoremstyle{plain}

\usepackage{txfonts,refstyle,xcolor}

\newref{con}{name = Conjecture\ }
\newref{prop}{name = Proposition\ }
\newref{def}{name = Definition\ }
\newref{sec}{name = Section\ }
\newref{sub}{name = Section\ }
\newref{thm}{name = Theorem\ }
\newref{lem}{name = Lemma\ }
\newref{cor}{name = Corollary\ }
\newref{fig}{name = Figure\ }

\usepackage{bbm}
\newcommand{\charf}{\mathbbm{1}}

\usepackage[all]{xy}

\newcommand{\xyR}[1]{%
     \makeatletter
     \xydef@\xymatrixrowsep@{#1}
     \makeatother
}

\newcommand{\xyC}[1]{%
     \makeatletter
     \xydef@\xymatrixcolsep@{#1}
     \makeatother
}

\newcommand{\ncol}[1]{\color{normalcolor}}

\makeatother

\addto\captionsamerican{\renewcommand{\corollaryname}{Corollary}}
\addto\captionsamerican{\renewcommand{\definitionname}{Definition}}
\addto\captionsamerican{\renewcommand{\lemmaname}{Lemma}}
\addto\captionsamerican{\renewcommand{\propositionname}{Proposition}}
\addto\captionsamerican{\renewcommand{\remarkname}{Remark}}
\addto\captionsamerican{\renewcommand{\theoremname}{Theorem}}
\addto\captionsenglish{\renewcommand{\corollaryname}{Corollary}}
\addto\captionsenglish{\renewcommand{\definitionname}{Definition}}
\addto\captionsenglish{\renewcommand{\lemmaname}{Lemma}}
\addto\captionsenglish{\renewcommand{\propositionname}{Proposition}}
\addto\captionsenglish{\renewcommand{\remarkname}{Remark}}
\addto\captionsenglish{\renewcommand{\theoremname}{Theorem}}
\providecommand{\corollaryname}{Corollary}
\providecommand{\definitionname}{Definition}
\providecommand{\lemmaname}{Lemma}
\providecommand{\propositionname}{Proposition}
\providecommand{\remarkname}{Remark}
\providecommand{\theoremname}{Theorem}
\begin{document}
\title{Bose particles in a box II. A convergent expansion of  the ground state of the Bogoliubov 
Hamiltonian  in the mean field limiting regime.} 
% \author{ A. Pizzo}
  \author{A. Pizzo \footnote{email: pizzo@mat.uniroma2.it}\\
 Dipartimento di Matematica, Universit\`a di Roma ``Tor Vergata",\\
 Via della Ricerca Scientifica 1, I-00133 Roma, Italy}
 % \author{ A. Pizzo}

\date{15/01/2017}

\maketitle

\abstract{In this paper we consider an interacting Bose gas at zero temperature, in a finite  box and in the mean field limiting regime. %{\color{red}(i.e., at fixed box volume $|\Lambda|$, for a number of particles, $N$, sufficiently large, and for a coupling constant inversely proportional to the particle density).} 
The $N$ gas particles interact through a pair potential of positive type and with an ultraviolet cut-off. Its (nonzero) Fourier components are sufficiently large with respect to the corresponding kinetic energies of the modes.  Using the multi-scale technique in the occupation numbers of particle states introduced in \cite{Pi1}, we provide a convergent expansion of the ground state of the \emph{particle number preserving} Bogoliubov Hamiltonian in terms of the bare operators. In the limit $N\to \infty$ the expansion is  up to any desired precision. 
% provided an ultraviolet cutoff has been imposed on the pair potential. 
%{\color{red}This procedure also provides an algorithm to expand the ground state vector in terms of the bare quantities up to any precision.}}
\\

\noindent
{\bf{Summary of contents}}
\begin{itemize}
\item In Section \ref{introduction},  the model of a Bose gas  is defined along with the notation used throughout the paper. In particular, the \emph{particle number preserving} Bogoliubov Hamiltonian (from now on Bogoliubov Hamiltonian) is defined.
\item In Section \ref{multiscale-HBog},  we review the main ideas and results of the multi-scale analysis in the occupation numbers of particle states implemented for the three-modes Bogoliubov Hamiltonian in \cite{Pi1}. In fact, the full Bogoliubov Hamiltonian can be thought of as a collection of three-modes systems, and we can iteratively apply the multi-scale analysis to them. The corresponding Feshbach-Schur flows are described informally in Section \ref{Feshbach}. 
\item In Section \ref{groundstateHBog},  in the mean field limiting regime (i.e., at fixed box volume $|\Lambda|$,  $N$ sufficiently large, and for a coupling constant  inversely proportional to the particle density) a convergent expansion of the ground state vector of the  Bogoliubov Hamiltonian is provided as a byproduct of subsequent Feshbach-Schur flows, each of them associated with a couple of (interacting) modes. 
%{\color{red}This also provides a convergent expansion of the vector up to any precision in terms of the bare quantities.}
%\item In Section \ref{new-proj}, ....
%\item
%Some of the proofs are deferred to the Appendix in Section \ref{appendix}.
\end{itemize}
\setcounter{equation}{0}

\section{Interacting Bose gas in a box}\label{introduction}

%\[
%a(x)=\frac{\sum_{\boldsymbol{n}}a_{\boldsymbol{n}}e^{ik_{\boldsymbol{n}}\cdot x}}{|\Lambda_{L'}|^{\frac{1}{2}}}
%\]
%where $|\Lambda_{L'}|=L'^{3}$and $k_{\boldsymbol{n}}=\frac{2\pi\boldsymbol{n}}{L'}$
%and $\boldsymbol{n}\in Z^{3}$. Commutation relations:

%\[
%[a_{\boldsymbol{n}},a_{\boldsymbol{n'}}^{*}]=\delta_{\boldsymbol{n},\boldsymbol{n}'}
%\]
%Hamiltonian
%\[
%H'=\int\frac{1}{2m}(\nabla a^{*})(\nabla a)(x)dx+\frac{\lambda}{2}\int\int(a^{*}(x)a(x)-\rho)\phi(x-y)(a^{*}(x)a(x)-\rho)dxdy
%\]
%where the integration is over $\Lambda_{L'}$ , and 
%\[
%\phi(x-y)=(2\pi)^{-\frac{3}{2}}\sum_{\boldsymbol{n}}\tilde{\phi}(k_{\boldsymbol{n}})\frac{e^{ik_{\boldsymbol{n}}\cdot(x-y)}}{|\Lambda_{L'}|^{\frac{1}{2}}}
%\]
%\\

We study a gas of (spinless) nonrelativistic Bose particles that, at zero temperature, are constrained to a $d$-dimensional box of side $L$ with $d\geq 1$. The particles interact through a pair potential with a coupling constant proportional to the inverse of the particle density  $\rho$.  The rigorous description of this system has many intriguing mathematical aspects not completely clarified yet. In spite of the progress in recent years, some important problems are still open to date, in particular in connection to the thermodynamic limit and the exact structure of the ground state vector.  In this paper we focus on the problem of the construction of the ground state vector for the gas system in the mean field limiting regime. For the convenience of the reader,  we recall some of the results closer to our present work.  
%and give references to the reader for the details. 

The low energy spectrum of the Hamiltonian in the mean field limit was predicted by Bogoliubov \cite{Bo1}, \cite{Bo2}. As for the ground state energy of a Bose gas, rigorous estimates have been provided for certain systems in \cite{LS1}, \cite{LS2}, \cite{ESY},\cite{YY}, and \cite{LSSY}.  Concerning  the excitation spectrum, in Bogoliubov theory it consists of elementary excitations whose energy is linear in the momentum for small momenta. 
After some important results restricted to one-dimensional models (see \cite{G}, \cite{LL}, \cite{L}), this conjecture was proven by Seiringer in \cite{Se1} (see also \cite{GS}) for the low-energy spectrum of an interacting Bose gas in a finite box and in the mean field limit,  where  the pair potential is of positive type. In  \cite{LNSS}  it has been extended to a more general class of potentials and the limiting behavior of the low energy eigenstates has been studied.  Later,  the result of \cite{Se1} has been proven to be valid in a sort of diagonal limit where the particle density and the box volume diverge according to a prescribed asymptotics; see  \cite{DN}. Recently, Bogoliubov's prediction on the energy spectrum in the mean field limit  has been shown to be valid also for the high energy eigenvalues (see \cite{NS}).

\noindent
These results are based on energy estimates starting from the spectrum of the corresponding Bogoliubov Hamiltonian. 

A different approach to studying a gas of Bose particles is based on renormalization group.  In this respect, we mention the paper by Benfatto, \cite{Be}, where he provided \emph{an order by order control} of the Schwinger  functions of this system in three dimensions and with an ultraviolet cut-off. His analysis holds at zero temperature in the infinite volume limit  and at finite particle density. Thus, it contains  a fully consistent treatment of the infrared divergences at a perturbative level. This program has been later developed in  \cite{CDPS1}, \cite{CDPS2}, and, more recently, in  \cite{C} and \cite{CG} by making use of \emph{Ward identities} to deal also with two-dimensional systems where some partial control of the renormalization flow has been provided; see \cite{C} for a detailed review of previous related results.

\noindent
Within the renormalization group approach, we also mention some results towards a rigorous construction of the functional integral for this system contained in \cite{BFKT1}, \cite{BFKT2}, and \cite{BFKT}.

Important  results concern  Bose-Einstein condensation  for a system of trapped Bose particles interacting each other via a potential of the type $N^{3\beta-1}\,v(N^{\beta}x)$,  with $0\leq\beta\leq1$, that for $\beta=1$ corresponds to the so called Gross-Pitaeveskii limiting regime: see \cite{LSY}, \cite{LS}, \cite{LSSY}, \cite{NRS}, and \cite{LNR}.  
%We also mention  the progress in the control of the dynamical properties of Bose gases. For references and for an update of the state of the art, we refer the reader to the introduction of \cite{DFPP}.

Both in the grand canonical and in the canonical ensemble  approach (see \cite{Se1}), starting from the Hamiltonian of the system one can define an approximated one, the Bogoliubov Hamiltonian. For a finite box and a large class of pair potentials, upon a  unitary transformation\footnote{In the canonical ensemble approach, the used unitary transformation yields the diagonalization of the (particle preserving) Bogoliubov Hamiltonian only in the mean field limit (see \cite{Se1}).} the Bogoliubov Hamiltonian  describes
a system of non-interacting bosons with a new energy dispersion law, which is in fact the correct description of the energy spectrum of the Bose particles system in the mean field limit.

%Somehow related to the content of this paper and of the companion papers,
%The literature on the Bose gas in general is vast and goes beyond the problems and the model addressed in this paper. However, it is surely .... 

%We also mention  the {\color{red}progress in the control} of the dynamical properties of Bose gases. For references and for an update of the state of the art we refer the reader to the introduction of \cite{DFPP}.
%\\
%\underline{For a survey of related results and the motivations underlying the multi-scale scheme  we refer} \underline{the reader to the introductory section of \cite{Pi1} and references therein.}
%\\

In the companion paper \cite{Pi1}, we construct the ground state of so called three-modes Bogoliubov Hamiltonians by means of a multi-scale analysis in the particle states occupation numbers.  Here, we explain how the new multi-scale scheme can be applied to a Bogoliubov Hamiltonian involving a finite number of interacting modes. More precisely, we implement the multi-scale scheme in the mean field limiting regime and under the \emph{strong interaction potential assumption} considered in \cite{Pi1}  where the interaction potential is strong with respect to the kinetic energy of the interacting modes; see Definition \ref{def-pot}. The results of the present paper are preliminary ingredients for the construction and expansion of the ground state of the complete Hamiltonian of the system in \cite{Pi3}.

We recall the Hamiltonian describing a gas of (spinless) nonrelativistic Bose particles that are constrained to a $d$-dimensional  box, $d\geq 1$,  and interact through a pair potential.  Though the number of particles is fixed we use the formalism of second quantization. The Hamiltonian corresponding to the pair potential $\phi(x-y)$ and to the coupling constant $\lambda>0$ is

\begin{equation}\label{initial-ham}
\mathscr{H}:=\int\frac{1}{2m}(\nabla a^{*})(\nabla a)(x)dx+\frac{\lambda}{2}\int\int a^{*}(x)a^*(y)\phi(x-y)a(x)a(y)dxdy\,,
\end{equation}
where: $\hbar$ has been set equal to $1$;
reference to the integration domain $\Lambda:=\{x\in \mathbb{R}^d\,|\,|x_i|\leq \frac{L}{2}\,,\,i=1,2,\dots, d\}$ is omitted and periodic boundary conditions are assumed; $dx$ is the Lebesgue measure in $d$ dimensions. Here, the operators $a^{*}(x)\:,a(x)$ are the usual operator-valued distributions on the bosonic Fock space$$\mathcal{F}:= \Gamma\left(L^{2}\left(\Lambda,\mathbb{C};dx\right)\right)\,$$  that satisfy the canonical
commutation relations (CCR)
\[
[a^{\#}(x),a^{\#}(y)]=0,\quad\quad[a(x),a^{*}(y)]=\delta(x-y)\charf,
\]
with $a^{\#}:=a$ or $a^{*}$. In terms of the field modes, they read
\[
a(x)=\sum_{\bold{j}\in \mathbb{Z}^d}\frac{a_{\bold{j}}e^{ik_{\bold{j}}\cdot x}}{|\Lambda|^{\frac{1}{2}}},\quad a^{*}(x)=\sum_{\bold{j}\in \mathbb{Z}^d}\frac{a_{\bold{j}}^*e^{-ik_{\bold{j}}\cdot x}}{|\Lambda|^{\frac{1}{2}}},\]
where $k_{\bold{j}}:=\frac{2\pi}{L} \bold{j}$, $\bold{j}=( j_1, \dots, j_d)$, $j_1, \dots, j_d \in \mathbb{Z}$, and $|\Lambda|=L^d$,  with CCR

\[
[a^{\#}_{\bold{j}},a^{\#}_{\bold{j}'}]=0,\quad\quad[a_{\bold{j}},a^{*}_{\bold{j}'}]=\delta_{\bold{j}\,,\,\bold{j}'}\,,\quad a^{\#}_{\bold{j}}=a_{\bold{j}}\,\,\text{or}\,\, a^{*}_{\bold{j}}.
\]
The (nondegenerate) vacuum vector of $\mathcal{F}$ is denoted by $\Omega$ ($\|\Omega\|=1$).

Given any function $\varphi\in L^{2}\left(\Lambda,\mathbb{C};dz\right)$, we express it in terms of its Fourier coefficients $\varphi_{\bold{j}}$, i.e.,
\begin{equation}
\varphi(z)=\frac{1}{|\Lambda|}\sum_{\bold{j}\in \mathbb{Z}^d}\varphi_{\bold{j}}e^{ik_{\bold{j}}z}\,.
\end{equation}
%and the Parseval identity reads
%\begin{equation}
%\int dz |\varphi|^2(z)=\frac{1}{|\Lambda|}\sum_{\bold{j}\in \mathbb{Z}^d}|\varphi_{\bold{j}}|^2<\infty\,.
%\end{equation}
\begin{definition}\label{def-pot}
The potential $\phi(x-y)$  is a bounded, real-valued function that is periodic, i.e., $\phi(z)=\phi(z+\bold{j}L)$  for $\bold{j}\in \mathbb{Z}^{d}$, and satisfies  the following conditions:
\begin{enumerate}
%\item
%$\phi(x-y)\geq 0$
%\item
%$\phi(x-y)=\phi(y-x)$, hence we write $\phi(x-y)=\frac{1}{|\Lambda|}\sum_{\bold{j}\in \mathbb{Z}^3}\phi_{\bold{j}}e^{ik_{\bold{j}}(x-y)}$
\item  $\phi(z)$ is an even function, in consequence $\phi_{\bold{j}}=\phi_{-\bold{j}}$. 
\item 
$\phi(z)$ is of positive type, i.e., the Fourier components $\phi_{\bold{j}}$ are nonnegative.
\item
The pair interaction has a fixed but arbitrarily large ultraviolet cutoff (i.e., the nonzero Fourier components $\phi_{\bold{j}}$  form a finite set $\{\phi_{\bold{0}}, \phi_{\pm\bold{j}_1},\dots, \phi_{\pm\bold{j}_M}\}$) with the requirement below to be satisfied:

 (\underline{Strong Interaction Potential Assumption})
The ratio $\epsilon_{\bold{j}}$ between the kinetic energy of the modes $\pm\bold{j}\neq \bold{0}$ and the corresponding Fourier component, $\phi_{\bold{j}}(\neq 0)$,  of the potential, i.e., $\frac{k_{\bold{j}}^2}{\phi_{\bold{j}}}=:\epsilon_{\bold{j}}$, is required to be small enough to ensure the estimates used in \cite{Pi1}. 
%Notice that $\epsilon_{\bold{j}}$ small corresponds either to a low energy mode  $\frac{2\pi\bold{j}}{L}$  or/and to a large potential component $\phi_{\bold{j}}$.
%\item [ii)] (\underline{Non-connected Frequencies Assumption}) For any $l,l'$ in the set $\{1,\dots, M\}$, the modes
%\begin{equation}
%\bold{j}_{l}\pm \bold{j}_{l'}\quad \text{and}\quad -(\bold{j}_{l}\pm \bold{j}_{l'})
%\end{equation}
%do not belong to the set $\{\pm \bold{j}_1,\dots,\pm \bold{j}_M\}$.
\end{enumerate}
\end{definition}
\begin{remark}
Other regimes for the ratios $\epsilon_{\bold{j}}$ can be explored with the same method by suitable modifications of some estimates (see \cite{CP}).
\end{remark}
%\begin{remark}
%The class of potentials chosen in Definition 1.1 is just to use only one method, i.e., the multi-scale analysis in the number of zero-mode particles using the Feshbach map. This method is quite effective with high potentials, because of the artificial gap induced by the chosen projection in the resolvents of the expansion: the higher is the potential with respect to the kinetic energy,  the higher is the gap. In fact,  the gap is the difference between the $\inf$ of the free Hamiltonian and the value of the spectral parameter $z\in \mathbb{R}, z<0$,  entering the Feshbach map. The procedure is expected to work for $z$ less or close to the Bogoliubov energy.
%\end{remark}
%\begin{remark}
%As it will be explained in the last section, in order to have an explicit expression (i.e., in terms of bare quantities) for the ground state --  we call this expression main term -- up to a remainder which is estimated in norm less than some fraction, say $1/16$, of the main term, we have to decompose the expansion into two steps (at least), by considering ``high" and ``low" components of the potential separately. Hopefully, at least in the case in which we have a bunch of high frequencies and another bunch of very low frequencies, the two sets of frequencies being well separated, this method should work.
%\end{remark}

We restrict $\mathscr{H}$ to the Fock subspace $\mathcal{F}^{N}$ of vectors with $N$ particles 
\begin{equation}
\mathscr{H}\upharpoonright_{\mathcal{F}^N}=\Big(\int\frac{1}{2m}(\nabla a^{*})(\nabla a)(x)dx+\frac{\lambda}{2}\int\int a^{*}(x)a^{*}(y) \phi(x-y)a(y)a(x)dxdy\Big)\upharpoonright_{\mathcal{F}^N}\,.
\end{equation}
From now on, we shall study the Hamiltonian
\begin{equation}
H:=\int\frac{1}{2m}(\nabla a^{*})(\nabla a)(x)dx+\frac{\lambda}{2}\int\int a^{*}(x)a^{*}(y)\phi(x-y)a(y)a(x)dxdy+c_N\charf 
\end{equation}
where $c_N:=\frac{\lambda\phi_{\bold{0}}}{2|\Lambda|}N-\frac{\lambda\phi_{\bold{0}}}{2|\Lambda|}N^2$ with $\bold{0}=(0,\dots,0)$. 
%{\color{red}The value $-c_N$ is in fact the leading order term of the ground state energy of $\mathscr{H}\upharpoonright_{\mathcal{F}^N}$.} 
Hereafter, the operator $H$ is meant to be restricted to the subspace $\mathcal{F}^N$. Note that $\mathscr{H}\upharpoonright_{\mathcal{F}^N}=(H-c_N \charf)\upharpoonright_{\mathcal{F}^N}\,.$

\subsection{The Hamiltonian $H$ and the Hamiltonian $H^{Bog}$}\label{hamiltonians}
\setcounter{equation}{0}
%In the next formulae of this section, we use the notation that will be necessary for a general potential, that means for a potential where more than three modes have corresponding components different from zero (and positive).

Using the definitions
\begin{equation}
\label{b in the strip}
a_{+}(x):=\sum_{\bold{j}\in\mathbb{Z}^d\setminus \{\bold{0}\} }\frac{a_{\bold{j}}}{|\Lambda|^{\frac{1}{2}}}e^{ik_{\bold{j}}\cdot x}\quad\text{and}\quad
a_{\bold{0}}(x):=\frac{a_{\bold{0}}}{|\Lambda|^{\frac{1}{2}}}\,,
\end{equation}
 the Hamiltonian $H$ reads
\begin{eqnarray}
H& = & \sum_{\bold{j}\in  \mathbb{Z}^d}\frac{k^2_{\bold{j}}}{2m}a_{\bold{j}}^{*}a_{\bold{j}}\\
% &  & +\frac{\lambda}{2}\int \int b^*|_{n,l+1}^{n,l}(x)b^*|_{n,l}^{0,0}(y)\phi(x-y)b|_{n,l+1}^{n,l}(x)b|_{n,l}^{0,0}(y)dxdy\\
& &+\frac{\lambda}{2}\int\int a^{*}_{+}(x)a^*_{+}(y)\phi(x-y)a_{+}(x)a_{+}(y)dxdy \\
& &+\lambda\int \int \{ a^*_{+}(x)a^*_{+}(y)\phi(x-y)a_{+}(x)a_{\bold{0}}(y)
 %&  & +\frac{\lambda}{2}\int \int a^*_{+}(x)a^*_{+}(y)\phi(x-y)a_{\bold{0}}(x)a_{+}(y)dxdy
  +h.c.\}dxdy\\
% &  & +\frac{\lambda}{2}\int \int a^*_{\bold{0}}(x)a^*_{+}(y)\phi(x-y)a_{+}(x)a_{+}(y)dxdy\\
 &  & +\frac{\lambda}{2}\int \int \{ a^*_{\bold{0}}(x)a^*_{\bold{0}}(y)\phi(x-y)a_{+}(x)a_{+}(y)
 + h.c.\}dxdy\\
 &  & +\lambda\int \int a^*_{\bold{0}}(x)a^*_{+}(y)\phi(x-y)a_{\bold{0}}(x)a_{+}(y)dxdy\\
 &  & +\lambda\int \int a^*_{\bold{0}}(x)a^*_{+}(y)\phi(x-y)a_{\bold{0}}(y)a_{+}(x)dxdy\\
& & +\frac{\lambda}{2}\int\int a^{*}_{\bold{0}}(x)a^*_{\bold{0}}(y)\phi(x-y)a_{\bold{0}}(x)a_{\bold{0}}(y)dxdy  \\
& &+c_N\charf \,.\label{constant}
\end{eqnarray}
Because of the implicit restriction to $\mathcal{F}^N$ and the subtraction of the constant $-c_N$ in (\ref{constant}), it turns out that
\begin{eqnarray}
H& = & \sum_{\bold{j}\in  \mathbb{Z}^d}\frac{k^2_{\bold{j}}}{2m}a_{\bold{j}}^{*}a_{\bold{j}}\\
% &  & +\frac{\lambda}{2}\int \int b^*|_{n,l+1}^{n,l}(x)b^*|_{n,l}^{0,0}(y)\phi(x-y)b|_{n,l+1}^{n,l}(x)b|_{n,l}^{0,0}(y)dxdy\\
%& &+ \frac{\lambda\phi_{\bold{0}}}{2|\Lambda|^2}\int\int a^{*}_{+}(x)a_{+}(y)dxdy
& &+\frac{\lambda}{2}\int\int a^{*}_{+}(x)a^*_{+}(y)\phi_{(\neq 0)}(x-y)a_{+}(x)a_{+}(y)dxdy  \\
& &+\lambda\int \int \{a^*_{+}(x)a^*_{+}(y)\phi_{(\neq 0)}(x-y)a_{+}(x)a_{\bold{0}}(y)
 %&  & +\frac{\lambda}{2}\int \int a^*_{+}(x)a^*_{+}(y)\phi(x-y)a_{\bold{0}}(x)a_{+}(y)dxdy\\
 + h.c.\}dxdy\\
% &  & +\frac{\lambda}{2}\int \int a^*_{\bold{0}}(x)a^*_{+}(y)\phi(x-y)a_{+}(x)a_{+}(y)dxdy\\
 &  & +\frac{\lambda}{2}\int \int \{a^*_{\bold{0}}(x)a^*_{\bold{0}}(y)\phi_{(\neq 0)}(x-y)a_{+}(x)a_{+}(y)
 + h.c.\}dxdy\quad\quad\quad\quad\\
% &  & +\lambda\int \int a^*_{\bold{0}}(x)a^*_{+}(y)\phi(x-y)a_{\bold{0}}(x)a_{+}(y)dxdy\\
 &  & +\lambda\int \int a^*_{\bold{0}}(x)a^*_{+}(y)\phi_{(\neq 0)}(x-y)a_{\bold{0}}(y)a_{+}(x)dxdy
%& & +\frac{\lambda}{2}\int\int a^{*}_{\bold{0}}(x)a^*_{\bold{0}}(y)\phi(x-y)a_{\bold{0}}(x)a_{\bold{0}}(y)dxdy  \\
%& &+c_N
\end{eqnarray}
where $\phi_{(\neq 0)}(x-y):=\phi (x-y)-\phi_{(0)}(x-y)$ with $\phi_{(0)}(x-y):=\frac{\phi_{\bold{0}}}{|\Lambda|}$.

Next, we define the \emph{particle number preserving} Bogoliubov Hamiltonian
 \begin{eqnarray}
H^{Bog} &:=  &\sum_{\bold{j}\in  \mathbb{Z}^d}\frac{k^2_{\bold{j}}}{2m}a_{\bold{j}}^{*}a_{\bold{j}}\\
& &+ \frac{\lambda}{2}\int \int a^*_{\bold{0}}(x)a^*_{\bold{0}}(y)\phi_{(\neq 0)}(x-y)a_{+}(x)a_{+}(y)dxdy\label{quartic-two-one}\\
& &+\frac{\lambda}{2}\int \int a^*_{+}(x)a^*_{+}(y)\phi_{(\neq 0)}(x-y)a_{\bold{0}}(x)a_{\bold{0}}(y)dxdy\label{quartic-two-two}\\
  &  & +\lambda\int \int a^*_{\bold{0}}(x)a^*_{+}(y)\phi_{(\neq 0)}(x-y)a_{\bold{0}}(y)a_{+}(x)dxdy\,
 \end{eqnarray}
that we can express in terms of the field modes,
\begin{eqnarray}
H^{Bog}
&=&\sum_{\bold{j}\in\mathbb{Z}^d\setminus \{\bold{0}\}} (\frac{k^2_{\bold{j}}}{2m}+\lambda\frac{\phi_{\bold{j}}}{|\Lambda|}a^*_{\bold{0}}a_{\bold{0}})a_{\bold{j}}^{*}a_{\bold{j}}+\frac{\lambda}{2}\sum_{\bold{j}\in \mathbb{Z}^d\setminus \{\bold{0}\}}\frac{\phi_{\bold{j}}}{|\Lambda|}\,\Big\{a^*_{\bold{0}}a^*_{\bold{0}}a_{\bold{j}}a_{-\bold{j}}+a^*_{\bold{j}}a^*_{-\bold{j}}a_{\bold{0}}a_{\bold{0}}\Big\}\,.
\end{eqnarray}
Hence, the Hamiltonian $H$ corresponds to
\begin{equation}\label{complete-H}
H=H^{Bog}+V\,
\end{equation}
with
\begin{eqnarray}
V& :=&\lambda\int \int a^*_{+}(x)a^*_{\bold{0}}(y)\phi_{(\neq 0)}(x-y)a_{+}(x)a_{+}(y)dxdy \\
& &+\lambda\int \int a^*_{+}(x)a^*_{+}(y)\phi_{(\neq 0)}(x-y)a_{+}(x)a_{\bold{0}}(y)dxdy \\
& & +\frac{\lambda}{2}\int\int a^{*}_{+}(x)a^*_{+}(y)\phi_{(\neq 0)}(x-y)a_{+}(x)a_{+}(y)dxdy\,.
\end{eqnarray}

\noindent
Following the convention of  \cite{Pi1}, we set 
\begin{equation}\label{definitions}
\lambda=\frac{1}{\rho}\quad,\quad m=\frac{1}{2}\quad,\quad N=\rho |\Lambda|\,\,\, \text{and}\,\,\text{even},
\end{equation}
where $\rho>0$ is the particle density. 

In this paper we are interested in the mean field limiting regime. This means that we keep the box fixed
and let the particle density $\rho$ be (arbitrarily)  large. Consequently, in the mean field limiting regime the number of particles $N$ is independent of $L$.\\

The main result of the paper is the recursive formula in (\ref{gs-Hm-start})-(\ref{gs-Hm-fin})  by which  in (\ref{construction-ground}) we construct the groundstate of the Hamiltonian $H^{Bog}$ assuming the conditions on the potential specified in Definition \ref{def-pot}. In Corollary \ref{expansion} we re-expand the operators entering the formula in (\ref{gs-Hm-start})-(\ref{gs-Hm-fin}), and obtain, in turn, an expansion of the groundstate vector in terms of the interaction terms $W^*_{\bold{j}_l}+W_{\bold{j}_l}$ and of the resolvents $\frac{1}{\hat{H}^0_{\bold{j}_l}-E^{Bog}_{\bold{j}_l}}$ (see (\ref{interaction-terms}) and (\ref{H0j}), respectively) applied to $\eta$ (the state with all the particles in the zero mode).

The paper relies on some results obtained in \cite{Pi1} that are listed in Section \ref{multiscale-HBog} along with an overview of the multi-scale technique applied to three-modes Bogoliubov Hamiltonians.  Before providing a rigorous proof of the algorithm for the construction of the groundstate in Theorem \ref{induction-many-modes}  (Section \ref{rigorousHbog}), in Section \ref{groundstateHBog}  we introduce the definitions entering the scheme and  outline the procedure leading to formula  (\ref{gs-Hm-start})-(\ref{gs-Hm-fin}).
\\

\noindent
{\bf{Notation}}

\begin{enumerate}
\item
The symbol $\charf$ stands for the identity operator. If helpful we specify the Hilbert space where it acts, e.g., $\charf_{\mathcal{F}^N}$. For $c-$number operators, e.g., $z\charf$, we may omit the symbol $\charf$.
\item
The symbol $\langle\,\,,\,\,\rangle$ stands for the scalar product in $\mathcal{F}^N$.
\item
The word mode is used for the wavelength $\frac{2\pi}{L}\bold{j}$ (or simply for $\bold{j}$) when we refer to the field mode associated with it.
\item
The symbol $\mathcal{O}(\alpha)$ stands for a quantity bounded in absolute value by a constant times $\alpha$ ($\alpha>0$). The symbol $o(\alpha)$ stands for a quantity such that $o(\alpha)/\alpha \to 0$ as $\alpha \to 0$. Throughout the paper the implicit multiplicative constants are always independent of $N$. 
\item
In some cases we use explicit constants, e.g., $C_I$, if the same quantity is used in later proofs. Unless otherwise specified, or unless it is obvious from the context,  the explicit constants may depend on the size of the box and on the details of the potential, in particular on the number, $M$,  of couples of nonzero frequency components in the Fourier expansion of the pair potential.
%, {\color{red}and we also mention the dependence on parameters of the system (eg., $\Delta_0:\min \{(k_{\bold{j}})^2\,;\,\bold{j}\in \mathbb{Z}^d\setminus \{\bold{0}\}\}$).}
%\item
%The implicit multiplicative constant is always independent of $N$.
\item
The symbol $|\psi \rangle \langle \psi|$, with $\|\psi\|=1$,  stands for the one-dimensional projection onto the state $\psi$.
%We use the word mode for the wavelength $\frac{2\pi}{L}\bold{j}$ (or simply for $\bold{j}$)  to refer to the field mode associated with it.
\item
Theorems and lemmas from the companion paper \cite{Pi1} are underlined, quoted in italic, and with the numbering that they have in the corresponding paper; e.g., \emph{\underline{Theorem 3.1} of \cite{Pi1}}.
\end{enumerate}

\section{Multi-scale analysis in the particle states occupation numbers for a three-modes Bogoliubov  Hamiltonian: Review of results}\label{multiscale-HBog}
\setcounter{equation}{0}
%Since the momentum operator $ \sum_{\bold{j}\in  \mathbb{Z}^3}k_{\bold{j}}a_{\bold{j}}^{*}a_{\bold{j}}$ commutes with $H$ and $H^{Bog}$, it is convenient to consider the decomposition of $\mathcal{F}^N$ into sectors $\mathcal{F}^N_{P}$ where $P$ is the sum of a collection of $k_{\bold{j}}$. Then, we consider the Hamiltonians $H_P$, $H^{Bog}_P$, and the interaction $\Delta H_P$ at any fixed total momentum $P$. 
In \cite{Pi1} we consider the \emph{three-modes Bogoliubov Hamiltonian}
\begin{eqnarray}
H^{Bog}_{\bold{j}_{*}}
&:=&\sum_{\bold{j}\in\mathbb{Z}^d\setminus \{\bold{0}\,;\, \pm\bold{j}_{*} \}} k^2_{\bold{j}}a_{\bold{j}}^{*}a_{\bold{j}}+\hat{H}^{Bog}_{\bold{j}_{*}}
%+\phi_{\bold{j}_{*}}\frac{a^*_{\bold{0}}a_{\bold{0}}}{N}a_{\bold{j}_{*}}^{*}a_{\bold{j}_{*}}+\phi_{\bold{j}_{*}}\frac{a^*_{\bold{0}}a_{\bold{0}}}{N}a_{-\bold{j}_{*}}^{*}a_{-\bold{j}_{*}}+\frac{\phi_{\bold{j}}}{N}\,\Big\{a^*_{\bold{0}}a^*_{\bold{0}}a_{\bold{j}_{*}}a_{-\bold{j}_{*}}+a^*_{\bold{j}_{*}}a^*_{-\bold{j}_{*}}a_{\bold{0}}a_{\bold{0}}\Big\}\,\quad
\end{eqnarray}
where 
\begin{equation}\label{check-HBogj}
\hat{H}^{Bog}_{\bold{j}_{*}}:=\hat{H}^0_{\bold{j} _{*}} +W_{\bold{j}_{*}}+W^*_{\bold{j}_{*}}\,
\end{equation} 
involves the three modes $\bold{j}=\bold{0},\pm\bold{j}_{*}$ only ($\bold{j}_*\neq \bold{0}$); see the definitions in (\ref{H0j})-(\ref{interaction-terms}). Therefore, $H^{Bog}_{\bold{j}_{*}}$ is the sum of:
\begin{itemize}
\item
The operator
 \begin{equation}\label{H0j}
\hat{H}^0_{\bold{j}_{*}}:=(k^2_{\bold{j} _{*}}+\phi_{\bold{j} _{*}}\frac{a^*_{\bold{0}}a_{\bold{0}}}{N})a_{\bold{j} _{*}}^{*}a_{\bold{j} _{*}}+(k^2_{\bold{j} _{*}}+\phi_{\bold{j} _{*}}\frac{a^*_{\bold{0}}a_{\bold{0}}}{N})a_{-\bold{j} _{*}}^{*}a_{-\bold{j} _{*}}
\end{equation}
commuting with each number operator $a^*_{\bold{j}}a_{\bold{j}}$;
\item
The interaction terms
\begin{equation}\label{interaction-terms}
\phi_{\bold{j} _{*}}\frac{a^*_{\bold{0}}a^*_{\bold{0}}a_{\bold{j} _{*}}a_{-\bold{j} _{*}}}{N}=:W_{\bold{j} _{*}}\,\quad,\quad
\phi_{\bold{j} _{*}}\frac{a_{\bold{0}}a_{\bold{0}}a^*_{\bold{j} _{*}}a^*_{-\bold{j} _{*}}}{N}=:W^*_{\bold{j} _{*}}\,
\end{equation} 
that change the number of particles in the three modes $\bold{j}=\bold{0},\pm\bold{j} _{*} $;
\item
The kinetic energy $\sum_{\bold{j}\in\mathbb{Z}^d\setminus \{ \pm\bold{j}_{*} \}} k^2_{\bold{j}}a_{\bold{j}}^{*}a_{\bold{j}}$ of the noninteracting modes.
\end{itemize}
The following identity follows from the definitions above:
\begin{equation}
%H_0=\frac{1}{2}\sum_{\bold{j}\in\mathbb{Z}^3\setminus \{\bold{0}\}}\hat{H}^{0}_{\bold{j}}\,\quad,\quad
H^{Bog}=\frac{1}{2}\sum_{\bold{j}\in\mathbb{Z}^{d}\setminus \{\bold{0}\}}\hat{H}^{Bog}_{\bold{j}}\,.
\end{equation}

%Now, we focus on a three-modes system where $\phi_{\bold{j}}\neq 0$ only for $\bold{j}\equiv \pm \bold{j}_{*}$ and $\bold{j}=0$, and we construct the ground state of the corresponding Bogoliubov Hamiltonian:
%Notice however that because of the selection rules on the particle momentum, the operators $V$, $V^*$, and $U$ are identically zero if the Hilbert space contains only the modes $ \{\bold{0};\bold{j};-\bold{j} \}$.  
%The treatment of the Hamiltonian $H$ in Section \ref{multiscale-H} is very similar but the algebra is slightly more involved. 
%\begin{eqnarray}
%H^{Bog}
%&=&\sum_{\bold{j}\in\mathbb{Z}^3\setminus \{\bold{0}\}} (\frac{k^2_{\bold{j}}}{2m}+\lambda\frac{\phi_{\bold{j}}}{|\Lambda|}a^*_{\bold{0}}a_{\bold{0}})a_{\bold{j}}^{*}a_{\bold{j}}+\frac{\lambda}{2}\sum_{\bold{j}\in \mathbb{Z}^3\setminus \{\bold{0}\}}\frac{\phi_{\bold{j}}}{|\Lambda|}\,\Big\{a^*_{\bold{0}}a^*_{\bold{0}}a_{\bold{j}}a_{-\bold{j}}+a^*_{\bold{j}}a^*_{-\bold{j}}a_{\bold{0}}a_{\bold{0}}\Big\}\,.
%\end{eqnarray}

\begin{remark}
Notice that $H^{Bog}_{\bold{j}_{*}}$ contains the kinetic energy corresponding to all the modes whereas $\hat{H}^{Bog}_{\bold{j}_{*}}$ contains the kinetic energy associated with the interacting modes only.
\end{remark}
\subsection{Feshbach-Schur projections and Feshbach-Schur Hamiltonians associated with $H^{Bog}_{\bold{j}_*}$}\label{Feshbach}
%The motivations of the strategy explained in this section are given (a posteriori) in Remark \ref{motivations}. 

In this section, we briefly summarize the main results of the strategy introduced in \cite{Pi1} where we construct the ground state of  $H^{Bog}_{\bold{j}_*}$ by implementing a multi-scale analysis in the occupation numbers of the modes $\pm\bold{j}_{*}$. The multi-scale analysis relies on a novel application of Feshbach-Schur map that we outline in the next lines. We recall the use of the Feshbach-Schur map for the spectral  analysis of quantum field theory systems starting with the seminal work by V. Bach, J. Fr\"ohlich, and I.M. Sigal, \cite{BFS},  followed by refinements of the technique and variants (see \cite{BCFS} and  \cite{GH}). In those papers, the use of the Feshbach-Schur map is in the spirit of the functional renormalization group, and the  projections ($\mathscr{P}$, $\overline{\mathscr{P}}$) (see the paragraph below equation (\ref{s})) are directly related to energy subspaces of the free Hamiltonian.
\\

For the Hamiltonian $H^{Bog}_{\bold{j}_*}$ applied to $\mathcal{F}^{N}$ we define:
\begin{itemize}
\item
$Q^{(0,1)}_{\bold{j}_*}:=$
the projection (in $\mathcal{F}^{N}$) onto the subspace generated by vectors with $N-0=N$ or $N-1$ particles in the modes $\bold{j}_{*}$ and $-\bold{j}_{*}$, i.e., the operator $a^*_{\bold{j}_*}a_{\bold{j}_*}+a^*_{-\bold{j}_*}a_{-\bold{j}_*}$ has eigenvalues $N$ and $N-1$ when restricted to $Q^{(0,1)}_{\bold{j}_*}\mathcal{F}^N$;
\item
$Q^{(>1)}_{\bold{j}_*}:=$ the projection onto the orthogonal complement of $Q^{(0,1)}_{\bold{j}_*}\mathcal{F}^{N}$ in $\mathcal{F}^{N}$.
\end{itemize}
 Therefore, we have $$Q^{(0,1)}_{\bold{j}_*}+Q^{(>1)}_{\bold{j}_*}=\charf_{\mathcal{F}^{N}}\,.$$ 
 
Analogously, starting from $i=2$ up to $i=N-2$ with $i$ even, we define 
$Q^{(i, i+1)}_{\bold{j}_*}$
the projection onto the subspace of $Q^{(>1)}_{\bold{j}_*}\mathcal{F}^{N}$ spanned by the vectors with $N-i$ or $N-i-1$ particles in the modes $\bold{j}_{*}$ and $-\bold{j}_{*}$.  Furthermore, $Q^{(>i+1)}_{\bold{j}_*}$ is the projection onto the orthogonal complement of $Q^{(i, i+1)}_{\bold{j}_*}Q^{(>i-1)}_{\bold{j}_*}\mathcal{F}^{N}$ in $Q^{(>i-1)}_{\bold{j}_*}\mathcal{F}^{N}$, i.e.,
\begin{equation}\label{s}
Q^{(>i+1)}_{\bold{j}_*}+Q^{(i,i+1)}_{\bold{j}_*}=Q^{(>i-1)}_{\bold{j}_*}\,.
\end{equation}
%\begin{remark}
%Notice that if we denote by $\mathcal{F}^{N}_{\{\bold{0};\bold{j}_{*};-\bold{j}_{*}\}}$ the subspace of vectors that contain only particles in the modes $\{\bold{0};\bold{j}_{*};-\bold{j}_{*}\}$, then $Q^{(>N-1)}\mathcal{F}^{N}_{\{\bold{0};\bold{j}_{*};-\bold{j}_{*}\}}\equiv Q^{(N)}\mathcal{F}^{N}_{\{\bold{0};\bold{j}_{*};-\bold{j}_{*}\}}$ where the R-H-S is the one-dimensional subspace generated by the state $$\eta:=\frac{1}{\sqrt{N!}}a_{\bold{0}}^*\dots a_{\bold{0}}^*\Omega$$ with all the $N$ particles in the zero-mode state. 
%\end{remark}

 We recall that given the (separable) Hilbert space $\mathcal{H}$,  the projections $\mathscr{P}$, $\overline{\mathscr{P}}$ ($\mathscr{P}=\mathscr{P}^2$, $\overline{\mathscr{P}}=\overline{\mathscr{P}}^2$) where $\mathscr{P}+\overline{\mathscr{P}}=\charf_{\mathcal{H}}$, and a closed operator $K-z $, $z$ in a subset of $ \mathbb{C}$,  acting on $\mathcal{H}$,  the Feshbach-Schur map associated with the couple $\mathscr{P}\,,\,\overline{\mathscr{P}}$ maps $K-z$ to the operator $\mathscr{F}(K-z)$ acting on $\mathscr{P}\mathcal{H}$ where  (formally)
\begin{equation}
\mathscr{F}(K-z):=\mathscr{P}(K-z)\mathscr{P}-\mathscr{P}K\overline{\mathscr{P}}\frac{1}{\overline{\mathscr{P}}(K-z)\overline{\mathscr{P}}}\overline{\mathscr{P}}K\mathscr{P}\,.
\end{equation}

We iterate the Feshbach-Schur map starting from $i=0$ up to $i=N-2$ with $i$ even, using the projections $\mathscr{P}^{(i)}$ and $\overline{{\mathscr{P}}^{(i)}}$  for the \emph{i-th}  step\footnote{We use this notation though the number of steps is in fact $i/2+1$ being $i$ an even number.} of the iteration where
\begin{equation}\label{projections}
\mathscr{P}^{(i)}:= Q^{(>i+1)}_{\bold{j}_*}\quad,\quad \overline{{\mathscr{P}}^{(i)}}:= Q^{(i,i+1)}_{\bold{j}_*}\,.
\end{equation}
We denote by  $\mathscr{F}^{(i)}$ the Feshbach-Schur map at the \emph{i-th} step ($i$ even) of the iteration. We start applying $\mathscr{F}^{(0)}$ to $H^{Bog}_{\bold{j}_*}-z$ where $z(\in \mathbb{R})$ ranges in the interval $(-\infty, z_{max})$ with $z_{max}$ larger but very close to
\begin{equation}
E^{Bog}_{\bold{j}_*}:=-\Big[k^2_{\bold{j}_*}+\phi_{\bold{j}_*}-\sqrt{(k^2_{\bold{j}_*})^2+2\phi_{\bold{j}}k^2_{\bold{j}_*}}\Big]\,.
\end{equation}
Hence, we define
\begin{eqnarray}
\mathscr{K}^{Bog\,(0)}_{\bold{j}_*}(z)&:=&\mathscr{F}^{(0)}(H^{Bog}_{\bold{j}_*}-z)\\
&=&Q^{(>1)}_{\bold{j}_*}(H^{Bog}_{\bold{j}_*}-z)Q^{(>1)}_{\bold{j}_*}-Q^{(>1)}_{\bold{j}_*}H^{Bog}_{\bold{j}_*}Q^{(0,1)}_{\bold{j}_*}\frac{1}{Q^{(0,1)}_{\bold{j}_*}(H^{Bog}_{\bold{j}_*}-z)Q^{(0,1)}_{\bold{j}_*}}Q^{(0,1)}_{\bold{j}_*}H^{Bog}_{\bold{j}_*}Q^{(>1)}_{\bold{j}_*}\quad\quad\quad\\
&=&Q^{(>1)}_{\bold{j}_*}(H^{Bog}_{\bold{j}_*}-z)Q^{(>1)}_{\bold{j}_*}-Q^{(>1)}_{\bold{j}_*}W_{\bold{j}_*}Q^{(0,1)}_{\bold{j}_*}\frac{1}{Q^{(0,1)}_{\bold{j}_*}(H^{Bog}_{\bold{j}_*}-z)Q^{(0,1)}_{\bold{j}_*}}Q^{(0,1)}_{\bold{j}_*}W^*_{\bold{j}_*}Q^{(>1)}_{\bold{j}_*}\,. \label{KappaBog0}
\end{eqnarray}
Next, by recursion we define
\begin{equation}\label{definition-kappa-operators}
\mathscr{K}^{Bog\,(i)}_{\bold{j}_*}(z):=\mathscr{F}^{(i)}(\mathscr{K}^{Bog\,(i-2)}_{\bold{j}_*}(z))\quad,\quad i=2,\dots,N-2\,\quad{i}\,\, \text{even},
\end{equation}
that acts on $Q^{(>i+1)}_{\bold{j}_*}\mathcal{F}^N$. In the derivation of the Feshbach-Schur Hamiltonians  $\mathscr{K}^{Bog\,(i)}_{\bold{j}_*}(z)$ we employ the following convenient notation $$W_{\bold{j}_*\,;\,i,i'}:=Q^{(i,i+1)}_{\bold{j}_*}W_{\bold{j}_*}Q^{(i',i'+1)}_{\bold{j}_*}\quad,\quad W^*_{\bold{j}_*\,;\,i,i'}:=Q^{(i,i+1)}_{\bold{j}_*}W^*_{\bold{j}_*}Q^{(i',i'+1)}_{\bold{j}_*},$$ and  $$
R^{Bog}_{\bold{j}_*\,;\,i,i}(z):=Q^{(i,i+1)}_{\bold{j}_*}\frac{1}{Q^{(i,i+1)}_{\bold{j}_*}(H^{Bog}_{\bold{j}_*}-z)Q^{(i,i+1)}_{\bold{j}_*}}Q^{(i,i+1)}_{\bold{j}_*}\,.$$
Due to the selection rules of the operators $W_{\bold{j}_*}$ and $W^*_{\bold{j}_*}$, the Feshbach-Schur Hamiltonian at the \emph{i-th} step is 
\begin{eqnarray}
& &\mathscr{K}^{Bog\,(i)}_{\bold{j}_*}(z)\\
&=&Q^{(>i+1)}_{\bold{j}_*}(H^{Bog}_{\bold{j}_*}-z)Q^{(>i+1)}_{\bold{j}_*}\\
& &-\sum_{l_i=0}^{\infty}Q^{(>i+1)}_{\bold{j}_*}W_{\bold{j}_*}Q^{(i,i+1)}_{\bold{j}_*}R^{Bog}_{\bold{j}_*\,;\,i,i}(z)\Big[W_{\bold{j}_*\,;\,i,i-2}\,R^{Bog}_{\bold{j}_*\,;\,i-2,i-2}(z)\times \\
& &\quad\quad\quad\times \sum_{l_{i-2}=0}^{\infty}\Big[W_{\bold{j}_*\,;\,i-2,i-4}\,\dots W^*_{\bold{j}_*\,;\,i-4,i-2}R^{Bog}_{\bold{j}_*\,;\,i-2,i-2}(z)\Big]^{l_{i-2}}W^*_{\bold{j}_*\,;\,i-2,i}R^{Bog}_{\bold{j}_*\,;\,i,i}(z)\Big]^{l_i}Q^{(i,i+1)}_{\bold{j}_*}W^*_{\bold{j}_*}Q^{(>i+1)}_{\bold{j}_*}\quad\quad\quad \label{dots-2}
\end{eqnarray}
where  the expressions corresponding to $\dots$ in (\ref{dots-2}) are made precise in \emph{\underline{Theorem 3.1} of \cite{Pi1}} reported below. \\

\begin{remark}\label{remark-cut}
The ratio $\epsilon_{\bold{j}_{*}}$ between the kinetic energy of the modes $\pm\bold{j}_{*}\neq \bold{0}$ and the corresponding Fourier component, $\phi_{\bold{j}_{*}}$,  of the potential, i.e., $\frac{k_{\bold{j}_{*}}^2}{\phi_{\bold{j}_{*}}}=:\epsilon_{\bold{j}_*}$, is required to be small to ensure the estimates used in \cite{Pi1}. In particular, $\epsilon_{\bold{j}_{*}}$ is assumed small enough so that $E^{Bog}_{\bold{j}_*}+ \phi_{\bold{j}_*}\sqrt{\epsilon_{\bold{j}_*}^2+2\epsilon_{\bold{j}_*}}<0$ and, likewise,  the spectral variable $z\in \mathbb{R}$ is restricted to negative values. Notice that $\epsilon_{\bold{j}_*}$ small corresponds either to a low energy mode  $\pm\frac{2\pi\bold{j}_{*}}{L}$  or/and to a large potential $\phi_{\bold{j}_{*}}$. 
\end{remark}
%Notice that \begin{equation}\frac{E^{Bog}_{\bold{j}_{*}}}{\phi_{\bold{j}_{*}}}=-\Big[\epsilon_{\bold{j}_*}+1-\sqrt{\epsilon_{\bold{j}_*}^2+2\epsilon_{\bold{j}_*}}\,\Big]\end{equation}

%\begin{remark}
%For our final purpose, i.e., the construction of the ground state of the Hamiltonian $H$, we need to have the result in the lemma below for values $z$ strictly above $E^{Bog}$. This is not a problem thanks to the gap of the unperturbed Hamiltonian. In other words, for fixed potential one can find a $\Delta$ sufficiently small such that the proof works also for $z<E^{Bog}+\Delta$ with a $\delta^{Bog}_N$ still larger than $1$.
%\end{remark}

%\begin{remark}
%.....TO BE CHANGED OR ERASED.....In order to have an explicit expression of the ground state,  and a self contained paper where we prove that the ground state energy is $E^{Bog}$ plus corrections that vanish as $N\to \infty$, we have to proceed differently as follows. By splitting the potential in high and low components, at first one construct the ground state for the high components according to the multi-scale analysis in the number of particles. Then, one uses ordinary perturbation theory for the low components. This way, without invoking the result obtained with different methods,  it should be possible to determine the ground state energy and then to have an explicit expression of the ground state energy up to small corrections (but not arbitrarily small!)
%\end{remark}
%\begin{remark}
%Notice that $\epsilon_{\bold{j}_*}\to 0$ as $\phi_{\bold{j}_{*}}\to \infty$.
%\end{remark}
Here, we recall the main result from \cite{Pi1} regarding the control of the flow of Feshbach-Schur Hamiltonians  up to the step $i=N-2$, along with some related tools. \\

\noindent
\emph{\underline{Theorem 3.1} of \cite{Pi1}}

\noindent
\emph{For \begin{equation}
z\leq E^{Bog}_{\bold{j}_*}+ (\delta-1)\phi_{\bold{j}_*}\sqrt{\epsilon_{\bold{j}_*}^2+2\epsilon_{\bold{j}_*}}\,(<0)
\end{equation} with $\delta\leq  1+\sqrt{\epsilon_{\bold{j}_*}}$, $\frac{1}{N}\leq \epsilon^{\nu}_{\bold{j}_*}$ for some $\nu >\frac{11}{8}$, and $\epsilon_{\bold{j}_*}$ sufficiently small, the operators $\mathscr{K}^{Bog\,(i)}_{\bold{j}_*}(z)$, $0\leq i\leq N-2$ and even,  are well defined.  For $i=0$, it is given in (\ref{KappaBog0}). For $i=2,4,6,\dots,N-2$ they correspond to
\begin{eqnarray}\label{KappaBog-i}
\mathscr{K}^{Bog\,(i)}_{\bold{j}_*}(z)&=&Q^{(>i+1)}_{\bold{j}_*}(H^{Bog}_{\bold{j}_*}-z)Q^{(>i+1)}_{\bold{j}_*}\\
%& &-\sum_{l_{i-1}=0}^{\infty}Q^{(>i)}_{\bold{j}_*}W_{\bold{j}_*}\,Q^{(i-1)}_{\bold{j}_*}R^{Bog}_{\bold{j}_*\,;\,i-1,i-1}(z)\,\Big[\Gamma^{Bog\,}_{\bold{j}_*\,;\,i-1,i-1}(z)R^{Bog}_{\bold{j}_*\,;\,i-1,i-1}(z)\Big]^{l_{i-1}}\,Q^{(i-1)}_{\bold{j}_*}W^*_{\bold{j}_*}Q^{(>i)}_{\bold{j}_*} \nonumber\\
& &-Q^{(>i+1)}_{\bold{j}_*}W_{\bold{j}_*}\,R^{Bog}_{\bold{j}_*\,;\,i,i}(z)\sum_{l_i=0}^{\infty}\Big[\Gamma^{Bog\,}_{\bold{j}_*\,;\,i,i}(z)R^{Bog}_{\bold{j}_*\,;\,i,i}(z)\Big]^{l_i}W^*_{\bold{j}_*}Q^{(>i+1)}_{\bold{j}_*} \nonumber
\end{eqnarray}
where:
\begin{itemize}
\item \begin{equation}\label{GammaBog-2}
\Gamma^{Bog\,}_{\bold{j}_*\,;\,2,2}(z):=W_{\bold{j}_*\,;\,2,0}\,R_{\bold{j}_*\,;\,0,0}^{Bog}(z)W_{\bold{j}_*\,;\,0,2}^*
\end{equation}
\item
 for $N-2\geq i\geq 4$,
%\begin{equation}
%\Gamma^{Bog\,}_{i-1,i-1}(z):=W_{i-1,i-3}\,R^{Bog}_{i-3,i-3}(z) \sum_{l_{i-3}=0}^{\infty}\Big[\Gamma^{Bog\,}_{i-3,i-3}(z)R^{Bog}_{i-3,i-3}(z)\Big]^{l_{i-3}}W^*_{i-3,i-1}
%\end{equation}
\begin{eqnarray}\label{GammaBog-i}
\Gamma^{Bog\,}_{\bold{j}_*\,;\,i,i}(z)&:=&W_{\bold{j}_*\,;\,i,i-2}\,R^{Bog}_{\bold{j}_*\,;\,i-2,i-2}(z) \sum_{l_{i-2}=0}^{\infty}\Big[\Gamma^{Bog}_{\bold{j}_*\,;\,i-2,i-2}(z)R^{Bog}_{\bold{j}_*\,;\,i-2,i-2}(z)\Big]^{l_{i-2}}W^*_{\bold{j}_*\,;\,i-2,i}\\
&=&W_{\bold{j}_*\,;\,i,i-2}\,(R^{Bog}_{\bold{j}_*\,;\,i-2,i-2}(z))^{\frac{1}{2}} \sum_{l_{i-2}=0}^{\infty}\Big[(R^{Bog}_{\bold{j}_*\,;\,i-2,i-2}(z))^{\frac{1}{2}}\Gamma^{Bog}_{\bold{j}_*\,;\,i-2,i-2}(z)(R^{Bog}_{\bold{j}_*\,;\,i-2,i-2}(z))^{\frac{1}{2}}\Big]^{l_{i-2}}\times \label{neumann}\quad\quad\quad\\
& &\quad\quad\quad \times (R^{Bog}_{\bold{j}_*\,;\,i-2,i-2}(z))^{\frac{1}{2}}W^*_{\bold{j}_*\,;\,i-2,i}\,.\nonumber
\end{eqnarray}
\end{itemize}
}

%\begin{remark} We defer the last step corresponding $i=N-1$ to the next section.
%\end{remark}
%\begin{equation}\label{GammaBog-4}
%\Gamma^{Bog\,}_{4,4}(z):=W_{4,2}\,R^{Bog}_{2,2}(z)\sum_{l_2=0}^{\infty}\Big[W_{2,0}\,R_{0,0}^{Bog}(z)W_{0,2}^*R^{Bog}_{2,2}(z)
%\Big]^{l_2}W^*_{2,4}\,.
%\end{equation}

Hence, \emph{\underline{Theorem 3.1} of \cite{Pi1}} states that the flow of  Feshbach-Schur Hamiltonians (up to the index value $N-2$) can be defined for spectral values $z$ up to $  E^{Bog}_{\bold{j}_*}+ (\delta-1)\phi_{\bold{j}^*}\sqrt{\epsilon_{\bold{j}_*}^2+2\epsilon_{\bold{j}_*}}$ with $\delta>1$ but very close to $1$. 
%In this respect, we observe that the first excited energy level is  known (see \cite{Se1})  to be located  asymptotically close to $E^{Bog}_{\bold{j}_*}+\phi_{\bold{j}^*}\sqrt{\epsilon_{\bold{j}_*}^2+2\epsilon_{\bold{j}_*}}$ as $N\to \infty$, that corresponds to $\delta=2$. 
The Hamiltonian $\mathscr{K}^{Bog\,(N-2)}_{\bold{j}_*}(z)$ acts on the Hilbert space $Q^{(>N-1)}_{\bold{j}_*}\mathcal{F}^N$.
\\

\noindent
The key estimates to prove \emph{\underline{Theorem 3.1} of \cite{Pi1}} are derived in \emph{\underline{Lemma 3.4}}  and \emph{\underline{Lemma 3.6}}  of \cite{Pi1}.  In fact, the Feshbach-Schur map is implementable as long as
\begin{equation}
\|(R^{Bog}_{\bold{j}_*\,;\,i-2,i-2}(z))^{\frac{1}{2}} \sum_{l_{i-2}=0}^{\infty}\Big[(R^{Bog}_{\bold{j}_*\,;\,i-2,i-2}(z))^{\frac{1}{2}}\Gamma^{Bog}_{\bold{j}_*\,;\,i-2,i-2}(z)(R^{Bog}_{\bold{j}_*\,;\,i-2,i-2}(z))^{\frac{1}{2}}\Big]^{l_{i-2}}(R^{Bog}_{\bold{j}_*\,;\,i-2,i-2}(z))^{\frac{1}{2}}\|<\infty\,. \label{key-object}
\end{equation}
The norm in (\ref{key-object}) can be related to the inverse of the $(N-i+2)-th$ term of the sequence studied in:
\\

\noindent
\emph{\underline{Lemma 3.6} of \cite{Pi1}}

\noindent
\emph{Assume $\epsilon>0$ sufficiently small.  Consider  for $j\in \mathbb{N}_{0}$ the sequence defined by
\begin{eqnarray}\label{def-X}
X_{2j+2}&:=&1-\frac{1}{4(1+a_{\epsilon}-\frac{2b_{\epsilon}}{N-2j-1}-\frac{1-c_{\epsilon}}{(N-2j-1)^2})X_{2j}}\,,
%x_{2j+3}&:=&1-\frac{1}{4(1+a_{\epsilon}-\frac{2b_{\epsilon}}{N-2j-1}-\frac{1-c_{\epsilon}}{(N-2j-1)^2})x_{2j+1}}
\end{eqnarray}
with initial condition $X_{0}=1$, up to $X_{2j=N-2}$ where $N(\geq 2)$ is even.  Here, 
\begin{equation}
a_{\epsilon}:=2\epsilon+\mathcal{O}(\epsilon^{\nu})\,,\quad \nu > \frac{11}{8}\,,
\end{equation}
\begin{equation}
b_{\epsilon}:=(1+\epsilon)\delta \, \chi_{[0,2)}(\delta)\sqrt{\epsilon^2+2\epsilon}\,\Big|_{\delta=1+\sqrt{\epsilon}}\,,
\end{equation}
and
\begin{equation}
c_{\epsilon}:=-(1-\delta^2 \, \chi_{[0,2)}(\delta))(\epsilon^2+2\epsilon)\,\Big|_{\delta=1+\sqrt{\epsilon}}\,
\end{equation}
with $\chi_{[0,2)}$ the characteristic function of the interval $[0,2)$.
Then, the following estimate holds true for $2\leq N-2j\leq  N $,
\begin{equation}\label{bound-even}
X_{2j}\geq\frac{1}{2}\Big[1+\sqrt{\eta a_{\epsilon}}-\frac{b_{\epsilon}/\sqrt{\eta a_{\epsilon}}}{N-2j-\epsilon^{\Theta}}\Big](>0)\,
\end{equation}
with $\eta=1-\sqrt{\epsilon}$, where $\Theta:=\min\{2(\nu-\frac{11}{8})\,;\,\frac{1}{4}\}$. }
\\

Indeed, the inequality 
\begin{equation}\label{ineq-Feshbach}
\|\sum_{l_{i-2}=0}^{\infty}\Big[(R^{Bog}_{\bold{j}_*\,;\,i-2,i-2}(z))^{\frac{1}{2}}\Gamma^{Bog}_{\bold{j}_*\,;\,i-2,i-2}(z)(R^{Bog}_{\bold{j}_*\,;\,i-2,i-2}(z))^{\frac{1}{2}}\Big]^{l_{i-2}}\|\leq \frac{1}{X_{i-2}}
\end{equation}  is used to control the Feshbach-Schur flow. It relies on the crucial estimate derived in the next lemma.
\\

\noindent
\emph{\underline{Lemma 3.4} of \cite{Pi1}}

\noindent
\emph{Let 
\begin{equation}
z\leq  E^{Bog}_{\bold{j}_*}+ (\delta-1)\phi_{\bold{j}^*}\sqrt{\epsilon_{\bold{j}_*}^2+2\epsilon_{\bold{j}_*}}\,(<0)
\end{equation} with $\delta< 2$,  $\frac{1}{N}\leq \epsilon^{\nu}_{\bold{j}_*}$ for some $\nu >1$, and $\epsilon_{\bold{j}_*}$ sufficiently small. Then
%\footnote{Notice that $$\frac{\epsilon_{\bold{j}_{*}}+1+\delta\sqrt{\epsilon^2_{\bold{j}_{*}}+2\epsilon_{\bold{j}_{*}}}}{N-i+1}\leq\frac{1+\epsilon_{\bold{j}_{*}}+\mathcal{O}(\epsilon_{\bold{j}_{*}}^{\frac{1}{2}})}{3}$$ because $N-i\geq 2$ and $\epsilon_{\bold{j}_{*}}>0$. Therefore, the denominator in (\ref{main-estimate-intermediate-a}) is strictly positive for $\epsilon_{\bold{j}_{*}}$ sufficiently small.}
\begin{eqnarray}\label{estimate-main-lemma-Bog}
%& &\frac{\sqrt{\frac{(N-l-1)}{(N-l-2)}}}{\Big[1+\frac{N}{N-l-2}(\epsilon+\frac{1+\epsilon-\sqrt{\epsilon^2+2\epsilon}}{l})\Big]^{\frac{1}{2}} \Big[1+\frac{N}{N-l}(\epsilon-\frac{1+\epsilon+\sqrt{\epsilon^2+2\epsilon}}{l})\Big]^{\frac{1}{2}}}\\
& &\| \Big[R^{Bog}_{\bold{j}_*\,;\,i,i}(z)\Big]^{\frac{1}{2}}\,W_{\bold{j}_*\,;\,i,i-2}\,\Big[R^{Bog}_{\bold{j}_*\,;\,i-2,i-2}(z)\Big]^{\frac{1}{2}}\|\,\|\Big[R^{Bog}_{\bold{j}_*\,;\,i-2,i-2}(z)\Big]^{\frac{1}{2}}\,W^*_{\bold{j}_*\,;\,i-2,i}\,\Big[R^{Bog}_{\bold{j}_*\,;\,i,i}(z)\Big]^{\frac{1}{2}}\|\quad\\
%%%%
%& &\|\Big[R^{Bog}_{i-2,i-2}(z)\Big]^{\frac{1}{2}}\,W_{i-2,i}\Big[R^{Bog}_{i,i}(z)\Big]^{\frac{1}{2}}\|\,\|\Big[R^{Bog}_{i,i}(z)\Big]^{\frac{1}{2}}W^*_{i,i-2}\,\Big[R^{Bog}_{i-2,i-2}(z)\Big]^{\frac{1}{2}}\|\\
%&\leq&\frac{1}{4\Big[1+\epsilon+\frac{\epsilon+1-\sqrt{\epsilon^2+2\epsilon}}{(N-i+2)}\Big]}\frac{1}{ \Big[1+\epsilon-\frac{\epsilon+1+\sqrt{\epsilon^2+2\epsilon}}{(N-i+2)}\Big]}+CN^{-\eta}\\
%%%%
%&\leq&\frac{1}{4\Big[1+\epsilon_{\bold{j}_*}+o(\epsilon)+\frac{\epsilon_{\bold{j}_*}+1-\delta\sqrt{\epsilon_{\bold{j}_*}^2+2\epsilon_{\bold{j}_*}}}{(N-i+1)}\Big]}\frac{1}{ \Big[1+\epsilon_{\bold{j}_*}+o(\epsilon)-\frac{\epsilon_{\bold{j}_*}+1+\delta\sqrt{\epsilon_{\bold{j}_*}^2+2\epsilon_{\bold{j}_*}}}{(N-i+1)}\Big]}\label{main-estimate-intermediate-a}\\
%%%
%\frac{1}{2\Big[1+(\epsilon+\frac{1+\epsilon-\sqrt{\epsilon^2+2\epsilon}}{N-i})-\mathcal{O}(\epsilon^2)\Big]^{\frac{1}{2}} \Big[1+(\epsilon-\frac{1+\epsilon+\sqrt{\epsilon^2+2\epsilon}}{N-i})-\mathcal{O}(\epsilon^2)\Big]^{\frac{1}{2}}}\\
%%%
&\leq&\frac{1}{4(1+a_{\epsilon_{\bold{j}_*}}-\frac{2b^{(\delta)}_{\epsilon_{\bold{j}_*}}}{N-i+1}-\frac{1-c^{(\delta)}_{\epsilon_{\bold{j}_*}}}{(N-i+1)^2})}\quad\label{def-deltabog}
\end{eqnarray}
holds for all $2\leq i\leq N-2$ ($i$ even). Here,  \begin{equation}\label{a}
a_{\epsilon_{\bold{j}_{*}}}:=2\epsilon_{\bold{j}_{*}}+\mathcal{O}(\epsilon^{\nu}_{\bold{j}_{*}})\,,
\end{equation}
\begin{equation}\label{b}
b^{(\delta)}_{\epsilon_{\bold{j}_{*}}}:=(1+\epsilon_{\bold{j}_{*}})\delta\, \chi_{[0,2)}\sqrt{\epsilon_{\bold{j}_{*}}^2+2\epsilon_{\bold{j}_{*}}}\,,
\end{equation}
and
\begin{equation}\label{c}
c^{(\delta)}_{\epsilon_{\bold{j}_{*}}}:=-(1-\delta^2\, \chi_{[0,2)})(\epsilon_{\bold{j}_{*}}^2+2\epsilon_{\bold{j}_{*}})\,,
\end{equation}
with $\chi_{[0,2)}$ the characteristic function of the interval $[0,2)$.
}
\\

A posteriori, one can observe that the choice of the Feshbach-Schur projections yields a rather refined control of the Neumann expansion in (\ref{neumann}) thanks to \emph{\underline{Lemma 3.4} of \cite{Pi1}}.  The price to be paid is the iteration from $i=0$ up to $i=N-2$ and the nontrivial control of the sequence $\{X_{2j}\}$ studied in \emph{\underline{Lemma 3.6} of \cite{Pi1}}.

\subsection{Fixed point, construction of the ground state of $H^{Bog}_{\bold{j}_*}$, and algorithm for the re-expansion}\label{groundstate}

We implement the Feshbach-Schur flow described in the previous section with the purpose to get a simple final effective Hamiltonian, namely a multiple of the one-dimensional projection $|\eta \rangle \langle \eta |$ where $\eta$ is the state with all the $N$ particles in the zero mode, i.e.,
\begin{equation}
	\eta=\frac{1}{\sqrt{N!}} a_{\bold{0}}^* \dots a_{\bold{0}}^*\Omega\,. \label{eta}
\end{equation}
This way we can reconstruct the ground state of $H_{\bold{j}_*}^{Bog}$ starting from $\eta$, by means of Feshbach-Schur theory. Consequently, for the last step of the Feshbach-Schur flow we employ the couple of projections: $\mathscr{P}_{\eta}:=|\eta \rangle \langle \eta |$ and $\overline{\mathscr{P}_{\eta}}$ such that
\begin{equation}
\mathscr{P}_{\eta}+\overline{\mathscr{P}_{\eta}}=\charf_{Q^{(>N-1)}_{\bold{j}_*}\mathcal{F}^N}\,.
\end{equation}
We remind that for $i=N-2$ the projection $Q^{(>i+1\equiv N-1)}_{\bold{j}_*}$ coincides with the projection onto the subspace where less than $N-i+1=N-N+1=1$ particles in the modes $\bold{j}_{*}$ and $-\bold{j}_{*}$ are present, i.e., where no particle in the modes $\bold{j}_{*}$ or $-\bold{j}_{*}$ is present.  

%{\color{red}TO BE ERASED---Including this last step,  the flow of Feshbach-Schur Hamiltonians is well defined as long as the spectral parameter $z(\in \mathbb{R})$ is strictly below the first excited eigenvalue of the energy spectrum of the Hamiltonian $H^{Bog}_{\bold{j}_*}$.---}
%In fact, the estimates allow $z$ to range between $-\infty$ and $z_{\#}$ where $z_{\#}$ is very close but still above the Bogoliubov energy 
%Therefore, for the three-modes system we have
%\begin{equation}
%E^{Bog}_{\bold{j}_{*}}:=-\Big[k^2_{\bold{j}_*}+\phi_{\bold{j}_*}-\sqrt{(k^2_{\bold{j}_*})^2+2\phi_{\bold{j}_*}k^2_{\bold{j}_*}}\Big]\,.
%\end{equation}

Starting from the formal expression
\begin{eqnarray}
& &\mathscr{K}^{Bog\,(N)}_{\bold{j}_*}(z) \\
&:=&\mathscr{F}^{(N)}(\mathscr{K}^{Bog\,(N-2)}_{\bold{j}_*}(z))\\
&=&\mathscr{P}_{\eta}(H^{Bog}_{\bold{j}_*}-z)\mathscr{P}_{\eta}\label{K-last-step-0}\\
& &-\mathscr{P}_{\eta}W_{\bold{j}_*}\,R^{Bog}_{\bold{j}_*\,;\,N-2,N-2}(z)\sum_{l_{N-2}=0}^{\infty}[\Gamma^{Bog}_{\bold{j}_*\,;\,N-2,N-2}(z) R^{Bog}_{\bold{j}_*\,;,N-2,N-2}(z)]^{l_{N-2}}\, W^*_{\bold{j}_*}\mathscr{P}_{\eta}\quad \nonumber\\
& &-\mathscr{P}_{\eta}W_{\bold{j}_*}\,\overline{\mathscr{P}_{\eta}}\,\frac{1}{\overline{\mathscr{P}_{\eta}}\mathscr{K}^{Bog\,(N-2)}_{\bold{j}_*}(z)\overline{\mathscr{P}_{\eta}}}\overline{\mathscr{P}_{\eta}}W^*_{\bold{j}_*}\mathscr{P}_{\eta}\,, \nonumber
\end{eqnarray}
 in \cite{Pi1} we study the invertibility of 
\begin{equation}
\overline{\mathscr{P}_{\eta}}\mathscr{K}^{Bog\,(N-2)}_{\bold{j}_*}(z)\overline{\mathscr{P}_{\eta}}\upharpoonright_{\overline{\mathscr{P}_{\eta}}\mathcal{F}^N}
\end{equation} 
and show that the R-H-S of (\ref{K-last-step-0}) is well defined   for $$z<\min\,\Big\{ z_{*}+\frac{\Delta_0}{2}\,;\,E^{Bog}_{\bold{j}_*}+ \sqrt{\epsilon_{\bold{j}_*}}\phi_{\bold{j}_*}\sqrt{\epsilon_{\bold{j}_*}^2+2\epsilon_{\bold{j}_*}}\Big\}\,,$$
where $\Delta_{0}:=\min\, \Big\{(k_{\bold{j}})^2\,|\,\bold{j}\in \mathbb{Z}^d \setminus \{\bold{0}\}\Big\}$ and $z_*$ is the unique solution of $f_{\bold{j}_*}(z)=0$ with
\begin{eqnarray}\label{function-f}
f_{\bold{j}_*}(z)&:=&-z-\langle \eta\,,\,W_{\bold{j}_*}\,R^{Bog}_{\bold{j}_*\,;\,N-2,N-2}(z)\sum_{l_{N-2}=0}^{\infty}[\Gamma^{Bog}_{\bold{j}_*\,;\,N-2,N-2}(z) R^{Bog}_{\bold{j}_*\,;\,N-2,N-2}(z)]^{l_{N-2}}\,W^*_{\bold{j}_*}\eta\rangle\,.\quad \quad  \label{fp-function}
\end{eqnarray}
%and $(1-\frac{N^{\mu}\phi_{\bold{j}_{*}}}{\Delta_0}\frac{1}{N})> \frac{1}{2}$ by assumption.

%But this also implies that  the expression in (\ref{absent}) is identically zero because $\overline{\mathscr{P}_{\eta}}W^*_{\bold{j}_*}\mathscr{P}_{\eta}=0$. 
\noindent
% \footnote{In \cite{Pi1} the fixed point problem $f_{\bold{j}_*}(z)=0$ is actually solved first and then it is deduced that  the inverse of $\overline{\mathscr{P}_{\eta}}\mathscr{K}^{Bog\,(N-2)}_{\bold{j}_*}(z)\overline{\mathscr{P}_{\eta}}$ is well defined in the given domain for $z$. }
%The latter trivially implies $\mathscr{K}^{Bog\,(N-2)}_{\bold{j}_*}(z)=f_{\bold{j}_*}(z)|\eta \rangle \langle \eta |$.} 
 In order to determine $z_*$, in {\emph{\underline{Section 4.1.1.} of \cite{Pi1}}, 
 %for a three-modes Bogoliubov Hamiltonian with interacting modes $\pm \bold{j}_{*}$ and 
  we derive the identity (for $z\leq E^{Bog}_{\bold{j}_*}+ \sqrt{\epsilon_{\bold{j}_*}}\phi_{\bold{j}^*}\sqrt{\epsilon_{\bold{j}_*}^2+2\epsilon_{\bold{j}_*}}<0$)
\begin{eqnarray}
& &\langle \eta\,,\,W_{\bold{j}_*}\,R^{Bog}_{\bold{j}_*\,;\,N-2,N-2}(z)\sum_{l_{N-2}=0}^{\infty}[\Gamma^{Bog}_{\bold{j}_*\,;\,N-2,N-2}(z) R^{Bog}_{\bold{j}_*\,;\,N-2,N-2}(z)]^{l_{N-2}}\,W^*_{\bold{j}_*}\eta\rangle\\
&=& (1-\frac{1}{N})\frac{\phi_{\bold{j}_{*}}}{2\epsilon_{\bold{j}_*}-\frac{4}{N}+2-\frac{z}{\phi_{\bold{j}_{*}}}}\check{\mathcal{G}}_{\bold{j}_{*}\,;\,N-2,N-2}(z)\label{correct-G}
\end{eqnarray}
where $\check{\mathcal{G}}_{\bold{j}_{*}\,;\,i,i}(z)$ is defined (for $i$ even and $N-2\geq i\geq 2$)  by
\begin{equation}\label{in-formula-G}
\check{\mathcal{G}}_{\bold{j}_{*}\,;\,i,i}(z):=\sum_{l_{i}=0}^{\infty}[\mathcal{W}_{\bold{j}_{*}\,;i,i-2}(z)\mathcal{W}^*_{\bold{j}_*\,;i-2,i}(z)\check{\mathcal{G}}_{\bold{j}_{*}\,;\,i-2,i-2}(z)]^{l_i}\quad,\quad \check{\mathcal{G}}_{\bold{j}_{*}\,;\,0,0}(z)=1\,,
\end{equation}
and
\begin{eqnarray}
\mathcal{W}_{\bold{j}_{*}\,;\,i,i-2}(z)\mathcal{W}^*_{\bold{j}_{*}\,;\,i-2,i}(z)\label{def-WW*}
&:= &\frac{(n_{\bold{j}_0}-1)n_{\bold{j}_0}}{N^2}\,\phi^2_{\bold{j}_{*}}\,\frac{ (n_{\bold{j}_{*}}+1)(n_{-\bold{j}_{*}}+1)}{\Big[(\frac{n_{\bold{j}_0}}{N}\phi_{\bold{j}_{*}}+(k_{\bold{j}_{*}}^2))(n_{\bold{j}_{*}}+n_{-\bold{j}_{*}})-z\Big]}\times\quad \\
& &\quad\quad\quad \times\frac{1}{\Big[(\frac{(n_{\bold{j}_0}-2)}{N}\phi_{\bold{j}_{*}}+(k_{\bold{j}_{*}}^2))(n_{\bold{j}_{*}}+n_{-\bold{j}_{*}}+2)-z\Big]}
%&:=& \frac{(n_{\bold{j}_0}+2)(n_{\bold{j}_0}+1)}{N^2}\,\phi^2_{\bold{j}_{*}}\,\frac{ (n_{\bold{j}_{*}}+1)(n_{-\bold{j}_{*}}+1)}{\Big[(\frac{n_{\bold{j}_0}}{N}\phi_{\bold{j}_{*}}+(k_{\bold{j}_{*}}^2))(n_{\bold{j}_{*}}+n_{-\bold{j}_{*}})-z\Big]}\times\\
%& &\times \frac{1}{\Big[(\frac{(n_{\bold{j}_0}+2)}{N}\phi_{\bold{j}_{*}}+(k_{\bold{j}_{*}}^2))(n_{\bold{j}_{*}}+n_{-\bold{j}_{*}})-2(\frac{(n_{\bold{j}_0}+2)}{N}\phi_{\bold{j}_{*}}+(k_{\bold{j}_{*}}^2))-z\Big]}\quad\quad\quad
\end{eqnarray} 
with 
\begin{equation}
n_{\bold{j}_{*}}+n_{-\bold{j}_{*}}=N-i\quad;\quad n_{\bold{j}_{*}}=n_{-\bold{j}_{*}}\quad;\quad n_{\bold{j}_0}=i \,.\label{fin-formula-G}
\end{equation}
\begin{remark}\label{nondecreasing}
In \cite{Pi1}, by induction  $\check{\mathcal{G}}_{\bold{j}_{*}\,;\,i,i}(z)$ is shown to be nondecreasing in $z$.
\end{remark}
Finally,
%Let $\eta$ be the (normalized) vector where all the $N$ particles are in the zero-mode state. We want to construct the Feshbach Hamiltonian obtained from $\mathscr{K}^{Bog\,(N-1)}(z)$ using the projection $\mathscr{P}_{\eta}:=|\eta \rangle \langle \eta |$ and the projection $\overline{\mathscr{P}_{\eta}}$ such that
%\begin{equation}
%\mathscr{P}_{\eta}+\overline{\mathscr{P}_{\eta}}=\charf_{Q^{(>N-1)}\mathcal{F}^N}
%\end{equation}
the identities $$\mathscr{P}_{\eta}(H^{Bog}_{\bold{j}_*}-z)\mathscr{P}_{\eta}=-z\mathscr{P}_{\eta}\quad,\quad \overline{\mathscr{P}_{\eta}}W^*_{\bold{j}_*}\mathscr{P}_{\eta}=0\,,$$ 
imply
\begin{equation}
\mathscr{K}^{Bog\,(N)}_{\bold{j}_*}(z)=f_{\bold{j}_*}(z)|\eta \rangle \langle \eta |\,.
\end{equation}

In \emph{\underline{Corollary 4.6} of \cite{Pi1}} we show that   in the limit $N\to \infty$ the ground state energy $z_*$ tends to $E^{Bog}_{\bold{j}_*}$ with a spectral gap that (in the mean field limit) can be estimated\footnote{In \cite{Se1} the results concerning the spectrum provide a much more accurate estimate of the spectral gap.} larger than $\frac{\Delta_0}{2}$ (where $\Delta_{0}:=\min\, \Big\{(k_{\bold{j}})^2\,|\,\bold{j}\in \mathbb{Z}^d \setminus \{\bold{0}\}\Big\}$). More importantly,  the ground state vector of the Hamiltonian $H^{Bog}_{\bold{j}_*}$ is derived exploiting Feshbach-Schur theory:
\begin{eqnarray}
& &\psi^{Bog}_{\bold{j}_*}\\
&:=&\Big[Q^{(>1)}_{\bold{j}_*}-\frac{1}{Q^{(0,1)}_{\bold{j}_*}(H^{Bog}_{\bold{j}_*}-z_*)Q^{(0,1)}_{\bold{j}_*}}Q^{(0,1)}_{\bold{j}_*}(H^{Bog}_{\bold{j}_*}-z_*)Q^{(>1)}_{\bold{j}_*}\Big]\times \label{gs-vector}\\
& &\quad\quad\quad\times \Big\{ \prod_{i=0\,,\, i\,\,\text{even}}^{N-4}\Big[Q^{(>i+3)}_{\bold{j}_*}-\frac{1}{Q^{(i+2,i+3)}_{\bold{j}_*}\mathscr{K}^{Bog\,(i)}_{\bold{j}_*}(z_*)Q^{(i+2,i+3)}_{\bold{j}_*}}Q^{(i+2,i+3)}_{\bold{j}_*}\mathscr{K}^{Bog\,(i)}_{\bold{j}_*}(z_*)Q^{(>i+3)}_{\bold{j}_*}\Big]\Big\}\eta \,\nonumber
\end{eqnarray}
that, taking the selection rules of $W_{\bold{j}_*}$ into account, can be also written
\begin{eqnarray}
\psi^{Bog}_{\bold{j}_*}
&=&\eta \label{gs-1}\\
& &-\frac{1}{Q^{(N-2,N-1)}_{\bold{j}_*}\mathscr{K}^{Bog\,(N-4)}_{\bold{j}_*}(z_*)Q^{(N-2,N-1)}_{\bold{j}_*}}Q^{(N-2,N-1)}_{\bold{j}_*}W^*_{\bold{j}_*}\eta \label{gs-1-bis}\\
& &-\sum_{j=2}^{N/2}\Big\{\prod^{2}_{r=j}\Big[-\frac{1}{Q^{(N-2r,N-2r+1)}_{\bold{j}_*}\mathscr{K}^{Bog\,(N-2r-2)}_{\bold{j}_*}(z_*)Q^{(N-2r,N-2r+1)}_{\bold{j}_*}}W^*_{\bold{j}_*\,;\,N-2r,N-2r+2}\Big]\Big\}\times \quad\quad\quad \label{gs-2}\\
& &\quad\quad\quad\quad\quad \quad\quad\times \frac{1}{Q^{(N-2,N-1)}_{\bold{j}_*}\mathscr{K}^{Bog\,(N-4)}_{\bold{j}_*}(z_*)Q^{(N-2,N-1)}_{\bold{j}_*}}Q^{(N-2, N-1)}_{\bold{j}_*}W^*_{\bold{j}_*}\eta
\nonumber
\end{eqnarray}
where $\mathscr{K}^{Bog\,(-2)}_{\bold{j}_*}(z_*):=H^{Bog}_{\bold{j}_*}-z_*$.\\

\begin{remark}\label{expansions}
The sum in (\ref{gs-2})  is controlled in norm by a multiple of
\begin{eqnarray}\label{convergent-series}
& &\sum_{j=2}^{\infty}c_j:=\sum_{j=2}^{\infty}\Big\{\prod^{2}_{l=j}\frac{1}{\Big[1+\sqrt{\eta a_{\epsilon_{\bold{j}_*}}}-\frac{b_{\epsilon_{\bold{j}_*}}/ \sqrt{\eta a_{\epsilon_{\bold{j}_*}}}}{2l-\epsilon^{\Theta}_{\bold{j}_*}}\Big]\Big[1+a_{\epsilon_{\bold{j}_*}}-\frac{2b_{\epsilon_{\bold{j}_*}}}{2l-1}-\frac{1-c_{\epsilon_{\bold{j}_*}}}{(2l-1)^2}\Big]^{\frac{1}{2}}}\Big\}
\end{eqnarray}
which is convergent for $\epsilon_{\bold{j}_*}>0$ because $\frac{c_{j+1}}{c_j}<1$ for $j$ sufficiently large. The series diverges in the limit $\epsilon_{\bold{j}_*} \to 0$. Hence, for any $\epsilon_{\bold{j}_*}>0$  sufficiently small there is a convergent expansion of $\psi^{Bog}_{\bold{j}_*}$ controlled by  the parameter $\Sigma_{\epsilon_{\bold{j}_*}}:=\frac{1}{1+\sqrt{\epsilon_{\bold{j}_*}}+o(\sqrt{\epsilon_{\bold{j}_*}})}$. 
%On the contrary, the expansion of the R-H-S of (\ref{fp-function}) is not divergent as $\epsilon_{\bold{j}_*}$ tends to zero.
\end{remark}

 In {\emph{\underline{Section 4} of \cite{Pi1}} we also show how to expand (for a gas in a fixed finite box) the vector $\psi^{Bog}_{{\bold{j}_*}}$ in terms of finite sums of finite products of the operators 
 %\footnote{The operator $H_0:=\frac{1}{2}\sum_{\bold{j}\in\mathbb{Z}^d\setminus \{\bold{0}\}}\hat{H}^{0}_{\bold{j}}$ is defined starting from $\hat{H}^{0}_{\bold{j}}$ in (\ref{H0j}).}      
 $W_{\bold{j}_*}, W^*_{\bold{j}_*}$, and $\frac{1}{\hat{H}^{0}_{\bold{j}_*}-E^{Bog}_{\bold{j}_*}}$ (see (\ref{H0j})) applied to the vector $\eta$, up to any desired precision in the limit $N\to \infty$ at fixed box size.
 
\begin{remark}\label{same-gs}
We observe that $\psi^{Bog}_{{\bold{j}_*}}$ is also eigenvector of $\hat{H}^{Bog}_{{\bold{j}_*}}$ (see the definition in (\ref{check-HBogj})) with the same eigenvalue $z_*$.
\end{remark}

\section{Ground state of $H^{Bog}$: Outline of the proof}\label{groundstateHBog} 
\setcounter{equation}{0}

From now on we consider the system in the mean field limiting regime: fixed box  $\Lambda$ and a number of particles $N$ independent of $|\Lambda|=L^d$. Since we have assumed that an ultraviolet cut-off is imposed on the interaction  potential,  there are $M$ couples of (nonzero) interacting modes with $M<\infty$. The strategy to construct the ground state of $H^{Bog}$ consists in three operations:
\begin{enumerate}
\item We define intermediate Bogoliubov Hamiltonians obtained by adding  (to the interaction Hamiltonian) a couple of modes, $\{\bold{j}_m\,,\,-\bold{j}_m\}$ with $1\leq m \leq M$,  at a time;
\item At each step, i.e., for each intermediate Hamiltonian,  we use the Feshbach-Schur map flow described in Section \ref{multiscale-HBog} and
associated with the new couple of modes, $\{\bold{j}_m\,,\,-\bold{j}_m\}$,  that has been considered;
\item We use the projection onto the ground state of the intermediate Bogoliubov  Hamiltonian at the $(m-1)-th$ step as the final projection $\mathscr{P}$ of the Feshbach-Schur map flow at the $m-th$ step.
\end{enumerate} 
In Section \ref{informal} we outline the procedure, and in Section \ref{rigorousHbog} we provide the results that are needed to make the construction rigorous.

\subsection{The Feshbach-Schur flows associated with $H^{Bog}$: The intermediate Hamiltonians $H^{Bog}_{\bold{j}_1,\,\dots, \bold{j}_m}$}\label{informal}

\noindent
We start from $H^{Bog}_{\bold{j}_1}$ and construct
\begin{eqnarray}
\mathscr{K}_{\bold{j}_1}^{Bog\,(N)}(z)
&= &\mathscr{P}_{\eta}(H^{Bog}_{\bold{j}_1}-z)\mathscr{P}_{\eta}\\
& &-\mathscr{P}_{\eta}W_{\bold{j}_1}\sum_{l_{N-2}=0}^{\infty}R^{Bog}_{\bold{j}_1;N-2,N-2}(z)\,\Big[\Gamma^{Bog\,}_{\bold{j}_1\,;\,N-2,N-2}(z)R^{Bog}_{\bold{j}_1;N-2,N-2}(z)\Big]^{l_{N-2}}W_{\bold{j}_1}^*\mathscr{P}_{\eta}\,. \nonumber
%& &-\sum_{l_i=0}^{\infty}Q^{(>i)}W\,Q^{(i)}R^{Bog}_{i,i}(z)\Big[\Gamma^{Bog\,}_{i,i}(z)R^{Bog}_{i,i}(z)\Big]^{l_i}Q^{(i)}W^*Q^{(>i)} \nonumber
\end{eqnarray}
Next, we determine the ground state energy, $z_{\bold{j}_1}^{Bog}$, of $H^{Bog}_{\bold{j}_1}$ by imposing
\begin{eqnarray}
z^{Bog}_{\bold{j}_1}=\langle \eta\,, \,W_{\bold{j}_1}\sum_{l_{N-2}=0}^{\infty}R^{Bog}_{\bold{j}_1;\,N-2,N-2}(z^{Bog}_{\bold{j}_1})\,\Big[\Gamma^{Bog\,}_{\bold{j}_1;\,N-2,N-2}(z^{Bog}_{\bold{j}_1})R^{Bog}_{\bold{j}_1;\,N-2,N-2}(z^{Bog}_{\bold{j}_1})\Big]^{l_{N-2}}W_{\bold{j}_1}^*\eta \rangle\,.\quad\quad
%&=&\langle \eta\,, \,W_{\bold{j}_1}\,Q_{\bold{j}_1}^{(N-2,N-1)}[R^{Bog}_{\bold{j}_1;\,N-2,N-2}(z_{\bold{j}_1})]^{\frac{1}{2}}\,\sum_{l_{N-2}=0}^{\infty}\Big[[R^{Bog}_{\bold{j}_1;\,N-2,N-2}(z_{\bold{j}_1})]^{\frac{1}{2}}\Gamma^{Bog\,}_{\bold{j}_1;\,N-2,N-2}(z_{\bold{j}_1})[R^{Bog}_{\bold{j}_1;\,N-2,N-2}(z_{\bold{j}_1})]^{\frac{1}{2}}\Big]^{l_{N-2}}\times \nonumber\\
%& &\quad\quad\quad\quad \times [R^{Bog}_{\bold{j}_1;\,N-2,N-2}(z_{\bold{j}_1})]^{\frac{1}{2}}\,Q_{\bold{j}_1}^{(N-2, N-1)}W_{\bold{j}_1}^*\eta \rangle
%& &-\sum_{l_i=0}^{\infty}Q^{(>i)}W\,Q^{(i)}R^{Bog}_{i,i}(z)\Big[\Gamma^{Bog\,}_{i,i}(z)R^{Bog}_{i,i}(z)\Big]^{l_i}Q^{(i)}W^*Q^{(>i)} \nonumber
\end{eqnarray}
Hence, the ground state of $H^{Bog}_{\bold{j}_1}$ is given by (\ref{gs-1})-(\ref{gs-2}) replacing $\bold{j}_*$ with $\bold{j}_1$.
%\begin{eqnarray}
%\psi^{Bog}_{\bold{j}_1}
%&:=&
%\Big[Q^{(>0)}-\frac{1}{Q^{(0)}(H^{Bog}-z_*)Q^{(0)}}Q^{(0)}(H^{Bog}-z_*)Q^{(>0)}\Big]\times\\
%& &\quad\quad\quad\times \Big\{ \prod_{i=0}^{N-2}\Big[Q^{(>i+1)}-\frac{1}{Q^{(i+1)}\mathscr{K}^{Bog\,(i)}(z_*)Q^{(i+1)}}Q^{(i+1)}\mathscr{K}^{Bog\,(i)}(z_*)Q^{(>i+1)}\Big]\Big\}\eta \quad \\
%&= &\eta \\
%& &-\frac{1}{Q^{(N-2)}\mathscr{K}^{Bog\,(N-3)}(z_*)Q^{(N-2)}}Q^{(N-2)}W^*\eta \\
%& &+\frac{1}{Q^{(N-4)}\mathscr{K}^{Bog\,(N-5)}(z_*)Q^{(N-4)}}W^*_{N-4,N-2}\frac{1}{Q^{(N-2)}\mathscr{K}^{Bog\,(N-3)}(z_*)Q^{(N-2)}}Q^{(N-2)}W^*\eta \quad \nonumber \\
%& &-\frac{1}{Q^{(N-6)}\mathscr{K}^{Bog\,(N-7)}(z_*)Q^{(N-6)}}W^*_{N-6,N-4}\frac{1}{Q^{(N-4)}\mathscr{K}^{Bog\,(N-5)}(z_*)Q^{(N-4)}}Q^{(N-4)}W^*Q^{(N-2)}\times \quad \nonumber \\
%& &\quad \times\frac{1}{Q^{(N-2)}\mathscr{K}^{Bog\,(N-3)}(z_*)Q^{(N-2)}}Q^{(N-2)}W^*\eta \\
%& &+\dots \\
%\eta \\
%& &-\frac{1}{Q_{\bold{j}_1}^{(N-2,N-1)}\mathscr{K}_{\bold{j}_1}^{Bog\,(N-4)}(z^{Bog}_{\bold{j}_1})Q_{\bold{j}_1}^{(N-2, N-1)}}Q_{\bold{j}_1}^{(N-2,N-1)}W_{\bold{j}_1}^*\eta\\
%& &-\sum_{j=2}^{N/2}\prod^{2}_{i=j}\Big[-\frac{1}{Q_{\bold{j}_1}^{(N-2i, N-2i+1)}\mathscr{K}_{\bold{j}_1}^{Bog\,(N-2i-2)}(z^{Bog}_{\bold{j}_1})Q_{\bold{j}_1}^{(N-2i, N-2i+1)}}W^*_{\bold{j}_1;\, N-2i,N-2i+2}\Big]\times \quad\quad\\
%& &\quad\quad\quad\quad\quad\quad \times \frac{1}{Q_{\bold{j}_1}^{(N-2,N-1)}\mathscr{K}_{\bold{j}_1}^{Bog\,(N-4)}(z^{Bog}_{\bold{j}_1})Q_{\bold{j}_1}^{(N-2, N-1)}}Q_{\bold{j}_1}^{(N-2, N-1)}W_{\bold{j}_1}^*\eta\nonumber
%\end{eqnarray}
This first step just requires the results of \cite{Pi1} that have been summarized in Section \ref{Feshbach}.
\\

In the next step, we consider the intermediate  Hamiltonian
\begin{equation}\label{ham-12}
H^{Bog}_{\bold{j}_1,\, \bold{j}_2}:=\sum_{\bold{j}\in\mathbb{Z}^d\setminus \{ \pm\bold{j}_{1}\,;\, \pm\bold{j}_{2} \}} k^2_{\bold{j}}a_{\bold{j}}^{*}a_{\bold{j}}+\hat{H}^{Bog}_{\bold{j}_1,\, \bold{j}_2}:=\sum_{\bold{j}\in\mathbb{Z}^d\setminus \{\pm\bold{j}_{1}\,;\, \pm\bold{j}_{2} \}} k^2_{\bold{j}}a_{\bold{j}}^{*}a_{\bold{j}}+\sum_{l=1}^{2}\hat{H}^{Bog}_{\bold{j}_l}\end{equation}
and construct the Feshbach-Schur Hamiltonians,
\begin{eqnarray}
& &\mathscr{K}_{\bold{j}_1,\,\bold{j}_2}^{Bog\,(0)}(z^{Bog}_{\bold{j}_1}+z)\\
&:=&Q_{\bold{j}_2}^{(>1)}(H^{Bog} _{\bold{j}_1,\, \bold{j}_2}-z^{Bog}_{\bold{j}_1}-z)Q^{(>1)}_{\bold{j}_2}\quad\quad\quad\\
& &-Q^{(>1)}_{\bold{j}_2}W_{\bold{j}_2}\,R^{Bog}_{\bold{j}_1,\bold{j}_2\,;\,0,0}(z^{Bog}_{\bold{j}_1}+z)W^*_{\bold{j}_2}Q^{(>1)}_{\bold{j}_2} \nonumber
%&=&Q_{\bold{j}_m}^{(>i)}(H^{Bog} _{\bold{j}_1,\, \bold{j}_2}-z_{\bold{j}_1}-z)Q^{(>i)}_{\bold{j}_m}\quad\quad\quad\\
%& &-Q^{(>i+1)}_{\bold{j}_2}\Gamma^{Bog}_{\bold{j}_1,\bold{j}_2\,;\,N,N}(z_{\bold{j}_1}+z))Q^{(>i+1)}_{\bold{j}_2} \nonumber
%&=&\mathscr{P}_{\psi^{Bog}_{\bold{j}_1}}(H^{Bog}_{\bold{j}_1,\, \bold{j}_2}-z_{\bold{j}_1}-z)\mathscr{P}_{\psi^{Bog}_{\bold{j}_1}}\\
%& &-\mathscr{P}_{\psi^{Bog}_{\bold{j}_1}}W_{\bold{j}_2}\,Q_{\bold{j}_2}^{(N-2, N-1)}\sum_{l_{N-2}=0}^{\infty}R^{Bog}_{\bold{j}_1,\bold{j}_2;\,N-2,N-2}(z_{\bold{j}_1}+z)\,\Big[\Gamma^{Bog\,}_{\bold{j}_1,\bold{j}_2\,;\,N-2,N-2}(z_{\bold{j}_1}+z)R^{Bog}_{\bold{j}_1,\bold{j}_2;\,N-2,N-2}(z_{\bold{j}_1}+z)\Big]^{l_{N-2}}\times \nonumber \\
%& &\quad\quad\quad \times Q_{\bold{j}_2}^{(N-2, N-1)}W_{\bold{j}_2}^*\mathscr{P}_{\psi^{Bog}_{\bold{j}_1}} \nonumber\\
%& &-\sum_{l_i=0}^{\infty}Q^{(>i)}W\,Q^{(i)}R^{Bog}_{i,i}(z)\Big[\Gamma^{Bog\,}_{i,i}(z)R^{Bog}_{i,i}(z)\Big]^{l_i}Q^{(i)}W^*Q^{(>i)} \nonumber
%&=&\mathscr{P}_{\psi^{Bog}_{\bold{j}_1}}(H^{Bog}_{\bold{j}_2}-z)\mathscr{P}_{\psi^{Bog}_{\bold{j}_1}}
\end{eqnarray}
and, for $2\leq i \leq N-2$ and even,
\begin{eqnarray}
& &\mathscr{K}_{\bold{j}_1,\,\bold{j}_2}^{Bog\,(i)}(z^{Bog}_{\bold{j}_1}+z)\\
&:=&Q_{\bold{j}_2}^{(>i+1)}(H^{Bog} _{\bold{j}_1,\, \bold{j}_2}-z^{Bog}_{\bold{j}_1}-z)Q^{(>i+1)}_{\bold{j}_2}\quad\quad\quad\\
& &-Q^{(>i+1)}_{\bold{j}_2}W_{\bold{j}_2}\,R^{Bog}_{\bold{j}_1,\,\bold{j}_2\,;\,i,i}(z^{Bog}_{\bold{j}_1}+z)\sum_{l_i=0}^{\infty}\Big[\Gamma^{Bog\,}_{\bold{j}_1,\,\bold{j}_2\,;\,i,i}(z^{Bog}_{\bold{j}_1}+z)\,R^{Bog}_{\bold{j}_1,\,\bold{j}_2\,;\,i,i}(z^{Bog}_{\bold{j}_1}+z)\Big]^{l_i}W^*_{\bold{j}_2}Q^{(>i+1)}_{\bold{j}_2} \nonumber
%&=&Q_{\bold{j}_m}^{(>i)}(H^{Bog} _{\bold{j}_1,\, \bold{j}_2}-z_{\bold{j}_1}-z)Q^{(>i)}_{\bold{j}_m}\quad\quad\quad\\
%& &-Q^{(>i+1)}_{\bold{j}_2}\Gamma^{Bog}_{\bold{j}_1,\bold{j}_2\,;\,N,N}(z_{\bold{j}_1}+z))Q^{(>i+1)}_{\bold{j}_2} \nonumber
%&=&\mathscr{P}_{\psi^{Bog}_{\bold{j}_1}}(H^{Bog}_{\bold{j}_1,\, \bold{j}_2}-z_{\bold{j}_1}-z)\mathscr{P}_{\psi^{Bog}_{\bold{j}_1}}\\
%& &-\mathscr{P}_{\psi^{Bog}_{\bold{j}_1}}W_{\bold{j}_2}\,Q_{\bold{j}_2}^{(N-2, N-1)}\sum_{l_{N-2}=0}^{\infty}R^{Bog}_{\bold{j}_1,\bold{j}_2;\,N-2,N-2}(z_{\bold{j}_1}+z)\,\Big[\Gamma^{Bog\,}_{\bold{j}_1,\bold{j}_2\,;\,N-2,N-2}(z_{\bold{j}_1}+z)R^{Bog}_{\bold{j}_1,\bold{j}_2;\,N-2,N-2}(z_{\bold{j}_1}+z)\Big]^{l_{N-2}}\times \nonumber \\
%& &\quad\quad\quad \times Q_{\bold{j}_2}^{(N-2, N-1)}W_{\bold{j}_2}^*\mathscr{P}_{\psi^{Bog}_{\bold{j}_1}} \nonumber\\
%& &-\sum_{l_i=0}^{\infty}Q^{(>i)}W\,Q^{(i)}R^{Bog}_{i,i}(z)\Big[\Gamma^{Bog\,}_{i,i}(z)R^{Bog}_{i,i}(z)\Big]^{l_i}Q^{(i)}W^*Q^{(>i)} \nonumber
%&=&\mathscr{P}_{\psi^{Bog}_{\bold{j}_1}}(H^{Bog}_{\bold{j}_2}-z)\mathscr{P}_{\psi^{Bog}_{\bold{j}_1}}
\end{eqnarray}
where we use the definitions:
\begin{itemize}
\item
\begin{equation}\label{rBog-ii}
R^{Bog}_{\bold{j}_1,\,\bold{j}_2\,;\,i,i}(z^{Bog}_{\bold{j}_1}+z):=Q^{(i, i+1)}_{\bold{j}_2}{\frac{1}{Q^{(i, i+1)}_{\bold{j}_2}(H^{Bog}_{\bold{j}_1,\, \bold{j}_2}-z^{Bog}_{\bold{j}_1}-z)Q^{(i,i+1)}_{\bold{j}_2}}Q^{(i, i+1)}_{\bold{j}_2}}\,;
\end{equation}
\item
\begin{equation}
\Gamma^{Bog\,}_{\bold{j}_1,\,\bold{j}_2\,;\,2,2}(z^{Bog}_{\bold{j}_1}+z):=W_{\bold{j}_2\,;\,2,0}R^{Bog}_{\bold{j}_1,\,\bold{j}_2\,;\,0,0}(z^{Bog}_{\bold{j}_1}+z)W_{\bold{j}_2\,;\,0,2}^*
\end{equation}
%\begin{equation}
%\Gamma^{Bog\,}_{\bold{j}_1,\,\bold{j}_2\,;\,3,3}(z_{\bold{j}_1}+z):=W_{\bold{j}_*\,;\,3,1}R^{Bog}_{\bold{j}_1,\,\bold{j}_2\,;\,1,1}(z_{\bold{j}_1}+z)W_{\bold{j}_*\,;\,1,3}^*\label{gamma-12}
%\end{equation}
and,  for $N-2\geq i\geq 4$ and even,
\begin{eqnarray}
& &\Gamma^{Bog\,}_{\bold{j}_1,\,\bold{j}_2\,;\,i,i}(z^{Bog}_{\bold{j}_1}+z)\\
&:=&W_{\bold{j}_2\,;\,i,i-2}\,R^{Bog}_{\bold{j}_1,\,\bold{j}_2\,;\,i-2,i-2}(z^{Bog}_{\bold{j}_1}+z) \sum_{l_{i-2}=0}^{\infty}\Big[\Gamma^{Bog}_{\bold{j}_1,\,\bold{j}_2\,;\,i-2,i-2}(z^{Bog}_{\bold{j}_1}+z)R^{Bog}_{\bold{j}_1,\,\bold{j}_2\,;\,\,i-2,i-2}(z^{Bog}_{\bold{j}_1}+z)\Big]^{l_{i-2}}W^*_{\bold{j}_2\,;\,i-2,i}\,.\quad\quad\quad\label{gamma-12}
\end{eqnarray}
%\item
%%\check{R}^{Bog}_{\bold{j}_1,\,\bold{j}_2\,;\,i,i}(z_{\bold{j}_1}+z):=\sum_{l_i=0}^{\infty}R^{Bog}_{\bold{j}_1,\bold{j}_2\,;\,i,i}(z_{\bold{j}_1}+z)\Big[\Gamma^{Bog\,}_{\bold{j}_1,\bold{j}_2\,;\,i,i}(z_{\bold{j}_1}+z)R^{Bog}_{\bold{j}_1,\bold{j}_2\,;\,i,i}(z_{\bold{j}_1}+z)\Big]^{l_i}
%\end{equation}
\end{itemize}
In the last implementation of the Feshbach-Schur map we use the projections
\begin{equation}\label{projection-step-1}
\mathscr{P}_{\psi^{Bog}_{\bold{j}_1}}:=|\frac{ \psi^{Bog}_{\bold{j}_1} }{\| \psi^{Bog}_{\bold{j}_1} \|}\rangle \langle\frac{ \psi^{Bog}_{\bold{j}_1} }{\| \psi^{Bog}_{\bold{j}_1} \|}|\quad,\quad \overline{\mathscr{P}_{\psi^{Bog}_{\bold{j}_1}}}:=\charf_{Q^{(>N-2)}_{\bold{j}_2}\mathcal{F}^N}-\mathscr{P}_{\psi^{Bog}_{\bold{j}_1}}
\end{equation}
where $Q^{(>N-1)}_{\bold{j}_2}\mathcal{F}^N$ is the  subspace of states in $\mathcal{F}^N$ with no particles in the modes $\pm \bold{j}_2$,
and we define
\begin{eqnarray}
& &\Gamma^{Bog\,}_{\bold{j}_1,\,\bold{j}_2\,;\,N,N}(z^{Bog}_{\bold{j}_1}+z)\\
&:=&W_{\bold{j}_2}\,R^{Bog}_{\bold{j}_1,\,\bold{j}_2\,;\,N-2,N-2}(z^{Bog}_{\bold{j}_1}+z) \sum_{l_{N-2}=0}^{\infty}\Big[\Gamma^{Bog}_{\bold{j}_1,\,\bold{j}_2\,;\,N-2,N-2}(z^{Bog}_{\bold{j}_1}+z)R^{Bog}_{\bold{j}_1,\,\bold{j}_2\,;\,\,N-2,N-2}(z^{Bog}_{\bold{j}_1}+z)\Big]^{l_{N-2}}W^*_{\bold{j}_2}\,.\quad\quad\quad\label{gamma-NN}
\end{eqnarray}
For the derivation of $\mathscr{K}_{\bold{j}_1,\,\bold{j}_2}^{Bog\,(N)}(z^{Bog}_{\bold{j}_1}+z)$, we point out that (see Remark \ref{same-gs})
$$(\hat{H}^{Bog}_{\bold{j}_1}-z^{Bog}_{\bold{j}_1})\mathscr{P}_{\psi^{Bog}_{\bold{j}_1}}=0\quad,\quad \sum_{\bold{j}\in \mathbb{Z}^d\setminus \{\bold{0},\pm \bold{j}_1\}}a^*_{\bold{j}}a_{\bold{j}}\mathscr{P}_{\psi^{Bog}_{\bold{j}_1}}=0\,,$$
and
$$\mathscr{P}_{\psi^{Bog}_{\bold{j}_1}}(H^{Bog}_{\bold{j}_1,\, \bold{j}_2}-z^{Bog}_{\bold{j}_1}-z)\mathscr{P}_{\psi^{Bog}_{\bold{j}_1}}=\mathscr{P}_{\psi^{Bog}_{\bold{j}_1}}(\hat{H}^{Bog}_{\bold{j}_2}-z)\mathscr{P}_{\psi^{Bog}_{\bold{j}_1}}=\mathscr{P}_{\psi^{Bog}_{\bold{j}_1}}(W_{\bold{j}_2}+W^*_{\bold{j}_2}-z)\mathscr{P}_{\psi^{Bog}_{\bold{j}_1}}=-z\mathscr{P}_{\psi^{Bog}_{\bold{j}_1}}\,,$$
 $$\mathscr{P}_{\psi^{Bog}_{\bold{j}_1}}(H^{Bog}_{\bold{j}_1,\, \bold{j}_2}-z^{Bog}_{\bold{j}_1}-z)\overline{\mathscr{P}_{\psi^{Bog}_{\bold{j}_1}}}=\mathscr{P}_{\psi^{Bog}_{\bold{j}_1}}(\hat{H}^{Bog}_{\bold{j}_2}-z)\overline{\mathscr{P}_{\psi^{Bog}_{\bold{j}_1}}}=\mathscr{P}_{\psi^{Bog}_{\bold{j}_1}}(W_{\bold{j}_2}+W^*_{\bold{j}_2})\overline{\mathscr{P}_{\psi^{Bog}_{\bold{j}_1}}}=0\,. $$ These identities follow from  the definitions of $\mathscr{P}_{\psi^{Bog}_{\bold{j}_1}}$, $\overline{\mathscr{P}_{\psi^{Bog}_{\bold{j}_1}}}$,  and $H^{Bog}_{\bold{j}_1,\, \bold{j}_2}$ combined with the fact that $Q^{(>N-1)}_{\bold{j}_2}$ is the projection onto the subspace of states with no particles in the modes $\pm \bold{j}_2$.
Formally, we get
\begin{eqnarray}
& &\mathscr{K}_{\bold{j}_1,\,\bold{j}_2}^{Bog\,(N)}(z^{Bog}_{\bold{j}_1}+z)\\
&:=&\mathscr{P}_{\psi^{Bog}_{\bold{j}_1}}(H^{Bog}_{\bold{j}_1,\, \bold{j}_2}-z^{Bog}_{\bold{j}_1}-z)\mathscr{P}_{\psi^{Bog}_{\bold{j}_1}}\\
& &-\mathscr{P}_{\psi^{Bog}_{\bold{j}_1}}\Gamma^{Bog}_{\bold{j}_1,\,\bold{j}_2;\,N,N}(z^{Bog}_{\bold{j}_1}+z)\mathscr{P}_{\psi^{Bog}_{\bold{j}_1}} \nonumber\\
& &-\mathscr{P}_{\psi^{Bog}_{\bold{j}_1}}\Gamma^{Bog}_{\bold{j}_1,\,\bold{j}_2;\,N,N}(z^{Bog}_{\bold{j}_1}+z)\overline{\mathscr{P}_{\psi^{Bog}_{\bold{j}_1}}}\times\\
& &\quad\quad\times \frac{1}{\overline{\mathscr{P}_{\psi^{Bog}_{\bold{j}_1}}}\mathscr{K}_{\bold{j}_1,\,\bold{j}_2}^{Bog\,(N-2)}(z^{Bog}_{\bold{j}_1}+z)\overline{\mathscr{P}_{\psi^{Bog}_{\bold{j}_1}}}}\overline{\mathscr{P}_{\psi^{Bog}_{\bold{j}_1}}}\Gamma^{Bog}_{\bold{j}_1,\bold{j}_2;\,N,N}(z^{Bog}_{\bold{j}_1}+z)\mathscr{P}_{\psi^{Bog}_{\bold{j}_1}}\\
%& &-\sum_{l_i=0}^{\infty}Q^{(>i)}W\,Q^{(i)}R^{Bog}_{i,i}(z)\Big[\Gamma^{Bog\,}_{i,i}(z)R^{Bog}_{i,i}(z)\Big]^{l_i}Q^{(i)}W^*Q^{(>i)} \nonumber
%&=&\mathscr{P}_{\psi^{Bog}_{\bold{j}_1}}(\hat{H}^{Bog}_{\bold{j}_2}-z)\mathscr{P}_{\psi^{Bog}_{\bold{j}_1}}\\
%& &-\mathscr{P}_{\psi^{Bog}_{\bold{j}_1}}\Gamma^{Bog}_{\bold{j}_1,\bold{j}_2;\,N,N}(z_{\bold{j}_1}+z)\mathscr{P}_{\psi^{Bog}_{\bold{j}_1}} \nonumber\\
%& &-\mathscr{P}_{\psi^{Bog}_{\bold{j}_1}}{\color{red}\Gamma^{Bog}_{\bold{j}_1,\bold{j}_2;\,N,N}(z_{\bold{j}_1}+z)}\overline{\mathscr{P}_{\psi^{Bog}_{\bold{j}_1}}}\times\\
%& &\quad\quad\times \frac{1}{\overline{\mathscr{P}_{\psi^{Bog}_{\bold{j}_1}}}\mathscr{K}_{\bold{j}_1,\,\bold{j}_2}^{Bog\,(N-2)}(z_{\bold{j}_1}+z)\overline{\mathscr{P}_{\psi^{Bog}_{\bold{j}_1}}}}\overline{\mathscr{P}_{\psi^{Bog}_{\bold{j}_1}}}\Gamma^{Bog}_{\bold{j}_1,\bold{j}_2;\,N,N}(z_{\bold{j}_1}+z)\mathscr{P}_{\psi^{Bog}_{\bold{j}_1}}\nonumber \\
&=&-z\mathscr{P}_{\psi^{Bog}_{\bold{j}_1}}\\
& &-\mathscr{P}_{\psi^{Bog}_{\bold{j}_1}}\Gamma^{Bog}_{\bold{j}_1,\,\bold{j}_2;\,N,N}(z^{Bog}_{\bold{j}_1}+z)\mathscr{P}_{\psi^{Bog}_{\bold{j}_1}} \nonumber\\
& &-\mathscr{P}_{\psi^{Bog}_{\bold{j}_1}}\Gamma^{Bog}_{\bold{j}_1,\,\bold{j}_2;\,N,N}(z^{Bog}_{\bold{j}_1}+z)\overline{\mathscr{P}_{\psi^{Bog}_{\bold{j}_1}}}\times\\
& &\quad\quad\times \frac{1}{\overline{\mathscr{P}_{\psi^{Bog}_{\bold{j}_1}}}\mathscr{K}_{\bold{j}_1,\,\bold{j}_2}^{Bog\,(N-2)}(z^{Bog}_{\bold{j}_1}+z)\overline{\mathscr{P}_{\psi^{Bog}_{\bold{j}_1}}}}\overline{\mathscr{P}_{\psi^{Bog}_{\bold{j}_1}}}\Gamma^{Bog}_{\bold{j}_1,\,\bold{j}_2;\,N,N}(z^{Bog}_{\bold{j}_1}+z)\mathscr{P}_{\psi^{Bog}_{\bold{j}_1}}\,.\nonumber
\end{eqnarray}

We determine  the ground state energy, $z^{Bog}_{\bold{j}_1,\bold{j}_2}:=z^{Bog}_{\bold{j}_1}+z^{(2)}$,  of $H^{Bog}_{\bold{j}_1,\, \bold{j}_2}$   by imposing
\begin{eqnarray}
z^{(2)}&=&-\langle \, \frac{\psi^{Bog}_{\bold{j}_1}}{\|\psi^{Bog}_{\bold{j}_1}\|}, \Gamma^{Bog}_{\bold{j}_1,\,\bold{j}_2;\,N,N}(z^{Bog}_{\bold{j}_1}+z^{(2)})\frac{\psi^{Bog}_{\bold{j}_1}}{\|\psi^{Bog}_{\bold{j}_1}\|}\rangle \nonumber\\
& &-\langle \, \frac{\psi^{Bog}_{\bold{j}_1}}{\|\psi^{Bog}_{\bold{j}_1}\|},\Gamma^{Bog}_{\bold{j}_1,\,\bold{j}_2;\,N,N}(z^{Bog}_{\bold{j}_1}+z^{(2)})\overline{\mathscr{P}_{\psi^{Bog}_{\bold{j}_1}}}\times\\
& &\quad\quad\times \frac{1}{\overline{\mathscr{P}_{\psi^{Bog}_{\bold{j}_1}}}\mathscr{K}_{\bold{j}_1,\,\bold{j}_2}^{Bog\,(N-2)}(z^{Bog}_{\bold{j}_1}+z^{(2)})\overline{\mathscr{P}_{\psi^{Bog}_{\bold{j}_1}}}}\overline{\mathscr{P}_{\psi^{Bog}_{\bold{j}_1}}}\Gamma^{Bog}_{\bold{j}_1,\,\bold{j}_2;\,N,N}(z^{Bog}_{\bold{j}_1}+z^{(2)})\frac{\psi^{Bog}_{\bold{j}_1}}{\|\psi^{Bog}_{\bold{j}_1}\|}\rangle\,.\nonumber
%& &-\sum_{l_i=0}^{\infty}Q^{(>i)}W\,Q^{(i)}R^{Bog}_{i,i}(z)\Big[\Gamma^{Bog\,}_{i,i}(z)R^{Bog}_{i,i}(z)\Big]^{l_i}Q^{(i)}W^*Q^{(>i)} \nonumber
\end{eqnarray}
Hence,  the ground state vector of $H^{Bog}_{\bold{j}_1,\, \bold{j}_2}$ is (up to normalization)
\begin{eqnarray}
& &\psi^{Bog}_{\bold{j}_1,\bold{j}_2}\\
&:=&
%& &-\Big\{\frac{1}{Q^{(N-2, N-1)}_{\bold{j}_2}\mathscr{K}_{\bold{j}_1,\bold{j}_2}^{Bog\,(N-4)}(z^{Bog}_{\bold{j}_1,\bold{j}_2})Q^{(N-2)}_{\bold{j}_2}}Q^{(N-2, N-1)}_{\bold{j}_2}W_{\bold{j}_2}^*\\
%& &-\sum_{j=2}^{N/2}\prod^{2}_{r=j}\Big[-\frac{1}{Q_{\bold{j}_2}^{(N-2r, N-2r+1)}\mathscr{K}_{\bold{j}_1,\bold{j}_2}^{Bog\,(N-2r-2)}(z^{Bog}_{\bold{j}_1,\bold{j}_2})Q_{\bold{j}_2}^{(N-2r, N-2r+1)}}W^*_{\bold{j}_2;N-2r,N-2r+2}\Big] \,\psi^{Bog}_{\bold{j}_1} \quad\quad\quad \\
\Big\{\sum_{j=2}^{N/2}\Big[\prod^{2}_{r=j}\Big(-\frac{1}{Q_{\bold{j}_2}^{(N-2r, N-2r+1)}\mathscr{K}_{\bold{j}_1,\bold{j}_2}^{Bog\,(N-2r-2)}(z^{Bog}_{\bold{j}_1,\bold{j}_2})Q_{\bold{j}_2}^{(N-2r, N-2r+1)}}W^*_{\bold{j}_2;N-2r,N-2r+2}\Big)\Big]+\charf \Big\}\times\quad\quad\quad\\
%& &\quad\quad\quad\quad\quad \times\Big[ \frac{1}{Q_{\bold{j}_2}^{(N-2, N-1)}\mathscr{K}_{\bold{j}_1,\bold{j}_2}^{Bog\,(N-4)}(z^{Bog}_{\bold{j}_1,\bold{j}_2})Q^{(N-2, N-1)}_{\bold{j}_2}}Q^{(N-2, N-1)}_{\bold{j}_2}W_{\bold{j}_2}^*\Big]\times\\
& &\quad\quad\quad\quad\times \Big[Q^{(>N-1)}_{\bold{j}_2}-\frac{1}{Q^{(N-2,N-1)}_{\bold{j}_2}\mathscr{K}^{Bog\,(N-4)}_{\bold{j}_1,\bold{j}_2}(z^{Bog}_{\bold{j}_1,\bold{j}_2})Q^{(N-2,N-1)}_{\bold{j}_2}}Q^{(N-2, N-1)}_{\bold{j}_1,\bold{j}_2}W^*_{\bold{j}_2}\Big]\times\\
& &\quad\quad\quad\quad\quad\quad \times \Big[\mathscr{P}_{\psi^{Bog}_{\bold{j}_1}}-\frac{1}{\overline{\mathscr{P}_{\psi^{Bog}_{\bold{j}_1}}}\mathscr{K}^{Bog\,(N-2)}_{\bold{j}_1,\bold{j}_2}(z^{Bog}_{\bold{j}_1,\bold{j}_2})\overline{\mathscr{P}_{\psi^{Bog}_{\bold{j}_1}}}}
\overline{\mathscr{P}_{\psi^{Bog}_{\bold{j}_1}}}\mathscr{K}^{Bog\,(N-2)}_{\bold{j}_1,\bold{j}_2}(z^{Bog}_{\bold{j}_1,\bold{j}_2})\Big]\,\psi^{Bog}_{\bold{j}_1}\,\nonumber
\end{eqnarray}
where $\mathscr{K}^{Bog\,(-2)}_{\bold{j}_1,\bold{j}_2}(z^{Bog}_{\bold{j}_1,\bold{j}_2}):=H^{Bog}_{\bold{j}_1,\bold{j}_2}-z^{Bog}_{\bold{j}_1,\bold{j}_2}$.
\\

At the $m-th$ step, we define
\begin{equation}
H^{Bog}_{\bold{j}_1,\,\dots, \bold{j}_m}:=\sum_{\bold{j}\in\mathbb{Z}^d\setminus \{\pm\bold{j}_{1},\dots , \pm\bold{j}_{m} \}} k^2_{\bold{j}}a_{\bold{j}}^{*}a_{\bold{j}}+\sum_{l=1}^{m}\hat{H}^{Bog}_{\bold{j}_l}\,,\end{equation}
where the reader should note that the kinetic energy of the interacting nonzero modes, $\pm\bold{j}_{1}\,,\,\dots \,,\, \pm\bold{j}_{m}$,  is contained in $\sum_{l=1}^{m}\hat{H}^{Bog}_{\bold{j}_{l}}$. Then, we construct (for $0\leq i\leq N-2$ and even)
\begin{eqnarray}
& &\mathscr{K}_{\bold{j}_1,\dots,\bold{j}_m}^{Bog\,(i)}(z^{Bog}_{\bold{j}_1,\dots,\bold{j}_m}+z) \label{fesh-i-m}\\
&:=&Q_{\bold{j}_m}^{(>i+1)}(H^{Bog} _{\bold{j}_1,\dots,\bold{j}_m}-z^{Bog}_{\bold{j}_1,\dots,\bold{j}_m}-z)Q^{(>i+1)}_{\bold{j}_m}\quad\quad\quad\\
& &-Q^{(>i+1)}_{\bold{j}_m}W_{\bold{j}_m}\,R^{Bog}_{\bold{j}_1,\dots,\bold{j}_m\,;\,i,i}(z^{Bog}_{\bold{j}_1,\dots,\bold{j}_m}+z)\sum_{l_i=0}^{\infty}\Big[\Gamma^{Bog\,}_{\bold{j}_1,\dots,\bold{j}_m\,;\,i,i}(z^{Bog}_{\bold{j}_1,\dots,\bold{j}_m}+z)\,R^{Bog}_{\bold{j}_1,\dots,\bold{j}_m\,;\,i,i}(z^{Bog}_{\bold{j}_1,\dots,\bold{j}_m}+z)\Big]^{l_i}W^*_{\bold{j}_m}Q^{(>i+1)}_{\bold{j}_m} \nonumber
%&=&Q_{\bold{j}_m}^{(>i)}(H^{Bog} _{\bold{j}_1,\, \bold{j}_2}-z_{\bold{j}_1}-z)Q^{(>i)}_{\bold{j}_m}\quad\quad\quad\\
%& &-Q^{(>i+1)}_{\bold{j}_2}\Gamma^{Bog}_{\bold{j}_1,\bold{j}_2\,;\,N,N}(z_{\bold{j}_1}+z))Q^{(>i+1)}_{\bold{j}_2} \nonumber
%&=&\mathscr{P}_{\psi^{Bog}_{\bold{j}_1}}(H^{Bog}_{\bold{j}_1,\, \bold{j}_2}-z_{\bold{j}_1}-z)\mathscr{P}_{\psi^{Bog}_{\bold{j}_1}}\\
%& &-\mathscr{P}_{\psi^{Bog}_{\bold{j}_1}}W_{\bold{j}_2}\,Q_{\bold{j}_2}^{(N-2, N-1)}\sum_{l_{N-2}=0}^{\infty}R^{Bog}_{\bold{j}_1,\bold{j}_2;\,N-2,N-2}(z_{\bold{j}_1}+z)\,\Big[\Gamma^{Bog\,}_{\bold{j}_1,\bold{j}_2\,;\,N-2,N-2}(z_{\bold{j}_1}+z)R^{Bog}_{\bold{j}_1,\bold{j}_2;\,N-2,N-2}(z_{\bold{j}_1}+z)\Big]^{l_{N-2}}\times \nonumber \\
%& &\quad\quad\quad \times Q_{\bold{j}_2}^{(N-2, N-1)}W_{\bold{j}_2}^*\mathscr{P}_{\psi^{Bog}_{\bold{j}_1}} \nonumber\\
%& &-\sum_{l_i=0}^{\infty}Q^{(>i)}W\,Q^{(i)}R^{Bog}_{i,i}(z)\Big[\Gamma^{Bog\,}_{i,i}(z)R^{Bog}_{i,i}(z)\Big]^{l_i}Q^{(i)}W^*Q^{(>i)} \nonumber
%&=&\mathscr{P}_{\psi^{Bog}_{\bold{j}_1}}(H^{Bog}_{\bold{j}_2}-z)\mathscr{P}_{\psi^{Bog}_{\bold{j}_1}}
\end{eqnarray}
and, eventually,
\begin{eqnarray}
& &\mathscr{K}_{\bold{j}_1,\dots,\bold{j}_m}^{Bog\,(N)}(z+z^{Bog}_{\bold{j}_1,\dots,\bold{j}_{m-1}})\label{final-fesh-step-m}\\
&:=&-z\mathscr{P}_{\psi^{Bog}_{\bold{j}_1,\dots,\bold{j}_{m-1}}}\\
& &-\mathscr{P}_{\psi^{Bog}_{\bold{j}_1,\dots,\bold{j}_{m-1}}}\Gamma^{Bog}_{\bold{j}_1,\dots,\bold{j}_m ;N,N}(z+z^{Bog}_{\bold{j}_1,\dots,\bold{j}_{m-1}})\mathscr{P}_{\psi^{Bog}_{\bold{j}_1,\dots,\bold{j}_{m-1}}}\nonumber\\
& &-\mathscr{P}_{\psi^{Bog}_{\bold{j}_1,\dots,\bold{j}_{m-1}}}\Gamma^{Bog}_{\bold{j}_1,\dots,\bold{j}_m;\,N,N}(z+z^{Bog}_{\bold{j}_1,\dots,\bold{j}_{m-1}})\overline{\mathscr{P}_{\psi^{Bog}_{\bold{j}_1,\dots,\bold{j}_{m-1}}}}\times\\
& &\quad\quad\times \frac{1}{\overline{\mathscr{P}_{\psi^{Bog}_{\bold{j}_1,\dots,\bold{j}_{m-1}}}}\mathscr{K}_{\bold{j}_1,\dots,\bold{j}_m}^{Bog\,(N-2)}(z+z^{Bog}_{\bold{j}_1,\dots,\bold{j}_{m-1}})\overline{\mathscr{P}_{\psi^{Bog}_{\bold{j}_1,\dots,\bold{j}_{m-1}}}}}\overline{\mathscr{P}_{\psi^{Bog}_{\bold{j}_1,\dots,\bold{j}_{m-1}}}}\Gamma^{Bog}_{\bold{j}_1,\dots,\bold{j}_m ;N,N}(z+z^{Bog}_{\bold{j}_1,\dots,\bold{j}_{m-1}})\mathscr{P}_{\psi^{Bog}_{\bold{j}_1,\dots,\bold{j}_{m-1}}}\nonumber\\
&=:&f^{Bog}_{\bold{j}_1,\dots,\bold{j}_m}(z+z^{Bog}_{\bold{j}_1,\dots,\bold{j}_{m-1}})\mathscr{P}_{\psi^{Bog}_{\bold{j}_1,\bold{j}_2,\dots,\bold{j}_{m-1}}}\label{def-f}
\end{eqnarray}
with definitions analogous to (\ref{rBog-ii})-(\ref{gamma-NN}):
\begin{itemize}
\item
\begin{equation}\label{final-proj-Bog}
\mathscr{P}_{\psi^{Bog}_{\bold{j}_1,\dots,\bold{j}_{m-1}}}:=|\frac{\psi^{Bog}_{\bold{j}_1,\dots,\bold{j}_{m-1}}}{\| \psi^{Bog}_{\bold{j}_1,\dots,\bold{j}_{m-1}} \|}\rangle \langle\frac{\psi^{Bog}_{\bold{j}_1,\dots,\bold{j}_{m-1}} }{\|\psi^{Bog}_{\bold{j}_1,\dots,\bold{j}_{m-1}}\|}|\quad,\quad \overline{\mathscr{P}_{\psi^{Bog}_{\bold{j}_1,\dots,\bold{j}_{m-1}}}}:=\charf_{Q^{(>N-1)}_{\bold{j}_m}\mathcal{F}^N}-\mathscr{P}_{\psi^{Bog}_{\bold{j}_1,\dots,\bold{j}_{m-1}}};
\end{equation}
\item
\begin{equation}
R^{Bog}_{\bold{j}_1,\dots,\bold{j}_m\,;\,i,i}(z^{Bog}_{\bold{j}_1,\dots ,\bold{j}_{m-1}}+z):=Q^{(i, i+1)}_{\bold{j}_m}\frac{1}{Q^{(i, i+1)}_{\bold{j}_m}(H^{Bog}_{\bold{j}_1, \dots,\bold{j}_m}-z^{Bog}_{\bold{j}_1,\dots ,\bold{j}_{m-1}}-z)Q^{(i, i+1)}_{\bold{j}_m}}Q^{(i, i+1)}_{\bold{j}_m}\,;\,\label{Rbog-m}
\end{equation}
\item
\begin{equation}
\Gamma^{Bog\,}_{\bold{j}_1,\dots,\bold{j}_m\,;\,2,2}(z^{Bog}_{\bold{j}_1,\dots ,\bold{j}_{m-1}}+z):=W_{\bold{j}_m\,;\,2,0}\,R^{Bog}_{\bold{j}_1,\dots,\bold{j}_m\,;\,0,0}(z^{Bog}_{\bold{j}_1,\dots ,\bold{j}_{m-1}}+z)W_{\bold{j}_m\,;\,0,2}^*\,,
\end{equation}
%\begin{equation}
%\Gamma^{Bog\,}_{\bold{j}_1,\dots,\bold{j}_m\,;\,3,3}(z_{\bold{j}_1,\dots ,\bold{j}_{m-1}}+z):=W_{\bold{j}_m\,;\,3,1}\,R^{Bog}_{\bold{j}_1,\dots,\bold{j}_m\,;\,i,i}(z_{\bold{j}_1,\dots ,\bold{j}_{m-1}}+z)W_{\bold{j}_m\,;\,1,3}^*
%\end{equation}
for $N-2\geq i\geq 4$ and even
\begin{eqnarray}
& &\Gamma^{Bog\,}_{\bold{j}_1,\dots,\bold{j}_m\,;\,i,i}(z^{Bog}_{\bold{j}_1,\dots,\bold{j}_{m-1}}+z)\label{gamma-i-m}\\
&:=&W_{\bold{j}_m\,;\,i,i-2}\,R^{Bog}_{\bold{j}_1,\dots,\bold{j}_m\,;\,i-2,i-2}(z^{Bog}_{\bold{j}_1,\dots,\bold{j}_{m-1}}+z)\times\\
& &\quad\times \sum_{l_{i-2}=0}^{\infty}\Big[\Gamma^{Bog}_{\bold{j}_1,\dots,\bold{j}_m\,;\,i-2,i-2}(z^{Bog}_{\bold{j}_1,\dots,\bold{j}_{m-1}}+z)R^{Bog}_{\bold{j}_1,\dots,\bold{j}_m\,;\,\,i-2,i-2}(z^{Bog}_{\bold{j}_1,\dots,\bold{j}_{m-1}}+z)\Big]^{l_{i-2}}W^*_{\bold{j}_m\,;\,i-2,i}\,,\quad\quad\quad \label{Gammabog-m}
\end{eqnarray}
\begin{eqnarray}
& &\Gamma^{Bog\,}_{\bold{j}_1,\dots,\bold{j}_m\,;\,N,N}(z^{Bog}_{\bold{j}_1,\dots,\bold{j}_{m-1}}+z) \label{GammaNNBog}\\
&:=&W_{\bold{j}_m}\,R^{Bog}_{\bold{j}_1,\dots,\bold{j}_m\,;\,N-2,N-2}(z^{Bog}_{\bold{j}_1,\dots,\bold{j}_{m-1}}+z)\times \label{sum}\\
& &\quad\times \sum_{l_{N-2}=0}^{\infty}\Big[\Gamma^{Bog}_{\bold{j}_1,\dots,\bold{j}_m\,;\,N-2,N-2}(z^{Bog}_{\bold{j}_1,\dots,\bold{j}_{m-1}}+z)R^{Bog}_{\bold{j}_1,\dots,\bold{j}_m\,;\,\,N-2,N-2}(z^{Bog}_{\bold{j}_1,\dots,\bold{j}_{m-1}}+z)\Big]^{l_{N-2}}W^*_{\bold{j}_m}\,.\quad\quad\quad \nonumber
\end{eqnarray}
\end{itemize}

\noindent
We compute the ground state energy, $z^{Bog}_{\bold{j}_1,\dots,\bold{j}_m}:=z^{Bog}_{\bold{j}_1,\dots,\bold{j}_{m-1}}+z^{(m)}$, of $H^{Bog}_{\bold{j}_1,\dots, \bold{j}_m}$ by solving the equation in $z$:
\begin{equation}\label{fixed-p-eq}
f^{Bog}_{\bold{j}_1,\dots,\bold{j}_m}(z+z^{Bog}_{\bold{j}_1,\dots,\bold{j}_{m-1}})=0.
\end{equation}
% that is we determine $z_*$ such that
%\begin{eqnarray}
%z_*&=& \langle  \frac{\psi^{Bog}_{\bold{j}_1,\dots,\bold{j}_{m-1}}}{\|\psi^{Bog}_{\bold{j}_1,\dots,\bold{j}_{m-1}}\|}\,, \,W_{\bold{j}_m}\,Q_{\bold{j}_m}^{(N-2, N-1)}\times \label{fixed-point-m}\\
%& &\times \sum_{l_{N-2}=0}^{\infty}R^{Bog}_{\bold{j}_1,\dots,\bold{j}_m,;\,N-2,N-2}(z_{\bold{j}_1,\dots,\bold{j}_{m-1}}+z_*)\,\Big[\Gamma^{Bog\,}_{\bold{j}_1,\dots,\bold{j}_m\,,;\,N-2,N-2}(z_{\bold{j}_1,\dots,\bold{j}_{m-1}}+z_*)R^{Bog}_{\bold{j}_1,\dots,\bold{j}_m\,;\,N-2,N-2}(z_{\bold{j}_1,\dots,\bold{j}_{m-1}}+z_*)\Big]^{l_{N-2}}\times \nonumber\\
%& &\quad \times Q_{\bold{j}_m}^{(N-2, N-1)}W_{\bold{j}_m}^* \frac{\psi^{Bog}_{\bold{j}_1,\dots,\bold{j}_{m-1}}}{\|\psi^{Bog}_{\bold{j}_1,\dots,\bold{j}_{m-1}}\|} \rangle\,. \nonumber
%& &-\sum_{l_i=0}^{\infty}Q^{(>i)}W\,Q^{(i)}R^{Bog}_{i,i}(z)\Big[\Gamma^{Bog\,}_{i,i}(z)R^{Bog}_{i,i}(z)\Big]^{l_i}Q^{(i)}W^*Q^{(>i)} \nonumber
%\end{eqnarray}
Hence,  the ground state vector of $H^{Bog}_{\bold{j}_1,\dots, \bold{j}_m}$ is (up to normalization)
\begin{eqnarray}
& &\psi^{Bog}_{\bold{j}_1,\dots,\bold{j}_m} \label{gs-Hm-start}\\
&:=&
%& &+\sum_{j=2}^{N/2}\prod^{2}_{r=j}\Big[-\frac{1}{Q_{\bold{j}_m}^{(N-2r, N-2r+1)}\mathscr{K}_{\bold{j}_1,\dots ,\bold{j}_m}^{Bog\,(N-2r-2)}(z^{Bog}_{\bold{j}_1,\dots ,\bold{j}_m})Q_{\bold{j}_m}^{(N-2r, N-2r+1)}}W^*_{\bold{j}_m;N-2r,N-2r+2}\Big]\psi^{Bog}_{\bold{j}_1, \dots, \bold{j}_{m-1}}\quad\quad\quad\label{gs-Hm-fin} \\
\Big\{\sum_{j=2}^{N/2}\Big[\prod^{2}_{r=j}\Big(-\frac{1}{Q_{\bold{j}_m}^{(N-2r, N-2r+1)}\mathscr{K}_{\bold{j}_1,\dots ,\bold{j}_m}^{Bog\,(N-2r-2)}(z^{Bog}_{\bold{j}_1,\dots ,\bold{j}_m})Q_{\bold{j}_m}^{(N-2r, N-2r+1)}}W^*_{\bold{j}_m;N-2r,N-2r+2}\Big)\Big]+\charf\Big\}\times\quad\quad\quad \label{gs-Hm-fin}\\
& &\quad\quad\quad\times \Big[Q^{(>N-1)}_{\bold{j}_m}-\frac{1}{Q^{(N-2,N-1)}_{\bold{j}_m}\mathscr{K}^{Bog\,(N-4)}_{\bold{j}_1,\dots,\bold{j}_m}(z^{Bog}_{\bold{j}_1,\dots,\bold{j}_m})Q^{(N-2,N-1)}_{\bold{j}_m}}Q^{(N-2, N-1)}_{\bold{j}_m}W^*_{\bold{j}_m}\Big]\times \nonumber \\
& &\quad\quad\quad \times \Big[\mathscr{P}_{\psi^{Bog}_{\bold{j}_1\dots,\bold{j}_{m-1}}}-\frac{1}{\overline{\mathscr{P}_{\psi^{Bog}_{\bold{j}_1,\dots,\bold{j}_{m-1}}}}\mathscr{K}^{Bog\,(N-2)}_{\bold{j}_1,\dots,\bold{j}_m}(z^{Bog}_{\bold{j}_1,\dots,\bold{j}_m})\overline{\mathscr{P}_{\psi^{Bog}_{\bold{j}_1, \dots, \bold{j}_{m-1}}}}}
\overline{\mathscr{P}_{\psi^{Bog}_{\bold{j}_1\dots,\bold{j}_{m-1}}}}\mathscr{K}^{Bog\,(N-2)}_{\bold{j}_1,\dots,\bold{j}_m}(z^{Bog}_{\bold{j}_1,\dots,\bold{j}_m})\Big]\psi^{Bog}_{\bold{j}_1, \dots, \bold{j}_{m-1}}\,\nonumber \\
&=:&T_{m}\,\psi^{Bog}_{\bold{j}_1, \dots, \bold{j}_{m-1}}
\end{eqnarray}
where $\mathscr{K}^{Bog\,(-2)}_{\bold{j}_1,\dots,\bold{j}_m}(z^{Bog}_{\bold{j}_1,\dots,\bold{j}_m}):=H^{Bog}_{\bold{j}_1,\dots,\bold{j}_m}-z^{Bog}_{\bold{j}_1,\dots, \bold{j}_m}$. Thus, we have derived the formula
\begin{equation}\label{construction-ground}
\psi^{Bog}_{\bold{j}_1, \dots, \bold{j}_{M}}=T_{M}\dots T_{1}\eta\,.
\end{equation}
\begin{remark}\label{vectornorm}
We point out that by construction
\begin{equation}
\|\psi^{Bog}_{\bold{j}_1, \dots, \bold{j}_{M}}\|\geq \|\psi^{Bog}_{\bold{j}_1, \dots, \bold{j}_{M-1}}\|\geq \dots \geq \|\eta\|=1\,.
\end{equation}
\end{remark}

In Corollary \ref{expansion}, we show how to approximate   the vector $\psi^{Bog}_{\bold{j}_1, \dots, \bold{j}_{M}}$ constructed in (\ref{construction-ground}) up to any arbitrarily small error $\zeta$ (for $N$ sufficiently large) with a vector $(\psi^{Bog}_{\bold{j}_1, \dots, \bold{j}_{M}})_{\zeta}$  corresponding to a $\zeta-$dependent finite sum of finite products of the interaction terms $W^*_{\bold{j}_l}+W_{\bold{j}_l}$, and of the resolvents $\frac{1}{\hat{H}^0_{\bold{j}_l}-E^{Bog}_{\bold{j}_l}}$ (see (\ref{H0j})), $1\leq l \leq M$, applied to $\eta$.
\section{Rigorous construction of the Feshbach-Schur Hamiltonians $\mathscr{K}_{\bold{j}_1,\dots,\bold{j}_m}^{Bog\,(i)}(z)$}\label{rigorousHbog}
\setcounter{equation}{0}
The derivation in Section \ref{informal} can be made rigorous  if we show that  at the $m-th$ step (for $\epsilon_{\bold{j}_m}$ sufficiently small and $N$ sufficiently large):
\begin{itemize}
\item[\bf{$\mathfrak{U} 1)$}]  The Feshbach-Schur Hamiltonian $\mathscr{K}_{\bold{j}_1,\dots,\bold{j}_m}^{Bog\,(i)}(z+z^{Bog}_{\bold{j}_1,\dots,\bold{j}_{m-1}})$, $0\leq i \leq N-2$,  is well defined provided
 \begin{equation}
z\leq E^{Bog}_{\bold{j}_m}+ (\delta-1)\phi_{\bold{j}_m}\sqrt{\epsilon_{\bold{j}_m}^2+2\epsilon_{\bold{j}_m}}\,(<0)
\end{equation} with $\delta=1+\sqrt{\epsilon_{\bold{j}_m}}$.
\item[$\mathfrak{U} 2)$] The expression $f^{Bog}_{\bold{j}_1,\dots,\bold{j}_m}(z+z^{Bog}_{\bold{j}_1,\dots,\bold{j}_{m-1}})$ is well defined and the
final Feshbach-Schur Hamiltonian is 
\begin{equation}
\mathscr{K}_{\bold{j}_1,\dots,\bold{j}_m}^{Bog\,(N)}(z+z^{Bog}_{\bold{j}_1,\dots,\bold{j}_{m-1}})=f^{Bog}_{\bold{j}_1,\dots,\bold{j}_m}(z+z^{Bog}_{\bold{j}_1,\dots,\bold{j}_{m-1}})\mathscr{P}_{\psi^{Bog}_{\bold{j}_1,\dots,\bold{j}_{m-1}}}\,
\end{equation} 
provided
 \begin{equation}\label{range}
 z\leq  z_{m}+\gamma \Delta_{m-1}-\mathcal{O}(\frac{1}{(\ln N)^{\frac{1}{2}}})<E^{Bog}_{\bold{j}_m}+ \sqrt{\epsilon_{\bold{j}_m}}\phi_{\bold{j}_m}\sqrt{\epsilon_{\bold{j}_m}^2+2\epsilon_{\bold{j}_m}}<0\quad,\quad \gamma= \frac{1}{2}\,,
 \end{equation}
 where  $z_{m}$ is the ground state energy of $H^{Bog}_{\bold{j}_m}$ and $\Delta_{m-1}>0$ is a lower bound to the spectral gap above the ground state energy of $H^{Bog}_{\bold{j}_1,\dots,\bold{j}_{m-1}}$. Both $\Delta_{m-1}$ and $\mathcal{O}(\frac{1}{(\ln N)^{\frac{1}{2}}})$ are specified later in Theorem \ref{induction-many-modes}. 
 
\noindent 
Furthermore,  there is a (unique) fixed point, $z^{(m)}$, in the given range of $z$  (see (\ref{range})) that solves $f^{Bog}_{\bold{j}_1,\dots,\bold{j}_m}(z^{(m)}+z^{Bog}_{\bold{j}_1,\dots,\bold{j}_{m-1}})=0$. 
% and which is close to $E^{Bog}_{\bold{j}_m}$ and $z_m$. %Indeed, in principle the ground state energy of $H^{Bog}_{\bold{j}_1,\dots, \bold{j}_m}$ might be larger than
%\begin{equation}
%{\color{red}....z_{\bold{j}_1,\dots,\bold{j}_{m-1}}+E^{Bog}_{\bold{j}_m}+ (\delta-1)\phi_{\bold{j}_m}\sqrt{\epsilon_{\bold{j}_m}^2+2\epsilon_{\bold{j}_m}}...}\,.
%\end{equation}
 \end{itemize}
 
Concerning requirement $\mathfrak{U} 1)$ we recall that
\begin{equation}
 H^{Bog}_{\bold{j}_1, \dots,\bold{j}_m}-z^{Bog}_{\bold{j}_1,\dots ,\bold{j}_{m-1}}-z=\sum_{\bold{j}\in\mathbb{Z}^d\setminus \{\pm\bold{j}_{1},\dots ,\pm\bold{j}_{m} \}} k^2_{\bold{j}}a_{\bold{j}}^{*}a_{\bold{j}}+\sum_{l=1}^{m-1}\hat{H}^{Bog}_{\bold{j}_l}-z^{Bog}_{\bold{j}_1,\dots ,\bold{j}_{m-1}}+\hat{H}^{Bog}_{\bold{j}_m}-z\,.
\end{equation}
If we assume
\begin{equation}\label{checkBog-gs}
\text{infspec}\,[\sum_{l=1}^{m-1}\hat{H}^{Bog}_{\bold{j}_l}]-z^{Bog}_{\bold{j}_1,\dots ,\bold{j}_{m-1}}\geq -\mathcal{O}(\frac{1}{(\ln N)^{\frac{1}{8}}})\,
\end{equation}
we can reproduce a result analogous to \emph{\underline{Lemma 3.4} of \cite{Pi1}} (reported in Section \ref{Feshbach}) for the Hamiltonian $H^{Bog}_{\bold{j}_1, \dots,\bold{j}_m}$; see Corollary \ref{main-lemma-H}. Indeed, thanks to this input, the operator norm estimate (\ref{estimate-main-lemma-H}) in Corollary \ref{main-lemma-H} can be derived as if the modes $\pm\bold{j}_1,\dots,\pm\bold{j}_{m-1}$ were absent. Consequently, the counterpart of \emph{\underline{Theorem 3.1} of \cite{Pi1}} (see Section \ref{Feshbach})  can be proven for $\mathscr{K}_{\bold{j}_1,\dots,\bold{j}_m}^{Bog\,(i)}(z+z^{Bog}_{\bold{j}_1,\dots,\bold{j}_{m-1}})$:
%{\color{red}Furthermore, the same argument of Section \ref{last-projection} applies with $\eta$ replaced by $\psi^{Bog}_{\bold{j}_1,\bold{j}_2,\dots,\bold{j}_{m-1}}$.} 
%Following the procedure of \cite{Pi1},  for the last implementation of the Feshbach map, i.e., for the construction of  $\mathscr{K}_{\bold{j}_1,\bold{j}_2,\dots,\bold{j}_m}^{Bog\,(N)}(z+z_{\bold{j}_1,\dots,\bold{j}_{m-1}})$,  we have first to solve the fixed point equation  in (\ref{fixed-p-eq}). 
\begin{thm}\label{Feshbach-Hbog}
Assume condition a) of Corollary \ref{main-lemma-H}.  Then, for \begin{equation}
z\leq E^{Bog}_{\bold{j}_m}+ (\delta-1)\phi_{\bold{j}_m}\sqrt{\epsilon_{\bold{j}_m}^2+2\epsilon_{\bold{j}_m}}\,(<0)
\end{equation} with $\delta=  1+\sqrt{\epsilon_{\bold{j}_m}}$,  $\epsilon_{\bold{j}_m}$ sufficiently small and $N$ sufficiently large, the
% and \begin{equation}
%z\leq E^{Bog}_{\bold{j}_m}+ (\delta-1)\phi_{\bold{j}_m}\sqrt{\epsilon_{\bold{j}_m}^2+2\epsilon_{\bold{j}_m}}
%\end{equation} with $\delta\leq  1+\sqrt{\epsilon_{\bold{j}_m}}$ and $\epsilon_{\bold{j}_m}$ sufficiently small. 
operators $\mathscr{K}_{\bold{j}_1,\dots,\bold{j}_m}^{\, Bog (i)}(z+z^{Bog}_{\bold{j}_1,\dots,\bold{j}_{m-1}})$, $0\leq i\leq N-2$ and even,  are well defined \footnote{$\mathscr{K}^{Bog\,(i)}_{\bold{j}_1,\dots,\bold{j}_m}(z)$ is  self-adjoint on the domain of the operator $Q^{(>i+1)}_{\bold{j}_m}(H^{Bog}_{\bold{j}_1,\dots,\bold{j}_m}-z)Q^{(>i+1)}_{\bold{j}_m}$.}. For $i=2,4,6,\dots,N-2$  they correspond to
\begin{eqnarray}\label{KappaBog-i}
& &\mathscr{K}_{\bold{j}_1,\dots,\bold{j}_m}^{Bog\,(i)}(z+z^{Bog}_{\bold{j}_1,\dots,\bold{j}_{m-1}})\\
&=&Q^{(>i+1)}_{\bold{j}_m}(H^{Bog}_{\bold{j}_1,\dots,\bold{j}_m}-z-z^{Bog}_{\bold{j}_1,\dots,\bold{j}_{m-1}})Q^{(>i+1)}_{\bold{j}_m}\\
%& &-\sum_{l_{i-1}=0}^{\infty}Q^{(>i)}_{\bold{j}_*}W_{\bold{j}_*}\,Q^{(i-1)}_{\bold{j}_*}R^{Bog}_{\bold{j}_*\,;\,i-1,i-1}(z)\,\Big[\Gamma^{Bog\,}_{\bold{j}_*\,;\,i-1,i-1}(z)R^{Bog}_{\bold{j}_*\,;\,i-1,i-1}(z)\Big]^{l_{i-1}}\,Q^{(i-1)}_{\bold{j}_*}W^*_{\bold{j}_*}Q^{(>i)}_{\bold{j}_*} \nonumber\\
& &-Q^{(>i+1)}_{\bold{j}_m}W_{\bold{j}_m}\,R^{Bog}_{\bold{j}_1,\dots,\bold{j}_m\,;\,i,i}(z+z^{Bog}_{\bold{j}_1,\dots,\bold{j}_{m-1}})\sum_{l_i=0}^{\infty}\Big[\Gamma^{Bog\,}_{\bold{j}_1,\dots,\bold{j}_m\,;\,i,i}(z+z^{Bog}_{\bold{j}_1,\dots,\bold{j}_{m-1}})R^{Bog}_{\bold{j}_1,\dots,\bold{j}_m\,;\,i,i}(z+z^{Bog}_{\bold{j}_1,\dots,\bold{j}_{m-1}})\Big]^{l_i}W^*_{\bold{j}_m}Q^{(>i+1)}_{\bold{j}_m} \nonumber
\end{eqnarray}
where $R^{Bog}_{\bold{j}_1,\dots,\bold{j}_m\,;\,i,i}(z+z^{Bog}_{\bold{j}_1,\dots,\bold{j}_{m-1}})$ and $\Gamma^{Bog\,}_{\bold{j}_1,\dots,\bold{j}_m\,;\,i,i}(z+z^{Bog}_{\bold{j}_1,\dots,\bold{j}_{m-1}})$ are defined in (\ref{Rbog-m})-(\ref{Gammabog-m}). 

\noindent
The following estimates hold true for $2\leq i \leq N-2$ and even:
\begin{equation}\label{est-gammatilde}
\|\check{\Gamma}^{Bog\,}_{\bold{j}_1,\dots,\bold{j}_m\,;\,i,i}(z+z^{Bog}_{\bold{j}_1,\dots,\bold{j}_{m-1}})\|\leq \frac{1}{X_{i}}|_{\epsilon\equiv \epsilon_{\bold{j}_*}}\label{Gamma-ineq}
\end{equation}
where 
\begin{equation}
\check{\Gamma}^{Bog\,}_{\bold{j}_1,\dots,\bold{j}_m\,;\,i,i}(z+z^{Bog}_{\bold{j}_1,\dots,\bold{j}_{m-1}}):=\sum_{l_i=0}^{\infty}\Big[(R^{Bog}_{\bold{j}_1,\dots,\bold{j}_m\,;\,i,i}(z+z^{Bog}_{\bold{j}_1,\dots,\bold{j}_{m-1}}))^{\frac{1}{2}}\Gamma^{Bog\,}_{\bold{j}_1,\dots,\bold{j}_m\,;\,i,i}(z+z^{Bog}_{\bold{j}_1,\dots,\bold{j}_{m-1}})(R^{Bog}_{\bold{j}_1,\dots,\bold{j}_m\,;\,i,i}(z+z^{Bog}_{\bold{j}_1,\dots,\bold{j}_{m-1}}))^{\frac{1}{2}}\Big]^{l_i}
\end{equation}
and $X_i$ is defined in \underline{Lemma 3.6} of \cite{Pi1} (that is reported in Section \ref{Feshbach}) and fulfills the bound
\begin{equation}\label{x-ineq}
X_{2j}\geq\frac{1}{2}\Big[1+\sqrt{\eta a_{\epsilon}}-\frac{b_{\epsilon}/\sqrt{\eta a_{\epsilon}}}{N-2j-\epsilon^{\Theta}}\Big]\,(>0)\,
\end{equation}
with $\eta=1-\sqrt{\epsilon}$,  $\Theta:=\min\{2(\nu-\frac{11}{8})\,;\,\frac{1}{4}\}$, and $a_{\epsilon}$, $b_{\epsilon}$ are those defined in Corollary \ref{main-lemma-H}.

\end{thm}

\noindent
\emph{Proof}

The proof is identical to \emph{\underline{Theorem 3.1} of \cite{Pi1}} using the estimates provided in Corollary \ref{main-lemma-H}.
\qed
\\

Requirement $\mathfrak{U} 2)$  has been proven to hold for the first step (i.e., for the Hamiltonian  $H^{Bog}_{\bold{j}_1}$) in \emph{Corollary 4.6} of \cite{Pi1} (see Section \ref{Feshbach})  with $\Delta_0:=\min \{(k_{\bold{j}})^2\,;\,\bold{j}\in \mathbb{Z}^d \setminus \{\bold{0}\}\}$. The proof for the successive steps (i.e., for the Hamiltonians  $ H^{Bog}_{\bold{j}_1, \dots,\bold{j}_m}$) is not straightforward,  rather it requires an inductive procedure implemented in Theorem \ref{induction-many-modes}. To better understand the strategy of Theorem \ref{induction-many-modes} we must explain the new difficulties in some detail. 

In the first step (i.e., for $H^{Bog}_{\bold{j}_1}$) the function 
\begin{equation}
f^{Bog}_{\bold{j}_1}(z):=-z-\langle \eta\,,\,W_{\bold{j}_*}\,R^{Bog}_{\bold{j}_1\,;\,N-2,N-2}(z)\sum_{l_{N-2}=0}^{\infty}[\Gamma^{Bog}_{\bold{j}_1,;\,N-2,N-2}(z) R^{Bog}_{\bold{j}_1\,;\,N-2,N-2}(z)]^{l_{N-2}}\,W^*_{\bold{j}_1}\eta\rangle \quad\quad
\end{equation}
is well defined as a result of the construction of $\mathscr{K}_{\bold{j}_1}^{Bog\,(N-2)}(z)$. Next,  according to the scheme used in \cite{Pi1}, starting from the existence of the (unique) solution $z^{(1)}\equiv z_1<E^{Bog}_{\bold{j}_1}+ \sqrt{\epsilon_{\bold{j}_1}}\phi_{\bold{j}_1}\sqrt{\epsilon_{\bold{j}_1}^2+2\epsilon_{\bold{j}_1}}$ of $f_{\bold{j}_1}(z)=0$ we show the invertibility on $\overline{\mathscr{P}_{\eta}}\mathcal{F}^N$ of the operator
\begin{equation}
\overline{\mathscr{P}_{\eta}}\mathscr{K}_{\bold{j}_1}^{Bog\,(N-2)}(z)\overline{\mathscr{P}_{\eta}}
\end{equation}
for $z\leq \min\,\Big\{z_1+ \frac{\Delta_0}{2}\,;\,E^{Bog}_{\bold{j}_1}+ \sqrt{\epsilon_{\bold{j}_1}}\phi_{\bold{j}_1}\sqrt{\epsilon_{\bold{j}_1}^2+2\epsilon_{\bold{j}_1}}\Big\}$.
On the contrary, starting from $m=2$ the definition of $f^{Bog}_{\bold{j}_1,\dots,\bold{j}_m}(z+z_{\bold{j}_1,\dots,\bold{j}_{m-1}})$ (see (\ref{def-f})) requires the existence of 
\begin{equation}
\overline{\mathscr{P}_{\psi^{Bog}_{\bold{j}_1,\dots,\bold{j}_{m-1}}}}\frac{1}{\overline{\mathscr{P}_{\psi^{Bog}_{\bold{j}_1,\dots,\bold{j}_{m-1}}}}\mathscr{K}_{\bold{j}_1,\dots,\bold{j}_m}^{Bog\,(N-2)}(z+z^{Bog}_{\bold{j}_1,\dots,\bold{j}_{m-1}})\overline{\mathscr{P}_{\psi^{Bog}_{\bold{j}_1,\dots,\bold{j}_{m-1}}}}}\overline{\mathscr{P}_{\psi^{Bog}_{\bold{j}_1,\dots,\bold{j}_{m-1}}}}\,.\label{perp-ham}
\end{equation}
The latter is proven  in Lemma \ref{invertibility-new} for $z$ in the interval specified in (\ref{range}). Then, we can define the 
 function $f^{Bog}_{\bold{j}_1,\dots,\bold{j}_m}(z+z^{Bog}_{\bold{j}_1,\dots,\bold{j}_{m-1}})$.
 % in the same interval of the spectral variable $z$. 
 \\
 
In order to determine the solution, $z^{(m)}$, to the equation in $z$}
\begin{equation}f^{Bog}_{\bold{j}_1,\dots,\bold{j}_m}(z+z^{Bog}_{\bold{j}_1,\dots,\bold{j}_{m-1}})=0\,, \label{fixed-point-m}
\end{equation} we exploit that the fixed point problem at the $m-th$ step boils down to the fixed point problem for a three-modes system if in the formula that defines  $f^{Bog}_{\bold{j}_1,\dots,\bold{j}_m}(z+z^{Bog}_{\bold{j}_1,\dots,\bold{j}_{m-1}})$  the vector  $\psi^{Bog}_{\bold{j}_1,\dots,\bold{j}_{m-1}}$ is replaced  with $\eta$ and
\begin{equation}\label{def-tilde-0}
R^{Bog}_{\bold{j}_1,\dots,\bold{j}_m;\,N-2,N-2}(z^{Bog}_{\bold{j}_1,\dots,\bold{j}_{m-1}}+z)\,\sum_{l_{N-2}=0}^{\infty}\Big[\Gamma^{Bog\,}_{\bold{j}_1,\dots,\bold{j}_m\,;\,N-2,N-2}(z^{Bog}_{\bold{j}_1,\dots,\bold{j}_{m-1}}+z)R^{Bog}_{\bold{j}_1,\dots,\bold{j}_m\,;\,N-2,N-2}(z^{Bog}_{\bold{j}_1,\dots,\bold{j}_{m-1}}+z)\Big]^{l_{N-2}}
\end{equation}  
is replaced with
\begin{equation}
\tilde{R}^{Bog}_{\bold{j}_1,\dots,\bold{j}_m;\,N-2,N-2}(z)\,\sum_{l_{N-2}=0}^{\infty}\Big[\tilde{\Gamma}^{Bog\,}_{\bold{j}_1,\dots,\bold{j}_m\,;\,N-2,N-2}(z)\tilde{R}^{Bog}_{\bold{j}_1,\dots,\bold{j}_m\,;\,N-2,N-2}(z)\Big]^{l_{N-2}}\label{def-tilde}
\end{equation}  
where $\,\tilde{}\,$ means that $$\sum_{l=1}^{m-1}\hat{H}^{Bog}_{\bold{j}_l}$$ is omitted in the resolvents $R^{Bog}_{\bold{j}_1,\dots,\bold{j}_m;\,N-2,N-2}(z)$ and in all the other resolvents entering the definition of the operators $\Gamma^{Bog\,}_{\bold{j}_1,\dots,\bold{j}_m\,;\,i,i}(z)$. 
In fact, with the help of Lemma \ref{main-relations-new}
we show that  
$$f^{Bog}_{\bold{j}_1,\dots,\bold{j}_m}(z+z^{Bog}_{\bold{j}_1,\dots,\bold{j}_{m-1}})=f^{Bog}_{\bold{j}_m}(z)+o_{N\to \infty}(1)\,.$$
%where the corrections vanish as $N\to \infty$.
%To this purpose, we need the results of the following lemma.

The result concerning $\mathfrak{U} 2)$ is finally proven in Theorem \ref{induction-many-modes}. To this end,  various preliminary ingredients are needed. Definition \ref{def-sums} and Proposition \ref{lemma-expansion-proof-0} deal with the re-expansion of $\Gamma^{Bog\,}_{\bold{j}_1,\dots,\bold{j}_m\,;\,N-2,N-2}(z^{Bog}_{\bold{j}_1,\dots,\bold{j}_{m-1}}+z)$ and are generalizations of structures already encountered in \cite{Pi1}. This re-expansion is crucial to implement the mechanisms behind the proofs of Lemma \ref{main-relations-new}
and Lemma \ref{invertibility-new}, that are explained in the \emph{{\bf{Outline of the proof}}} provided in this section for both.
%concern the definition of the fixed point equation and represent the main tools to estimate to which extent the system with many couples, $\{\pm \bold{j}_1, \dots, \pm \bold{j}_m\}$,  of interacting modes behaves effectively as a collection of independent couples of modes in the mean field limit. }
\\

To streamline formulae, in Definition \ref{def-sums}, Proposition \ref{lemma-expansion-proof-0}, and Remarks \ref{prod-control} and \ref{estimation-proc} below, we write $W_{j,j-2}$, $W^*_{j-2,j}$, $R^{Bog}_{j-2,j-2}(w)$, and $\Gamma^{Bog\,}_{j,j}(w)$ instead of $W_{\bold{j}_m\,;\,j,j-2}$, $W^*_{\bold{j}_m\,;\,j-2,j}$, $R^{Bog}_{\bold{j}_1,\dots,\bold{j}_m\,;\,j-2,j-2}(z^{Bog}_{\bold{j}_1,\dots,\bold{j}_{m-1}}+z)$, and $\Gamma^{Bog\,}_{\bold{j}_1,\dots,\bold{j}_m\,;\,j,j}(z^{Bog}_{\bold{j}_1,\dots,\bold{j}_{m-1}}+z)$, respectively.

\begin{definition} \label{def-sums}

\noindent
Let $h\in \mathbb{N}$, $h\geq 2$, and 
\begin{equation}
w\leq z^{Bog}_{\bold{j}_1,\dots,\bold{j}_{m-1}}+E^{Bog}_{\bold{j}_m}+ (\delta-1)\phi_{\bold{j}_m}\sqrt{\epsilon_{\bold{j}_m}^2+2\epsilon_{\bold{j}_m}}\,.
\end{equation}
with $\delta\leq 1+\sqrt{\epsilon_{\bold{j}_m}}$. Assume condition a) of Corollary \ref{main-lemma-H},  and let  $\epsilon_{\bold{j}_m}$ be sufficiently small and $N$ sufficiently large. We define:
\begin{enumerate}
\item For $N-2\geq  j \geq 4$ and even
\begin{equation}\label{initial-gamma-1}
[\Gamma^{Bog\,}_{j,j}(w)]_{(j-2, h_-)}:=[\Gamma^{Bog\,}_{j,j}(w)]^{(0)}_{(j-2, h_-)}+[\Gamma^{Bog\,}_{j,j}(w)]^{(>0)}_{(j-2, h_-)}
\end{equation}
where
\begin{eqnarray}\label{def-0}
& &[\Gamma^{Bog\,}_{j,j}(w)]^{(0)}_{(j-2, h_-)}:=W_{j,j-2}R^{Bog}_{j-2,j-2}(w) W^*_{j-2,j}\quad \text{for}\,\,j\geq 2
\end{eqnarray}
and
\begin{eqnarray}
& &[\Gamma^{Bog\,}_{j,j}(w)]^{(>0)}_{(j-2, h_-)}\label{initial-gamma-2}\\
&:=&W_{j,j-2}\,(R^{Bog}_{j-2,j-2}(w))^{\frac{1}{2}}\times \\
& &\quad \times  \sum_{l_{j-2}=1}^{h-1}\Big[(R^{Bog}_{j-2,j-2}(w))^{\frac{1}{2}}W_{j-2,j-4}\,R_{j-4,j-4}^{Bog}(w)W_{j-4,j-2}^*(R^{Bog}_{j-2,j-2}(w))^{\frac{1}{2}} \Big]^{l_{j-2}}(R^{Bog}_{j-2,j-2}(w))^{\frac{1}{2}} W^*_{j-2,j}\nonumber \\
&=&W_{j,j-2}\,(R^{Bog}_{j-2,j-2}(w))^{\frac{1}{2}}\times \label{initial-gamma-3}\\
& &\quad \times  \sum_{l_{j-2}=1}^{h-1}\Big[(R^{Bog}_{j-2,j-2}(w))^{\frac{1}{2}}[\Gamma^{Bog\,}_{j-2,j-2}(w)]^{(0)}_{(j-4, h_-)}(R^{Bog}_{j-2,j-2}(w))^{\frac{1}{2}} \Big]^{l_{j-2}}(R^{Bog}_{j-2,j-2}(w))^{\frac{1}{2}} W^*_{j-2,j}\,,\nonumber 
\end{eqnarray}
for $N-2\geq  j \geq 4$ and even
\begin{eqnarray}
[\Gamma^{Bog\,}_{j,j}(w)]_{(j-2, h_+)}&:=&W_{j,j-2}\,(R^{Bog}_{j-2,j-2}(w))^{\frac{1}{2}}\times \\
& &\quad \times  \sum_{l_{j-2}=h}^{\infty}\Big[(R^{Bog}_{j-2,j-2}(w))^{\frac{1}{2}}\Gamma^{Bog\,}_{j-2,j-2}(w)(R^{Bog}_{j-2,j-2}(w))^{\frac{1}{2}} \Big]^{l_{j-2}}\times \nonumber \\
& &\quad\quad\quad \times (R^{Bog}_{j-2,j-2}(w))^{\frac{1}{2}} W^*_{j-2,j}\,.\nonumber
\end{eqnarray}
\item
For $N-2\geq  j \geq 6$,  $2\leq l \leq j-4$ (both even numbers)
\begin{eqnarray}
& &[\Gamma^{Bog\,}_{j,j}(w)]_{(l,h_-; l+2,h_-;\dots;j-4,h_-;j-2,h_-)}\label{def-gamma-relation}\\
&:= &W_{j,j-2}\,(R^{Bog}_{j-2,j-2}(w))^{\frac{1}{2}} \check{\sum}_{l_{j-2}=1}^{h-1}\Big[(R^{Bog}_{j-2,j-2}(w))^{\frac{1}{2}}[\Gamma^{Bog\,}_{j-2,j-2}(w)]_{(l,h_-;l+2,h_-;\dots;j-4,h_-)}(R^{Bog}_{j-2,j-2}(w))^{\frac{1}{2}} \Big]^{l_{j-2}}\times \nonumber\\
& &\quad\quad\quad\quad \times(R^{Bog}_{j-2,j-2}(w))^{\frac{1}{2}} W^*_{j-2,j}\,. \label{collection}
\end{eqnarray}
Here, the  symbol $\check{\sum}^{h-1}_{l_{j-2}=1}$ stands for a sum of terms resulting from operations $\mathcal{A}1$ and $\mathcal{A}2$ below:
\begin{itemize}
\item[$\mathcal{A}1)$]
At fixed $1\leq l_{j-2} \leq h-1$ summing all the products
\begin{equation}
\Big[(R^{Bog}_{j-2,j-2}(w))^{\frac{1}{2}}\mathcal{X}(R^{Bog}_{j-2,j-2}(w))^{\frac{1}{2}} \Big]^{l_{j-2}}
\end{equation}
that are obtained by replacing  $\mathcal{X}$ for each factor with the operators (defined by iteration) of the type $[\Gamma^{Bog\,}_{j-2,j-2}(w)]_{(s,h_-;s+2,h_-;\dots;j-4,h_-)}$ with $l\leq s\leq j-4$ and even, with the constraint that  if $l\leq j-6$ then $\mathcal{X}$ is replaced with $[\Gamma^{Bog\,}_{j-2,j-2}(w)]_{(l,h_-;4,h_-;\dots;j-4,h_-)}$ in one factor at least, whereas if $l= j-4$ then  $\mathcal{X}$ is replaced with $[\Gamma^{Bog\,}_{j-2,j-2}(w)]^{(>0)}_{(j-4,h_-)}$ in one factor at least;
\item[$\mathcal{A}2)$]
Summing from $l_{j-2}=1$ up to $l_{j-2}=h-1$.
\end{itemize}
\item
For $N-2\geq  j \geq 6$,  $2\leq l \leq j-4$ and even
\begin{eqnarray}
& &[\Gamma^{Bog\,}_{j,j}(w)]_{(l,h_+; l+2,h_-;\dots;j-4,h_-;j-2,h_-)}\\
&:= &W_{j,j-2}\,(R^{Bog}_{j-2,j-2}(w))^{\frac{1}{2}} \check{\sum}_{l_{j-2}=1}^{h-1}\Big[(R^{Bog}_{j-2,j-2}(w))^{\frac{1}{2}}[\Gamma^{Bog\,}_{j-2,j-2}(w)]_{(l,h_+;l+2,h_-;\dots;j-4,h_-)}\times \nonumber\\
& &\quad\quad \quad\quad\quad\quad\quad\quad\quad\quad\quad\quad\times (R^{Bog}_{j-2,j-2}(w))^{\frac{1}{2}} \Big]^{l_{2}}(R^{Bog}_{j-2,j-2}(w))^{\frac{1}{2}} W^*_{j-2,j}\,.\label{collection-bis}
\end{eqnarray}
Here, the  symbol $\check{\sum}^{h-1}_{l_{j-2}=1}$ stands for a sum of terms resulting from operations $\mathcal{B}1$ and $\mathcal{B}2$ below:
\begin{itemize}
\item[$\mathcal{B}1)$]
At fixed $1\leq l_{j-2} \leq h-1$, summing all the products
\begin{equation}
\Big[(R^{Bog}_{j-2,j-2}(w))^{\frac{1}{2}}\mathcal{X}(R^{Bog}_{j-2,j-2}(w))^{\frac{1}{2}} \Big]^{l_{j-2}}
\end{equation}
that are obtained by replacing  $\mathcal{X}$ for each factor with the operators (iteratively defined) of the type $[\Gamma^{Bog\,}_{j-2,j-2}(z)]_{(s,h_+;s+2,h_-;\dots;j-4,h_-)}$ and $[\Gamma^{Bog\,}_{j-2,j-2}(z)]_{(s',h_-;s'+2,h_-;\dots;j-4,h_-)}$  with $l\leq s\leq j-4$ and $2\leq s'\leq j-4$ where $s$ and $s'$ are even, and with the constraint   that $\mathcal{X}$ is replaced with $[\Gamma^{Bog\,}_{j-2,j-2}(z)]_{(l,h_+;l+2,h_-;\dots;j-4,h_-)}$ in one factor at least.
%that are obtained by replacing  $\mathcal{X}$ for each factor with the operators (defined by iteration) of the type  $[\Gamma^{Bog\,}_{j-2,j-2}(w)]_{(l,h_+;l+2,h_-;\dots;j-4,h_-)}$ and $[\Gamma^{Bog\,}_{j-2,j-2}(w)]_{(s,h_-;s+2,h_-;\dots;j-4,h_-)}$  with $l<s\leq j-4$ and even, with the constraint   that $\mathcal{X}$ is replaced with $[\Gamma^{Bog\,}_{j-2,j-2}(z)]_{(l,h_+;l+2,h_-;\dots;j-4,h_-)}$ in one factor at least.
\item[$\mathcal{B}2)$]
Summing from $l_{j-2}=1$ up to $h-1$.
\end{itemize}
\end{enumerate}
The definitions of above can be adapted in an obvious manner to the case $h=\infty$, in particular the terms $[\Gamma^{Bog\,}_{j,j}(w)]_{(l,h_+; l+2,h_-;\dots;j-4,h_-;j-2,h_-)}$ are absent.
\end{definition}
\begin{prop}\label{lemma-expansion-proof-0} Assume condition a) of Corollary \ref{main-lemma-H}, and let $\epsilon_{\bold{j}_m}\equiv \epsilon$ be sufficiently small and $N$ sufficiently large.
For any fixed $2\leq h \in \mathbb{N}$ and for $N-2\geq i\geq 4$ and even,  the splitting
\begin{eqnarray}
\Gamma^{Bog\,}_{i,i}(w)&=&\sum_{l=2,\,l\, even}^{i-2}[\Gamma^{Bog\,}_{i,i}(w)]_{(l,h_-;l+2,h_-;\dots ; i-2,h_-)}+\sum_{l=2\,,\,l\, even}^{i-2}[\Gamma^{Bog\,}_{i,i}(w)]_{(l,h_+; l+2,h_-;\dots; i-2,h_-)}\quad\quad\label{decomposition}
\end{eqnarray}
holds true for $w\leq z^{Bog}_{\bold{j}_1,\dots,\bold{j}_{m-1}}+E^{Bog}_{\bold{j}_m}+ (\delta -1)\phi_{\bold{j}_m}\sqrt{\epsilon_{\bold{j}_m}^2+2\epsilon_{\bold{j}_m}}$ and $\delta\leq 1+\sqrt{\epsilon_{\bold{j}_m}}$. Moreover, for $2\leq l \leq i-2$ and even,  the estimates
\begin{eqnarray}\label{gamma-exp-1-0}
& &\Big\|(R^{Bog}_{i,i}(w))^{\frac{1}{2}}[\Gamma^{Bog\,}_{i,i}(w)]_{(l,h_-;l+2, h_-;\, \dots \,; i-2,h_-)}(R^{Bog}_{i,i}(w))^{\frac{1}{2}}\Big\|\\
& &\leq  \prod_{f=l+2\,,\, f-l\,\text{even}}^{i}\frac{K_{f,\epsilon}}{(1-Z_{f-2,\epsilon})^2}\nonumber
%\Big(\frac{2}{3}+\mathcal{O}(\sqrt{\epsilon})\Big)^{\frac{i-l}{2}}\prod_{f=l+2\,,\, f-l\,\text{even}}^{i-2}(1+a_{\epsilon}-\frac{2b_{\epsilon}}{N-f-1}-\frac{1-c_{\epsilon}}{(N-f-1)^2})^{-1}\nonumber
\end{eqnarray}
and
%\begin{equation}
%\Big\|(R^{Bog}_{i,i}(z))^{\frac{1}{2}}[\Gamma^{Bog\,}_{i,i}(z)]_{(l,h_-;\dots ; j,h_-;j-4,h_+)}
%(R^{Bog}_{i,i}(z))^{\frac{1}{2}}\Big\|\leq {\color{red}...}
%\end{equation}
\begin{eqnarray}\label{gamma-exp-2-0}
& &\|(R^{Bog}_{i,i}(w))^{\frac{1}{2}}[\Gamma^{Bog\,}_{i,i}(w)]_{(l,h_+; 4,h_-;\dots;i-2,h_-)}(R^{Bog}_{i,i}(w))^{\frac{1}{2}}\|\\
&\leq& (Z_{l,\epsilon})^{h}\,\prod_{f=l+2\,,\, f-l\,\text{even}}^{i}\frac{K_{f,\epsilon}}{(1-Z_{f-2,\epsilon})^2} \nonumber
%& &\quad\times \prod_{f=l+2\,,\, f-l\,\text{even}}^{i-2}(1+a_{\epsilon}-\frac{2b_{\epsilon}}{N-f-1}-\frac{1-c_{\epsilon}}{(N-f-1)^2})^{-1}\nonumber
 \end{eqnarray}
hold true, where 
\begin{equation}\label{KZ-0}
K_{i,\epsilon}:=\frac{1}{4(1+a_{\epsilon}-\frac{2b_{\epsilon}}{N-i+1}-\frac{1-c_{\epsilon}}{(N-i+1)^2})}\quad,\quad Z_{i-2,\epsilon}:=\frac{1}{4(1+a_{\epsilon}-\frac{2b_{\epsilon}}{N-i+3}-\frac{1-c_{\epsilon}}{(N-i+3)^2})}\frac{2}{\Big[1+\sqrt{\eta a_{\epsilon}}-\frac{b_{\epsilon}/\sqrt{\eta a_{\epsilon}}}{N-i+4-\epsilon^{\Theta}}\Big]}\,.
\end{equation}
where $a_{\epsilon}, b_{\epsilon}, c_{\epsilon}$,  are those defined in Corollary \ref{main-lemma-H}, and $\Theta$ is defined in \underline{Lemma 3.6} of \cite{Pi1} (reported in Section \ref{Feshbach}). 
\end{prop}

\noindent
 \emph{Proof}
 
\noindent
The proof is a straighforward generalization of \emph{\underline{Proposition 4.10} of \cite{Pi1}},  which is possible thanks to estimate (\ref{estimate-main-lemma-H}) in Corollary \ref{main-lemma-H} that implies (\ref{Gamma-ineq}) through \emph{\underline{Lemma 3.6} of \cite{Pi1}} (reported in Section \ref{Feshbach}). \qed

\begin{remark}\label{prod-control}
From the definitions in (\ref{KZ-0}) and the $\epsilon-$dependence of $a_{\epsilon}$, $b_{\epsilon}$, and $c_{\epsilon}$ (see (\ref{a-bis})-(\ref{b-bis})-(\ref{c-bis})),  it is evident that there exist constants $C,c>0$ such that (assuming $\epsilon$ sufficiently small)
\begin{equation}\label{KZ}
\frac{K_{f,\epsilon}}{(1-Z_{f-2,\epsilon})^2}\leq \frac{1}{1+c\sqrt{\epsilon}}
\end{equation}
for $N-f>\frac{C}{\sqrt{\epsilon}}$\,.  With a similar computation, one can check that for $N-2\geq i> N-\frac{C}{\sqrt{\epsilon}}$ and some $c'>0$ 
\begin{equation}
\frac{K_{f,\epsilon}}{(1-Z_{f-2,\epsilon})^2}\leq (1+c'\frac{\sqrt{\epsilon}}{N-f}+\mathcal{O}(\frac{1}{(N-f)^2}))
\end{equation}
In consequence, for $N-2\geq i> N-\frac{C}{\sqrt{\epsilon}}$ (and assuming for simplicity that $N-\frac{C}{\sqrt{\epsilon}}$ is an even number) the inequality 
\begin{eqnarray}
& &\prod_{f=N-\frac{C}{\sqrt{\epsilon}}\,,\, f\,\text{even}}^{i}\frac{K_{f,\epsilon}}{(1-Z_{f-2,\epsilon})^2}\\
&\leq & \prod_{f=N-\frac{C}{\sqrt{\epsilon}}\,,\, f\,\text{even}}^{i}(1+c'\frac{\sqrt{\epsilon}}{N-f}+\mathcal{O}(\frac{1}{(N-f)^2}))\\
&=&\prod_{f=N-\frac{C}{\sqrt{\epsilon}}\,,\, f\,\text{even}}^{i}\exp[\ln\Big(1+c'\frac{\sqrt{\epsilon}}{N-f}+\mathcal{O}(\frac{1}{(N-f)^2})\Big)]\leq \mathcal{O}(1)\,
\end{eqnarray}
holds true. Therefore, we can conclude that:

\noindent
1) If $N-\frac{C}{\sqrt{\epsilon}}\geq i\geq l+2$
\begin{equation}
\prod_{f=l+2\,,\, f\,\text{even}}^{i}\frac{K_{f,\epsilon}}{(1-Z_{f-2,\epsilon})^2}\leq\mathcal{O}((\frac{1}{1+c\sqrt{\epsilon}})^{i-l-2})\quad ;  
\end{equation}
2) If $N-2\geq i>N-\frac{C}{\sqrt{\epsilon}}\geq l+2$, then
\begin{eqnarray}
& &\prod_{f=l+2\,,\, f\,\text{even}}^{i}\frac{K_{f,\epsilon}}{(1-Z_{f-2,\epsilon})^2}\\
& =&\Big[\prod_{f=l+2\,,\, f\,\text{even}}^{N-\frac{C}{\sqrt{\epsilon}}}\frac{K_{f,\epsilon}}{(1-Z_{f-2,\epsilon})^2}\Big]\,\Big[\prod_{f=N-\frac{C}{\sqrt{\epsilon}}+2\,,\, f\,\text{even}}^{i}\frac{K_{f,\epsilon}}{(1-Z_{f-2,\epsilon})^2} \Big]\\
&\leq & \mathcal{O}((\frac{1}{1+c\sqrt{\epsilon}})^{N-\frac{C}{\sqrt{\epsilon}}-l-2})\,.
\quad\end{eqnarray}
\end{remark}
\begin{remark}\label{estimation-proc}
In this remark we explain how to provide an estimate  of
\begin{equation}\label{estimate-E}
\|(R^{Bog}_{i,i}(w))^{\frac{1}{2}}\,\sum_{l=2,\,l\, even}^{i-2}[\Gamma^{Bog\,}_{i,i}(w)]_{(l,h_-; l+2,h_-;\dots ; i-2,h_-)}\,(R^{Bog}_{i,i}(w))^{\frac{1}{2}}\|\,,\quad i\leq N-2\,,
\end{equation}
without using  (\ref{gamma-exp-1-0}) and the computations in Remark \ref{prod-control}. Indeed, this would make the estimate worse.

\noindent
%We observe that the bound in (\ref{def-deltabog}) of Lemma \ref{main-lemma-Bog} can be employed to provide an upper bound to (\ref{estimate-E}) since 
The operator in (\ref{estimate-E}) can be expressed as a sum of products of operators of the type in (\ref{estimate-main-lemma-Bog}). We call ``blocks" the operators of the type in (\ref{estimate-main-lemma-Bog}) and define
\begin{equation}\label{E-est}
\mathcal{E}(\|(R^{Bog}_{i,i}(z))^{\frac{1}{2}}\,\sum_{l=2,\, l\, even}^{i-2}[\Gamma^{Bog\,}_{i,i}(w)]_{(r,h_-; r+2,h_-;\dots ; i-2,h_-)}\,(R^{Bog}_{i,i}(z))^{\frac{1}{2}}\|)\,
\end{equation}
the upper bound (to the operator norm in the argument of $\mathcal{E}$) obtained estimating the norm of the sum (of the operators) with the sum of the norms of the summands, and the norm of each operator product with the product of the norms of the blocks. The estimate of the norm of each block is provided by Corollary \ref{main-lemma-H} in the Appendix.
%{\color{red}\underline{Lemma 3.4} of \cite{Pi1} (reported in Section \ref{Feshbach}).}

%summing up the estimate of the norm of the operator products that follows from Lemma \ref{main-lemma-Bog}.

\noindent
Next, we point out that
\begin{itemize}
  \item by using the decomposition in (\ref{decomposition})  of Proposition \ref{lemma-expansion-proof-0}  for $h'\equiv \infty$, we get 
  % up to a remainder of arbitrarily small norm we can approximate 
\begin{eqnarray}
& &(R^{Bog}_{i,i}(w))^{\frac{1}{2}}\Gamma^{Bog\,}_{i,i}(w)(R^{Bog}_{i,i}(w))^{\frac{1}{2}}\label{complete-zero} \\
& =&(R^{Bog}_{i,i}(w))^{\frac{1}{2}}\sum_{l=2,\,l\, even}^{i-2}[\Gamma^{Bog\,}_{i,i}(w)]_{(l,h'_-; l+2,h'_-;\dots ; i-2,h'_-)}(R^{Bog}_{i,i}(w))^{\frac{1}{2}}\,,\label{partial-0-0}
 \end{eqnarray}
% by choosing an $h'$ sufficiently large. 
and
\begin{equation} (R^{Bog}_{i,i}(w))^{\frac{1}{2}}\sum_{l=2,\,l\, even}^{i-2}[\Gamma^{Bog\,}_{i,i}(w)]_{(l,h_-; l+2,h_-;\dots ; i-2,h_-)}(R^{Bog}_{i,i}(w))^{\frac{1}{2}}\,,\quad h<\infty , \, \label{partial-0}
\end{equation} is by construction a partial sum of the terms in (\ref{partial-0-0});
 \item
both for the estimate of  the norm of (\ref{complete-zero}) provided in Theorem  \ref{Feshbach-Hbog} and for $\mathcal{E}(\|(\ref{partial-0})\|)$ we use the same procedure: by Corollary \ref{main-lemma-H} in the Appendix
%\underline{Lemma 3.4} of \cite{Pi1} (reported in Section \ref{Feshbach})
 we estimate the operator norm of the  blocks and of the products of blocks;  then for each sum of products of blocks we sum up the (estimates of the) operator norms of the products. 
\end{itemize}
Hence, we can conclude that (for $i\leq N-2$)
\begin{eqnarray}\label{estimate-E-bis}
& &\mathcal{E}(\|(R^{Bog}_{i,i}(z))^{\frac{1}{2}}\,\sum_{l=2,\,l\, even}^{i-2}[\Gamma^{Bog\,}_{i,i}(w)]_{(l,h_-; l+2,h_-;\dots ; i-2,h_-)}\,(R^{Bog}_{i,i}(z))^{\frac{1}{2}}\|)\\
&\leq &\mathcal{E}(\|(R^{Bog}_{i,i}(z))^{\frac{1}{2}}\,\Gamma^{Bog\,}_{i,i}(w)\,(R^{Bog}_{i,i}(z))^{\frac{1}{2}}\|)\\
&\leq &\frac{4}{5}
\end{eqnarray}
where the last step follows  for $\epsilon $ sufficiently small from the identity
\begin{eqnarray}
& &(R^{Bog}_{i,i}(w))^{\frac{1}{2}}\Gamma^{Bog\,}_{i,i}(R^{Bog}_{i,i}(w))^{\frac{1}{2}}\label{gammacheck}\\
&=&(R^{Bog}_{i,i}(w))^{\frac{1}{2}}W_{i,i-2}\,(R^{Bog}_{i-2,i-2}(w))^{\frac{1}{2}} \times\\
& &\quad\quad\quad\times\sum_{l_{i-2}=0}^{\infty}\Big[(R^{Bog}_{i-2,i-2}(w))^{\frac{1}{2}}\Gamma^{Bog}_{i-2,i-2}(w)(R^{Bog}_{i-2,i-2}(w))^{\frac{1}{2}}\Big]^{l_{i-2}}\times \quad\quad\quad\label{Estimate-E-bis-bis}\\
& &\quad\quad\quad\quad\quad\quad\quad\quad\times (R^{Bog}_{i-2,i-2}(w))^{\frac{1}{2}}W^*_{i-2,i}(R^{Bog}_{i,i}(w))^{\frac{1}{2}}\,\nonumber
%&=&(R^{Bog}_{i,i}(z))^{\frac{1}{2}}W_{i,i-2}\,(R^{Bog}_{i-2,i-2}(z))^{\frac{1}{2}}\times\\
%& &\quad\times  \sum_{l_{i-2}=0}^{\infty}\Big[(R^{Bog}_{i-2,i-2}(z))^{\frac{1}{2}}\Gamma^{Bog}_{i-2,i-2}(z)R^{Bog}_{i-2,i-2}(z)\Big]^{l_{i-2}}\times \\
%& &\quad \times (R^{Bog}_{i-2,i-2}(z))^{\frac{1}{2}}W^*_{i-2,i}(R^{Bog}_{i,i}(z))^{\frac{1}{2}}\,.
\end{eqnarray}
and from estimates (\ref{main-estimate-intermediate}),(\ref{Gamma-ineq}), and (\ref{x-ineq}).
\end{remark}
In the next lemma we develop some tools for Theorem \ref{induction-many-modes}. More precisely, the results in Lemma \ref{main-relations-new} are used later to prove that
$$f^{Bog}_{\bold{j}_1,\dots,\bold{j}_m}(z+z^{Bog}_{\bold{j}_1,\dots,\bold{j}_{m-1}})=f^{Bog}_{\bold{j}_m}(z)+o_{N\to \infty}(1)\,.$$
\begin{lemma}\label{main-relations-new}
Let $2\leq m \leq M$ and assume that the Hamiltonian $H^{Bog}_{\bold{j}_1,\dots,\bold{j}_{m-1}}$ has ground state vector $\psi^{Bog}_{\bold{j}_1,\dots,\bold{j}_{m-1}}$ (see (\ref{gs-Hm-start})-(\ref{gs-Hm-fin})) with ground state energy $z^{Bog}_{\bold{j}_1,\dots,\bold{j}_{m-1}}$. 
%Define $$\Delta_{0}:=\min\, \Big\{\epsilon_{\bold{j}}\,|\,\bold{j}\in \mathbb{Z}^3 \Big\}\,,$$  and, for $m\geq 2$, assume that there exists $\Delta_{m-1}{\color{red}>0}$ such that
%\begin{equation}
%\text{Gap}\,\Big[\hat{H}^{Bog}_{\bold{j}_1, \dots,\bold{j}_{m-1}}+\sum_{\bold{j}\neq \{\pm\bold{j}_1, \dots, \pm\bold{j}_{m}\}}(k_{\bold{j}})^2a^*_{\bold{j}}a_{\bold{j}}\Big]\geq \Delta_{m-1}
%\end{equation}
%(Notice that $$\hat{H}^{Bog}_{\bold{j}_1, \dots,\bold{j}_{m-1}}+\sum_{\bold{j}\neq \{\pm\bold{j}_1, \dots, \pm\bold{j}_{m}\}}(k_{\bold{j}})^2a^*_{\bold{j}}a_{\bold{j}}=H^{Bog}_{\bold{j}_1, \dots,\bold{j}_{m-1}}-(k_{\bold{j}_m})^2(a^*_{\bold{j}_m}a_{\bold{j}_m}+a^*_{-\bold{j}_m}a_{-\bold{j}_m})\,,$$ i.e., the kinetic energy associated with the modes $\pm \bold{j}_m$ is absent). 
Furthermore, assume:

\begin{enumerate}
\item
\begin{equation}\label{assumption-flow-0}
\text{infspec}\,[\sum_{l=1}^{m-1}\hat{H}^{Bog}_{\bold{j}_l}]-z^{Bog}_{\bold{j}_1,\dots ,\bold{j}_{m-1}}\geq -\frac{(m-1)}{(\ln N)^{\frac{1}{8}}}
%\hat{H}^{Bog}_{\bold{j}_1, \dots,\bold{j}_{m}}\geq z^{Bog}_{\bold{j}_1,\dots,\bold{j}_{m}}-\mathcal{O}(\frac{1}{(\ln N)^{\frac{1}{8}}})
\end{equation}
%\begin{equation}
%(H^{\#}_{\bold{j}_{1},\dots,\bold{j}_{m-1}})_{\xi}-(1-\xi)T_{\bold{j}=\{\pm\bold{j}_m\}}\geq z^{\#}_{\bold{j}_1,\dots\,\bold{j}_{m-1}}-\frac{(m-1)\xi^{\frac{1}{2}}}{M}\,,\label{ass-1-main-lemma}
%\end{equation}
where $z^{Bog}_{\bold{j}_1,\dots\,\bold{j}_{m-1}}$ is the ground state energy of $H^{Bog}_{\bold{j}_1,\dots,\bold{j}_{m-1}}$ for $m\geq 2$, and $z^{Bog}_{\bold{j}_1,\dots\,\bold{j}_{m-1}}|_{m=1}\equiv 0$;
\item
there exist $0\leq\tilde{C}_{m-1}\leq \tilde{C}_{M-1}<\infty$ such that
 \begin{equation}\label{ass-number-0-new}
\langle \frac{\psi^{Bog}_{\bold{j}_1,\dots,\bold{j}_{m-1}}}{\|\psi^{Bog}_{\bold{j}_1,\dots,\bold{j}_{m-1}}\|}\,,\,\sum_{\bold{j}\in\mathbb{Z}^d\setminus \{\bold{0}\}} a_{\bold{j}}^{*}a_{\bold{j}}\,\frac{\psi^{Bog}_{\bold{j}_1,\dots,\bold{j}_{m-1}}}{\|\psi^{Bog}_{\bold{j}_1,\dots,\bold{j}_{m-1}}\|}\rangle \leq \tilde{C}_{m-1}\,.
\end{equation}
\end{enumerate}
Let $\epsilon_{\bold{j}_m}$ be sufficiently small  and $N$ sufficiently large. 
%such that Theorem \ref{Feshbach-Hbog} can be applied. 
%the Feshbach flow associated with $H^{Bog}_{\bold{j}_1,\dots,\bold{j}_{m}}$ and the couple of modes $\pm \bold{j}_m$ is well defined for $$z\leq E^{Bog}_{\bold{j}_*}+ \sqrt{\epsilon_{\bold{j}_m}}\phi_{\bold{j}_m}\sqrt{\epsilon_{\bold{j}_m}^2+2\epsilon_{\bold{j}_m}}\,.$$
 Then, for $z$ in the interval $$z\leq E^{Bog}_{\bold{j}_m}+ \sqrt{\epsilon_{\bold{j}_m}}\phi_{\bold{j}_m}\sqrt{\epsilon_{\bold{j}_m}^2+2\epsilon_{\bold{j}_m}}-\frac{(\lfloor (\ln N)^{\frac{1}{2}} \rfloor +4)\phi_{\bold{j}_m}}{N}\,(<0)\,$$
the following estimates hold true
\begin{eqnarray}
& &\Big|\langle  \frac{\psi^{Bog}_{\bold{j}_1,\dots,\bold{j}_{m-1}}}{\|\psi^{Bog}_{\bold{j}_1,\dots,\bold{j}_{m-1}}\|}\,,\,\Gamma^{Bog}_{\bold{j}_1,\dots,\bold{j}_m ;N,N}(z+z^{Bog}_{\bold{j}_1,\dots,\bold{j}_{m-1}}) \frac{\psi^{Bog}_{\bold{j}_1,\dots,\bold{j}_{m-1}}}{\|\psi^{Bog}_{\bold{j}_1,\dots,\bold{j}_{m-1}}\|}\rangle- \langle  \eta\,,\,\tilde{\Gamma}^{Bog}_{\bold{j}_1,\dots,\bold{j}_m ;N,N}(z) \eta \rangle \Big| \leq \frac{C_I}{(\ln N)^{\frac{1}{4}}}\,,\quad \quad\quad \label{primero-1-1-new}
%&= &\langle  \frac{\psi^{Bog}_{\bold{j}_1,\dots,\bold{j}_{m-1}}}{\|\psi^{Bog}_{\bold{j}_1,\dots,\bold{j}_{m-1}}\|}\,,\,W_{\bold{j}_2}\,R^{Bog}_{\bold{j}_1,\,\bold{j}_2\,;\,N-2,N-2}(z_{\bold{j}_1}+z)\times\\
%& &\quad\quad \times \sum_{l_{N-2}=0}^{\infty}\Big[\Gamma^{Bog}_{\bold{j}_1,\,\bold{j}_2\,;\,N-2,N-2}(z_{\bold{j}_1}+z)R^{Bog}_{\bold{j}_1,\,\bold{j}_2\,;\,\,N-2,N-2}(z_{\bold{j}_1}+z)\Big]^{l_{N-2}}W^*_{\bold{j}_2}\, \frac{\psi^{Bog}_{\bold{j}_1,\dots,\bold{j}_{m-1}}}{\|\psi^{Bog}_{\bold{j}_1,\dots,\bold{j}_{m-1}}\|}\rangle \quad\quad\quad\\
%&= &\langle  \eta\,,\,\tilde{\Gamma}^{Bog}_{\bold{j}_1,\dots,\bold{j}_m ;N,N}(z) \eta \rangle +{\color{red}\frac{C_I}{N^{\frac{1}{16}}}}\label{A-remainder}
\end{eqnarray}
\begin{eqnarray}
& &\|\mathscr{P}_{\psi^{Bog}_{\bold{j}_1,\dots,\bold{j}_{m-1}}}\Gamma^{Bog}_{\bold{j}_1,\dots,\bold{j}_m;\,N,N}(z+z^{Bog}_{\bold{j}_1,\dots,\bold{j}_{m-1}})\,\overline{\mathscr{P}_{\psi^{Bog}_{\bold{j}_1,\dots,\bold{j}_{m-1}}}}\|\leq \frac{C_{II}}{(\ln N)^{\frac{1}{4}}} \label{primero-3-1-new}
\end{eqnarray}
for some $0< C_{I}, C_{II}<\infty $, where $\tilde{\Gamma}^{Bog}_{\bold{j}_1,\dots,\bold{j}_m ;N,N}(z)$ is defined starting from  $\Gamma^{Bog}_{\bold{j}_1,\dots,\bold{j}_m ;N,N}(z+z^{Bog}_{\bold{j}_1,\dots,\bold{j}_{m-1}})$ (see (\ref{GammaNNBog})) by replacing (\ref{def-tilde-0}) with (\ref{def-tilde}).
\end{lemma}

\noindent
\emph{{\bf{Outline of the proof.}}} The proof is deferred to Lemma \ref{main-relations} in the Appendix, where at the beginning we show the result contained in (\ref{primero-1-1-new}). For (\ref{primero-3-1-new}) the argument is partially the same and is provided afterwards. Here, we outline the steps leading to the inequality in (\ref{primero-1-1-new}). They are implemented and explained in full detail in Lemma 
\ref{main-relations}:
\begin{itemize}
%\item[{\bf{I)}}]
% by truncating the sum on the R-H-S of (\ref{fixed-point-m-bis});
%\item[{\bf{STEP I)}}]
\item[{\bf{STEP I)}}]
In the expression
\begin{equation}\label{expansion-z}
\langle  \frac{\psi^{Bog}_{\bold{j}_1,\dots,\bold{j}_{m-1}}}{\|\psi^{Bog}_{\bold{j}_1,\dots,\bold{j}_{m-1}}\|}\,,\,\Gamma^{Bog}_{\bold{j}_1,\dots,\bold{j}_m ;N,N}(w) \frac{\psi^{Bog}_{\bold{j}_1,\dots,\bold{j}_{m-1}}}{\|\psi^{Bog}_{\bold{j}_1,\dots,\bold{j}_{m-1}}\|}\rangle\,,\quad w=z+z^{Bog}_{\bold{j}_1,\dots,\bold{j}_{m-1}}\,,
\end{equation}
we isolate a first remainder by truncating the sum in (\ref{sum}).
\item[{\bf{STEP II)}}]
By using the expansion of $$\Gamma^{Bog\,}_{\bold{j}_1,\dots,\bold{j}_m\,,;\,N-2,N-2}(w)$$ discussed in Proposition \ref{lemma-expansion-proof-0} we isolate a second remainder.  
%Thus,  we isolate a second remainder;
\item[{\bf{STEP III)}}]
From {\bf{Step II)}}, an expression consisting of a finite sum of (finite) products with inner factors of the type 
\begin{equation}\label{typical}
(R^{Bog}_{\bold{j}_1,\dots,\bold{j}_m;\,i,i}(w))^{\frac{1}{2}}W_{\bold{j}_m\,;\,i,i-2}(R^{Bog}_{\bold{j}_1,\dots,\bold{j}_m ;\,i-2,i-2}(w))^{\frac{1}{2}}\quad,\quad (R^{Bog}_{\bold{j}_1,\dots,\bold{j}_m;\,i-2,i-2}(w))^{\frac{1}{2}}W^*_{\bold{j}_m\,;\,i-2,i}(R^{Bog}_{\bold{j}_1,\dots,\bold{j}_m ;\,i,i}(w))^{\frac{1}{2}}
\end{equation}
and the two outer factors
\begin{equation}\label{outers}
W_{\bold{j}_m}(R^{Bog}_{\bold{j}_1,\dots,\bold{j}_m ;\,N-2,N-2}(w))^{\frac{1}{2}}\quad,\quad (R^{Bog}_{\bold{j}_1,\dots,\bold{j}_m ;\,N-2,N-2}(w))^{\frac{1}{2}}W^*_{\bold{j}_m}\,,
\end{equation}
is left. Then, we make use of the main mechanism of the proof that we highlight below: 
\\

\noindent
\emph{After the truncations in {\bf{Steps I)}} and {\bf{II)}}, in the resulting expression the resolvents $R^{Bog}_{\bold{j}_1,\dots,\bold{j}_m;\,i,i}(w)$ are sufficiently ``close" to the vector $\psi^{Bog}_{\bold{j}_1,\dots,\bold{j}_{m-1}}$,  hence they can be replaced with  $\tilde{R}^{Bog}_{\bold{j}_1,\dots,\bold{j}_m;\,i,i}(z)$ (see the definition after (\ref{def-tilde})) up to a small remainder, by exploiting the identity  $$\sum_{l=1}^{m-1}\hat{H}^{Bog}_{\bold{j}_l}\psi^{Bog}_{\bold{j}_1,\dots,\bold{j}_{m-1}}=z^{Bog}_{\bold{j}_1,\dots,\bold{j}_{m-1}}\psi^{Bog}_{\bold{j}_1,\dots,\bold{j}_{m-1}}.$$}
\item[{\bf{STEP IV)}}] In this step the resolvents $\tilde{R}^{Bog}_{\bold{j}_1,\dots,\bold{j}_m;\,i,i}(z)$ are replaced with $z$-dependent $c-numbers$ and the vector $\psi^{Bog}_{\bold{j}_1,\dots,\bold{j}_{m-1}}$ is replaced with $\eta$. One more remainder term is produced.
\end{itemize}

Next lemma deals with the invertibility of $$\overline{\mathscr{P}_{\psi^{Bog}_{\bold{j}_1,\dots,\bold{j}_{m-1}}}}\mathscr{K}_{\bold{j}_1,\dots,\bold{j}_m}^{Bog\,(N-2)}(z+z^{Bog}_{\bold{j}_1,\dots,\bold{j}_{m-1}})\overline{\mathscr{P}_{\psi^{Bog}_{\bold{j}_1,\dots,\bold{j}_{m-1}}}}$$  on $\overline{\mathscr{P}_{\psi^{Bog}_{\bold{j}_1,\dots,\bold{j}_{m-1}}}}\mathcal{F}^{N}$.
\begin{lemma}\label{invertibility-new}
Let $2\leq m\leq M$ and assume that the Hamiltonian $H^{Bog}_{\bold{j}_1,\dots,\bold{j}_{m-1}}$ has nondegenerate ground state energy $z^{Bog}_{\bold{j}_1,\dots,\bold{j}_{m-1}}$ and ground state vector $\psi^{Bog}_{\bold{j}_1,\dots,\bold{j}_{m-1}}$ with the property in (\ref{ass-number-0-new}). 

\noindent
Furthermore, assume that:
\begin{enumerate}
\item
There exists
 $\Delta_0\geq \Delta_{m-1}>0$ such that
\begin{equation}\label{assumption-gap-new}
\text{infspec}\,\Big[\Big(\hat{H}^{Bog}_{\bold{j}_1, \dots,\bold{j}_{m-1}}+\sum_{\bold{j}\in \mathbb{Z}^d \setminus \{\pm\bold{j}_1, \dots, \pm\bold{j}_{m}\}}(k_{\bold{j}})^2a^*_{\bold{j}}a_{\bold{j}}\Big)\upharpoonright_{(Q^{(>N-1)}_{\bold{j}_m}\mathcal{F}^N)\ominus \{\mathbb{C}\psi^{Bog}_{\bold{j}_1, \dots,\bold{j}_{m-1}}\}}\Big]-z^{Bog}_{\bold{j}_1,\dots,\bold{j}_{m-1}}\geq \Delta_{m-1}
\end{equation}
where $Q^{(>N-1)}_{\bold{j}_m}\mathcal{F}^N$ is the subspace of states in $\mathcal{F}^N$ with no particles in the modes $\pm \bold{j}_m$, and $\{\mathbb{C}\psi^{Bog}_{\bold{j}_1, \dots,\bold{j}_{m-1}}\}$ is the subspace generated by the vector $\psi^{Bog}_{\bold{j}_1, \dots,\bold{j}_{m-1}}$.
(Notice that $$\hat{H}^{Bog}_{\bold{j}_1, \dots,\bold{j}_{m-1}}+\sum_{\bold{j}\in \mathbb{Z}^d \setminus \{\pm\bold{j}_1, \dots, \pm\bold{j}_{m}\}}(k_{\bold{j}})^2a^*_{\bold{j}}a_{\bold{j}}=H^{Bog}_{\bold{j}_1, \dots,\bold{j}_{m-1}}-(k_{\bold{j}_m})^2(a^*_{\bold{j}_m}a_{\bold{j}_m}+a^*_{-\bold{j}_m}a_{-\bold{j}_m})\,,$$ i.e., the kinetic energy associated with the modes $\pm \bold{j}_m$ is absent.)
\item
\begin{equation}\label{lower-bound-spec}
\text{infspec}\,[\sum_{l=1}^{m-1}\hat{H}^{Bog}_{\bold{j}_l }]-z^{Bog}_{\bold{j}_1,\dots ,\bold{j}_{m-1}}\geq -\frac{m-1}{(\ln N)^{\frac{1}{8}}}\,.
\end{equation}
\end{enumerate}

\noindent
Let $\epsilon_{\bold{j}_m}$ be sufficiently small and $N$ sufficiently large such that: 

\noindent
a) for
$$z\leq E^{Bog}_{\bold{j}_m}+ \sqrt{\epsilon_{\bold{j}_m}}\phi_{\bold{j}_m}\sqrt{\epsilon_{\bold{j}_m}^2+2\epsilon_{\bold{j}_m}}\,(<0)$$the Feshbach-Schur flow  associated with the couple of modes $\pm \bold{j}_m$ is well defined (see Theorem \ref{Feshbach-Hbog}) up to the index value $i=N-2$; 

\noindent
b) \begin{equation}\label{def-cjm-new}
 \frac{\ln N}{N}\ll1 \quad \text{and} \quad \frac{U_{\bold{j}_m}}{\sqrt{N}}<\frac{\Delta_{m-1}}{2}\quad \text{where}\quad U_{\bold{j}_m}:=k^2_{\bold{j}_m}+\phi_{\bold{j}_m}\,;
 \end{equation}

\noindent
c)\footnote{This condition holds for $N$ sufficiently large because in \emph{\underline{Corollary 4.6} of \cite{Pi1}} we show that   in the limit $N\to \infty$ the ground state energy, $z_*$, of $H_{\bold{j}_*}^{Bog}$ tends to $E^{Bog}_{\bold{j}_*}$ (see Section \ref{groundstate}),  and because $\gamma\Delta_{m-1}\leq \frac{\Delta_0}{2}$.}  \begin{equation}
 z_{m}+\gamma \Delta_{m-1}<E^{Bog}_{\bold{j}_m}+\frac{1}{2} \sqrt{\epsilon_{\bold{j}_m}}\phi_{\bold{j}_m}\sqrt{\epsilon_{\bold{j}_m}^2+2\epsilon_{\bold{j}_m}}\quad,\quad\gamma= \frac{1}{2}\,,
 \end{equation}
where  $z_{m}$ is the ground state energy of $H^{Bog}_{\bold{j}_m}$.

\noindent
Then, there exists a constant $C^{\perp}>0$ such that for  
\begin{equation}
z\leq  z_{m}-\frac{C^{\perp}}{(\ln N)^{\frac{1}{2}}}+\gamma \Delta_{m-1} \label{bound-z-0-new}
\end{equation}
the following estimate holds true:
 %\begin{eqnarray}
%& &\overline{\mathscr{P}_{\psi^{Bog}_{\bold{j}_1,\dots,\bold{j}_{m-1}}}}\Gamma^{Bog}_{\bold{j}_1,\dots,\bold{j}_m;\,N,N}(z+z^{Bog}_{\bold{j}_1,\dots,\bold{j}_{m-1}})\,\overline{\mathscr{P}_{\psi^{Bog}_{\bold{j}_1,\dots,\bold{j}_{m-1}}}}\leq -z_m\overline{\mathscr{P}_{\psi^{Bog}_{\bold{j}_1,\dots,\bold{j}_{m-1}}}}\label{ineq-z_m}
%\end{eqnarray} 
%and 
\begin{equation}\label{invert-est-new}
\overline{\mathscr{P}_{\psi^{Bog}_{\bold{j}_1,\dots,\bold{j}_{m-1}}}}\mathscr{K}_{\bold{j}_1,\dots,\bold{j}_m}^{Bog\,(N-2)}(z+z^{Bog}_{\bold{j}_1,\dots,\bold{j}_{m-1}})\overline{\mathscr{P}_{\psi^{Bog}_{\bold{j}_1,\dots,\bold{j}_{m-1}}}}\geq (1-\gamma)\Delta_{m-1}\overline{\mathscr{P}_{\psi^{Bog}_{\bold{j}_1,\dots,\bold{j}_{m-1}}}}\,.
\end{equation}
%in the  $z$ given in (\ref{bound-z-0}).
\end{lemma}

\noindent
\emph{{\bf{Outline of the proof.}}} Like for Lemma \ref{main-relations-new}, the detailed proof is deferred to the Appendix; see Lemma \ref{invertibility}. Here, we outline the procedure. We start observing that due to the definition in (\ref{final-proj-Bog})
\begin{eqnarray}
& &\overline{\mathscr{P}_{\psi^{Bog}_{\bold{j}_1,\dots,\bold{j}_{m-1}}}}\mathscr{K}_{\bold{j}_1,\dots,\bold{j}_m}^{Bog\,(N-2)}(z+z_{\bold{j}_1,\dots,\bold{j}_{m-1}})\overline{\mathscr{P}_{\psi^{Bog}_{\bold{j}_1,\dots,\bold{j}_{m-1}}}}\\
%& =&\overline{\mathscr{P}_{\psi^{Bog}_{\bold{j}_1,\dots,\bold{j}_{m-1}}}}\mathscr{K}_{\bold{j}_1,\dots,\bold{j}_m}^{Bog\,(N-2)}(z+z_{\bold{j}_1,\dots,\bold{j}_{m-1}})\overline{\mathscr{P}_{\psi^{Bog}_{\bold{j}_1,\dots,\bold{j}_{m-1}}}}\\
%&=&\overline{\mathscr{P}_{\psi^{Bog}_{\bold{j}_1,\dots,\bold{j}_{m-1}}}}(H^{Bog} _{\bold{j}_1,\dots, \bold{j}_m}-z-z_{\bold{j}_1,\dots,\bold{j}_{m-1}})\overline{\mathscr{P}_{\psi^{Bog}_{\bold{j}_1,\dots,\bold{j}_{m-1}}}}\quad\quad\quad\\
%& &-\overline{\mathscr{P}_{\psi^{Bog}_{\bold{j}_1,\dots,\bold{j}_{m-1}}}}\Gamma^{Bog\,}_{\bold{j}_1,\dots,\bold{j}_m\,,;\,N,N}(z_{\bold{j}_1,\dots,\bold{j}_{m-1}}+z)\overline{\mathscr{P}_{\psi^{Bog}_{\bold{j}_1,\dots,\bold{j}_{m-1}}}}\\
&=&\overline{\mathscr{P}_{\psi^{Bog}_{\bold{j}_1,\dots,\bold{j}_{m-1}}}}(H^{Bog} _{\bold{j}_1,\dots, \bold{j}_{m-1}}-z-z_{\bold{j}_1,\dots,\bold{j}_{m-1}})\overline{\mathscr{P}_{\psi^{Bog}_{\bold{j}_1,\dots,\bold{j}_{m-1}}}}\quad\quad\quad\\
& &-\overline{\mathscr{P}_{\psi^{Bog}_{\bold{j}_1,\dots,\bold{j}_{m-1}}}}\Gamma^{Bog\,}_{\bold{j}_1,\dots,\bold{j}_m\,;\,N,N}(z_{\bold{j}_1,\dots,\bold{j}_{m-1}}+z)\overline{\mathscr{P}_{\psi^{Bog}_{\bold{j}_1,\dots,\bold{j}_{m-1}}}}\\
&=&\overline{\mathscr{P}_{\psi^{Bog}_{\bold{j}_1,\dots,\bold{j}_{m-1}}}}(\hat{H}^{Bog}_{\bold{j}_1, \dots,\bold{j}_{m-1}}+\sum_{\bold{j}\neq \{\pm\bold{j}_1, \dots, \pm\bold{j}_{m}\}}(k_{\bold{j}})^2a^*_{\bold{j}}a_{\bold{j}}-z_{\bold{j}_1,\dots,\bold{j}_{m-1}}-z)\overline{\mathscr{P}_{\psi^{Bog}_{\bold{j}_1,\dots,\bold{j}_{m-1}}}}\label{-4}\quad\quad\quad\\
& &-\overline{\mathscr{P}_{\psi^{Bog}_{\bold{j}_1,\dots,\bold{j}_{m-1}}}}\Gamma^{Bog\,}_{\bold{j}_1,\dots,\bold{j}_m\,;\,N,N}(z_{\bold{j}_1,\dots,\bold{j}_{m-1}}+z)\overline{\mathscr{P}_{\psi^{Bog}_{\bold{j}_1,\dots,\bold{j}_{m-1}}}}\label{-5}\,.
%& &-\frac{\phi_{\bold{j}_{m}}}{2\epsilon_{\bold{j}_m}+2-\frac{z-\Delta_{m-1}(1-\frac{\phi_{\bold{j}_m}\lfloor (\ln N)^{\frac{1}{2}}\rfloor}{N\Delta_{m-1}})}{\phi_{\bold{j}_{m}}}}\check{\mathcal{G}}_{\bold{j}_{m}\,;\,N-2,N-2}(z-\Delta_{m-1}(1-\frac{\phi_{\bold{j}_m}\lfloor (\ln N)^{\frac{1}{2}} \rfloor}{N\Delta_{m-1}}))\overline{\mathscr{P}_{\psi^{Bog}_{\bold{j}_1,\dots,\bold{j}_{m-1}}}}-\frac{C^{\perp}}{(\ln N)^{\frac{1}{2}}}\overline{\mathscr{P}_{\psi^{Bog}_{\bold{j}_1,\dots,\bold{j}_{m-1}}}} \nonumber
\end{eqnarray}
Hence, the result in (\ref{invert-est-new}) is proven if 
\begin{eqnarray}
& &\overline{\mathscr{P}_{\psi^{Bog}_{\bold{j}_1,\dots,\bold{j}_{m-1}}}}(\hat{H}^{Bog}_{\bold{j}_1, \dots,\bold{j}_{m-1}}+\sum_{\bold{j}\neq \{\pm\bold{j}_1, \dots, \pm\bold{j}_{m}\}}(k_{\bold{j}})^2a^*_{\bold{j}}a_{\bold{j}}-z_{\bold{j}_1,\dots,\bold{j}_{m-1}}-z)\overline{\mathscr{P}_{\psi^{Bog}_{\bold{j}_1,\dots,\bold{j}_{m-1}}}}\quad\quad\quad\\
& &-\overline{\mathscr{P}_{\psi^{Bog}_{\bold{j}_1,\dots,\bold{j}_{m-1}}}}\Gamma^{Bog\,}_{\bold{j}_1,\dots,\bold{j}_m\,;\,N,N}(z_{\bold{j}_1,\dots,\bold{j}_{m-1}}+z)\overline{\mathscr{P}_{\psi^{Bog}_{\bold{j}_1,\dots,\bold{j}_{m-1}}}}\\
&\geq&(1-\gamma)\Delta_{m-1}\overline{\mathscr{P}_{\psi^{Bog}_{\bold{j}_1,\dots,\bold{j}_{m-1}}}}\,.\label{ineq-lemma4.4}
%& &-\frac{\phi_{\bold{j}_{m}}}{2\epsilon_{\bold{j}_m}+2-\frac{z-\Delta_{m-1}(1-\frac{\phi_{\bold{j}_m}\lfloor (\ln N)^{\frac{1}{2}}\rfloor}{N\Delta_{m-1}})}{\phi_{\bold{j}_{m}}}}\check{\mathcal{G}}_{\bold{j}_{m}\,;\,N-2,N-2}(z-\Delta_{m-1}(1-\frac{\phi_{\bold{j}_m}\lfloor (\ln N)^{\frac{1}{2}} \rfloor}{N\Delta_{m-1}}))\overline{\mathscr{P}_{\psi^{Bog}_{\bold{j}_1,\dots,\bold{j}_{m-1}}}}-\frac{C^{\perp}}{(\ln N)^{\frac{1}{2}}}\overline{\mathscr{P}_{\psi^{Bog}_{\bold{j}_1,\dots,\bold{j}_{m-1}}}} \nonumber
\end{eqnarray}
Assuming $\epsilon_{\bold{j}_m}$ sufficiently small and $N$ sufficiently large, if $z<-4\phi_{\bold{j}_m}$ the inequality in (\ref{ineq-lemma4.4})  follows easily from the assumption in (\ref{assumption-gap-new}), 
the estimates in (\ref{est-gammatilde}) and (\ref{x-ineq}), and the norm bound
\begin{equation}
\|W_{\bold{j}_m}\,(R^{Bog}_{\bold{j}_1,\dots,\bold{j}_m\,;\,N-2,N-2}(z^{Bog}_{\bold{j}_1,\dots,\bold{j}_{m-1}}+z))^{\frac{1}{2}}\|< \frac{\sqrt{\phi_{\bold{j}_m}}}{2}.
\end{equation}
Assume that for 
\begin{equation}\label{interval}
-4\phi_{\bold{j}_m}\leq z< z_{m}+\gamma \Delta_{m-1}\,\Big(< E^{Bog}_{\bold{j}_m}+ \frac{1}{2}\sqrt{\epsilon_{\bold{j}_m}}\phi_{\bold{j}_m}\sqrt{\epsilon_{\bold{j}_m}^2+2\epsilon_{\bold{j}_m}}\,<0\Big) \,
\end{equation}
%in the intersection of two intervals in  (\ref{interval}) and (\ref{bound-z-0}), respectively,  
we can show
%As far as  (\ref{ineq-z_m}) is concerned, ssuming 
%\begin{eqnarray}
%& &(\charf -\mathscr{P}_{\psi^{Bog}_{\bold{j}_1,\dots,\bold{j}_{m-1}}})\Big(\hat{H}^{Bog}_{\bold{j}_1, \dots,\bold{j}_{m-1}}+\sum_{\bold{j}\neq \{\pm\bold{j}_1, \dots, \pm\bold{j}_{m}\}}(k_{\bold{j}})^2a^*_{\bold{j}}a_{\bold{j}}-z_{\bold{j}_1,\dots,\bold{j}_{m-1}}\Big)(\charf-\mathscr{P}_{\psi^{Bog}_{\bold{j}_1,\dots,\bold{j}_{m-1}}})\quad\quad \\
%&\geq & \Delta_{m-1} (\charf -\mathscr{P}_{\psi^{Bog}_{\bold{j}_1,\dots,\bold{j}_{m-1}}})
%\end{eqnarray}
\begin{eqnarray}
& &\|\overline{\mathscr{P}_{\psi^{Bog}_{\bold{j}_1,\dots,\bold{j}_{m-1}}}}\Gamma^{Bog}_{\bold{j}_1,\dots,\bold{j}_m;\,N,N}(z+z^{Bog}_{\bold{j}_1,\dots,\bold{j}_{m-1}})\,\overline{\mathscr{P}_{\psi^{Bog}_{\bold{j}_1,\dots,\bold{j}_{m-1}}}}\|\label{GammaPerp}\\
%& =&\|\overline{\mathscr{P}_{\psi^{Bog}_{\bold{j}_1,\dots,\bold{j}_{m-1}}}}\Gamma^{Bog\,}_{\bold{j}_1,\,\bold{j}_2\,;\,N,N}(z_{\bold{j}_1}+z)\\
&=&\|\overline{\mathscr{P}_{\psi^{Bog}_{\bold{j}_1,\dots,\bold{j}_{m-1}}}}W_{\bold{j}_m}\,R^{Bog}_{\bold{j}_1,\dots,\bold{j}_m\,;\,N-2,N-2}(z^{Bog}_{\bold{j}_1,\dots,\bold{j}_{m-1}}+z)\times \label{outer-1}\\
& &\quad\quad \times \sum_{l_{N-2}=0}^{\infty}\Big[\Gamma^{Bog}_{\bold{j}_1,\dots,\bold{j}_m\,;\,N-2,N-2}(z^{Bog}_{\bold{j}_1,\dots,\bold{j}_{m-1}}+z)R^{Bog}_{\bold{j}_1,\dots,\bold{j}_m\,;\,\,N-2,N-2}(z^{Bog}_{\bold{j}_1,\dots,\bold{j}_{m-1}}+z)\Big]^{l_{N-2}}W^*_{\bold{j}_m}\,\overline{\mathscr{P}_{\psi^{Bog}_{\bold{j}_1,\dots,\bold{j}_{m-1}}}}\|\,\quad\quad\quad \label{outer-2}\\
&\leq &(1-\frac{1}{N})\frac{\phi_{\bold{j}_{m}}}{2\epsilon_{\bold{j}_m}+2-\frac{4}{N}-\frac{z-\Delta_{m-1}+\frac{U_{\bold{j}_m}}{\sqrt{N}}}{\phi_{\bold{j}_{m}}}}\check{\mathcal{G}}_{\bold{j}_{m}\,;\,N-2,N-2}(z-\Delta_{m-1}+\frac{U_{\bold{j}_m}}{\sqrt{N}})+\mathcal{O}(\frac{1}{(\ln N)^{\frac{1}{2}}})\label{primero-2-new}\,
\end{eqnarray}
where $\check{\mathcal{G}}_{\bold{j}_{m}\,;\,N-2,N-2}(z)$ is introduced in (\ref{in-formula-G})-(\ref{fin-formula-G})  (with $\bold{j}_*\equiv \bold{j}_m$) and $U_{\bold{j}_m}$ is defined in (\ref{def-cjm-new}).  Then, 
%following the same  procedure of \emph{\underline{Lemma 4.4} of \cite{Pi1}}, 
we can readily set a positive constant $C^{\perp}$ such that also for $z$ in the intersection of the two intervals in  (\ref{interval}) and (\ref{bound-z-0-new}), 
\begin{eqnarray}
& &(\ref{-4})+(\ref{-5})\\
%& &\overline{\mathscr{P}_{\psi^{Bog}_{\bold{j}_1,\dots,\bold{j}_{m-1}}}}\mathscr{K}_{\bold{j}_1,\dots,\bold{j}_m}^{Bog\,(N-2)}(z+z_{\bold{j}_1,\dots,\bold{j}_{m-1}})\overline{\mathscr{P}_{\psi^{Bog}_{\bold{j}_1,\dots,\bold{j}_{m-1}}}}\\
%& =&\overline{\mathscr{P}_{\psi^{Bog}_{\bold{j}_1,\dots,\bold{j}_{m-1}}}}\mathscr{K}_{\bold{j}_1,\dots,\bold{j}_m}^{Bog\,(N-2)}(z+z_{\bold{j}_1,\dots,\bold{j}_{m-1}})\overline{\mathscr{P}_{\psi^{Bog}_{\bold{j}_1,\dots,\bold{j}_{m-1}}}}\\
%&=&\overline{\mathscr{P}_{\psi^{Bog}_{\bold{j}_1,\dots,\bold{j}_{m-1}}}}(H^{Bog} _{\bold{j}_1,\dots, \bold{j}_m}-z-z_{\bold{j}_1,\dots,\bold{j}_{m-1}})\overline{\mathscr{P}_{\psi^{Bog}_{\bold{j}_1,\dots,\bold{j}_{m-1}}}}\quad\quad\quad\\
%& &-\overline{\mathscr{P}_{\psi^{Bog}_{\bold{j}_1,\dots,\bold{j}_{m-1}}}}\Gamma^{Bog\,}_{\bold{j}_1,\dots,\bold{j}_m\,,;\,N,N}(z_{\bold{j}_1,\dots,\bold{j}_{m-1}}+z)\overline{\mathscr{P}_{\psi^{Bog}_{\bold{j}_1,\dots,\bold{j}_{m-1}}}}\\
%&=&\overline{\mathscr{P}_{\psi^{Bog}_{\bold{j}_1,\dots,\bold{j}_{m-1}}}}(H^{Bog} _{\bold{j}_1,\dots, \bold{j}_{m-1}}-z-z_{\bold{j}_1,\dots,\bold{j}_{m-1}})\overline{\mathscr{P}_{\psi^{Bog}_{\bold{j}_1,\dots,\bold{j}_{m-1}}}}\quad\quad\quad\\
%& &-\overline{\mathscr{P}_{\psi^{Bog}_{\bold{j}_1,\dots,\bold{j}_{m-1}}}}\Gamma^{Bog\,}_{\bold{j}_1,\dots,\bold{j}_m\,;\,N,N}(z_{\bold{j}_1,\dots,\bold{j}_{m-1}}+z)\overline{\mathscr{P}_{\psi^{Bog}_{\bold{j}_1,\dots,\bold{j}_{m-1}}}}\\
& \geq&\overline{\mathscr{P}_{\psi^{Bog}_{\bold{j}_1,\dots,\bold{j}_{m-1}}}}(\hat{H}^{Bog}_{\bold{j}_1, \dots,\bold{j}_{m-1}}+\sum_{\bold{j}\neq \{\pm\bold{j}_1, \dots, \pm\bold{j}_{m}\}}(k_{\bold{j}})^2a^*_{\bold{j}}a_{\bold{j}}-z_{\bold{j}_1,\dots,\bold{j}_{m-1}}-z)\overline{\mathscr{P}_{\psi^{Bog}_{\bold{j}_1,\dots,\bold{j}_{m-1}}}}\label{-4-bis}\quad\quad\quad\\
%& &-\frac{\phi_{\bold{j}_{m}}}{2\epsilon_{\bold{j}_m}+2-\frac{z-\Delta_{m-1}+\frac{C_{\bold{j}_m}}{\sqrt{N}}}{\phi_{\bold{j}_{m}}}}\check{\mathcal{G}}_{\bold{j}_{m}\,;\,N-2,N-2}(z-\Delta_{m-1}+\frac{C_{\bold{j}_m}}{\sqrt{N}})\overline{\mathscr{P}_{\psi^{Bog}_{\bold{j}_1,\dots,\bold{j}_{m-1}}}}-\frac{C^{\perp}}{(\ln N)^{\frac{1}{2}}}\overline{\mathscr{P}_{\psi^{Bog}_{\bold{j}_1,\dots,\bold{j}_{m-1}}}} \nonumber\\
& &-(1-\frac{1}{N})\frac{\phi_{\bold{j}_{m}}}{2\epsilon_{\bold{j}_m}+2-\frac{4}{N}-\frac{z-\Delta_{m-1}+\frac{U_{\bold{j}_m}}{\sqrt{N}}}{\phi_{\bold{j}_{m}}}}\,\check{\mathcal{G}}_{\bold{j}_{m}\,;\,N-2,N-2}(z-\Delta_{m-1}+\frac{U_{\bold{j}_m}}{\sqrt{N}}))\overline{\mathscr{P}_{\psi^{Bog}_{\bold{j}_1,\dots,\bold{j}_{m-1}}}}-\frac{C^{\perp}}{(\ln N)^{\frac{1}{2}}}\overline{\mathscr{P}_{\psi^{Bog}_{\bold{j}_1,\dots,\bold{j}_{m-1}}}} \nonumber\\
&\geq&(\Delta_{m-1}-z)\overline{\mathscr{P}_{\psi^{Bog}_{\bold{j}_1,\dots,\bold{j}_{m-1}}}}+(z_m-\frac{C^{\perp}}{(\ln N)^{\frac{1}{2}}})\overline{\mathscr{P}_{\psi^{Bog}_{\bold{j}_1,\dots,\bold{j}_{m-1}}}}\label{-2}\\
&\geq &(1-\gamma)\Delta_{m-1}\overline{\mathscr{P}_{\psi^{Bog}_{\bold{j}_1,\dots,\bold{j}_{m-1}}}}\label{ineq-3}
%& &5555-\overline{\mathscr{P}_{\psi^{Bog}_{\bold{j}_1,\dots,\bold{j}_{m-1}}}}\,\tilde{\Gamma}^{Bog}_{\bold{j}_1,\dots,\bold{j}_m ;N,N}(z-\Delta_{m-1})\,\overline{\mathscr{P}_{\psi^{Bog}_{\bold{j}_1,\dots,\bold{j}_{m-1}}}}-{\color{red}\mathcal{\sigma}}\mathcal{O}(...)5555
\end{eqnarray}
where in the steps from (\ref{-4-bis}) to (\ref{-2}) and from (\ref{-2}) to (\ref{ineq-3}) we have used the following ingredients:
\begin{enumerate}
\item
$\check{\mathcal{G}}_{\bold{j}_{m}\,;\,N-2,N-2}(z)$ is nondecreasing (see Remark \ref{nondecreasing})  in the given range of $z$ , 
%and $$-\Delta_{m-1}+\frac{C_{\bold{j}_m}}{\sqrt{N}}<-\frac{\Delta_{m-1}}{2}$$ due to the assumption in (\ref{def-cjm}), 
therefore
\begin{eqnarray}
& &-(1-\frac{1}{N})\frac{\phi_{\bold{j}_{m}}}{2\epsilon_{\bold{j}_m}+2-\frac{4}{N}-\frac{z-\Delta_{m-1}+\frac{U_{\bold{j}_m}}{\sqrt{N}}}{\phi_{\bold{j}_{m}}}}\,\check{\mathcal{G}}_{\bold{j}_{m}\,;\,N-2,N-2}(z-\Delta_{m-1}+\frac{U_{\bold{j}_m}}{\sqrt{N}}))\quad\\
%& &-\frac{\phi_{\bold{j}_{m}}}{2\epsilon_{\bold{j}_m}+2-\frac{z-\Delta_{m-1}+\frac{C_{\bold{j}_m}}{\sqrt{N}}}{\phi_{\bold{j}_{m}}}}\check{\mathcal{G}}_{\bold{j}_{m}\,;\,N-2,N-2}(z-\Delta_{m-1}+\frac{C_{\bold{j}_m}}{\sqrt{N}})\quad\quad\quad\quad\\
&\geq &-(1-\frac{1}{N})\frac{\phi_{\bold{j}_{m}}}{2\epsilon_{\bold{j}_m}+2-\frac{4}{N}-\frac{z_m}{\phi_{\bold{j}_{m}}}}\check{\mathcal{G}}_{\bold{j}_{m}\,;\,N-2,N-2}(z_m)\quad
\end{eqnarray}
because $z-\Delta_{m-1}+\frac{U_{\bold{j}_m}}{\sqrt{N}}<z-\frac{\Delta_{m-1}}{2}< z_m$ (see (\ref{def-cjm-new}) and (\ref{interval}));
\item By definition
\begin{equation}
f_{\bold{j}_m}(z):=-z-(1-\frac{1}{N})\frac{\phi_{\bold{j}_{m}}}{2\epsilon_{\bold{j}_m}+2-\frac{4}{N}-\frac{z}{\phi_{\bold{j}_{m}}}}\check{\mathcal{G}}_{\bold{j}_{m}\,;\,N-2,N-2}(z)
\end{equation} 
and $z_m$ solves the fixed point equation $f_{\bold{j}_m}(z)=0$, therefore
\begin{equation}
-(1-\frac{1}{N})\frac{\phi_{\bold{j}_{m}}}{2\epsilon_{\bold{j}_m}+2-\frac{4}{N}-\frac{z_m}{\phi_{\bold{j}_{m}}}}\check{\mathcal{G}}_{\bold{j}_{m}\,;\,N-2,N-2}(z_m)=z_m\,;
\end{equation}
\item Due to the assumption in (\ref{assumption-gap-new}),
\begin{equation}
\overline{\mathscr{P}_{\psi^{Bog}_{\bold{j}_1,\dots,\bold{j}_{m-1}}}}(\hat{H}^{Bog}_{\bold{j}_1, \dots,\bold{j}_{m-1}}+\sum_{\bold{j}\neq \{\pm\bold{j}_1, \dots, \pm\bold{j}_{m}\}}(k_{\bold{j}})^2a^*_{\bold{j}}a_{\bold{j}}-z_{\bold{j}_1,\dots,\bold{j}_{m-1}})\overline{\mathscr{P}_{\psi^{Bog}_{\bold{j}_1,\dots,\bold{j}_{m-1}}}}\geq \Delta_{m-1}\overline{\mathscr{P}_{\psi^{Bog}_{\bold{j}_1,\dots,\bold{j}_{m-1}}}}\,;
\end{equation}
due to (\ref{bound-z-0-new}),  $-z+z_m -\frac{C^{\perp}}{(\ln N)^{\frac{1}{2}}}\geq -\frac{\Delta_{m-1}}{2}$.
\end{enumerate}
In oder to prove the key inequality in (\ref{primero-2-new}),  in Lemma \ref{invertibility} we implement an $h$-dependent truncation\footnote{Following the argument of  \emph{\underline{Corollary  5.9} of \cite{Pi1}} the second term in (\ref{5.22-bis}) is $\mathcal{O}(\frac{1}{\sqrt{\epsilon_{\bold{j}_m}}}\,(\frac{1}{1+c\sqrt{\epsilon_{\bold{j}_m}}})^{h})$. However, in the present paper the multiplicative constants may depend on the size of the box and the details of the potential.}  of the sum over $l_{N-2}$ and of $\Gamma^{Bog}_{\bold{j}_1,\dots,\bold{j}_m\,;\,N-2,N-2}(z+z^{Bog}_{\bold{j}_1,\dots,\bold{j}_{m-1}})$,
\begin{eqnarray}
& &(\ref{GammaPerp})\\
&=&\|\overline{\mathscr{P}_{\psi^{Bog}_{\bold{j}_1,\dots,\bold{j}_{m-1}}}}W_{\bold{j}_m}\,R^{Bog}_{\bold{j}_1,\dots,\bold{j}_m\,;\,N-2,N-2}(z+z^{Bog}_{\bold{j}_1,\dots,\bold{j}_{m-1}})\times \label{express-new}\\
& &\quad\quad \times \sum_{l_{N-2}=0}^{h-1}\Big[[\Gamma^{Bog}_{\bold{j}_1,\dots,\bold{j}_m\,;\,N-2,N-2}(z+z^{Bog}_{\bold{j}_1,\dots,\bold{j}_{m-1}})]_{\tau_h}\,R^{Bog}_{\bold{j}_1,\dots,\bold{j}_m\,;\,\,N-2,N-2}(z+z^{Bog}_{\bold{j}_1,\dots,\bold{j}_{m-1}})\Big]^{l_{N-2}}W^*_{\bold{j}_m}\,\overline{\mathscr{P}_{\psi^{Bog}_{\bold{j}_1,\dots,\bold{j}_{m-1}}}}\|\nonumber \label{step-3objiects}\\
& &+\mathcal{O}((\frac{4}{5})^{h})+\mathcal{O}(\,(\frac{1}{1+c\sqrt{\epsilon_{\bold{j}_m}}})^h)\,\label{5.22-bis}
\end{eqnarray}
where $[\Gamma^{Bog}_{\bold{j}_1,\dots,\bold{j}_m\,;\,N-2,N-2}(z+z^{Bog}_{\bold{j}_1,\dots,\bold{j}_{m-1}})]_{\tau_h}$ is defined in Lemma \ref{main-relations}; see (\ref{identity-Gammas}). This truncation relies on the results of Proposition \ref{lemma-expansion-proof-0}.
%\begin{remark}\label{remark-4.5}
Next, we exploit the following mechanism:
\\

\noindent
\emph{
For $h$ sufficiently ``small"  with respect to $N$ all the resolvents in expression (\ref{express-new}) are sufficiently ``close" to $\overline{\mathscr{P}_{\psi^{Bog}_{\bold{j}_1,\dots,\bold{j}_{m-1}}}}$ so that up to a small error  the operator
\begin{equation}
\sum_{\bold{j}\in \mathbb{Z}^d \setminus \{\pm\bold{j}_1, \dots, \pm\bold{j}_{m}\}}(k_{\bold{j}})^2a^*_{\bold{j}}a_{\bold{j}}+\hat{H}^{Bog}_{\bold{j}_1,\dots,\bold{j}_{m-1}}-z^{Bog}_{\bold{j}_1,\dots,\bold{j}_{m-1}}
\end{equation}
contained in each resolvent is bounded below by
\begin{equation}
%infspec \overline{\mathscr{P}_{\psi^{Bog}_{\bold{j}_1,\dots,\bold{j}_{m-1}}}}(\hat{H}^{Bog}_{\bold{j}_1,\dots,\bold{j}_{m-1}}-z^{Bog}_{\bold{j}_1,\dots,\bold{j}_{m-1}})\overline{\mathscr{P}_{\psi^{Bog}_{\bold{j}_1,\dots,\bold{j}_{m-1}}}}\geq 
\Delta_{m-1} \overline{\mathscr{P}_{\psi^{Bog}_{\bold{j}_1,\dots,\bold{j}_{m-1}}}}
\end{equation}
in a sense specified in the detailed proof (Lemma \ref{invertibility}).  The implementation requires some lengthy technical steps; see $\bf{a)}$, $\bf{b}$, $\bf{c)}$, $\bf{d)}$, $\bf{e)}$, and  $\bf{f)}$ in Lemma \ref{invertibility}.}
\\
\\

The construction of the Feshbach-Schur flow and of the ground state for each Hamiltonian $H^{Bog}_{\bold{j}_1,\dots,\bold{j}_m}$, $1\leq m\leq M$, is the content of Theorem \ref{induction-many-modes} below that concerns five properties proven by induction. For the convenience of the reader, we list the five properties and outline the structure of the proof. 

\noindent
Property 1. ensures the construction of the Feshbach-Schur flow $\{\mathscr{K}^{Bog\,(i)}_{\bold{j}_1,\,\dots, \bold{j}_{m}}(z+z^{Bog}_{\bold{j}_1,\dots,\bold{j}_{m-1}})\,|\,0\leq  i\,\, \text{and even}\,\}$ up to $i=N$. 

\noindent
Property 2. provides the existence of the unique solution of the fixed point equation associated with the Feshbach-Schur Hamiltonian $\mathscr{K}^{Bog\,(N)}_{\bold{j}_1,\,\dots, \bold{j}_{m}}(z+z^{Bog}_{\bold{j}_1,\dots,\bold{j}_{m-1}})$  defined with Property 1. Hence, the (non-degenerate) ground state energy of $H^{Bog}_{\bold{j}_1, \dots,\bold{j}_{m}}$ and the corresponding ground state vector are determined.

\noindent
Property 3. is concerned with the spectral gap of the Hamiltonian $$ \Big(\hat{H}^{Bog}_{\bold{j}_1, \dots,\bold{j}_{m}}+\sum_{\bold{j}\in \mathbb{Z}^d\setminus \{\pm\bold{j}_1, \dots, \pm\bold{j}_{m+1}\}}(k_{\bold{j}})^2a^*_{\bold{j}}a_{\bold{j}}\Big)\upharpoonright_{Q^{(>N-1)}_{\bold{j}_{m+1}}\mathcal{F}^N}\,.$$ The gap condition at step $m$ provided by Property 3. is needed to apply Lemma \ref{invertibility-new} and derive Property 1. and 2. at step $m+1$. 

\noindent
Property 4. provides the information on $$\text{infspec}\,[\sum_{l=1}^{m}\hat{H}^{Bog}_{\bold{j}_l }]-z^{Bog}_{\bold{j}_1,\dots ,\bold{j}_{m}}$$ that is assumed in Corollary \ref{main-lemma-H} and Lemma \ref{invertibility-new}. Then,  Property 4. is needed to derive Property 1. at step $m+1$. Thanks to this input, the operator norm estimate (\ref{estimate-main-lemma-H}) in Corollary \ref{main-lemma-H} can be derived as if the modes $\pm\bold{j}_1,\dots,\pm\bold{j}_{m-1}$ were absent.

\noindent
Property 5. provides the bound on the expectation value of the number operator $\mathcal{N}_+:=\sum_{\bold{j}\in\mathbb{Z}^d\setminus \{\bold{0}\}} a_{\bold{j}}^{*}a_{\bold{j}}$ in the ground state of $H^{Bog}_{\bold{j}_1,\dots,\bold{j}_{m}}$. This information is needed to control the fixed point equation associated with $H^{Bog}_{\bold{j}_1,\dots,\bold{j}_{m+1}}$.

\begin{thm}\label{induction-many-modes}
Let $max_{1\leq m \leq M}\,\epsilon_{\bold{j}_m}$ be sufficiently small and $N$ sufficiently large. 
%Set $\Delta_0\equiv \min\, \Big\{\epsilon_{\bold{j}}\,|\,\bold{j}\in \mathbb{Z}^3 \Big\}\,$.  
Then the following properties hold true for all $1\leq m \leq M$:
\begin{enumerate}
\item[1)]The Feshbach-Schur Hamiltonian $\mathscr{K}_{\bold{j}_1,\dots,\bold{j}_m}^{Bog\,(N)}(z+z^{Bog}_{\bold{j}_1,\dots,\bold{j}_{m-1}})$ in (\ref{final-fesh-step-m})-(\ref{def-f}) is well defined for \begin{equation}
 z \leq z_{m}+\gamma\Delta_{m-1} -\frac{C^{\perp}}{(\ln N)^{\frac{1}{2}}} < E^{Bog}_{\bold{j}_m}+ \frac{1}{2}\sqrt{\epsilon_{\bold{j}_m}}\phi_{\bold{j}_m}\sqrt{\epsilon_{\bold{j}_m}^2+2\epsilon_{\bold{j}_m}} \,(<0)\quad,\quad \gamma=\frac{1}{2},\label{inter-def}
 \end{equation}
 where:
 \begin{itemize}
 \item $z^{Bog}_{\bold{j}_1,\dots,\bold{j}_{m-1}}$ is the ground state energy of $H^{Bog}_{\bold{j}_1,\dots,\bold{j}_{m-1}}$ and is defined recursively in point 2) below;
 \item
   $z_m$ is the ground state energy of $H^{Bog}_{\bold{j}_m}$;
   \item $\Delta_{m-1}$  is defined recursively by $\Delta_{0}:=\min\, \Big\{(k_{\bold{j}})^2\,|\,\bold{j}\in \mathbb{Z}^d \setminus \{\bold{0}\}\Big\}$ and
  \begin{eqnarray}
\Delta_m&:= &\gamma\Delta_{m-1}-\frac{C^{\perp}}{(\ln N)^{\frac{1}{2}}}-(\frac{2}{\gamma})^m\frac{C_{III}}{(\ln N)^{\frac{1}{4}}}\,\,(>0) \,\quad \quad\quad\quad \label{Deltagap}
%&=&\gamma \Delta_{m-1}-{\color{red}(\frac{2}{\gamma})^m}\frac{C_{III}}{(\ln N)^{\frac{1}{4}}}
 \end{eqnarray}
 with $C_{III}:=C_{I}+\frac{C_{II}^2}{(1-\gamma)\Delta_{0}}$ where the constants $C_{I},C_{II}$ are introduced in Lemma \ref{main-relations-new}, and $C^{\perp}$ is introduced in Lemma \ref{invertibility-new}.
\end{itemize} 
 \item[2)]
 For $z$ as in (\ref{inter-def}), there exists a unique value $z^{(m)}$ such that $$f^{Bog}_{\bold{j}_1,\dots,\bold{j}_m}(z+z^{Bog}_{\bold{j}_1,\dots,\bold{j}_{m-1}})|_{z\equiv z^{(m)}}=0\,.$$  The inequality 
 \begin{equation}\label{difference-tra-z}
 |z^{(m)}-z_m|\leq (\frac{2}{\gamma})^m\frac{C_{III}}{(\ln N)^{\frac{1}{4}}}
 \end{equation}
  holds true.  
 
 \noindent
 The Hamiltonian $H^{Bog}_{\bold{j}_1,\dots,\bold{j}_{m}}$ has (nondegenerate) ground state energy $$z^{Bog}_{\bold{j}_1,\dots,\bold{j}_{m}}:=z^{Bog}_{\bold{j}_1,\dots,\bold{j}_{m-1}}+z^{(m)}$$ where $z^{Bog}_{\bold{j}_1,\dots,\bold{j}_{m-1}}|_{m=1}\equiv 0$. The corresponding eigenvector is given in  (\ref{gs-Hm-start})-(\ref{gs-Hm-fin}).
 
\item[3)] 
 The spectral gap of the two operators
\begin{equation}
H^{Bog}_{\bold{j}_1,\dots,\bold{j}_{m}}\quad, \quad \Big(\hat{H}^{Bog}_{\bold{j}_1, \dots,\bold{j}_{m}}+\sum_{\bold{j}\in \mathbb{Z}^d\setminus \{\pm\bold{j}_1, \dots, \pm\bold{j}_{m+1}\}}(k_{\bold{j}})^2a^*_{\bold{j}}a_{\bold{j}}\Big)\upharpoonright_{Q^{(>N-1)}_{\bold{j}_{m+1}}\mathcal{F}^N}\,
\end{equation}
 above the (common) ground state energy $z^{Bog}_{\bold{j}_1,\dots,\bold{j}_{m}}$  is larger or equal to $\Delta_{m}$.

%\item The ground state $\psi^{Bog}_{\bold{j}_1,\bold{j}_2,\dots,\bold{j}_{m}}$ fulfills the expansion:
%\begin{equation}
%\psi^{Bog}_{\bold{j}_1,\bold{j}_2,\dots,\bold{j}_{m}}=\sum_{l=2}^{m} T_{m}\dots T_{l+1}S_{l}\,\psi^{Bog}_{\bold{j}_1,\bold{j}_2,\dots,\bold{j}_{l-1}}+T_{m}\dots T_0\eta
%\end{equation}
%\begin{equation}
%S_n:=\quad \|S_n\|\leq \frac{C^n}{N}
%\end{equation}
\item[4)]
The lower bound 
\begin{equation}\label{lower}
\text{infspec}\,[\sum_{l=1}^{m}\hat{H}^{Bog}_{\bold{j}_l }]-z^{Bog}_{\bold{j}_1,\dots ,\bold{j}_{m}}\geq -\frac{m}{(\ln N)^{\frac{1}{8}}}\,
\end{equation}
holds true.
 \item[5)] For $\tilde{C}_m=\frac{\sum_{l=1}^{m}\phi_{\bold{j}_l}}{\Delta_0}$, the upper bound
 \begin{equation}
\langle \frac{\psi^{Bog}_{\bold{j}_1,\dots,\bold{j}_{m}}}{\|\psi^{Bog}_{\bold{j}_1,\dots,\bold{j}_{m}}\|}\,,\,\sum_{\bold{j}\in\mathbb{Z}^d\setminus \{\bold{0}\}} a_{\bold{j}}^{*}a_{\bold{j}}\,\frac{\psi^{Bog}_{\bold{j}_1,\dots,\bold{j}_{m}}}{\|\psi^{Bog}_{\bold{j}_1,\dots,\bold{j}_{m}}\|}\rangle \leq \tilde{C}_m\,
\end{equation}
holds true.
 \end{enumerate}
\end{thm}

\noindent
\emph{Proof}

For $m=1$, by construction $z_1\equiv z^{Bog}_{\bold{j}_1}$ and Property 1) and 2) have been proven in \emph{\underline{Corollary 4.6} of \cite{Pi1}} that is reported in Section \ref{groundstate}. As far as Property 3)  is concerned, we observe that the vector $\psi^{Bog}_{\bold{j}_1}$ belongs to  $Q^{(>N-1)}_{\bold{j}_2}\mathcal{F}^N$ and is also eigenvector of $\Big(\hat{H}^{Bog}_{\bold{j}_1}+\sum_{\bold{j}\in \mathbb{Z}^d\setminus \{\pm\bold{j}_1, \pm\bold{j}_{2}\}}(k_{\bold{j}})^2a^*_{\bold{j}}a_{\bold{j}}\Big)\upharpoonright_{Q^{(>N-1)}_{\bold{j}_2}\mathcal{F}^N}$. Thus,  for any $\psi\in Q^{(>N-1)}_{\bold{j}_2}\mathcal{F}^N $, $\|\psi\|=1$,  that is orthogonal to $\psi^{Bog}_{\bold{j}_1}$ we can derive
\begin{eqnarray}
& &\langle \psi\,,\,\Big(\hat{H}^{Bog}_{\bold{j}_1}+\sum_{\bold{j}\in \mathbb{Z}^d \setminus \{\pm\bold{j}_1, \pm\bold{j}_{2}\}}(k_{\bold{j}})^2a^*_{\bold{j}}a_{\bold{j}}\Big)\psi \rangle\\
&=&\langle \psi\,,\,\Big(\hat{H}^{Bog}_{\bold{j}_1}+\sum_{\bold{j}\in \mathbb{Z}^d \setminus \{\pm\bold{j}_1\}}(k_{\bold{j}})^2a^*_{\bold{j}}a_{\bold{j}}\Big)\psi \rangle\\
&=&\langle \psi\,,\,H^{Bog}_{\bold{j}_1}\,\psi \rangle\\
&\geq &z^{Bog}_{\bold{j}_1}+\Delta_1\,
\end{eqnarray}
where the last step follows from the result in \emph{\underline{Corollary 4.6} of \cite{Pi1}}.

\noindent
Concerning Property 4), we recall that $\psi^{Bog}_{{\bold{j}_1}}$ is also eigenstate of $\hat{H}^{Bog}_{{\bold{j}_1}}$  with the same eigenvalue $z^{Bog}_1$. Furthermore, we can restrict $\hat{H}^{Bog}_{{\bold{j}_1}}$ to any subspace $[\mathcal{F}^N]_j$ of $\mathcal{F}^N$ with a fixed number of particles, $j$,  in the modes different from $\pm\bold{j}_1$ and $\bold{0}$. We assume that  $j$ is less than $N-2$ (if $j=N-2$ the property is trivially proven) and even, but the final results hold also for $j$ odd\footnote{In this case we re-define the first couple of projections $\overline{\hat{\mathscr{P}}}^{(j)}:=\hat{Q}_{\bold{j}_1}^{(j)}:=Q_{\bold{j}_1}^{(j)}\charf_{[\mathcal{F}^N]_j}\quad\text{and}\quad\hat{\mathscr{P}}^{(j)}:=\hat{Q}_{\bold{j}_1}^{(>j)}:=\charf_{[\mathcal{F}^N]_j}-\hat{Q}^{(j)}_{\bold{j}_1}\,,$ where $Q_{\bold{j}_1}^{(j)}$ is the projection onto the subspace of $\mathcal{F}^N$ with $N-j$ particles in the modes $\pm \bold{j}_1$. Then, we proceed for $i\geq j+1$ (and even) up to $N-2$ according to the definitions in (\ref{hat-proj}). }. By adapting Corollary \ref{main-lemma-H} and Theorem \ref{Feshbach-Hbog} to the Hamiltonian $\hat{H}^{Bog}_{\bold{j}_1 }$, the Feshbach-Schur flow can be implemented in the same way it has been done for $H^{Bog}_{{\bold{j}_1}}-w$ starting from the projections 
\begin{equation}\overline{\hat{\mathscr{P}}}^{(j)}:=\hat{Q}_{\bold{j}_1}^{(j,j+1)}:=Q_{\bold{j}_1}^{(j,j+1)}\charf_{[\mathcal{F}^N]_j}\quad\text{and}\quad\hat{\mathscr{P}}^{(j)}:=\hat{Q}_{\bold{j}_1}^{(>j+1)}:=\charf_{[\mathcal{F}^N]_j}-\hat{Q}^{(j,j+1)}_{\bold{j}_1}\,,
\end{equation}
and proceeding for $i>j$ (and even) up to $i=N-2$ with the definitions
\begin{equation}\label{hat-proj}
\overline{\hat{\mathscr{P}}}^{(i)}:=\hat{Q}_{\bold{j}_1}^{(i,i+1)}:=Q_{\bold{j}_1}^{(i,i+1)}\hat{Q}_{\bold{j}_1}^{(>i-1)}\quad\text{and}\quad\hat{\mathscr{P}}^{(i)}:=\hat{Q}_{\bold{j}_1}^{(>i+1)}:=\hat{Q}_{\bold{j}_1}^{(>i-1)}-\hat{Q}^{(i,i+1)}_{\bold{j}_1}\,.
\end{equation}
We call $\hat{\mathscr{K}}^{Bog\,(i)}_{\bold{j}_1}(w)$ the Feshbach-Schur Hamiltonians so defined. Note that $\hat{Q}_{\bold{j}_1}^{(>N-1)}$  is the projection onto the subspace of states with $N-j$ particles in the zero mode state and $j$ particles in the modes different from $\pm\bold{j}_1$ and $\bold{0}$. Setting $w\equiv z$ with $z$ in the range defined in (\ref{inter-def}),  we obtain
\begin{eqnarray}
& &\hat{\mathscr{K}}^{Bog\,(N-2)}_{\bold{j}_1}(z)\\
%&=&\mathscr{P}_{\eta}(H^{Bog}_{\bold{j}_*}-z)\mathscr{P}_{\eta}\\
%& &-\mathscr{P}_{\eta}W_{\bold{j}_*}\,Q^{(N-2)}_{\bold{j}_*}\,R^{Bog}_{\bold{j}_*\,;\,N-2,N-2}(z)\sum_{l_{N-2}=0}^{\infty}[\Gamma^{Bog}_{\bold{j}_*\,;\,N-2,N-2}(z) R^{Bog}_{\bold{j}_*\,;\,N-2,N-2}(z)]^{l_{N-2}}\, Q^{(N-2)}_{\bold{j}_*}W^*_{\bold{j}_*}\mathscr{P}_{\eta}\nonumber\\
&=&-z\hat{Q}_{\bold{j}_1}^{(>N-1)}\\
& &-\hat{Q}_{\bold{j}_1}^{(>N-1)}W_{\bold{j}_1}\,\hat{R}^{Bog}_{\bold{j}_1\,;\,N-2,N-2}(z)\sum_{l_{N-2}=0}^{\infty}[\hat{\Gamma}^{Bog}_{\bold{j}_1\,;\,N-2,N-2}(z) \hat{R}^{Bog}_{\bold{j}_1\,;\,N-2,N-2}(z)]^{l_{N-2}}\, W^*_{\bold{j}_1}\hat{Q}_{\bold{j}_1}^{(>N-1)}\quad\quad\quad\label{restriction},
%& &-Q^{(>N-1)}W\,Q^{(N-2)}\,R^{Bog}_{N-2,N-2}(z)\sum_{l_{N-2}=0}^{\infty}[\Gamma^{Bog}_{N-2,N-2}(z) R^{Bog}_{N-2,N-2}(z)]^{l_{N-2}}\, Q^{(N-2)}W^*Q^{(>N-1)}\quad \quad\quad\quad\label{first-term-last-Bog}\\
%& &-\mathscr{V}^{(N-1)}(z)\,R_{N-1,N-1}(z)\sum_{l_{N-1}=0}^{\infty}[\Gamma_{N-1,N-1}(z) R_{N-1,N-1}(z)]^{l_{N-1}}\,(\mathscr{V}^{(N-1)}(z) )^*\quad\quad \label{last-term-last-Bog}
\end{eqnarray}
where $\hat{R}^{Bog}_{\bold{j}_1\,;\,i,i}(z)$ and $\hat{\Gamma}^{Bog}_{\bold{j}_1\,;\,i,i}(z)$ have the same definition of $R^{Bog}_{\bold{j}_1\,;\,i,i}(z)$ and $\Gamma^{Bog}_{\bold{j}_1\,;\,i,i}(z)$ but in terms of $\hat{H}^{Bog}_{{\bold{j}_1}}$ and of the new projections.  
%As last couple of projections we consider $\hat{\mathscr{P}}^{(N-j)}:=|\eta_j\rangle \langle \eta_j|$ and $\overline{\hat{\mathscr{P}}}^{(N-j)}:=\hat{Q}_{\bold{j}_1}^{(>N-j-1)}-|\eta_j\rangle \langle \eta_j|$.  
Next, we observe that since $\hat{Q}_{\bold{j}_1}^{(>N-1)}$ projects onto a subspace of states with no particles in the modes $\pm \bold{j}_1$  the operators 
\begin{equation}
\Big[\hat{R}^{Bog}_{\bold{j}_1\,;\,i,i}(z)\Big]^{\frac{1}{2}}\,W_{\bold{j}_1\,;\,i,i-2}\,\Big[\hat{R}^{Bog}_{\bold{j}_1\,;\,i-2,i-2}(z)\Big]^{\frac{1}{2}}
\end{equation}
appearing in (\ref{restriction}) and entering the definition of $\hat{\Gamma}^{Bog}_{\bold{j}_1\,;\,N-2,N-2}(z)$  are in fact restricted to states with an even number of particles in the modes $\pm \bold{j}_1$.
Then, we use a procedure analogous to  Lemma \ref{invertibility} by implementing {\bf{Steps}} {\bf{a)}}, {\bf{b)}}, {\bf{d)}} , and {\bf{e)}}; see Remark \ref{invert}.
%\begin{equation}
%\|\Big[R^{Bog}_{\bold{j}_1\,;\,i,i}(z)\Big]^{\frac{1}{2}}\,W_{\bold{j}_1\,;\,i,i-2}\,\Big[R^{Bog}_{\bold{j}_1\,;\,i-2,i-2}(z)\Big]^{\frac{1}{2}}\|^2\leq \mathcal{W}_{\bold{j}_1\,;\,i,i-2}(z)\mathcal{W}^*_{\bold{j}_1\,;\,i-2,i}(z)
%\end{equation}
%where the R-H-S is defined in (\ref{def-WW*}) (for $\bold{j}_*\equiv \bold{j}_1$) and enters the definition of $\check{\mathcal{G}}_{\bold{j}_1\,;\,N-2,N-2}(z)$ through the formula in (\ref{in-formula-G}).
We  conclude that for $N$ sufficiently large
\begin{eqnarray}
& &\|\hat{Q}_{\bold{j}_1}^{(>N-1)}W_{\bold{j}_1}\,\hat{R}^{Bog}_{\bold{j}_1\,;\,N-2,N-2}(z)\sum_{l_{N-2}=0}^{\infty}[\hat{\Gamma}^{Bog}_{\bold{j}_1\,;\,N-2,N-2}(z) \hat{R}^{Bog}_{\bold{j}_1\,;\,N-2,N-2}(z)]^{l_{N-2}}\, W^*_{\bold{j}_1}\hat{Q}_{\bold{j}_1}^{(>N-1)}\|\nonumber\\
&\leq&(1-\frac{1}{N}) \frac{\phi_{\bold{j}_{1}}}{2\epsilon_{\bold{j}_1}+2-\frac{4}{N}-\frac{z+\frac{U_{\bold{j}_1}}{\sqrt{N}}}{\phi_{\bold{j}_{1}}}}\check{\mathcal{G}}_{\bold{j}_1\,;\,N-2,N-2}(z+\frac{U_{\bold{j}_1}}{\sqrt{N}})+\mathcal{O}(\frac{1}{(\ln N)^{\frac{1}{2}}})\,.\label{ineq-inf-Hcapp-0-0}
\end{eqnarray}
Note that, for $N$ sufficiently large and $z$ in the range  (\ref{inter-def}), we have $$z+\frac{U_{\bold{j}_1}}{\sqrt{N}}+\frac{1}{2(\ln N)^{\frac{1}{8}}}<E^{Bog}_{\bold{j}_m}+\sqrt{\epsilon_{\bold{j}_m}}\phi_{\bold{j}_m}\sqrt{\epsilon_{\bold{j}_m}^2+2\epsilon_{\bold{j}_m}}\,,$$ therefore $\check{\mathcal{G}}_{\bold{j}_1\,;\,N-2,N-2}(z+\frac{U_{\bold{j}_1}}{\sqrt{N}})$  and $\check{\mathcal{G}}_{\bold{j}_1\,;\,N-2,N-2}(z+\frac{U_{\bold{j}_1}}{\sqrt{N}}+\frac{1}{2(\ln N)^{\frac{1}{8}}})$ are well defined; see (\ref{function-f})-(\ref{fin-formula-G}). Since $\check{\mathcal{G}}_{\bold{j}_1\,;\,N-2,N-2}(z)$ is nondecreasing  (see Remark \ref{nondecreasing}) the inequality in (\ref{ineq-inf-Hcapp-0-0}) implies
\begin{eqnarray}
& &\|\hat{Q}_{\bold{j}_1}^{(>N-1)}W_{\bold{j}_1}\,\hat{R}^{Bog}_{\bold{j}_1\,;\,N-2,N-2}(z)\sum_{l_{N-2}=0}^{\infty}[\hat{\Gamma}^{Bog}_{\bold{j}_1\,;\,N-2,N-2}(z) \hat{R}^{Bog}_{\bold{j}_1\,;\,N-2,N-2}(z)]^{l_{N-2}}\, W^*_{\bold{j}_1}\hat{Q}_{\bold{j}_1}^{(>N-1)}\|\nonumber\\
&\leq& (1-\frac{1}{N})\frac{\phi_{\bold{j}_{1}}}{2\epsilon_{\bold{j}_1}+2-\frac{4}{N}-\frac{z+\frac{U_{\bold{j}_1}}{\sqrt{N}}+\frac{1}{2(\ln N)^{\frac{1}{8}}}}{\phi_{\bold{j}_{1}}}}\check{\mathcal{G}}_{\bold{j}_1\,;\,N-2,N-2}(z+\frac{U_{\bold{j}_1}}{\sqrt{N}}+\frac{1}{2(\ln N)^{\frac{1}{8}}})+\mathcal{O}(\frac{1}{(\ln N)^{\frac{1}{2}}})\quad\quad\quad\quad\label{ineq-inf-Hcapp-0}
\end{eqnarray}
and, consequently,
\begin{eqnarray}
& &-z-\|\hat{Q}_{\bold{j}_1}^{(>N-j-1)}W_{\bold{j}_1}\,\hat{R}^{Bog}_{\bold{j}_1\,;\,N-2,N-2}(z)\sum_{l_{N-2}=0}^{\infty}[\hat{\Gamma}^{Bog}_{\bold{j}_1\,;\,N-2,N-2}(z) \hat{R}^{Bog}_{\bold{j}_1\,;\,N-2,N-2}(z)]^{l_{N-2}}\, W^*_{\bold{j}_1}\hat{Q}_{\bold{j}_1}^{(>N-1)}\| \nonumber\\
%&\geq &-z-\frac{\phi_{\bold{j}_{1}}}{2\epsilon_{\bold{j}_1}+2-\frac{z+\frac{\phi_{\bold{j}_1}N^{\mu}}{N}}{\phi_{\bold{j}_{1}}}}\check{\mathcal{G}}_{\bold{j}_1\,;\,N-2,N-2}(z+\frac{\phi_{\bold{j}_1}N^{\mu}}{N})+\mathcal{O}(\frac{1}{(\ln N)^{\frac{1}{2}}})\label{step-inf-pen}\\
&\geq&\frac{U_{\bold{j}_1}}{\sqrt{N}}+\frac{1}{2(\ln N)^{\frac{1}{8}}} -(z+\frac{U_{\bold{j}_1}}{\sqrt{N}}+\frac{1}{2(\ln N)^{\frac{1}{8}}})\label{step-inf-ult-0}\\
& &-(1-\frac{1}{N})\frac{\phi_{\bold{j}_{1}}}{2\epsilon_{\bold{j}_1}+2-\frac{4}{N}-\frac{z+\frac{U_{\bold{j}_1}}{\sqrt{N}}+\frac{1}{2(\ln N)^{\frac{1}{8}}}}{\phi_{\bold{j}_{1}}}}\check{\mathcal{G}}_{\bold{j}_1\,;\,N-2,N-2}(z+\frac{U_{\bold{j}_1}}{\sqrt{N}}+\frac{1}{2(\ln N)^{\frac{1}{8}}})+\mathcal{O}(\frac{1}{(\ln N)^{\frac{1}{2}}})\,.\quad\quad\quad\quad\,.\label{step-inf-ult}
\end{eqnarray}
%where in the step from (\ref{step-inf-pen}) to (\ref{step-inf-ult}) we have exploited that $\frac{\phi_{\bold{j}_{1}}}{2\epsilon_{\bold{j}_1}+2-\frac{w}{\phi_{\bold{j}_{1}}}}\check{\mathcal{G}}_{\bold{j}_1\,;\,N-2,N-2}(w)$ is non decreasing in $w$ (see \emph{\underline{Remark 4.1} of \cite{Pi1}}). 
Since, due to Remark \ref{nondecreasing}, the derivative with respect to $z$ of
\begin{equation}
z+ (1-\frac{1}{N})\frac{\phi_{\bold{j}_{1}}}{2\epsilon_{\bold{j}_1}+2-\frac{4}{N}-\frac{z}{\phi_{\bold{j}_{1}}}}\check{\mathcal{G}}_{\bold{j}_1\,;\,N-2,N-2}(z)\equiv -f^{Bog}_{\bold{j}_1}(z)
\end{equation}
is not smaller than $1$,  for $N$ large enough and  $$z< z^{Bog}_1-\frac{1}{(\ln N)^{\frac{1}{8}}}< z^{Bog}_1-\frac{U_{\bold{j}_1}}{\sqrt{N}}-\frac{1}{2(\ln N)^{\frac{1}{8}}}\quad \Rightarrow \quad z+\frac{U_{\bold{j}_1}}{\sqrt{N}}+\frac{1}{2(\ln N)^{\frac{1}{8}}}< z^{Bog}_1 $$ we deduce that
\begin{eqnarray}
& & -(z+\frac{U_{\bold{j}_1}}{\sqrt{N}}+\frac{1}{2(\ln N)^{\frac{1}{8}}})\\
& &-(1-\frac{1}{N})\frac{\phi_{\bold{j}_{1}}}{2\epsilon_{\bold{j}_1}+2-\frac{4}{N}-\frac{z+\frac{U_{\bold{j}_1}}{\sqrt{N}}+\frac{1}{2(\ln N)^{\frac{1}{8}}}}{\phi_{\bold{j}_{1}}}}\check{\mathcal{G}}_{\bold{j}_1\,;\,N-2,N-2}(z+\frac{U_{\bold{j}_1}}{\sqrt{N}}+\frac{1}{2(\ln N)^{\frac{1}{8}}})\\
&\geq&f^{Bog}_{\bold{j}_1}(z^{Bog}_{\bold{j}_1})\\
&=&0\,.
\end{eqnarray}
Hence, $(\ref{step-inf-ult-0})+(\ref{step-inf-ult})>0$ and, consequently,  the operator $\hat{\mathscr{K}}^{Bog\,(N-2)}_{\bold{j}_1}(z)$ acting on $\hat{Q}_{\bold{j}_1}^{(>N-1)}\mathcal{F}^N$ is bounded invertible for $z< z^{Bog}_1-\frac{1}{(\ln N)^{\frac{1}{8}}}$.
Since this holds for any subspace $[\mathcal{F}^N]_j$ of $\mathcal{F}^N$, and due to the isospectrality of the Feshbach-Schur map, also $\hat{H}^{Bog}_{\bold{j}_1}-z$ is bounded invertible for $z < z^{Bog}_1-\frac{1}{(\ln N)^{\frac{1}{8}}}$. 
%But since $z^{Bog}_1$ is an eigenvalue of $\hat{H}^{Bog}_{\bold{j}_1}$, it is the ground state energy.

\noindent
 Property 5) follows from the identity below
 \begin{eqnarray}
\hat{H}^{Bog}_{\bold{j}_1} &=&\sum_{\bold{j}=\pm \bold{j}_1}k^2_{\bold{j}}a_{\bold{j}}^{*}a_{\bold{j}}+\Big\{\frac{\phi_{\bold{j}_1}}{N}(a^*_{\bold{0}}a_{\bold{j}_1}+a_{\bold{0}}a^*_{-\bold{j}_1})(a_{\bold{0}}a^*_{\bold{j}_1}+a^*_{\bold{0}}a_{-\bold{j}_1})-\frac{\phi_{\bold{j}_1}}{N}[a^*_{-\bold{j}_1}a_{-\bold{j}_1}+a^*_{\bold{0}}a_{\bold{0}}]\Big\} \,.\quad\quad\quad\quad \label{id-1}
%&=&\langle \frac{\psi^{Bog}_{\bold{j}_1}}{\|\psi^{Bog}_{\bold{j}_1}\|}\,,\,k^2_{\bold{j}_1}\sum_{\bold{j}=\pm \bold{j}_1}a_{\bold{j}}^{*}a_{\bold{j}}\,\frac{\psi^{Bog}_{\bold{j}_1}}{\|\psi^{Bog}_{\bold{j}_1}\|}\rangle\\
%& &+\langle \frac{\psi^{Bog}_{\bold{j}_1}}{\|\psi^{Bog}_{\bold{j}_1}\|}\,,\,\Big\{\frac{\phi_{\bold{j}_1}}{N}(a^*_{\bold{0}}a^*_{\bold{j}_1}+a_{\bold{0}}a_{-\bold{j}_1})(a^*_{\bold{0}}a^*_{-\bold{j}_1}+a_{\bold{0}}a_{\bold{j}_1})-\frac{\phi_{\bold{j}_1}}{N}[a^*_{-\bold{j}_1}a_{-\bold{j}_1}+a^*_{\bold{0}}a_{\bold{0}}+1]\Big\}\,\frac{\psi^{Bog}_{\bold{j}_1}}{\|\psi^{Bog}_{\bold{j}_1}\|}\rangle\quad \nonumber\\
%& &+\langle \frac{\psi^{Bog}_{\bold{j}_1}}{\|\psi^{Bog}_{\bold{j}_1}\|}\,,\,\Big\{\frac{\phi_{\bold{j}_1}}{N}(a^*_{\bold{0}}a_{\bold{j}_1}+a_{\bold{0}}a^*_{-\bold{j}_1})(a_{\bold{0}}a^*_{\bold{j}_1}+a^*_{\bold{0}}a_{-\bold{j}_1})-\frac{\phi_{\bold{j}_1}}{N}[a^*_{-\bold{j}_1}a_{-\bold{j}_1}+a^*_{\bold{0}}a_{\bold{0}}]\Big\}\,\frac{\psi^{Bog}_{\bold{j}_1}}{\|\psi^{Bog}_{\bold{j}_1}\|}\rangle\,.\quad\quad\quad
\end{eqnarray}
Indeed, since 
\begin{equation}\label{energy-est-in}
(a^*_{\bold{0}}a_{\bold{j}_1}+a_{\bold{0}}a^*_{-\bold{j}_1})(a_{\bold{0}}a^*_{\bold{j}_1}+a^*_{\bold{0}}a_{-\bold{j}_1})\geq 0
%(a^*_{\bold{0}}a^*_{\bold{j}_1}+a_{\bold{0}}a_{-\bold{j}_1})(a^*_{\bold{0}}a^*_{-\bold{j}_1}+a_{\bold{0}}a_{\bold{j}_1})\geq 0
\end{equation}
and
\begin{equation}\label{energy-est-in-0}
a^*_{-\bold{j}_1}a_{-\bold{j}_1}+a^*_{\bold{0}}a_{\bold{0}}\leq N\,,
\end{equation}
from 
\begin{eqnarray}
& &z^{Bog}_{\bold{j}_1}\\
&=&\langle \frac{\psi^{Bog}_{\bold{j}_1}}{\|\psi^{Bog}_{\bold{j}_1}\|}\,,\,k^2_{\bold{j}_1}\sum_{\bold{j}=\pm \bold{j}_1}a_{\bold{j}}^{*}a_{\bold{j}}\,\frac{\psi^{Bog}_{\bold{j}_1}}{\|\psi^{Bog}_{\bold{j}_1}\|}\rangle\\
%& &+\langle \frac{\psi^{Bog}_{\bold{j}_1}}{\|\psi^{Bog}_{\bold{j}_1}\|}\,,\,\Big\{\frac{\phi_{\bold{j}_1}}{N}(a^*_{\bold{0}}a^*_{\bold{j}_1}+a_{\bold{0}}a_{-\bold{j}_1})(a^*_{\bold{0}}a^*_{-\bold{j}_1}+a_{\bold{0}}a_{\bold{j}_1})-\frac{\phi_{\bold{j}_1}}{N}[a^*_{-\bold{j}_1}a_{-\bold{j}_1}+a^*_{\bold{0}}a_{\bold{0}}+1]\Big\}\,\frac{\psi^{Bog}_{\bold{j}_1}}{\|\psi^{Bog}_{\bold{j}_1}\|}\rangle\quad \nonumber\\
& &+\langle \frac{\psi^{Bog}_{\bold{j}_1}}{\|\psi^{Bog}_{\bold{j}_1}\|}\,,\,\Big\{\frac{\phi_{\bold{j}_1}}{N}(a^*_{\bold{0}}a_{\bold{j}_1}+a_{\bold{0}}a^*_{-\bold{j}_1})(a_{\bold{0}}a^*_{\bold{j}_1}+a^*_{\bold{0}}a_{-\bold{j}_1})-\frac{\phi_{\bold{j}_1}}{N}[a^*_{-\bold{j}_1}a_{-\bold{j}_1}+a^*_{\bold{0}}a_{\bold{0}}]\Big\}\,\frac{\psi^{Bog}_{\bold{j}_1}}{\|\psi^{Bog}_{\bold{j}_1}\|}\rangle\,\quad\quad\quad
\end{eqnarray}
we readily derive
\begin{equation}
\langle \frac{\psi^{Bog}_{\bold{j}_1}}{\|\psi^{Bog}_{\bold{j}_1}\|}\,,\,\sum_{\bold{j}=\pm \bold{j}_1}a_{\bold{j}}^{*}a_{\bold{j}}\,\frac{\psi^{Bog}_{\bold{j}_1}}{\|\psi^{Bog}_{\bold{j}_1}\|}\rangle\leq \frac{z^{Bog}_{\bold{j}_1}+\frac{\phi_{\bold{j}_1}N}{N}}{k^2_{\bold{j}_1}}\leq \frac{\phi_{\bold{j}_1}}{k^2_{\bold{j}_1}}\label{energy-est-fin}
\end{equation}
where we have used that $z^{Bog}_{\bold{j}_1}<0$. Furthermore, by construction of $\psi^{Bog}_{\bold{j}_1}$
\begin{equation}
\langle \frac{\psi^{Bog}_{\bold{j}_1}}{\|\psi^{Bog}_{\bold{j}_1}\|}\,,\,\sum_{\bold{j}\in \mathbb{Z}^d\setminus \{\pm \bold{j}_1,\bold{0}\}}a_{\bold{j}}^{*}a_{\bold{j}}\,\frac{\psi^{Bog}_{\bold{j}_1}}{\|\psi^{Bog}_{\bold{j}_1}\|}\rangle=0\,.
\end{equation}
\\
 
 Now, we assume that Properties 1), 2), 3), 4), and 5) hold for $1\leq m-1\leq M-1$ and show that they are true also for $m$.
 \\
 
 \noindent
 \emph{Property 1)}  
Given Property 4) at step $m-1$, we can apply Corollary \ref{main-lemma-H}. Then, Theorem \ref{Feshbach-Hbog} ensures that for $z\leq E^{Bog}_{\bold{j}_m}+ \sqrt{\epsilon_{\bold{j}_m}}\phi_{\bold{j}_m}\sqrt{\epsilon_{\bold{j}_m}^2+2\epsilon_{\bold{j}_m}}$ the Feshbach-Schur Hamiltonian $\mathscr{K}_{\bold{j}_1,\dots,\bold{j}_m}^{Bog\,(N-2)}(z+z^{Bog}_{\bold{j}_1,\dots,\bold{j}_{m-1}})$ is well defined. Furthermore, thanks to Properties 3) and 4) at step $m-1$, we can apply Lemma \ref{invertibility-new}  and conclude that for $N$ sufficiently large and
 \begin{equation}
z\leq  z_{m}-\frac{C^{\perp}}{(\ln N)^{\frac{1}{2}}}+\gamma \Delta_{m-1}<E^{Bog}_{\bold{j}_m}+ \frac{1}{2}\sqrt{\epsilon_{\bold{j}_m}}\phi_{\bold{j}_m}\sqrt{\epsilon_{\bold{j}_m}^2+2\epsilon_{\bold{j}_m}}\,(<0)\quad,\quad \gamma = \frac{1}{2}\,,\label{bound-z-0-bis-primo}
\end{equation}
the operator $$\overline{\mathscr{P}_{\psi^{Bog}_{\bold{j}_1,\dots,\bold{j}_{m-1}}}}\mathscr{K}_{\bold{j}_1,\dots,\bold{j}_m}^{Bog\,(N-2)}(w)\overline{\mathscr{P}_{\psi^{Bog}_{\bold{j}_1,\dots,\bold{j}_{m-1}}}}\quad,\quad w=z+z^{Bog}_{\bold{j}_1,\dots,\bold{j}_{m-1}},$$ is bounded invertible on $\overline{\mathscr{P}_{\psi^{Bog}_{\bold{j}_1,\dots,\bold{j}_{m-1}}}}\mathcal{F}^{N}$. Note that the second inequality in (\ref{bound-z-0-bis-primo}) holds because from \emph{\underline{Lemma 5.5} of \cite{Pi1}} we can estimate $|z_m- E^{Bog}_{\bold{j}_m}|\leq \mathcal{O}(\frac{1}{N^{\beta}})$ for any $0<\beta<1$. Thus, for $z$ in the interval given in (\ref{bound-z-0-bis-primo}) we can define ($w=z+z^{Bog}_{\bold{j}_1,\dots,\bold{j}_{m-1}}$)
\begin{eqnarray}
& &\mathscr{K}_{\bold{j}_1,\dots,\bold{j}_m}^{Bog\,(N)}(w)\label{final-fesh-step-m-proof}\\
&=&-z\mathscr{P}_{\psi^{Bog}_{\bold{j}_1,\dots,\bold{j}_{m-1}}}\\
& &-\mathscr{P}_{\psi^{Bog}_{\bold{j}_1,\dots,\bold{j}_{m-1}}}\Gamma^{Bog}_{\bold{j}_1,\dots,\bold{j}_m ;N,N}(w)\mathscr{P}_{\psi^{Bog}_{\bold{j}_1,\dots,\bold{j}_{m-1}}}\nonumber\\
& &-\mathscr{P}_{\psi^{Bog}_{\bold{j}_1,\dots,\bold{j}_{m-1}}}\Gamma^{Bog}_{\bold{j}_1,\dots,\bold{j}_m;\,N,N}(w)\overline{\mathscr{P}_{\psi^{Bog}_{\bold{j}_1,\dots,\bold{j}_{m-1}}}}\times\\
& &\quad\quad\times \frac{1}{\overline{\mathscr{P}_{\psi^{Bog}_{\bold{j}_1,\dots,\bold{j}_{m-1}}}}\mathscr{K}_{\bold{j}_1,\dots,\bold{j}_m}^{Bog\,(N-2)}(w)\overline{\mathscr{P}_{\psi^{Bog}_{\bold{j}_1,\dots,\bold{j}_{m-1}}}}}\overline{\mathscr{P}_{\psi^{Bog}_{\bold{j}_1,\dots,\bold{j}_{m-1}}}}\Gamma^{Bog}_{\bold{j}_1,\dots,\bold{j}_m ;N,N}(w)\mathscr{P}_{\psi^{Bog}_{\bold{j}_1,\dots,\bold{j}_{m-1}}}\nonumber\\
&=:&f^{Bog}_{\bold{j}_1,\dots,\bold{j}_m}(w)\mathscr{P}_{\psi^{Bog}_{\bold{j}_1,\dots,\bold{j}_{m-1}}}\,.\label{def-f-proof}
\end{eqnarray}

\noindent
\emph{Property 2)}
The Hamiltonian $\mathscr{K}_{\bold{j}_1,\dots,\bold{j}_m}^{Bog\,(N)}(z+z^{Bog}_{\bold{j}_1,\dots,\bold{j}_{m-1}})$ has eigenvalue zero if there is a solution, $z\equiv z^{(m)},$ to $f^{Bog}_{\bold{j}_1,\dots,\bold{j}_m}(z+z^{Bog}_{\bold{j}_1,\dots,\bold{j}_{m-1}})=0$ in the range given in (\ref{bound-z-0-bis-primo}). This equation can also be written
 \begin{eqnarray}
%& &\mathscr{K}_{\bold{j}_1,\dots,\bold{j}_m}^{Bog\,(N)}(z+z_{\bold{j}_1,\dots,\bold{j}_{m-1}})\label{final-fesh-step-m}\\
z&= &- \langle  \frac{\psi^{Bog}_{\bold{j}_1,\dots,\bold{j}_{m-1}}}{\|\psi^{Bog}_{\bold{j}_1,\dots,\bold{j}_{m-1}}\|}\,,\,\Gamma^{Bog}_{\bold{j}_1,\dots,\bold{j}_m ;N,N}(z+z^{Bog}_{\bold{j}_1,\dots,\bold{j}_{m-1}}) \frac{\psi^{Bog}_{\bold{j}_1,\dots,\bold{j}_{m-1}}}{\|\psi^{Bog}_{\bold{j}_1,\dots,\bold{j}_{m-1}}\|}\rangle \label{leading-fp}\\
& &- \langle  \frac{\psi^{Bog}_{\bold{j}_1,\dots,\bold{j}_{m-1}}}{\|\psi^{Bog}_{\bold{j}_1,\dots,\bold{j}_{m-1}}\|}\,,\,\Gamma^{Bog}_{\bold{j}_1,\dots,\bold{j}_m;\,N,N}(z+z^{Bog}_{\bold{j}_1,\dots,\bold{j}_{m-1}})\overline{\mathscr{P}_{\psi^{Bog}_{\bold{j}_1,\dots,\bold{j}_{m-1}}}}\times \label{fin-fp}\\
& &\quad\quad\times \frac{1}{\overline{\mathscr{P}_{\psi^{Bog}_{\bold{j}_1,\dots,\bold{j}_{m-1}}}}\mathscr{K}_{\bold{j}_1,\dots,\bold{j}_m}^{Bog\,(N-2)}(z+z^{Bog}_{\bold{j}_1,\dots,\bold{j}_{m-1}})\overline{\mathscr{P}_{\psi^{Bog}_{\bold{j}_1,\dots,\bold{j}_{m-1}}}}}\times \nonumber \\
& &\quad\quad\quad\quad \times\overline{\mathscr{P}_{\psi^{Bog}_{\bold{j}_1,\dots,\bold{j}_{m-1}}}}\Gamma^{Bog}_{\bold{j}_1,\dots,\bold{j}_m ;N,N}(z+z^{Bog}_{\bold{j}_1,\dots,\bold{j}_{m-1}}) \frac{\psi^{Bog}_{\bold{j}_1,\dots,\bold{j}_{m-1}}}{\|\psi^{Bog}_{\bold{j}_1,\dots,\bold{j}_{m-1}}\|}\rangle\,. \nonumber
%&=:&f_{\bold{j}_1,\bold{j}_2,\dots,\bold{j}_m}(z+z_{\bold{j}_1,\dots,\bold{j}_{m-1}})\mathscr{P}_{\psi^{Bog}_{\bold{j}_1,\bold{j}_2,\dots,\bold{j}_{m-1}}}\label{def-f}
\end{eqnarray}
Properties 1), 2), 4) and 5) at step $m-1$ enable us to derive estimates (\ref{primero-1-1-new}), (\ref{primero-3-1-new}) in Lemma \ref{main-relations-new}. 
%{\color{red}In the previous point} (see \emph{Property 1)}) we have derived the 
We recall that Properties 3) and 4) at step $m-1$ yield the inequality in (\ref{invert-est-new}).  Consequently, combining (\ref{primero-1-1-new})-(\ref{primero-3-1-new}) with (\ref{invert-est-new}), we can rewrite the fixed point equation in (\ref{leading-fp})-(\ref{fin-fp}) in the following form
\begin{equation}\label{eq-dollar-bis}
z=-\langle \eta \,,\,\Gamma^{Bog}_{\bold{j}_m ;N,N}(z)\,\eta \rangle +\mathcal{Y}(z)
\end{equation}
with 
\begin{eqnarray}\
|\mathcal{Y}(z)|&\leq& \frac{C_I}{(\ln N)^{\frac{1}{4}}}+\frac{C_{II}^2}{(\ln N)^{\frac{1}{4}}(1-\gamma)\Delta_{m-1}}\\
&\leq & \frac{C_I}{(\ln N)^{\frac{1}{4}}}+(\frac{2}{\gamma})^m\frac{C_{II}^2}{(\ln N)^{\frac{1}{4}}(1-\gamma) \Delta_{0}}\label{bound-Y}
\end{eqnarray}
where we have used that for $1\leq m\leq M$ and $N$ large enough
$$\Delta_{m-1}\geq  \Delta_0 (\frac{\gamma}{2})^m\,;$$
see the definition of $\Delta_m$ in (\ref{Deltagap}).
% $${\color{red}(\frac{2}{\gamma})^m}\frac{C_{III}}{(\ln N)^{\frac{1}{4}}}\leq \gamma\frac{\Delta_{m-1}}{2}\,.$$
Consequently, we have reduced the fixed point equation $$f^{Bog}_{\bold{j}_1, \dots,\bold{j}_m}(z+z^{Bog}_{\bold{j}_1,\dots,\bold{j}_{m-1}})=0$$ to the fixed point equation, $f^{Bog}_{\bold{j}_m}(z)=0$, of a three-modes system, up to the small error $\mathcal{Y}(z)$. Therefore, the same argument of \emph{\underline{Theorem 4.1} of \cite{Pi1}} implies that for $N$ large enough there exists a $z^{(m)}$ in the range (\ref{bound-z-0-bis-primo}) that solves the equation in (\ref{eq-dollar-bis}). Now, we show the inequality 
\begin{equation}\label{diff-z}
 |z^{(m)}-z_m|\leq (\frac{2}{\gamma})^m\frac{C_{III}}{(\ln N)^{\frac{1}{4}}}\,\quad,\quad C_{III}:=C_I+\frac{C_{II}^2}{(1-\gamma)\Delta_{0}}\,.\,
\end{equation}
Since, by construction,
\begin{equation}
z_m=-\langle \eta \,,\,\Gamma^{Bog}_{\bold{j}_m ;N,N}(z_m)\,\eta \rangle\,, 
\end{equation}
after subtracting the equation in (\ref{eq-dollar-bis}) we get
\begin{equation}\label{diff-z-bis}
z_m+\langle \eta \,,\,\Gamma^{Bog}_{\bold{j}_m ;N,N}(z_m)\,\eta \rangle -z^{(m)}-\langle \eta \,,\,\Gamma^{Bog}_{\bold{j}_m ;N,N}(z^{(m)})\,\eta \rangle=-\mathcal{Y}(z^{(m)})\,.
\end{equation}
Hence, the inequality in (\ref{diff-z}) follows from (\ref{bound-Y}) and the mean value theorem applied to the L-H-S of (\ref{diff-z-bis}),  because the derivative with respect to $z$  of $$z+\langle \eta \,,\,\Gamma^{Bog}_{\bold{j}_m ;N,N}(z)\eta \rangle=z+(1-\frac{1}{N})\frac{\phi_{\bold{j}_{m}}}{2\epsilon_{\bold{j}_m}+2-\frac{4}{N}-\frac{z}{\phi_{\bold{j}_{m}}}}\check{\mathcal{G}}_{\bold{j}_m\,;\,N-2,N-2}(z)$$ is not smaller than $1$; see Remark \ref{nondecreasing}. 

\noindent
Using the isospectrality of the Feshbach-Schur map, we deduce that $H^{Bog}_{\bold{j}_1,\dots,\bold{j}_{m}}$ has the eigenvalue $$z^{(m)}+z^{Bog}_{\bold{j}_1,\dots,\bold{j}_{m-1}}=:z^{Bog}_{\bold{j}_1,\dots,\bold{j}_{m}}\,.$$
The corresponding eigenvector is given in  (\ref{gs-Hm-start})-(\ref{gs-Hm-fin}).

\noindent 
The uniqueness of $z^{(m)}$ in the considered range of $z$ holds because for any other value, $(z^{(m)})'$, solving the fixed point problem we can state the inequality 
$$ |(z^{(m)})'-z_m|\leq \mathcal{O}(\frac{1}{(\ln N)^{\frac{1}{4}}})\,$$
by the same argument used to prove the inequality in (\ref{diff-z}). Now, assuming that there are two distinct eigenvalues $z^{Bog}_{\bold{j}_1,\dots,\bold{j}_{m-1}}+z^{(m)}$ and $z^{Bog}_{\bold{j}_1,\dots,\bold{j}_{m-1}}+(z^{(m)})'$, the corresponding eigenvectors, $\psi^{Bog}_{\bold{j}_1, \dots, \bold{j}_m}$ and $(\psi^{Bog}_{\bold{j}_1, \dots, \bold{j}_m})'$, are given by the formula in  (\ref{gs-Hm-start})-(\ref{gs-Hm-fin}) replacing $z^{Bog}_{\bold{j}_1,\dots,\bold{j}_{m}}$ with $z^{Bog}_{\bold{j}_1,\dots,\bold{j}_{m-1}}+z^{(m)}$ and $z^{Bog}_{\bold{j}_1,\dots,\bold{j}_{m-1}}+(z^{(m)})'$, respectively. Next, starting from the expression of $\mathscr{K}^{Bog\,(N-2r-2)}_{\bold{j}_1,\dots,\bold{j}_{m-1}}(w)$ (see (\ref{fesh-i-m})) and using the relation in (\ref{gamma-i-m})-(\ref{Gammabog-m}), we exploit the expansion
\begin{eqnarray}
& &\frac{1}{Q^{(N-2r,N-2r+1)}_{\bold{j}_m}\mathscr{K}^{Bog\,(N-2r-2)}_{\bold{j}_1,\dots,\bold{j}_{m-1}}(w)Q^{(N-2r,N-2r+1)}_{\bold{j}_*}}\\
&=&\sum_{l_{N-2r}=0}^{\infty}R^{Bog}_{\bold{j}_1,\dots,\bold{j}_{m-1}\,;\,N-2r,N-2r}(w)\Big[\Gamma^{Bog\,}_{\bold{j}_1,\dots,\bold{j}_{m-1}\,;\,N-2r,N-2r}(w)\,R^{Bog}_{\bold{j}_1,\dots,\bold{j}_{m-1}\,;\,N-2r,N-2r}(w)\Big]^{l_{N-2r}}\quad\quad
\end{eqnarray}
where $w=z^{Bog}_{\bold{j}_1,\dots,\bold{j}_{m-1}}+z^{(m)}$ or $w=z^{Bog}_{\bold{j}_1,\dots,\bold{j}_{m-1}}+(z^{(m)})'$ depending on the considered eigenvalue. Due to the control of the series in (\ref{convergent-series}) and by means of
a procedure analogous to {\bf{STEP I}} and {\bf{II}} in Lemma \ref{main-relations-new},  we can re-expand the operators $\Gamma^{Bog\,}_{\bold{j}_1,\dots,\bold{j}_{m-1}\,;\,N-2r,N-2r}(w)$ and estimate (with a suitable choice of $h$)
\begin{equation}
\|\psi^{Bog}_{\bold{j}_1, \dots, \bold{j}_m}-(\psi^{Bog}_{\bold{j}_1, \dots, \bold{j}_m})'\|\leq  \mathcal{O}(\frac{1}{[\ln (\ln N)]^{\frac{1}{2}}})
\end{equation}
because $|(z^{(m)})'-z^{(m)}|\leq  |(z^{(m)})'-z_m|+|(z^{(m)})-z_m|\leq \mathcal{O}(\frac{1}{(\ln N)^{\frac{1}{4}}})$ (see section 0.1 in \emph{supporting-file-Bose2.pdf}). Hence, since $$\|\psi^{Bog}_{\bold{j}_1, \dots, \bold{j}_m}\|\,,\,\|(\psi^{Bog}_{\bold{j}_1, \dots, \bold{j}_m})'\|\geq 1$$  by construction (see Remark \ref{vectornorm}), for $N$ large enough the two eigenvectors are not orthogonal and  the corresponding eigenvalues must coincide.
%The uniqueness of $z^{(m)}$ follows from the fact that the derivative of $$z+\langle \eta \,,\,\Gamma^{Bog}_{\bold{j}_m ;N,N}(z)\eta \rangle-\mathcal{Y}(z)$$ is larger than $\frac{1}{2}$ for $N$ sufficiently large; see Lemma \ref{appendix-lemma-1}.

\noindent
\emph{Property 3)}   In Property  2) we have derived that in the interval
 \begin{equation}
z\leq  z_{m}-\frac{C^{\perp}}{(\ln N)^{\frac{1}{2}}}+\gamma \Delta_{m-1}<E^{Bog}_{\bold{j}_m}+ \frac{1}{2}\sqrt{\epsilon_{\bold{j}_m}}\phi_{\bold{j}_m}\sqrt{\epsilon_{\bold{j}_m}^2+2\epsilon_{\bold{j}_m}}\,(<0)\quad,\quad \gamma = \frac{1}{2}\,,\label{bound-z-0-bis}
\end{equation}
the Hamiltonian $\mathscr{K}_{\bold{j}_1,\dots,\bold{j}_m}^{Bog\,(N)}(z+z^{Bog}_{\bold{j}_1,\dots,\bold{j}_{m-1}})$ is bounded invertible except for $z\equiv z^{(m)}$.  This together with (\ref{diff-z}) imply that for $N$ large enough
 \begin{eqnarray}
& &\text{infspec}\,\Big[ H^{Bog}_{\bold{j}_1, \dots,\bold{j}_{m}}\upharpoonright_{\mathcal{F}^N\ominus \{\mathbb{C}\psi^{Bog}_{\bold{j}_1, \dots,\bold{j}_{m}}\}}\Big]-z^{Bog}_{\bold{j}_1, \dots,\bold{j}_{m}}\\
&\geq  &\Delta_m:=\gamma\Delta_{m-1}-\frac{C^{\perp}}{(\ln N)^{\frac{1}{2}}}-(\frac{2}{\gamma})^m\frac{C_{III}}{(\ln N)^{\frac{1}{4}}}\,.\quad\quad\quad
 \end{eqnarray}
By means of the same argument used for $m=1$, we deduce that
\begin{equation}
\text{infspec}\,\Big[\Big(\hat{H}^{Bog}_{\bold{j}_1, \dots,\bold{j}_{m}}+\sum_{\bold{j}\notin \{\pm\bold{j}_1, \dots, \pm\bold{j}_{m+1}\}}(k_{\bold{j}})^2a^*_{\bold{j}}a_{\bold{j}}\Big)\upharpoonright_{(Q^{(>N-1)}_{\bold{j}_{m+1}}\mathcal{F}^N)\ominus \{\mathbb{C}\psi^{Bog}_{\bold{j}_1, \dots,\bold{j}_{m}}\}}\Big]-z^{Bog}_{\bold{j}_1, \dots,\bold{j}_{m}}\geq\Delta_m\,.
%\text{Gap}\,\Big\{\hat{H}^{Bog}_{\bold{j}_1, \dots,\bold{j}_{m}}+\sum_{\bold{j}\neq \{\pm\bold{j}_1, \dots, \pm\bold{j}_{m+1}\}}(k_{\bold{j}})^2a^*_{\bold{j}}a_{\bold{j}}\Big\}\geq\Delta_m 
 \end{equation}

\noindent
\emph{Property 4)} The argument is analogous to the case $m=1$, making use of the restriction to subspaces with fixed number, $j$, (that we assume less than $N-2$ and even) of particles in the modes different from $\pm\bold{j}_1\,,\dots \,,\pm\bold{j}_m$ and $\bold{0}$. Invoking  Property 4) at step $m-1$ we can adapt Corollary \ref{main-lemma-H} and Theorem \ref{Feshbach-Hbog} to the Hamiltonian $\sum_{l=1}^{m}\hat{H}^{Bog}_{\bold{j}_l }$. Hence, we define the Feshbach-Schur Hamiltonian ($w\equiv z+z^{Bog}_{\bold{j}_1,\dots ,\bold{j}_{m-1}}$) 

\begin{eqnarray}
& &\hat{\mathscr{K}}^{Bog\,(N-2)}_{\bold{j}_1,\dots,\bold{j}_m}(w)\\
%&=&\mathscr{P}_{\eta}(H^{Bog}_{\bold{j}_*}-z)\mathscr{P}_{\eta}\\
%& &-\mathscr{P}_{\eta}W_{\bold{j}_*}\,Q^{(N-2)}_{\bold{j}_*}\,R^{Bog}_{\bold{j}_*\,;\,N-2,N-2}(z)\sum_{l_{N-2}=0}^{\infty}[\Gamma^{Bog}_{\bold{j}_*\,;\,N-2,N-2}(z) R^{Bog}_{\bold{j}_*\,;\,N-2,N-2}(z)]^{l_{N-2}}\, Q^{(N-2)}_{\bold{j}_*}W^*_{\bold{j}_*}\mathscr{P}_{\eta}\nonumber\\
&=&\hat{Q}_{\bold{j}_m}^{(>N-1)}(\sum_{l=1}^{m}\hat{H}^{Bog}_{\bold{j}_l }-z^{Bog}_{\bold{j}_1,\dots ,\bold{j}_{m-1}}-z)\hat{Q}_{\bold{j}_m}^{(>N-1)}\label{firstline}\\
& &-\hat{Q}_{\bold{j}_m}^{(>N-1)}W_{\bold{j}_m}\,\hat{R}^{Bog}_{\bold{j}_1,\dots,\bold{j}_m\,;\,N-2,N-2}(w)\times \label{invert-G}\\
& &\quad\quad\quad\quad \times \sum_{l_{N-2}=0}^{\infty}[\hat{\Gamma}^{Bog}_{\bold{j}_1,\dots,\bold{j}_m\,;\,N-2,N-2}(w) \hat{R}^{Bog}_{\bold{j}_1,\dots,\bold{j}_m\,;\,N-2,N-2}(w)]^{l_{N-2}}\, W^*_{\bold{j}_m}\hat{Q}_{\bold{j}_m}^{(>N-1)}\,.\quad\quad\quad\nonumber
\end{eqnarray}
Due to Property 4) at step $m-1$, by implementing {\bf{Steps}} {\bf{a)}}, {\bf{b)}}, {\bf{d)}} , and {\bf{e)}} of Lemma \ref{invertibility} (see Remark \ref{invert}),  we can estimate for $w\equiv z+z^{Bog}_{\bold{j}_1,\dots ,\bold{j}_{m-1}}$ with $z$ in the interval defined in (\ref{inter-def})
\begin{eqnarray}
& &\|\hat{Q}_{\bold{j}_m}^{(>N-1)}W_{\bold{j}_m}\,\hat{R}^{Bog}_{\bold{j}_1,\dots,\bold{j}_m\,;\,N-2,N-2}(w)\sum_{l_{N-2}=0}^{\infty}[\hat{\Gamma}^{Bog}_{\bold{j}_1,\dots,\bold{j}_m\,;\,N-2,N-2}(w) \hat{R}^{Bog}_{\bold{j}_1,\dots,\bold{j}_m\,;\,N-2,N-2}(w)]^{l_{N-2}}\, W^*_{\bold{j}_m}\hat{Q}_{\bold{j}_m}^{(>N-1)}\|\nonumber\\
&\leq& (1-\frac{1}{N})\frac{\phi_{\bold{j}_{m}}}{2\epsilon_{\bold{j}_m}+2-\frac{4}{N}-\frac{1}{\phi_{\bold{j}_{m}}}(z+\frac{U_{\bold{j}_m}}{\sqrt{N}}+\frac{m-1+\frac{1}{2}}{(\ln N)^{\frac{1}{8}}})}\check{\mathcal{G}}_{\bold{j}_m\,;\,N-2,N-2}(z+\frac{U_{\bold{j}_m}}{\sqrt{N}}+\frac{m-1+\frac{1}{2}}{(\ln N)^{\frac{1}{8}}})\label{ineq-inf-Hcapp}\\
& &+\mathcal{O}(\frac{1}{(\ln N)^{\frac{1}{2}}})\,
\end{eqnarray}
where we have also made use of the property $$\check{\mathcal{G}}_{\bold{j}_m\,;\,N-2,N-2}(z+\frac{U_{\bold{j}_m}}{\sqrt{N}}+\frac{m-1}{(\ln N)^{\frac{1}{8}}})\leq \check{\mathcal{G}}_{\bold{j}_m\,;\,N-2,N-2}(z+\frac{U_{\bold{j}_m}}{\sqrt{N}}+\frac{m-1+\frac{1}{2}}{(\ln N)^{\frac{1}{8}}});$$ see Remark \ref{nondecreasing}.
Using Property 4) at step $m-1$ one more time in order to estimate the infimum of (\ref{firstline}), we deduce that the spectrum of the Feshbach-Schur Hamiltonian $\hat{\mathscr{K}}^{Bog\,(N-2)}_{\bold{j}_1,\dots,\bold{j}_m}(w)$ is bounded from below by
\begin{eqnarray}
%& &-Q^{(>N-1)}W\,Q^{(N-2)}\,R^{Bog}_{N-2,N-2}(z)\sum_{l_{N-2}=0}^{\infty}[\Gamma^{Bog}_{N-2,N-2}(z) R^{Bog}_{N-2,N-2}(z)]^{l_{N-2}}\, Q^{(N-2)}W^*Q^{(>N-1)}\quad \quad\quad\quad\label{first-term-last-Bog}\\
%& &-\mathscr{V}^{(N-1)}(z)\,R_{N-1,N-1}(z)\sum_{l_{N-1}=0}^{\infty}[\Gamma_{N-1,N-1}(z) R_{N-1,N-1}(z)]^{l_{N-1}}\,(\mathscr{V}^{(N-1)}(z) )^*\quad\quad \label{last-term-last-Bog}
&  &-z-\frac{m-1}{(\ln N)^{\frac{1}{8}}}\\
& &-\|\hat{Q}_{\bold{j}_m}^{(>N-1)}W_{\bold{j}_m}\,\hat{R}^{Bog}_{\bold{j}_1,\dots,\bold{j}_m\,;\,N-2,N-2}(w)\sum_{l_{N-2}=0}^{\infty}[\hat{\Gamma}^{Bog}_{\bold{j}_1,\dots,\bold{j}_m\,;\,N-2,N-2}(w) \hat{R}^{Bog}_{\bold{j}_1,\dots,\bold{j}_m\,;\,N-2,N-2}(w)]^{l_{N-2}}\, W^*_{\bold{j}_m}\hat{Q}_{\bold{j}_m}^{(>N-1)}\| \nonumber \\
&\geq  &\frac{1}{2(\ln N)^{\frac{1}{8}}}+\frac{U_{\bold{j_m}}}{\sqrt{N}}+\mathcal{O}(\frac{1}{(\ln N)^{\frac{1}{2}}})\label{final-cappuccio-1}\\
& &-(z+\frac{U_{\bold{j_m}}}{\sqrt{N}}+\frac{m-1+\frac{1}{2}}{(\ln N)^{\frac{1}{8}}})\label{final-cappuccio-2}\\
& &-(1-\frac{1}{N})\frac{\phi_{\bold{j}_{m}}}{2\epsilon_{\bold{j}_m}+2-\frac{4}{N}-\frac{1}{\phi_{\bold{j}_{m}}}(z+\frac{U_{\bold{j_m}}}{\sqrt{N}}+\frac{m-1+\frac{1}{2}}{(\ln N)^{\frac{1}{8}}})}\check{\mathcal{G}}_{\bold{j}_m\,;\,N-2,N-2}(z+\frac{U_{\bold{j_m}}}{\sqrt{N}}+\frac{m-1+\frac{1}{2}}{(\ln N)^{\frac{1}{8}}})\,.\label{final-cappuccio-3}
%&=&\frac{\phi_{\bold{j}_m}N^{\mu}}{N}+\mathcal{O}(\frac{1}{(\ln N)^{\frac{1}{2}}})\label{final-cappuccio-1}\\
%& &-\frac{\phi_{\bold{j}_{m}}}{2\epsilon_{\bold{j}_m}+2-\frac{1}{\phi_{\bold{j}_{m}}}(z+\frac{\phi_{\bold{j}_m}N^{\mu}}{N}+\frac{m-1}{(\ln N)^{\frac{1}{8}}})}\check{\mathcal{G}}_{\bold{j}_m\,;\,N-2,N-2}(z+\frac{\phi_{\bold{j}_m}N^{\mu}}{N}+\frac{m-1}{(\ln N)^{\frac{1}{8}}})-(z+\frac{\phi_{\bold{j}_m}N^{\mu}}{N}+\frac{m-1}{(\ln N)^{\frac{1}{8}}})\,.\label{final-cappuccio}
\end{eqnarray}
Next, we observe that, for $N$ sufficiently large and  $$z+\frac{m-1+\frac{1}{2}}{(\ln N)^{\frac{1}{8}}}+\frac{U_{\bold{j_m}}}{\sqrt{N}}\leq z_m\quad \Rightarrow \quad z\leq z_m-\frac{m-\frac{1}{2}}{(\ln N)^{\frac{1}{8}}}-\frac{U_{\bold{j_m}}}{\sqrt{N}}\,,$$ the inequalities   $(\ref{final-cappuccio-1})>0$ and $(\ref{final-cappuccio-2})+(\ref{final-cappuccio-3})\geq 0$ hold true. Hence, the isospectrality of the Feshbach-Schur map implies that $$\sum_{l=1}^{m}\hat{H}^{Bog}_{\bold{j}_l }-w$$ is bounded invertible if (recall $z^{(m)}+z^{Bog}_{\bold{j}_1,\dots,\bold{j}_{m-1}}\equiv z^{Bog}_{\bold{j}_1,\dots,\bold{j}_{m}}$) $$w=z+z^{Bog}_{\bold{j}_1,\dots,\bold{j}_{m-1}}\leq z_m-\frac{m-\frac{1}{2}}{(\ln N)^{\frac{1}{8}}}-\frac{U_{\bold{j_m}}}{\sqrt{N}}-z^{(m)}+z^{(m)}+z^{Bog}_{\bold{j}_1,\dots,\bold{j}_{m-1}}=(z_m-z^{(m)})+z^{Bog}_{\bold{j}_1,\dots,\bold{j}_{m}}-\frac{m-\frac{1}{2}}{(\ln N)^{\frac{1}{8}}}-\frac{U_{\bold{j_m}}}{\sqrt{N}}\,.$$ The difference $|z_m-z^{(m)}|$ is estimated in (\ref{difference-tra-z}),  from which we conclude that
$$|z_m-z^{(m)}|+\frac{U_{\bold{j_m}}}{\sqrt{N}}$$
 is bounded by $\frac{1}{2(\ln N)^{\frac{1}{8}}}$ for $N$ large. Thus, $\sum_{l=1}^{m}\hat{H}^{Bog}_{\bold{j}_l }-w$ is bounded invertible for $w< z^{Bog}_{\bold{j}_1,\dots,\bold{j}_{m}}-\frac{m}{(\ln N)^{\frac{1}{8}}}$.
\\

\noindent
\emph{Property 5.} The argument is analogous to the case $m=1$.\qed

The very last result of this section concerns the expansion of the ground state vector $\psi^{Bog}_{\bold{j}_1, \dots, \bold{j}_{M}}$ in terms of the bare quantities.
 
\begin{corollary}\label{expansion}
Assume  that the hypotheses of Theorem \ref{induction-many-modes}
%Proposition \ref{lemma-expansion-proof-0}, Lemma \ref{main-relations-new}, and Lemma \ref{invertibility-new}
are satisfied. Then, at fixed $M$, for any arbitrarily small $\zeta>0$,  there exist $N_{\zeta}<\infty$ and a vector $(\psi^{Bog}_{\bold{j}_1, \dots, \bold{j}_{m}})_{\zeta}$,  corresponding to a ($\zeta$-dependent) finite sum of (finite) products of the interaction terms $W^*_{\bold{j}_l}+W_{\bold{j}_l}$ and of the resolvents $\frac{1}{\hat{H}^0_{\bold{j}_l}-E^{Bog}_{\bold{j}_l}}$ applied to $\eta$,  such that 
$$\| \psi^{Bog}_{\bold{j}_1, \dots, \bold{j}_{m}}-(\psi^{Bog}_{\bold{j}_1, \dots, \bold{j}_{m}})_{\zeta}\|\leq \zeta\,,\quad 1\leq m \leq M\,,$$  
for $N>N_{\zeta}$.
\end{corollary}

\noindent
\emph{Proof}

\noindent
Making use of the formulae in (\ref{gs-Hm-start})-(\ref{gs-Hm-fin}) and  (\ref{construction-ground}), we implement the following operations on each $T_m \psi^{Bog}_{\bold{j}_1, \dots, \bold{j}_{m-1}}$ starting from $m\leq M$ down to $m=1$:
\begin{itemize}
\item[{\bf{i)}}] We replace 
\begin{equation}
 \Big[\mathscr{P}_{\psi^{Bog}_{\bold{j}_1\dots,\bold{j}_{m-1}}}-\frac{1}{\overline{\mathscr{P}_{\psi^{Bog}_{\bold{j}_1,\dots,\bold{j}_{m-1}}}}\mathscr{K}^{Bog\,(N-2)}_{\bold{j}_1,\dots,\bold{j}_m}(z^{Bog}_{\bold{j}_1,\dots,\bold{j}_m})\overline{\mathscr{P}_{\psi^{Bog}_{\bold{j}_1, \dots, \bold{j}_{m-1}}}}}
\overline{\mathscr{P}_{\psi^{Bog}_{\bold{j}_1\dots,\bold{j}_{m-1}}}}\mathscr{K}^{Bog\,(N-2)}_{\bold{j}_1,\dots,\bold{j}_m}(z^{Bog}_{\bold{j}_1,\dots,\bold{j}_m})\Big]\psi^{Bog}_{\bold{j}_1, \dots, \bold{j}_{m-1}}
\end{equation}
 with $\psi^{Bog}_{\bold{j}_1, \dots, \bold{j}_{m-1}}$ up to a remainder with norm less than $\mathcal{O}(\frac{1}{\Delta_{m-1} (\ln N)^{\frac{1}{4}}} )$ thanks to (\ref{primero-3-1-new}) in Lemma \ref{main-relations-new} and (\ref{invert-est-new})  in Lemma \ref{invertibility-new}.
\item[{\bf{ii)}}]
We truncate the sum in (\ref{gs-Hm-fin}) at some $\bar{j}$ using the convergence of the series in  (\ref{convergent-series}) with $\bold{j}_{*}$ replaced with $\bold{j}_m$, and where $a_{\epsilon_{\bold{j}_{*}}}, b_{\epsilon_{\bold{j}_{*}}}. c_{\epsilon_{\bold{j}_{*}}}$ with $\epsilon_{\bold{j}_*}\equiv\epsilon_{\bold{j}_m}$ are those defined in Corollary \ref{main-lemma-H} and $\Theta$ is defined in Theorem \ref{Feshbach-Hbog}. The remaining expression is (see the procedure in section 0.1 of \emph{supporting-file-Bose2.pdf}):
\begin{eqnarray}
& &\psi^{Bog}_{\bold{j}_1, \dots, \bold{j}_{m-1}}\\
& &+\Big[-\frac{1}{Q^{(N-2,N-1)}_{\bold{j}_m}\mathscr{K}^{Bog\,(N-4)}_{\bold{j}_1,\dots,\bold{j}_m}(z^{Bog}_{\bold{j}_1,\dots,\bold{j}_m})Q^{(N-2,N-1)}_{\bold{j}_m}}Q^{(N-2, N-1)}_{\bold{j}_m}W^*_{\bold{j}_m}\Big]\psi^{Bog}_{\bold{j}_1, \dots, \bold{j}_{m-1}}\label{gs-Hm-fin-bis-0}\\
%& &+\sum_{j=2}^{N/2}\prod^{2}_{r=j}\Big[-\frac{1}{Q_{\bold{j}_m}^{(N-2r, N-2r+1)}\mathscr{K}_{\bold{j}_1,\dots ,\bold{j}_m}^{Bog\,(N-2r-2)}(z^{Bog}_{\bold{j}_1,\dots ,\bold{j}_m})Q_{\bold{j}_m}^{(N-2r, N-2r+1)}}W^*_{\bold{j}_m;N-2r,N-2r+2}\Big]\psi^{Bog}_{\bold{j}_1, \dots, \bold{j}_{m-1}}\quad\quad\quad\label{gs-Hm-fin} \\
& &+\sum_{j=2}^{\bar{j}}\Big\{\prod^{2}_{r=j}\Big[-\frac{1}{Q_{\bold{j}_m}^{(N-2r, N-2r+1)}\mathscr{K}_{\bold{j}_1,\dots ,\bold{j}_m}^{Bog\,(N-2r-2)}(z^{Bog}_{\bold{j}_1,\dots ,\bold{j}_m})Q_{\bold{j}_m}^{(N-2r, N-2r+1)}}W^*_{\bold{j}_m;N-2r,N-2r+2}\Big]\times\quad\quad\quad\quad\quad  \label{gs-Hm-fin-bis}\\
& &\quad\quad\quad\times \Big[-\frac{1}{Q^{(N-2,N-1)}_{\bold{j}_m}\mathscr{K}^{Bog\,(N-4)}_{\bold{j}_1,\dots,\bold{j}_m}(z^{Bog}_{\bold{j}_1,\dots,\bold{j}_m})Q^{(N-2,N-1)}_{\bold{j}_m}}Q^{(N-2, N-1)}_{\bold{j}_m}W^*_{\bold{j}_m}\Big]\Big\}\psi^{Bog}_{\bold{j}_1, \dots, \bold{j}_{m-1}}\,.\,\nonumber 
\end{eqnarray}
The remainder can be estimated in norm less than $\mathcal{O}([\frac{1}{1+c\sqrt{\epsilon_{\bold{j}_m}}}]^{\bar{j}})$ for some $c>0$ (see (\ref{convergent-series})).
\item[{\bf{iii)}}]
For each factor  appearing in (\ref{gs-Hm-fin-bis-0}) and (\ref{gs-Hm-fin-bis})  we exploit the expansion (see the definition in (\ref{fesh-i-m}) and the relation in (\ref{gamma-i-m})-(\ref{Gammabog-m}))
\begin{eqnarray}
& &\frac{1}{Q^{(N-2r,N-2r+1)}_{\bold{j}_m}\mathscr{K}^{Bog\,(N-2r-2)}_{\bold{j}_1,\dots,\bold{j}_{m-1}}(w)Q^{(N-2r,N-2r+1)}_{\bold{j}_m}}\label{fake-res}\\
&=&\sum_{l_{N-2r}=0}^{\infty}R^{Bog}_{\bold{j}_1,\dots,\bold{j}_{m-1}\,;\,N-2r,N-2r}(w)\Big[\Gamma^{Bog\,}_{\bold{j}_1,\dots,\bold{j}_{m-1}\,;\,N-2r,N-2r}(w)\,R^{Bog}_{\bold{j}_1,\dots,\bold{j}_{m-1}\,;\,N-2r,N-2r}(w)\Big]^{l_{N-2r}}\,\label{fake-res2}\quad\quad\quad\quad
\end{eqnarray}
with $w\equiv z^{Bog}_{\bold{j}_1,\dots,\bold{j}_{m}}=z^{Bog}_{\bold{j}_1,\dots,\bold{j}_{m-1}}+z^{(m)}$. For each summand in (\ref{gs-Hm-fin-bis}) it is clear that we obtain an expression on which  we  can
implement a procedure analogous to  {\bf{Steps}} $\bf{I}$, $\bf{II}$, $\bf{III}$, and $\bf{IV}$ of Lemma \ref{main-relations} because we have just a finite number of factors of the type in  (\ref{fake-res})  and the complete product of operators is  applied to $\psi^{Bog}_{\bold{j}_1, \dots, \bold{j}_{m-1}}$. This way, for some (even) $h$ to be determined later,  we replace: a) each factor in (\ref{fake-res}) with a truncation of the sum  in (\ref{fake-res2}) at $l_{N-2r}=h$; b) each $\Gamma^{Bog\,}_{\bold{j}_1,\dots,\bold{j}_{m-1}\,;\,N-2r,N-2r}(w)$  with
$$[\tilde{\Gamma}^{Bog\,}_{\bold{j}_1,\dots,\bold{j}_m\,;\,N-2,N-2}(z^{(m)})]_{\mathcal{\tau}_h} \, $$
that is defined in {\bf{STEP III}} of Lemma \ref{main-relations};
c) each $R^{Bog}_{\bold{j}_1,\dots,\bold{j}_{m-1}\,;\,N-2r,N-2r}(w)$ with $\tilde{R}^{Bog}_{\bold{j}_1,\dots,\bold{j}_{m-1}\,;\,N-2r,N-2r}(z^{(m)})$. The norm of the remainder produced in this step is less than $\mathcal{O}\Big((\bar{j}^2)\,\frac{\,h(2h)^{\frac{h+2}{2}}\cdot h\,(2h)^{\frac{h+2}{2}}}{\sqrt{N}}\Big)$ (see an analogous procedure  in section 0.1 of \emph{supporting-file-Bose2.pdf}).
%With some further work one can derive bounds from above and below for $\check{\mathcal{G}}_{\bold{j}_*\,;\,N-2,N-2}(E^{Bog}_{\bold{j}_*})$. Then, the property mentioned in Remark \ref{increasing} implies a straightforward control of the difference between $E^{Bog}_{\bold{j}_*}$ and $z_*$.
%\item
%$\bar{j}$ and the expansion of the operators $\Gamma^{Bog}_{\bold{j}_*\,;\,i,i}(z)$ are chosen so that the sum of the remainder terms is an operator in norm less than $\delta$.
\item[{\bf{iv)}}]
In the leading term (that consists of a finite number of products) resulting from the previous operations, we replace $z^{(m)}$ with $E^{Bog}_{\bold{j}_m}$ up to a remainder term that can be estimated thanks to (\ref{diff-z}) combined with  \emph{\underline{Lemma 5.5} of \cite{Pi1}}. Indeed, this lemma provides the estimate  $|z_m- E^{Bog}_{\bold{j}_m}|\leq \mathcal{O}(\frac{1}{N^{\beta}})$ for any $0<\beta<1$. 
\end{itemize}
Hence, it is evident that repeating the procedure for each $m$ down to $m=1$, the leading term that is obtained is of the form described in the statement of the lemma, and the norm of the sum of all the remainder terms is less than any $\zeta>0$ by setting a suitable $\zeta$-dependence for $\bar{j}$,  $h$, and  $N$.
 \qed

\section{Appendix}
\setcounter{equation}{0}

\begin{corollary}\label{main-lemma-H}
For $M\geq m\geq 1$ assume:

\noindent
(a)  
%\begin{equation}
%(H^{Bog}_{\bold{j}_1,\dots,\bold{j}_{m-1}})_{\xi}\geq z_{\bold{j}_1,\dots\,\bold{j}_m}-\mathcal{O}(\xi)
%\end{equation}
\begin{equation}\label{assumption-flow}
\text{infspec}\,[\sum_{l=1}^{m-1}\hat{H}^{Bog}_{\bold{j}_l}]-z^{Bog}_{\bold{j}_1,\dots ,\bold{j}_{m-1}}\geq -\frac{(m-1)}{(\ln N)^{\frac{1}{8}}}
%\hat{H}^{Bog}_{\bold{j}_1, \dots,\bold{j}_{m}}\geq z^{Bog}_{\bold{j}_1,\dots,\bold{j}_{m}}-\mathcal{O}(\frac{1}{(\ln N)^{\frac{1}{8}}})
\end{equation}
%\begin{equation}
%(H^{\#}_{\bold{j}_{1},\dots,\bold{j}_{m-1}})_{\xi}-(1-\xi)T_{\bold{j}=\{\pm\bold{j}_m\}}\geq z^{\#}_{\bold{j}_1,\dots\,\bold{j}_{m-1}}-\frac{(m-1)\xi^{\frac{1}{2}}}{M}\,,\label{ass-1-main-lemma}
%\end{equation}
where $z^{Bog}_{\bold{j}_1,\dots\,\bold{j}_{m-1}}$ is the ground state energy of $H^{Bog}_{\bold{j}_1,\dots,\bold{j}_{m-1}}$ for $m\geq 2$ and $z^{Bog}_{\bold{j}_1,\dots\,\bold{j}_{m-1}}|_{m=1}\equiv 0$;\\
 % with $z^{\#}_{\bold{j}_1,\dots\,\bold{j}_{m-1}}\equiv 0$ if $m=1$.
%
%\noindent
%b) 
%\begin{equation}
%|z^{\#}_{\bold{j}_1,\dots\,\bold{j}_{m-1}}-z_{\bold{j}_1,\dots\,\bold{j}_{m-1}}|\leq \frac{(m-1)\xi^{\frac{1}{2}}}{3M}\,.
%\end{equation}

\noindent
(b)  
\begin{equation}
w:=z+z^{Bog}_{\bold{j}_1,\dots,\bold{j}_{m-1}}\leq  z^{Bog}_{\bold{j}_1,\dots,\bold{j}_{m-1}}+ E^{Bog}_{\bold{j}_m}+ (\delta-1)\phi_{\bold{j}_m}\sqrt{\epsilon_{\bold{j}_m}^2+2\epsilon_{\bold{j}_M}}
\end{equation}  with $\delta\leq 1+\sqrt{\epsilon_{\bold{j}_m}}$ and $\epsilon_{\bold{j}_m}$ sufficiently small.  

\noindent
Then, for 
 $N$ sufficiently large 
\begin{eqnarray}\label{estimate-main-lemma-H}
%& &\frac{\sqrt{\frac{(N-l-1)}{(N-l-2)}}}{\Big[1+\frac{N}{N-l-2}(\epsilon+\frac{1+\epsilon-\sqrt{\epsilon^2+2\epsilon}}{l})\Big]^{\frac{1}{2}} \Big[1+\frac{N}{N-l}(\epsilon-\frac{1+\epsilon+\sqrt{\epsilon^2+2\epsilon}}{l})\Big]^{\frac{1}{2}}}\\
& &\| \Big[R_{\bold{j}_1,\dots,\bold{j}_m\,;\,i,i}(w)\Big]^{\frac{1}{2}}\,W_{\bold{j}_m\,;i,i-2}\,\Big[R_{\bold{j}_1,\dots,\bold{j}_m\,;\,i-2,i-2}(w)\Big]^{\frac{1}{2}}\|\,\|\Big[R_{\bold{j}_1,\dots,\bold{j}_m\,;\,i-2,i-2}(w)\Big]^{\frac{1}{2}}W^*_{\bold{j}_m\,;i-2,i}\,\Big[R_{\bold{j}_1,\dots,\bold{j}_m\,;\,i,i}(w)\Big]^{\frac{1}{2}}\|\quad \quad\quad\quad \label{operator-blocks}\\
%%%%
%& &\|\Big[R^{Bog}_{i-2,i-2}(z)\Big]^{\frac{1}{2}}\,W_{i-2,i}\Big[R^{Bog}_{i,i}(z)\Big]^{\frac{1}{2}}\|\,\|\Big[R^{Bog}_{i,i}(z)\Big]^{\frac{1}{2}}W^*_{i,i-2}\,\Big[R^{Bog}_{i-2,i-2}(z)\Big]^{\frac{1}{2}}\|\\
%&\leq&\frac{1}{4\Big[1+\epsilon+\frac{\epsilon+1-\sqrt{\epsilon^2+2\epsilon}}{(N-i+2)}\Big]}\frac{1}{ \Big[1+\epsilon-\frac{\epsilon+1+\sqrt{\epsilon^2+2\epsilon}}{(N-i+2)}\Big]}+CN^{-\eta}\\
%%%%
%&\leq&\frac{1}{4\Big[1+\epsilon_{\bold{j}_m}+o(\epsilon_{\bold{j}_m})+\frac{\epsilon_{\bold{j}_m}+1-\delta\sqrt{\epsilon_{\bold{j}_m}^2+2\epsilon_{\bold{j}_m}}}{(N-i+2)}\Big]}\frac{1}{ \Big[1+\epsilon_{\bold{j}_m}+o(\epsilon_{\bold{j}_m})-\frac{\epsilon_{\bold{j}_m}+1+\delta\sqrt{\epsilon_{\bold{j}_m}^2+2\epsilon_{\bold{j}_m}}}{(N-i+2)}\Big]}\label{main-estimate-intermediate}\\
%%%
%\frac{1}{2\Big[1+(\epsilon+\frac{1+\epsilon-\sqrt{\epsilon^2+2\epsilon}}{N-i})-\mathcal{O}(\epsilon^2)\Big]^{\frac{1}{2}} \Big[1+(\epsilon-\frac{1+\epsilon+\sqrt{\epsilon^2+2\epsilon}}{N-i})-\mathcal{O}(\epsilon^2)\Big]^{\frac{1}{2}}}\\
%%%
&\leq &\frac{1}{4(1+a_{\epsilon_{\bold{j}_m}}-\frac{2b^{(\delta)}_{\epsilon_{\bold{j}_m}}}{N-i+2}-\frac{1-c^{(\delta)}_{\epsilon_{\bold{j}_m}}}{(N-i+2)^2})}\leq \frac{1}{4(1+a_{\epsilon_{\bold{j}_m}}-\frac{2b_{\epsilon_{\bold{j}_m}}}{N-i+2}-\frac{1-c_{\epsilon_{\bold{j}_m}}}{(N-i+2)^2})}\label{main-estimate-intermediate}
\end{eqnarray}
holds for $2\leq i\leq N-2$ and even. Here,  \begin{equation}\label{a-bis}
a_{\epsilon_{\bold{j}_{m}}}:=2\epsilon_{\bold{j}_{m}}+\mathcal{O}(\epsilon^{\nu}_{\bold{j}_{m}})\,,\quad \nu>\frac{11}{8}\,,
\end{equation}
\begin{equation}\label{b-bis}
b^{(\delta)}_{\epsilon_{\bold{j}_{m}}}:=(1+\epsilon_{\bold{j}_{m}})\delta \, \chi_{[0,2)}(\delta)\sqrt{\epsilon_{\bold{j}_{m}}^2+2\epsilon_{\bold{j}_{m}}}\quad,\quad b_{\epsilon_{\bold{j}_{m}}}:=(1+\epsilon_{\bold{j}_{m}})\delta \, \chi_{[0,2)}(\delta)\sqrt{\epsilon_{\bold{j}_{m}}^2+2\epsilon_{\bold{j}_{m}}}\,\Big|_{\delta=1+\sqrt{\epsilon_{\bold{j}_m}}}
\end{equation}
and
\begin{equation}\label{c-bis}
c^{(\delta)}_{\epsilon_{\bold{j}_{m}}}:=-(1-\delta^2 \, \chi_{[0,2)}(\delta))(\epsilon_{\bold{j}_{m}}^2+2\epsilon_{\bold{j}_{m}})\quad,\quad c_{\epsilon_{\bold{j}_{m}}}:=-(1-\delta^2 \, \chi_{[0,2)}(\delta))(\epsilon_{\bold{j}_{m}}^2+2\epsilon_{\bold{j}_{m}})\,\Big|_{\delta=1+\sqrt{\epsilon_{\bold{j}_m}}}
\end{equation}
%(Notice that for $\delta<2$$$\frac{\epsilon_{\bold{j}_{m}}+1+\delta\sqrt{\epsilon^2_{\bold{j}_{m}}+2\epsilon_{\bold{j}_{m}}}}{(N-i+2)}<1+\epsilon_{\bold{j}_{m}}$$ because $N-i\geq 1$ and $\epsilon_{\bold{j}_{m}}>0$. Therefore, the denominator in (\ref{main-estimate-intermediate}) is strictly positive.)
with $\chi_{[0,2)}$ the characteristic function of the interval $[0,2)$.
\end{corollary}

\noindent
\emph{Proof}

We recall that
\begin{equation}
R^{Bog}_{\bold{j}_1,\dots,\bold{j}_m\,;\,i,i}(w):=Q^{(i,i+1)}_{\bold{j}_m}\frac{1}{Q^{(i,i+1)}_{\bold{j}_m}[\, \hat{H}^{Bog}_{\bold{j}_1, \dots,\bold{j}_{m-1}}+\sum_{\bold{j}\in \mathbb{Z}^d \setminus \{\pm\bold{j}_1, \dots, \pm\bold{j}_{m}\}}(k_{\bold{j}})^2a^*_{\bold{j}}a_{\bold{j}}-z^{Bog}_{\bold{j}_1,\dots ,\bold{j}_{m-1}}+ \hat{H}^{Bog}_{\bold{j}_m}-z\,]Q^{(i,i+1)}_{\bold{j}_m}}Q^{(i,i+1)}_{\bold{j}_m}\,.
\end{equation}
Consider  $\epsilon_{\bold{j}_m}$ sufficiently small and $N$ sufficiently large so that $-z-\frac{(m-1)}{(\ln N)^{\frac{1}{8}}}>0$. Consequently, the operator 
\begin{equation}\label{S-operator}
S^{Bog}_{\bold{j}_m\,;\,i,i}(z):=Q^{(i,i+1)}_{\bold{j}_m}\frac{1}{Q^{(i,i+1)}_{\bold{j}_m}[\,\hat{H}^{Bog}_{\bold{j}_m}-z-\frac{(m-1)}{(\ln N)^{\frac{1}{8}}}\,]Q^{(i,i+1)}_{\bold{j}_m}}Q^{(i,i+1)}_{\bold{j}_m}\,
\end{equation}
is well defined.
Next, we observe that 
\begin{equation}
\|\Big[R^{Bog}_{\bold{j}_1,\dots,\bold{j}_m\,;\,i,i}(w)\Big]^{\frac{1}{2}}\,\frac{1}{\Big[S^{Bog}_{\bold{j}_m\,;\,i,i}(z)\Big]^{\frac{1}{2}}}\,\charf_{Q^{(i,i+1)}_{\bold{j}_m}\mathcal{F}^N}\|\leq 1\,
\end{equation}
due to the inequality in (\ref{assumption-flow}).
Using the procedure of \emph{\underline{Lemma 3.4} of \cite{Pi1}},  for $z\leq   E^{Bog}_{\bold{j}_m}+ (\delta-1)\phi_{\bold{j}_m}\sqrt{\epsilon_{\bold{j}_m}^2+2\epsilon_{\bold{j}_m}}$ and $N$ large and $\epsilon_{\bold{j}_m}$ small (recall that we have assumed $-z-\frac{(m-1)}{(\ln N)^{\frac{1}{8}}}>0$) we can bound 
\begin{eqnarray}
& &\| \Big[S_{\bold{j}_1,\dots,\bold{j}_m\,;\,i,i}(z)\Big]^{\frac{1}{2}}\,W_{\bold{j}_m\,;i,i-2}\,\Big[S_{\bold{j}_1,\dots,\bold{j}_m\,;\,i-2,i-2}(z)\Big]^{\frac{1}{2}}\|\,\|\Big[S_{\bold{j}_1,\dots,\bold{j}_m\,;\,i-2,i-2}(z)\Big]^{\frac{1}{2}}W^*_{\bold{j}_m\,;i-2,i}\,\Big[S_{\bold{j}_1,\dots,\bold{j}_m\,;\,i,i}(z)\Big]^{\frac{1}{2}}\|\nonumber \quad \\
&\leq &\sup_{2\leq n_{\bold{j}_0}\leq i}\frac{\,\frac{n_{\bold{j}_0}-1}{N}\phi_{\bold{j}_{m}}(N-i+1)}{4\Big[(N-i-1)\Big(\frac{n_{\bold{j}_0}}{N}\phi_{\bold{j}_{m}}+k_{\bold{j}_{m}}^2\Big)-z-\frac{(m-1)}{(\ln N)^{\frac{1}{8}}}\Big]}\,\frac{\frac{n_{\bold{j}_0}}{N}\phi_{\bold{j}_{m}}(N-i+1)}{ \Big[(N-i+1)\Big(\frac{n_{\bold{j}_0}-2}{N}\phi_{\bold{j}_{m}}+k_{\bold{j}_{m}}^2\Big)-z-\frac{(m-1)}{(\ln N)^{\frac{1}{8}}}\Big]}\quad \label{bound-S}\\
&= &\sup_{2\leq n_{\bold{j}_0}\leq i}\frac{\,\frac{n_{\bold{j}_0}-1}{N}\phi_{\bold{j}_{m}}}{4\Big[\Big(\frac{n_{\bold{j}_0}}{N}\phi_{\bold{j}_{m}}+k_{\bold{j}_{m}}^2\Big)(1-\frac{2}{N-i+1})-\frac{z+\frac{(m-1)}{(\ln N)^{1/8}}}{N-i+1}\Big]}\,\frac{\frac{n_{\bold{j}_0}}{N}\phi_{\bold{j}_{m}}}{ \Big[\Big(\frac{n_{\bold{j}_0}-2}{N}\phi_{\bold{j}_{m}}+k_{\bold{j}_{m}}^2\Big)-\frac{z+\frac{(m-1)}{(\ln N)^{1/8}}}{N-i+1}\Big]} \label{step-a}\\
%&=&\frac{1}{4\Big[(1+\frac{N\epsilon_{\bold{j}_*}+1}{n_{\bold{j}_0}-1})(1-\frac{2}{N-i+1})-\frac{N}{(N-i+1)(n_{\bold{j}_{0}}-1)}\frac{z}{\phi_{\bold{j}_{*}}}\Big]}\, \frac{1}{ \Big[1+\frac{N\epsilon_{\bold{j}_{*}}-2}{n_{\bold{j}_0}}-\frac{N}{n_{\bold{j}_0}(N-i+1)}\frac{z}{\phi_{\bold{j}_{*}}}\Big]}\quad\quad\label{fin}\\
&\leq &\frac{1}{4\Big[(1+\frac{N\epsilon_{\bold{j}_m}}{n_{\bold{j}_0}})(1-\frac{2}{N-i+1})-\frac{N}{(N-i+1)n_{\bold{j}_{0}}}\frac{z+\frac{(m-1)}{(\ln N)^{1/8}}}{\phi_{\bold{j}_{m}}}\Big]}\,\Big|_{n_{\bold{j}_0}=N}\, \frac{1}{ \Big[1+\frac{N\epsilon_{\bold{j}_{m}}-1}{n_{\bold{j}_0}-1}-\frac{N}{(n_{\bold{j}_0}-1)(N-i+1)}\frac{z+\frac{(m-1)}{(\ln N)^{1/8}}}{\phi_{\bold{j}_{m}}}\Big]}\Big|_{n_{\bold{j}_0}-1=N}\,.\quad\quad\quad\label{bound-S-1}
%&=&\frac{1}{4\Big[1+\frac{N\epsilon_{\bold{j}_{*}}-1}{n_{\bold{j}_0}+1}-\frac{1}{(N-i-1)}\frac{z}{\phi_{\bold{j}_{*}}}\Big]\,\Big[1+\frac{N}{n_{\bold{j}_0}+1}\epsilon_{\bold{j}_{*}}(1-\frac{2}{N-i-1})-\frac{2}{N-i-1}-\frac{1}{N-i-1}\frac{z}{\phi_{\bold{j}_{*}}}\Big]}\quad\quad\quad \label{fin}
\end{eqnarray}
Like in \emph{\underline{Lemma 3.4} of \cite{Pi1}}, (\ref{bound-S-1}) implies   the inequality in (\ref{main-estimate-intermediate})  where the $N-$dependent correction due to $-\frac{(m-1)}{(\ln N)^{\frac{1}{8}}}$ is hidden in the term $\mathcal{O}(\epsilon^{\nu}_{\bold{j}_m})$\,.
The second inequality in (\ref{main-estimate-intermediate}) holds because $-\frac{2b^{(\delta)}_{\epsilon_{\bold{j}_m}}}{N-i+1}-\frac{1-c^{(\delta)}_{\bold{j}_m}}{(N-i+1)^2}$ is nonincreasing as a function of $\delta$ in the considered  range $\delta\leq 1+\sqrt{\epsilon_{\bold{j}_m}}$ provided  $\epsilon_{\bold{j}_m}$ is sufficiently small. %\begin{equation}
%Q^{(i,i+1)}_{\bold{j}_m}(H^{Bog}_{\bold{j}_1,\dots,\bold{j}_{m-1}}-z^{Bog}_{\bold{j}_1,\dots,\bold{j}_{m-1}})Q^{(i,i+1)}_{\bold{j}_m}\geq 0\,.
%\end{equation} 
\qed
\begin{lemma}\label{main-relations}
Let $2\leq m \leq M$ and assume that the Hamiltonian $H^{Bog}_{\bold{j}_1,\dots,\bold{j}_{m-1}}$ has ground state vector $\psi^{Bog}_{\bold{j}_1,\dots,\bold{j}_{m-1}}$ (see (\ref{gs-Hm-start})-(\ref{gs-Hm-fin})) with ground state energy $z^{Bog}_{\bold{j}_1,\dots,\bold{j}_{m-1}}$. 
%Define $$\Delta_{0}:=\min\, \Big\{\epsilon_{\bold{j}}\,|\,\bold{j}\in \mathbb{Z}^3 \Big\}\,,$$  and, for $m\geq 2$, assume that there exists $\Delta_{m-1}{\color{red}>0}$ such that
%\begin{equation}
%\text{Gap}\,\Big[\hat{H}^{Bog}_{\bold{j}_1, \dots,\bold{j}_{m-1}}+\sum_{\bold{j}\neq \{\pm\bold{j}_1, \dots, \pm\bold{j}_{m}\}}(k_{\bold{j}})^2a^*_{\bold{j}}a_{\bold{j}}\Big]\geq \Delta_{m-1}
%\end{equation}
%(Notice that $$\hat{H}^{Bog}_{\bold{j}_1, \dots,\bold{j}_{m-1}}+\sum_{\bold{j}\neq \{\pm\bold{j}_1, \dots, \pm\bold{j}_{m}\}}(k_{\bold{j}})^2a^*_{\bold{j}}a_{\bold{j}}=H^{Bog}_{\bold{j}_1, \dots,\bold{j}_{m-1}}-(k_{\bold{j}_m})^2(a^*_{\bold{j}_m}a_{\bold{j}_m}+a^*_{-\bold{j}_m}a_{-\bold{j}_m})\,,$$ i.e., the kinetic energy associated with the modes $\pm \bold{j}_m$ is absent). 
Assume condition a) of Corollary \ref{main-lemma-H} and that there exist $0\leq\tilde{C}_{m-1}\leq \tilde{C}_{M-1}<\infty$ such that
 \begin{equation}\label{ass-number-0}
\langle \frac{\psi^{Bog}_{\bold{j}_1,\dots,\bold{j}_{m-1}}}{\|\psi^{Bog}_{\bold{j}_1,\dots,\bold{j}_{m-1}}\|}\,,\,\sum_{\bold{j}\in\mathbb{Z}^d\setminus \{\bold{0}\}} a_{\bold{j}}^{*}a_{\bold{j}}\,\frac{\psi^{Bog}_{\bold{j}_1,\dots,\bold{j}_{m-1}}}{\|\psi^{Bog}_{\bold{j}_1,\dots,\bold{j}_{m-1}}\|}\rangle \leq \tilde{C}_{m-1}\,.
\end{equation}
Let $\epsilon_{\bold{j}_m}$ be sufficiently small  and $N$ sufficiently large. 
%such that Theorem \ref{Feshbach-Hbog} can be applied. 
%the Feshbach flow associated with $H^{Bog}_{\bold{j}_1,\dots,\bold{j}_{m}}$ and the couple of modes $\pm \bold{j}_m$ is well defined for $$z\leq E^{Bog}_{\bold{j}_*}+ \sqrt{\epsilon_{\bold{j}_m}}\phi_{\bold{j}_m}\sqrt{\epsilon_{\bold{j}_m}^2+2\epsilon_{\bold{j}_m}}\,.$$
 Then, for $z$ in the interval $$z\leq E^{Bog}_{\bold{j}_m}+ \sqrt{\epsilon_{\bold{j}_m}}\phi_{\bold{j}_m}\sqrt{\epsilon_{\bold{j}_m}^2+2\epsilon_{\bold{j}_m}}-\frac{(\lfloor (\ln N)^{\frac{1}{2}} \rfloor +4)\phi_{\bold{j}_m}}{N}\,(<0)$$
the following estimates hold true
\begin{eqnarray}
& &\Big|\langle  \frac{\psi^{Bog}_{\bold{j}_1,\dots,\bold{j}_{m-1}}}{\|\psi^{Bog}_{\bold{j}_1,\dots,\bold{j}_{m-1}}\|}\,,\,\Gamma^{Bog}_{\bold{j}_1,\dots,\bold{j}_m ;N,N}(z+z^{Bog}_{\bold{j}_1,\dots,\bold{j}_{m-1}}) \frac{\psi^{Bog}_{\bold{j}_1,\dots,\bold{j}_{m-1}}}{\|\psi^{Bog}_{\bold{j}_1,\dots,\bold{j}_{m-1}}\|}\rangle- \langle  \eta\,,\,\tilde{\Gamma}^{Bog}_{\bold{j}_1,\dots,\bold{j}_m ;N,N}(z) \eta \rangle \Big| \leq \frac{C_I}{(\ln N)^{\frac{1}{4}}}\,,\quad \quad\quad \label{primero-1-1}
%&= &\langle  \frac{\psi^{Bog}_{\bold{j}_1,\dots,\bold{j}_{m-1}}}{\|\psi^{Bog}_{\bold{j}_1,\dots,\bold{j}_{m-1}}\|}\,,\,W_{\bold{j}_2}\,R^{Bog}_{\bold{j}_1,\,\bold{j}_2\,;\,N-2,N-2}(z_{\bold{j}_1}+z)\times\\
%& &\quad\quad \times \sum_{l_{N-2}=0}^{\infty}\Big[\Gamma^{Bog}_{\bold{j}_1,\,\bold{j}_2\,;\,N-2,N-2}(z_{\bold{j}_1}+z)R^{Bog}_{\bold{j}_1,\,\bold{j}_2\,;\,\,N-2,N-2}(z_{\bold{j}_1}+z)\Big]^{l_{N-2}}W^*_{\bold{j}_2}\, \frac{\psi^{Bog}_{\bold{j}_1,\dots,\bold{j}_{m-1}}}{\|\psi^{Bog}_{\bold{j}_1,\dots,\bold{j}_{m-1}}\|}\rangle \quad\quad\quad\\
%&= &\langle  \eta\,,\,\tilde{\Gamma}^{Bog}_{\bold{j}_1,\dots,\bold{j}_m ;N,N}(z) \eta \rangle +{\color{red}\frac{C_I}{N^{\frac{1}{16}}}}\label{A-remainder}
\end{eqnarray}
\begin{eqnarray}
& &\|\mathscr{P}_{\psi^{Bog}_{\bold{j}_1,\dots,\bold{j}_{m-1}}}\Gamma^{Bog}_{\bold{j}_1,\dots,\bold{j}_m;\,N,N}(z+z^{Bog}_{\bold{j}_1,\dots,\bold{j}_{m-1}})\,\overline{\mathscr{P}_{\psi^{Bog}_{\bold{j}_1,\dots,\bold{j}_{m-1}}}}\|\leq \frac{C_{II}}{(\ln N)^{\frac{1}{4}}} \label{primero-3-1}
\end{eqnarray}
for some $0< C_{I}, C_{II}<\infty $, where $\tilde{\Gamma}^{Bog}_{\bold{j}_1,\dots,\bold{j}_m ;N,N}(z)$ is defined starting from  $\Gamma^{Bog}_{\bold{j}_1,\dots,\bold{j}_m ;N,N}(z+z^{Bog}_{\bold{j}_1,\dots,\bold{j}_{m-1}})$ (see (\ref{GammaNNBog})) by replacing (\ref{def-tilde-0}) with (\ref{def-tilde}).
\end{lemma}

\noindent
\emph{Proof}

At first, we show the result contained in (\ref{primero-1-1}). For (\ref{primero-3-1}) the argument is partially the same and is provided afterwards.

\noindent
We implement {\bf{STEPS I-IV}} outlined in \emph{{\bf{Outline of the proof}}}; see Lemma \ref{main-relations-new}.
\\

\noindent
{\bf{STEP I)}}

\noindent
For $\epsilon_{\bold{j}_m}$ sufficiently small and $N$ sufficiently large, we can estimate (recall $w\equiv z+z^{Bog}_{\bold{j}_1,\dots,\bold{j}_{m-1}}$)
\begin{eqnarray}
& &(\ref{expansion-z})\\
&=& \langle  \frac{\psi^{Bog}_{\bold{j}_1,\dots,\bold{j}_{m-1}}}{\|\psi^{Bog}_{\bold{j}_1,\dots,\bold{j}_{m-1}}\|}\,, \,W_{\bold{j}_m}R^{Bog}_{\bold{j}_1,\dots,\bold{j}_m,;\,N-2,N-2}(w)\,\sum_{l_{N-2}=0}^{\bar{j}-1}\Big[\Gamma^{Bog\,}_{\bold{j}_1,\dots,\bold{j}_m\,,;\,N-2,N-2}(w)R^{Bog}_{\bold{j}_1,\dots,\bold{j}_m\,;\,N-2,N-2}(w)\Big]^{l_{N-2}}\times \quad\quad\quad\label{fixed-point-m-bis-bis}\\
& &\quad \times W_{\bold{j}_m}^* \frac{\psi^{Bog}_{\bold{j}_1,\dots,\bold{j}_{m-1}}}{\|\psi^{Bog}_{\bold{j}_1,\dots,\bold{j}_{m-1}}\|} \rangle \nonumber\\
& &+\mathcal{O}((\frac{4}{5})^{\bar{j}})
%& &-\sum_{l_i=0}^{\infty}Q^{(>i)}W\,Q^{(i)}R^{Bog}_{i,i}(z)\Big[\Gamma^{Bog\,}_{i,i}(z)R^{Bog}_{i,i}(z)\Big]^{l_i}Q^{(i)}W^*Q^{(>i)} \nonumber
\end{eqnarray}
where $\bar{j}\geq 2$ is fixed \emph{a posteriori}. This follows from the identity
\begin{eqnarray}
& &(R^{Bog}_{\bold{j}_1,\dots,\bold{j}_m\,;\,N-2,N-2}(w))^{\frac{1}{2}}\Gamma^{Bog\,}_{\bold{j}_1,\dots,\bold{j}_m\,;\,N-2,N-2}(R^{Bog}_{\bold{j}_1,\dots,\bold{j}_m\,;\,N-2,N-2}(w))^{\frac{1}{2}}\label{gammacheck}\\
&=&(R^{Bog}_{\bold{j}_1,\dots,\bold{j}_m\,;\,N-2,N-2}(w))^{\frac{1}{2}}W_{\bold{j}_m\,;\,N-2,N-4}\,(R^{Bog}_{\bold{j}_1,\dots,\bold{j}_m\,;\,N-4,N-4}(w))^{\frac{1}{2}} \times \label{initial}\\
& &\quad\quad\quad\times\sum_{l_{i-2}=0}^{\infty}\Big[(R^{Bog}_{\bold{j}_1,\dots,\bold{j}_m\,;\,N-4,N-4}(w))^{\frac{1}{2}}\Gamma^{Bog}_{\bold{j}_1,\dots,\bold{j}_m\,;\,N-4,N-4}(w)(R^{Bog}_{\bold{j}_1,\dots,\bold{j}_m\,;\,N-4,N-4}(w))^{\frac{1}{2}}\Big]^{l_{i-2}}\times \quad\quad\quad \label{middle}\\
& &\quad\quad\quad\quad\quad\quad\quad\quad\times (R^{Bog}_{\bold{j}_1,\dots,\bold{j}_m\,;\,N-4,N-4}(w))^{\frac{1}{2}}W^*_{\bold{j}_m\,;\,N-4,N-2}(R^{Bog}_{\bold{j}_1,\dots,\bold{j}_m\,;\,N-2,N-2}(w))^{\frac{1}{2}}\, \nonumber
%&=&(R^{Bog}_{i,i}(z))^{\frac{1}{2}}W_{i,i-2}\,(R^{Bog}_{i-2,i-2}(z))^{\frac{1}{2}}\times\\
%& &\quad\times  \sum_{l_{i-2}=0}^{\infty}\Big[(R^{Bog}_{i-2,i-2}(z))^{\frac{1}{2}}\Gamma^{Bog}_{i-2,i-2}(z)R^{Bog}_{i-2,i-2}(z)\Big]^{l_{i-2}}\times \\
%& &\quad \times (R^{Bog}_{i-2,i-2}(z))^{\frac{1}{2}}W^*_{i-2,i}(R^{Bog}_{i,i}(z))^{\frac{1}{2}}\,.
\end{eqnarray}
and the estimates in (\ref{main-estimate-intermediate}) and (\ref{x-ineq}) that imply $\|(\ref{initial})\|^2\leq \frac{4}{15}+o(1)$ and $\|(\ref{middle})\|\leq \frac{8}{3}+o(1)$\,.

\noindent
{\bf{STEP II)}}

\noindent
From the expansion of $\Gamma^{Bog\,}_{\bold{j}_1,\dots,\bold{j}_m\,,;\,N-2,N-2}(w)$ discussed in Proposition \ref{lemma-expansion-proof-0} where $h$ is an even natural number,  $N-4\geq h\geq 2$, that is fixed \emph{a posteriori}, we write
 \begin{eqnarray}
& &(\ref{expansion-z})\\
&=&\langle  \frac{\psi^{Bog}_{\bold{j}_1,\dots,\bold{j}_{m-1}}}{\|\psi^{Bog}_{\bold{j}_1,\dots,\bold{j}_{m-1}}\|}\,, \,W_{\bold{j}_m}\,\times  \label{second-term-fp-eq}\\
& &\quad\times \sum_{l_{N-2}=0}^{\bar{j}-1}R^{Bog}_{\bold{j}_1,\dots,\bold{j}_m,;\,N-2,N-2}(w)\,\Big[\sum_{l=N-2-h,\,l\, even}^{N-4}[\Gamma^{Bog\,}_{\bold{j}_1,\dots,\bold{j}_m;\,N-2,N-2}(w)]_{(l,h_-,\,l+2,h_-,\, \dots \,, N-4,h_-)}\times \nonumber\\
& & \quad\quad\quad\quad \quad\quad\quad\quad \times R^{Bog}_{\bold{j}_1,\dots,\bold{j}_m,;\,N-2,N-2}(w)\Big]^{l_{N-2}}W_{\bold{j}_m}^* \frac{\psi^{Bog}_{\bold{j}_1,\dots,\bold{j}_{m-1}}}{\|\psi^{Bog}_{\bold{j}_1,\dots,\bold{j}_{m-1}}\|} \rangle \nonumber\\
& &+\mathcal{O}((\frac{4}{5})^{\bar{j}})+\mathcal{O}((\frac{1}{1+c\sqrt{\epsilon_{\bold{j}_m}}})^{h})\label{5.22}
\end{eqnarray}
for some $c>0$
 thanks to the estimates in (\ref{gamma-exp-1-0}), (\ref{gamma-exp-2-0}),  and  the observations  in Remarks \ref{prod-control}  and \ref{estimation-proc}. For the details\footnote{Following the argument of  \emph{\underline{Corollary  5.9} of \cite{Pi1}} the second term in (\ref{5.22}) is $\mathcal{O}(\frac{1}{\sqrt{\epsilon_{\bold{j}_m}}}\,(\frac{1}{1+c\sqrt{\epsilon_{\bold{j}_m}}})^{h})$. However, in the present paper the multiplicative constants may depend on the size of the box and the details of the potential.}  we refer the reader to \emph{\underline{Corollary  5.9} of \cite{Pi1}} where a similar argument is implemented.
\\

\noindent
{\bf{ STEP III)}}

\noindent
Concerning the quantity in (\ref{second-term-fp-eq}), in Lemma \ref{truncation}
%begin{equation}\label{leading-step-2}
%\sum_{l={\color{red}N-2-h},\,l\, even}^{N-2}\langle  \frac{\psi^{Bog}_{\bold{j}_1,\dots,\bold{j}_{m-1}}}{\|\psi^{Bog}_{\bold{j}_1,\dots,\bold{j}_{m-1}}\|}\,, \,[\Gamma^{Bog\,}_{\bold{j}_1,\dots,\bold{j}_m\,,;\,N,N}(z_{\bold{j}_1,\dots,\bold{j}_{m-1}}+z)]_{(l,h_-;l+2,h_-;\dots ; N-2,h_-)} \frac{\psi^{Bog}_{\bold{j}_1,\dots,\bold{j}_{m-1}}}{\|\psi^{Bog}_{\bold{j}_1,\dots,\bold{j}_{m-1}}\|} \rangle\,,
%\end{equation} 
we show  that 
\begin{equation}\label{identity-Gammas}
\sum_{l=N-2-h,\,l\, even}^{N-4}[\Gamma^{Bog\,}_{\bold{j}_1,\dots,\bold{j}_m\,,;\,N-2,N-2}(w)]_{(l,h_-;l+2,h_-;\dots ; N-4,h_-)} \equiv [\Gamma^{Bog\,}_{\bold{j}_1,\dots,\bold{j}_m\,;\,N-2,N-2}(w)]_{\mathcal{\tau}_h}
\end{equation} 
where the R-H-S is defined recursively by 
\begin{eqnarray}\label{tau-exp}
& &[\Gamma^{Bog\,}_{\bold{j}_1,\dots,\bold{j}_m\,;\,i,i}(w)]_{\mathcal{\tau}_h}\\
&:=&W_{\bold{j}_m} R^{Bog}_{\bold{j}_1,\dots,\bold{j}_m\,;\,i-2,i-2}(w)\,\sum_{l_{i-2}=0}^{h-1}\Big\{[\Gamma^{Bog\,}_{\bold{j}_1,\dots,\bold{j}_m\,;\,i-2,i-2}(w)]_{\mathcal{\tau}_h}\,R^{Bog}_{\bold{j}_1,\dots,\bold{j}_m\,;\,i-2,i-2}(w)\Big\}^{l_{i-2}} W^*_{\bold{j}_m}\nonumber
\end{eqnarray}
for $N-2\geq i \geq N-h$, and
\begin{equation}\label{tau-exp-bis}
[\Gamma^{Bog\,}_{\bold{j}_1,\dots,\bold{j}_m\,;\,N-2-h,N-2-h}(w)]_{\mathcal{\tau}_h}
:=  W_{\bold{j}_m}\,R^{Bog}_{\bold{j}_1,\dots,\bold{j}_m\,;\, N-4-h, N-4-h}(w)W_{\bold{j}_m}^*\,.\quad\quad
\end{equation}
In the sequel, we make use of the quantity $[\tilde{\Gamma}^{Bog\,}_{\bold{j}_1,\dots,\bold{j}_m\,;\,i,i}(z)]_{\mathcal{\tau}_h}$ defined like $[\Gamma^{Bog\,}_{\bold{j}_1,\dots,\bold{j}_m\,;\,i,i}(w)]_{\mathcal{\tau}_h}$ in (\ref{tau-exp})-(\ref{tau-exp-bis}) but with each operators $R^{Bog}_{\bold{j}_1,\dots,\bold{j}_m\,;\,i,i}(w)$, $w=z+z^{Bog}_{\bold{j}_1,\dots,\bold{j}_{m-1}}$,  replaced with the corresponding  $\tilde{R}^{Bog}_{\bold{j}_1,\dots,\bold{j}_m\,;\,i,i}(z)$.

Next, we observe that:
\begin{itemize}
\item[$\bf{III_{1}\Big)}$]
The expansion in (\ref{tau-exp}) of the operator 
\begin{equation}\label{full-expression}
[\Gamma^{Bog\,}_{\bold{j}_1,\dots,\bold{j}_m\,;\,i,i}(w)]_{\mathcal{\tau}_h}
\end{equation}
 produces $\mathcal{O}(h^2)$ operators $[\Gamma^{Bog\,}_{\bold{j}_1,\dots,\bold{j}_m\,;\,i-2,i-2}(w)]_{\mathcal{\tau}_h}$
 . Then,  using iteratively (\ref{tau-exp}) from $i=N-2$ down to $i=N-h$ we get
 \begin{equation}\label{summands-rl}
[\Gamma^{Bog\,}_{\bold{j}_1,\dots,\bold{j}_m\,;\,N-2,N-2}(w)]_{\mathcal{\tau}_h}=: \sum_{r=1}^{\bar{r}}[\Gamma^{Bog\,}_{\bold{j}_1,\dots,\bold{j}_m\,;\,N-2,N-2}(w)]^{(r)}_{\mathcal{\tau}_h}
 \end{equation}
for some $h-$dependent  $\bar{r}<\infty$ and each summand $[\Gamma^{Bog\,}_{\bold{j}_1,\dots,\bold{j}_m\,;\,N-2,N-2}(w)]^{(r)}_{\mathcal{\tau}_h}$ corresponds to $W_{\bold{j}_m}$ multiplying on the right a finite product of ``RW-blocks", i.e., operators of the type
 \begin{equation}\label{blocks}
R^{Bog}_{\bold{j}_1,\dots,\bold{j}_m\,;\,i,i}(w)W_{\bold{j}_m}^*\quad,\quad R^{Bog}_{\bold{j}_1,\dots,\bold{j}_m\,;\,i,i}(w)W_{\bold{j}_m}\,,
\end{equation}
where $i$ is even and ranges\footnote{For the expression $R^{Bog}_{\bold{j}_1,\dots,\bold{j}_m\,;\,i,i}(w)W_{\bold{j}_m}$, the index $i $ ranges from $N-2-h$ to $N-4$.} from $N-4-h$ to $N-4$.  The number of the RW-blocks for each $[\Gamma^{Bog\,}_{\bold{j}_1,\dots,\bold{j}_m\,;\,N-2,N-2}(w)]^{(r)}_{\mathcal{\tau}_h}$ is bounded\footnote{It is enough to proceed by induction taking into account the maximum number of factors $[\Gamma^{Bog\,}_{\bold{j}_1,\dots,\bold{j}_m\,;\,i-2,i-2}(w)]_{\mathcal{\tau}_h}$  and  $R^{Bog}_{\bold{j}_1,\dots,\bold{j}_m\,;\,i-2,i-2}(w)$ in $W_{\bold{j}_m} R^{Bog}_{\bold{j}_1,\dots,\bold{j}_m\,;\,i-2,i-2}(w)\,\Big\{[\Gamma^{Bog\,}_{\bold{j}_1,\dots,\bold{j}_m\,;\,i-2,i-2}(w)]_{\mathcal{\tau}_h}\,R^{Bog}_{\bold{j}_1,\dots,\bold{j}_m\,;\,i-2,i-2}(w)\Big\}^{h} W^*_{\bold{j}_m}$ and  that $R^{Bog}_{\bold{j}_1,\dots,\bold{j}_m\,;\,N-h-4,N-h-4}(w)[\Gamma^{Bog\,}_{\bold{j}_1,\dots,\bold{j}_m\,;\,N-h-4,N-h-4}(w)]_{\mathcal{\tau}_h}$  contains two RW-blocks. }  by $\mathcal{O}((2h)^{\frac{h+2}{2}})$ (see section 0.2 in \emph{supporting-file-Bose2.pdf}). 
%The number of summands $\bar{r}$ can be easily bounded with a quantity $ {\color{red}\mathcal{O}(h^{2h})}$.
%{\color{red}Therefore, the full expression (\ref{full-expression}) contains  $\mathcal{O}(h^{2h})$ blocks of the type (\ref{blocks}), 
%at most. }
\item[$\bf{III_{2}\Big)}$]
We can write
\begin{eqnarray}
& & (\ref{second-term-fp-eq})\\
&=&\langle  \frac{\psi^{Bog}_{\bold{j}_1,\dots,\bold{j}_{m-1}}}{\|\psi^{Bog}_{\bold{j}_1,\dots,\bold{j}_{m-1}}\|}\,, \,W_{\bold{j}_m}\,\times  \\
& &\quad\times \sum_{l_{N-2}=0}^{\bar{j}-1}R^{Bog}_{\bold{j}_1,\dots,\bold{j}_m,;\,N-2,N-2}(w)\,\Big[ \sum_{r=1}^{\bar{r}}[\Gamma^{Bog\,}_{\bold{j}_1,\dots,\bold{j}_m\,;\,N-2,N-2}(w)]^{(r)}_{\mathcal{\tau}_h} R^{Bog}_{\bold{j}_1,\dots,\bold{j}_m,;\,N-2,N-2}(w)\Big]^{l_{N-2}}W_{\bold{j}_m}^* \frac{\psi^{Bog}_{\bold{j}_1,\dots,\bold{j}_{m-1}}}{\|\psi^{Bog}_{\bold{j}_1,\dots,\bold{j}_{m-1}}\|} \rangle \nonumber \\
%\end{eqnarray}
%Then (\ref{second-term-fp-eq}) can be written as a  sum of ${\color{red}\mathcal{O}((\bar{r}^{\bar{j}})\bar{j})}$ terms of the type
%\begin{eqnarray}
&= &\sum_{\bar{l}=0}^{\bar{j}-1} \sum_{r_1=1}^{\bar{r}}\dots  \sum_{r_{\,\bar{l}\,}=1}^{\bar{r}}\langle  \frac{\psi^{Bog}_{\bold{j}_1,\dots,\bold{j}_{m-1}}}{\|\psi^{Bog}_{\bold{j}_1,\dots,\bold{j}_{m-1}}\|}\,, \,W_{\bold{j}_m}\,\times \label{expr-406}\\
& &\quad\times\Big\{ R^{Bog}_{\bold{j}_1,\dots,\bold{j}_m,;\,N-2,N-2}(w)\,\prod_{l=1}^{\bar{l}}\Big[\, [\Gamma^{Bog\,}_{\bold{j}_1,\dots,\bold{j}_m\,;\,N-2,N-2}(w)]^{(r_l)}_{\mathcal{\tau}_h} R^{Bog}_{\bold{j}_1,\dots,\bold{j}_m,;\,N-2,N-2}(w)\Big] \Big\}W_{\bold{j}_m}^* \frac{\psi^{Bog}_{\bold{j}_1,\dots,\bold{j}_{m-1}}}{\|\psi^{Bog}_{\bold{j}_1,\dots,\bold{j}_{m-1}}\|} \rangle \nonumber
\end{eqnarray}
where each $[\Gamma^{Bog\,}_{\bold{j}_1,\dots,\bold{j}_m\,;\,N-2,N-2}(w)]^{(r_l)}_{\mathcal{\tau}_h}$ is a summand in (\ref{summands-rl}) and $\prod_{l=1}^{\bar{l}=0}[\dots]\equiv \charf$. The operator
\begin{equation}
W_{\bold{j}_m}\Big\{R^{Bog}_{\bold{j}_1,\dots,\bold{j}_m,;\,N-2,N-2}(w)\,\prod_{l=1}^{\bar{l}}\Big[ [\Gamma^{Bog\,}_{\bold{j}_1,\dots,\bold{j}_m\,;\,N-2,N-2}(w)]^{(r_l)}_{\mathcal{\tau}_h}  R^{Bog}_{\bold{j}_1,\dots,\bold{j}_m,;\,N-2,N-2}(w)\Big]\Big\} W_{\bold{j}_m}^* \label{second-term-fp-eq-bis}
\end{equation}
corresponds to $W_{\bold{j}_m}$ multiplying on the right a finite product of RW-blocks
 \begin{equation}
R^{Bog}_{\bold{j}_1,\dots,\bold{j}_m\,;\,i,i}(w)W_{\bold{j}_m}^*\quad,\quad R^{Bog}_{\bold{j}_1,\dots,\bold{j}_m\,;\,i,i}(w)W_{\bold{j}_m}\,,
\end{equation}
where $i$ is even and ranges\footnote{For the expression $R^{Bog}_{\bold{j}_1,\dots,\bold{j}_m\,;\,i,i}(w)W_{\bold{j}_m}$, the index $i $ ranges from $N-2-h$ to $N-2$.} from $N-4-h$ to $N-2$. The number of RW-blocks in (\ref{second-term-fp-eq-bis})
is  $\mathcal{O}(\bar{j}(2h)^{\frac{h+2}{2}})$ at most  because $0\leq \bar{l}\leq \bar{j}-1$. 
\item[$\bf{III_{3}\Big)}$]
We implement the procedure described below on the operator in (\ref{second-term-fp-eq-bis}) starting from the RW-block on the very right, i.e., $R^{Bog}_{\bold{j}_1,\dots,\bold{j}_m\,;\,N-2,N-2}(w)W_{\bold{j}_m}^*$: We make use of the resolvent equation \begin{eqnarray}
R^{Bog}_{\bold{j}_1,\dots,\bold{j}_m\,;\,i,i}(w)
&=&\tilde{R}^{Bog}_{\bold{j}_1,\dots,\bold{j}_m\,;\,i,i}(z) \\
& &-R^{Bog}_{\bold{j}_1,\dots,\bold{j}_m\,;\,i,i}(w)\{\hat{H}^{Bog}_{\bold{j}_1}+ \dots+\hat{H}^{Bog}_{\bold{j}_{m-1}}-z_{\bold{j}_1,\dots ,\bold{j}_{m-1}}\}\tilde{R}^{Bog}_{\bold{j}_1,\dots,\bold{j}_m\,;\,i,i}(z)\,\quad\quad
%&=&\Big(\charf -R^{Bog}_{\bold{j}_1,\dots,\bold{j}_m\,;\,N-2,N-2}(w)\{\hat{H}^{Bog}_{\bold{j}_1}+ \dots+\hat{H}^{Bog}_{\bold{j}_{m-1}}-z_{\bold{j}_1,\dots ,\bold{j}_{m-1}}\}\Big)\times \\
%& &\quad\quad\times  W_{\bold{j}_m}^*\tilde{R}^{Bog}_{\bold{j}_1,\dots,\bold{j}_m\,;\,N-2,N-2}(w+(k_{\bold{j}_m})^2+\frac{2a^*_{\bold{0}}a_{\bold{0}}}{N}\phi_{\bold{j}_m})\mathscr{P}^{(i)}_{\psi^{Bog}_{\bold{j}_1,\dots,\bold{j}_{m-1}}}\quad\quad\quad \label{res-eq-1}
\end{eqnarray}
and pull $ \hat{H}^{Bog}_{\bold{j}_1}+ \dots+\hat{H}^{Bog}_{\bold{j}_{m-1}}-z^{Bog}_{\bold{j}_1,\dots ,\bold{j}_{m-1}}$ to the right until it hits the vector $\psi^{Bog}_{\bold{j}_1,\dots,\bold{j}_{m-1}}$ in the expression
\begin{eqnarray}
%& &W_{\bold{j}_m}\Big\{ \prod_{l=1}^{h-1}\Big[R^{Bog}_{\bold{j}_1,\dots,\bold{j}_m,;\,N-2,N-2}\, [\Gamma^{Bog\,}_{\bold{j}_1,\dots,\bold{j}_m\,;\,N,N}]^{(r_l)}_{\mathcal{\tau}_h}\times \quad \label{gamma(r)} R^{Bog}_{\bold{j}_1,\dots,\bold{j}_m,;\,N-2,N-2}\Big](z_{\bold{j}_1,\dots,\bold{j}_{m-1}}+z) \Big\}W_{\bold{j}_m}^* \frac{\psi^{Bog}_{\bold{j}_1,\dots,\bold{j}_{m-1}}}{\|\psi^{Bog}_{\bold{j}_1,\dots,\bold{j}_{m-1}}\|}\,\nonumber\\
& &W_{\bold{j}_m}\Big\{R^{Bog}_{\bold{j}_1,\dots,\bold{j}_m,;\,N-2,N-2}(w)\,\prod_{l=1}^{\bar{l}}\Big[ [\Gamma^{Bog\,}_{\bold{j}_1,\dots,\bold{j}_m\,;\,N-2,N-2}(w)]^{(r_l)}_{\mathcal{\tau}_h}  R^{Bog}_{\bold{j}_1,\dots,\bold{j}_m,;\,N-2,N-2}(w)\Big] \Big\}W_{\bold{j}_m}^* \frac{\psi^{Bog}_{\bold{j}_1,\dots,\bold{j}_{m-1}}}{\|\psi^{Bog}_{\bold{j}_1,\dots,\bold{j}_{m-1}}\|}\,.\quad\quad\quad\label{gamma(r)}.
\end{eqnarray}
Recall that, by construction, 
\begin{equation}
(\hat{H}^{Bog}_{\bold{j}_1}+ \dots+\hat{H}^{Bog}_{\bold{j}_{m-1}}-z^{Bog}_{\bold{j}_1,\dots ,\bold{j}_{m-1}})\psi^{Bog}_{\bold{j}_1,\dots,\bold{j}_{m-1}}=0\,.
\end{equation}
Assuming that the resolvent $R^{Bog}_{\bold{j}_1,\dots,\bold{j}_m\,;\,i,i}(w)$  to be replaced with $\tilde{R}^{Bog}_{\bold{j}_1,\dots,\bold{j}_m\,;\,i,i}(z)$ is contained in a RW-block of the type
$$R^{Bog}_{\bold{j}_1,\dots,\bold{j}_m\,;\,i,i}(w)W_{\bold{j}_m}^*\,,$$
we consider the identity
\begin{eqnarray}
& &\{\sum_{l=1}^{m-1}\hat{H}^{Bog}_{\bold{j}_l}-z_{\bold{j}_1,\dots ,\bold{j}_{m-1}}\}\tilde{R}^{Bog}_{\bold{j}_1,\dots,\bold{j}_m\,;\,i,i}(z)W_{\bold{j}_m}^* \\
&=&\tilde{R}^{Bog}_{\bold{j}_1,\dots,\bold{j}_m\,;\,i,i}(z)W_{\bold{j}_m}^* \{\sum_{l=1}^{m-1}\hat{H}^{Bog}_{\bold{j}_l}-z_{\bold{j}_1,\dots ,\bold{j}_{m-1}}\}\\
& &-\tilde{R}^{Bog}_{\bold{j}_1,\dots,\bold{j}_m\,;\,i,i}(z)\Delta_{\mathcal{R}\,,\,\bold{j}_m}\tilde{R}^{Bog}_{\bold{j}_1,\dots,\bold{j}_m\,;\,i,i}(z)W_{\bold{j}_m}^* \quad\quad\label{DeltaR}\\
& &+\tilde{R}^{Bog}_{\bold{j}_1,\dots,\bold{j}_m\,;\,i,i}(z)\Delta_{\mathcal{W}^*\,,\,\bold{j}_m}\label{DeltaW}
\end{eqnarray}
where
\begin{equation}\label{comm-W}
\Delta_{\mathcal{W}^*\,,\,\bold{j}_m}:=[\sum_{l=1}^{m-1}\hat{H}^{Bog}_{\bold{j}_l}\,,\,W_{\bold{j}_m}^*]
\end{equation}
and
\begin{equation}\label{comm-R}
-\tilde{R}^{Bog}_{\bold{j}_1,\dots,\bold{j}_m\,;\,i,i}(z)\Delta_{\mathcal{R}\,,\,\bold{j}_m}\tilde{R}^{Bog}_{\bold{j}_1,\dots,\bold{j}_m\,;\,i,i}(z):=[\sum_{l=1}^{m-1}\hat{H}^{Bog}_{\bold{j}_l}\,,\,\tilde{R}^{Bog}_{\bold{j}_1,\dots,\bold{j}_m\,;\,i,i}(z)]\,.
\end{equation}
The terms proportional to (\ref{DeltaR}) and (\ref{DeltaW}) are the remainders produced by pulling the operator $\sum_{l=1}^{m-1}\hat{H}^{Bog}_{\bold{j}_{l}}-z^{Bog}_{\bold{j}_1,\dots ,\bold{j}_{m-1}}$ through the RW-block
$$\tilde{R}^{Bog}_{\bold{j}_1,\dots,\bold{j}_m\,;\,i,i}(z)W_{\bold{j}_m}^*\,.$$
An analogous procedure applies to a RW-block of the type $R^{Bog}_{\bold{j}_1,\dots,\bold{j}_m\,;\,i,i}(w)W_{\bold{j}_m}$.
\item[$\bf{III_{4}\Big)}$]
 We proceed with the computation of the commutators in  (\ref{comm-W}) and (\ref{comm-R}):
\begin{eqnarray}
& &[\sum_{l=1}^{m-1}\hat{H}^{Bog}_{\bold{j}_l}\,,\,W_{\bold{j}_m}^*]\label{comm-W-bis}\\
&=&\sum_{l=1}^{m-1}[\phi_{\bold{j}_l}\frac{a^*_{\bold{0}}a_{\bold{0}}}{N}(a^*_{\bold{j}_l}a_{\bold{j}_l}+a^*_{-\bold{j}_l}a_{-\bold{j}_l})+\phi_{\bold{j}_l}\frac{a_{\bold{0}}a_{\bold{0}}}{N}a_{\bold{j}_l}^{*}a^*_{-\bold{j}_l}+\phi_{\bold{j}_l}\frac{a^*_{\bold{0}}a^*_{\bold{0}}}{N}a_{\bold{j}_l}a_{-\bold{j}_l}\,,\,\phi_{\bold{j}_m}\frac{a_{\bold{0}}a_{\bold{0}}}{N}a_{\bold{j}_m}^{*}a^*_{-\bold{j}_m}]\nonumber \\
&=&\sum_{l=1}^{m-1}[\phi_{\bold{j}_l}\frac{a^*_{\bold{0}}a_{\bold{0}}}{N}(a^*_{\bold{j}_l}a_{\bold{j}_l}+a^*_{-\bold{j}_l}a_{-\bold{j}_l})\,,\,\phi_{\bold{j}_m}\frac{a_{\bold{0}}a_{\bold{0}}}{N}a_{\bold{j}_m}^{*}a^*_{-\bold{j}_m}]+\sum_{l=1}^{m-1}[\phi_{\bold{j}_l}\frac{a^*_{\bold{0}}a^*_{\bold{0}}}{N}a_{\bold{j}_l}a_{-\bold{j}_l}\,,\,\phi_{\bold{j}_m}\frac{a_{\bold{0}}a_{\bold{0}}}{N}a_{\bold{j}_m}^{*}a^*_{-\bold{j}_m}]\nonumber\\
&=&-2\sum_{l=1}^{m-1}\phi_{\bold{j}_l}\frac{a_{\bold{0}}a_{\bold{0}}}{N^2}(a^*_{\bold{j}_l}a_{\bold{j}_l}+a^*_{-\bold{j}_l}a_{-\bold{j}_l})\,\phi_{\bold{j}_m}a_{\bold{j}_m}^{*}a^*_{-\bold{j}_m}-\sum_{l=1}^{m-1}\phi_{\bold{j}_l}\frac{4a^*_{\bold{0}}a_{\bold{0}}+2}{N^2}a_{\bold{j}_l}a_{-\bold{j}_l}\,\phi_{\bold{j}_m}a_{\bold{j}_m}^{*}a^*_{-\bold{j}_m}\nonumber\\
&=&-2\phi_{\bold{j}_m}a_{\bold{j}_m}^{*}a^*_{-\bold{j}_m}\frac{a_{\bold{0}}a_{\bold{0}}}{N}\frac{1}{N}\sum_{l=1}^{m-1}\phi_{\bold{j}_l}(a^*_{\bold{j}_l}a_{\bold{j}_l}+a^*_{-\bold{j}_l}a_{-\bold{j}_l})\,\nonumber\\
& &-\phi_{\bold{j}_m}a_{\bold{j}_m}^{*}a^*_{-\bold{j}_m}\frac{a^*_{\bold{0}}a_{\bold{0}}}{N}\sum_{l=1}^{m-1}\phi_{\bold{j}_l}\frac{4}{N}a_{\bold{j}_l}a_{-\bold{j}_l}-\phi_{\bold{j}_m}a_{\bold{j}_m}^{*}a^*_{-\bold{j}_m}\sum_{l=1}^{m-1}\phi_{\bold{j}_l}\frac{2}{N^2}a_{\bold{j}_l}a_{-\bold{j}_l}\,\nonumber
%&=&-\phi_{\bold{j}_m}a_{\bold{j}_m}^{*}a^*_{-\bold{j}_m}\frac{a_{\bold{0}}a_{\bold{0}}}{N}\,\Big\{\frac{2}{N}\sum_{l=1}^{m-1}\phi_{\bold{j}_l}(a^*_{\bold{j}_l}a_{\bold{j}_l}+a^*_{-\bold{j}_l}a_{-\bold{j}_l})+\sum_{l=1}^{m-1}\phi_{\bold{j}_l}\frac{4}{N}a_{\bold{j}_l}a_{-\bold{j}_l}\Big\}\,\nonumber\\
%& &-\phi_{\bold{j}_m}a_{\bold{j}_m}^{*}a^*_{-\bold{j}_m}\sum_{l=1}^{m-1}\phi_{\bold{j}_l}\frac{2}{N^2}a_{\bold{j}_l}a_{-\bold{j}_l}\,\nonumber
%&=:&\Delta_{\mathcal{W}\,,\,\bold{j}_m}
\end{eqnarray}
and (using $[Q^{(i,i+1)}_{\bold{j}_m}\,,\,\sum_{l=1}^{m-1}\hat{H}^{Bog}_{\bold{j}_l}]=0$)
\begin{eqnarray}
& &[\sum_{l=1}^{m-1}\hat{H}^{Bog}_{\bold{j}_l}\,,\,\tilde{R}^{Bog}_{\bold{j}_1,\dots,\bold{j}_m\,;\,i,i}(z)]\\
&=&-\tilde{R}^{Bog}_{\bold{j}_1,\dots,\bold{j}_m\,;\,i,i}(z)[\sum_{l=1}^{m-1}\hat{H}^{Bog}_{\bold{j}_l}\,,\,\phi_{\bold{j}_m}\frac{a^*_{\bold{0}}a_{\bold{0}}}{N}(a_{\bold{j}_m}^{*}a_{\bold{j}_m}+a_{-\bold{j}_m}^{*}a_{-\bold{j}_m})]\tilde{R}^{Bog}_{\bold{j}_1,\dots,\bold{j}_m\,;\,i,i}(z)\quad\quad \nonumber
%&=&-\tilde{R}^{Bog}_{\bold{j}_1,\dots,\bold{j}_m\,;\,i,i}(w)\sum_{l=1}^{m-1}[\phi_{\bold{j}_l}\frac{a^*_{\bold{0}}a_{\bold{0}}}{N}(a^*_{\bold{j}_l}a_{\bold{j}_l}+a^*_{-\bold{j}_l}a_{-\bold{j}_l})+\phi_{\bold{j}_l}\frac{a_{\bold{0}}a_{\bold{0}}}{N}a_{\bold{j}_l}^{*}a^*_{-\bold{j}_l}+\phi_{\bold{j}_l}\frac{a^*_{\bold{0}}a^*_{\bold{0}}}{N}a_{\bold{j}_l}a_{-\bold{j}_l}\,,\,\phi_{\bold{j}_m}\frac{a^*_{\bold{0}}a_{\bold{0}}}{N}(a_{\bold{j}_m}^{*}a_{\bold{j}_m}+a_{-\bold{j}_m}^{*}a_{-\bold{j}_m})]\tilde{R}^{Bog}_{\bold{j}_1,\dots,\bold{j}_m\,;\,i,i}(w)\nonumber \\
\end{eqnarray}
where
\begin{eqnarray}
& &[\sum_{l=1}^{m-1}\hat{H}^{Bog}_{\bold{j}_l}\,,\,\phi_{\bold{j}_m}\frac{a^*_{\bold{0}}a_{\bold{0}}}{N}(a_{\bold{j}_m}^{*}a_{\bold{j}_m}+a_{-\bold{j}_m}^{*}a_{-\bold{j}_m})]\\
&=&\sum_{l=1}^{m-1}[\phi_{\bold{j}_l}\frac{a^*_{\bold{0}}a_{\bold{0}}}{N}(a^*_{\bold{j}_l}a_{\bold{j}_l}+a^*_{-\bold{j}_l}a_{-\bold{j}_l})\,,\,\phi_{\bold{j}_m}\frac{a^*_{\bold{0}}a_{\bold{0}}}{N}(a_{\bold{j}_m}^{*}a_{\bold{j}_m}+a_{-\bold{j}_m}^{*}a_{-\bold{j}_m})]\nonumber\\
& &+\sum_{l=1}^{m-1}[\phi_{\bold{j}_l}\frac{a^*_{\bold{0}}a^*_{\bold{0}}}{N}a_{\bold{j}_l}a_{-\bold{j}_l}+\phi_{\bold{j}_l}\frac{a_{\bold{0}}a_{\bold{0}}}{N}a_{\bold{j}_l}^{*}a^*_{-\bold{j}_l}\,,\,\phi_{\bold{j}_m}\frac{a^*_{\bold{0}}a_{\bold{0}}}{N}(a_{\bold{j}_m}^{*}a_{\bold{j}_m}+a_{-\bold{j}_m}^{*}a_{-\bold{j}_m})]\nonumber\\
&=&\sum_{l=1}^{m-1}[\phi_{\bold{j}_l}\frac{a^*_{\bold{0}}a^*_{\bold{0}}}{N}a_{\bold{j}_l}a_{-\bold{j}_l}+\phi_{\bold{j}_l}\frac{a_{\bold{0}}a_{\bold{0}}}{N}a_{\bold{j}_l}^{*}a^*_{-\bold{j}_l}\,,\,\phi_{\bold{j}_m}\frac{a^*_{\bold{0}}a_{\bold{0}}}{N}(a_{\bold{j}_m}^{*}a_{\bold{j}_m}+a_{-\bold{j}_m}^{*}a_{-\bold{j}_m})]\\
&=&-2\sum_{l=1}^{m-1}\phi_{\bold{j}_l}\frac{a^*_{\bold{0}}a^*_{\bold{0}}}{N^2}a_{-\bold{j}_l}a_{\bold{j}_l}\,\phi_{\bold{j}_m}(a_{\bold{j}_m}^{*}a_{\bold{j}_m}+a_{-\bold{j}_m}^{*}a_{-\bold{j}_m})+2\sum_{l=1}^{m-1}\phi_{\bold{j}_l}\frac{a_{\bold{0}}a_{\bold{0}}}{N^2}a^*_{-\bold{j}_l}a^*_{\bold{j}_l}\,\phi_{\bold{j}_m}(a_{\bold{j}_m}^{*}a_{\bold{j}_m}+a_{-\bold{j}_m}^{*}a_{-\bold{j}_m})\nonumber\\
&=&-2\frac{a^*_{\bold{0}}a^*_{\bold{0}}}{N^2}\,\phi_{\bold{j}_m}(a_{\bold{j}_m}^{*}a_{\bold{j}_m}+a_{-\bold{j}_m}^{*}a_{-\bold{j}_m})\sum_{l=1}^{m-1}\phi_{\bold{j}_l}a_{-\bold{j}_l}a_{\bold{j}_l}+2\frac{a_{\bold{0}}a_{\bold{0}}}{N^2}\,\phi_{\bold{j}_m}(a_{\bold{j}_m}^{*}a_{\bold{j}_m}+a_{-\bold{j}_m}^{*}a_{-\bold{j}_m})\sum_{l=1}^{m-1}\phi_{\bold{j}_l}a^*_{-\bold{j}_l}a^*_{\bold{j}_l}\,.\nonumber
%&=:&\Delta_{\mathcal{R}\,,\,\bold{j}_m}
\end{eqnarray}
\item[$\bf{III_{5}\Big)}$]
Using the previous mechanism, step by step  in expression (\ref{gamma(r)}) we replace each $R^{Bog}_{\bold{j}_1,\dots,\bold{j}_m\,;\,i,i}(w)$ with $\tilde{R}^{Bog}_{\bold{j}_1,\dots,\bold{j}_m\,;\,i,i}(z)$ by pulling the operator $\hat{H}^{Bog}_{\bold{j}_1,\dots, \bold{j}_{m-1}}-z_{\bold{j}_1,\dots ,\bold{j}_{m-1}}$ through  each block until it hits the vector $\psi^{Bog}_{\bold{j}_1,\dots,\bold{j}_{m-1}}$. This yields a (leading) product of blocks  denoted by
\begin{eqnarray}
& &W_{\bold{j}_m}\Big\{ \tilde{R}^{Bog}_{\bold{j}_1,\dots,\bold{j}_m,;\,N-2,N-2}(z)\,\prod_{l=1}^{\bar{l}}\Big[ [\tilde{\Gamma}^{Bog\,}_{\bold{j}_1,\dots,\bold{j}_m\,;\,N-2,N-2}(z)]^{(r_l)}_{\mathcal{\tau}_h} \tilde{R}^{Bog}_{\bold{j}_1,\dots,\bold{j}_m,;\,N-2,N-2}(z)\Big] \Big\}W_{\bold{j}_m}^* \frac{\psi^{Bog}_{\bold{j}_1,\dots,\bold{j}_{m-1}}}{\|\psi^{Bog}_{\bold{j}_1,\dots,\bold{j}_{m-1}}\|}\,\nonumber
\end{eqnarray}where each $R^{Bog}_{\bold{j}_1,\dots,\bold{j}_m\,;\,i,i}(w)$ has been replaced with the corresponding $\tilde{R}^{Bog}_{\bold{j}_1,\dots,\bold{j}_m\,;\,i,i}(z)$ and $[\tilde{\Gamma}^{Bog\,}_{\bold{j}_1,\dots,\bold{j}_m\,;\,i,i}(z)]^{(r_l)}_{\mathcal{\tau}_h}$ stands for  $[\Gamma^{Bog\,}_{\bold{j}_1,\dots,\bold{j}_m\,;\,i,i}(w)]^{(r_l)}_{\mathcal{\tau}_h}$ after these replacements. In this process, due to the estimate of the number of RW-blocks in (\ref{second-term-fp-eq-bis}), each RW-block in (\ref{gamma(r)}) produces $\mathcal{O}(\bar{j}(2h)^{\frac{h+2}{2}})$ remainder terms  at most. Thus,  the total number of remainder terms associated with (\ref{gamma(r)})
is bounded by $\mathcal{O}(\bar{j}(2h)^{\frac{h+2}{2}}\cdot \bar{j}(2h)^{\frac{h+2}{2}})$. 
\item[$\bf{III_{6}\Big)}$]
In general (unless $i'$ in (\ref{remain-part1}) is equal to $N-2$), each remainder term can be written as
\begin{eqnarray}
& &W_{\bold{j}_m}\underbrace{\Big\{R^{Bog}_{\bold{j}_1,\dots,\bold{j}_m,;\,N-2,N-2}(w)\,\prod_{l=1}^{\bar{l}}\Big[ [\Gamma^{Bog\,}_{\bold{j}_1,\dots,\bold{j}_m\,;\,N-2,N-2}(w)]^{(r_l)}_{\mathcal{\tau}_h}  R^{Bog}_{\bold{j}_1,\dots,\bold{j}_m,;\,N-2,N-2}(w)\Big] \Big\}}(-)R^{Bog}_{\bold{j}_1,\dots,\bold{j}_m\,;\,i',i'}(w)\times \quad\quad\quad\label{remain-part1}\\
& &\quad \quad\quad\quad\quad\quad\quad\quad \quad\quad\quad Left \nonumber\\
%& &\quad \times R^{Bog}_{\bold{j}_1,\dots,\bold{j}_m\,;\,N-2-h,N-2-h}(w)\tilde{R}^{Bog}_{\bold{j}_1,\dots,\bold{j}_m\,;\,N-2-h,N-2-h}(w)\Delta_{\mathcal{W}\,,\,\bold{j}_m}\,\times \\
& &\quad\quad \times \underbrace{\Big\{\tilde{R}^{Bog}_{\bold{j}_1,\dots,\bold{j}_m,;\,N-2,N-2}(z)\,\prod_{l=1}^{\bar{l}}\Big[ [\tilde{\Gamma}^{Bog\,}_{\bold{j}_1,\dots,\bold{j}_m\,;\,N-2,N-2}(z)]^{(r_l)}_{\mathcal{\tau}_h} \tilde{R}^{Bog}_{\bold{j}_1,\dots,\bold{j}_m,;\,N-2,N-2}(z)\Big] \Big\}W^*_{\bold{j}_m}}\frac{\psi^{Bog}_{\bold{j}_1,\dots,\bold{j}_{m-1}}}{\|\psi^{Bog}_{\bold{j}_1,\dots,\bold{j}_{m-1}}\|}  \quad\quad \label{remain-part2}\\
& & \quad\quad\quad\quad\quad\quad\quad\quad\quad\quad\quad\quad Right\quad\quad\quad\quad\quad\quad\quad \nonumber
\end{eqnarray} 
for some $i'$, $N-4-h\leq i' \leq N-2$, so that the original product of RW-blocks has been split into the product of three subproducts:  
\begin{enumerate}
\item
 the symbol ``Left" stands for a subproduct where the original RW-blocks are unchanged. Note that  if  $i'=N-2$ the ``Left" part is absent;
 \item
 the ``separating resolvent" $(-)R^{Bog}_{\bold{j}_1,\dots,\bold{j}_m\,;\,i',i'}(w)$  that separates the ``Left" part from the ``Right" part;
\item
the symbol ``Right" stands for a subproduct where,  with respect to the original expression, all the RW-blocks minus one --  the  ``exceptional RW-block" -- are unchanged except for the fact that the operators $R^{Bog}_{\bold{j}_1,\dots,\bold{j}_m\,;\,i,i}(w)$ have been replaced with $\tilde{R}^{Bog}_{\bold{j}_1,\dots,\bold{j}_m\,;\,i,i}(z)$. Assume that the exceptional RW-block is of the type $R^{Bog}_{\bold{j}_1,\dots,\bold{j}_m\,;\,i'',i''}(z)W^*_{\bold{j}_m}$. (The other case is analogous.) Then, it is replaced either with
 \begin{equation}\label{new-1}
\tilde{R}^{Bog}_{\bold{j}_1,\dots,\bold{j}_m\,;\,i'',i''}(z)\Delta_{\mathcal{W}^{*}\,,\,\bold{j}_m}
 \end{equation}
%or
% \begin{equation}\label{new-1-bis}
%\tilde{R}^{Bog}_{\bold{j}_1,\dots,\bold{j}_m\,;\,i'',i''}(z)\Delta_{\mathcal{W}\,,\,\bold{j}_m}
% \end{equation}
 or with
\begin{equation}\label{new-2}
-\tilde{R}^{Bog}_{\bold{j}_1,\dots,\bold{j}_m\,;\,i'',i''}(z)\Delta_{\mathcal{R}\,,\,\bold{j}_m}\tilde{R}^{Bog}_{\bold{j}_1,\dots,\bold{j}_m\,;\,i'',i''}(z)W_{\bold{j}_m}^{*}\,.
\end{equation}
%If  it is the first block on the right  after the separating RW-block, it is replaced with
%$$
%R^{Bog}_{\bold{j}_1,\dots,\bold{j}_m\,;\,i'',i''}(w)\tilde{R}^{Bog}_{\bold{j}_1,\dots,\bold{j}_m\,;\,i'',i''}(z)\Delta_{\mathcal{W}^{*}\,,\,\bold{j}_m}
%$$
%or with
%$$
%-R^{Bog}_{\bold{j}_1,\dots,\bold{j}_m\,;\,i'',i''}(w)\tilde{R}^{Bog}_{\bold{j}_1,\dots,\bold{j}_m\,;\,i'',i''}(z)\Delta_{\mathcal{R}\,,\,\bold{j}_m}\tilde{R}^{Bog}_{\bold{j}_1,\dots,\bold{j}_m\,;\,i'',i''}(z)W_{\bold{j}_m}^{*}\,.
%\tilde{R}^{Bog}_{\bold{j}_1,\dots,\bold{j}_m\,;\,i'',i''}(z)\Delta_{\mathcal{W}^{*}\,,\,\bold{j}_m}\,.
%$$
%or
%\begin{equation}\label{new-2-bis}
%-\tilde{R}^{Bog}_{\bold{j}_1,\dots,\bold{j}_m\,;\,i'',i''}(z)\Delta_{\mathcal{R}\,,\,\bold{j}_m}\tilde{R}^{Bog}_{\bold{j}_1,\dots,\bold{j}_m\,;\,i'',i''}(z)W_{\bold{j}_m}\,.
%\end{equation}
\end{enumerate}

Furthermore,  the exceptional RW-block splits the subproduct labeled by ``Right" into two subproducts of RW-blocks labeled by ``Right 1" and ``Right 2", respectively, unless: 1) $i'\equiv i''$, in this case the factor ``Right 1" is just the identity; 2) $i''\equiv N-2$, in this case the factor ``Right 2" is just the identity. Therefore, assuming for example that the replacement of the exceptional RW-block  is with (\ref{new-1}) (see also (\ref{comm-W}) and (\ref{comm-W-bis})) we get
\begin{eqnarray}
%&&[\Gamma^{Bog\,}_{\bold{j}_1,\dots,\bold{j}_m\,;\,N,N}(z_{\bold{j}_1,\dots,\bold{j}_{m-1}}+z)]^{(r)}_{\mathcal{\tau}_h}\\
& &W_{\bold{j}_m}\underbrace{\Big\{R^{Bog}_{\bold{j}_1,\dots,\bold{j}_m,;\,N-2,N-2}(w)\, \prod_{l=1}^{\bar{l}}\Big[ [\Gamma^{Bog\,}_{\bold{j}_1,\dots,\bold{j}_m\,;\,N-2,N-2}(w)]^{(r_l)}_{\mathcal{\tau}_h}  R^{Bog}_{\bold{j}_1,\dots,\bold{j}_m,;\,N-2,N-2}(w)\Big] \Big\}}(-)R^{Bog}_{\bold{j}_1,\dots,\bold{j}_m\,;\,i',i'}(w)\times\quad\quad \nonumber\\
& & \quad\quad\quad\quad\quad\quad\quad\quad\quad\quad\quad\quad\quad Left\quad\\
& &\times \underbrace{\Big\{\tilde{R}^{Bog}_{\bold{j}_1,\dots,\bold{j}_m,;\,N-2,N-2}(z)\, \prod_{l=1}^{\bar{l}}\Big[ [\tilde{\Gamma}^{Bog\,}_{\bold{j}_1,\dots,\bold{j}_m\,;\,N-2,N-2}(z)]^{(r_l)}_{\mathcal{\tau}_h} \tilde{R}^{Bog}_{\bold{j}_1,\dots,\bold{j}_m,;\,N-2,N-2}(z)\Big] \Big\}}\times \\
& & \quad\quad\quad\quad\quad\quad\quad\quad\quad\quad\quad\quad Right\,\,\, 1\\
& &\times \tilde{R}^{Bog}_{\bold{j}_1,\dots,\bold{j}_m\,;\,i'',i''}(z) \phi_{\bold{j}_m}\frac{a_{\bold{j}_m}^{*}a^*_{-\bold{j}_m}a_{\bold{0}}a_{\bold{0}}}{N}\times \label{excep-1}\\
& &\quad\times\, \underbrace{\Big\{ \tilde{R}^{Bog}_{\bold{j}_1,\dots,\bold{j}_m,;\,N-2,N-2}(z)\,\prod_{l=1}^{\bar{l}}\Big[ [\tilde{\Gamma}^{Bog\,}_{\bold{j}_1,\dots,\bold{j}_m\,;\,N-2,N-2}(z)]^{(r_l)}_{\mathcal{\tau}_h} \tilde{R}^{Bog}_{\bold{j}_1,\dots,\bold{j}_m,;\,N-2,N-2}(z)\Big] \Big\}W^*_{\bold{j}_m}}\times \\
& &\quad\quad \quad\quad\quad\quad\quad\quad\quad\quad\quad\quad Right\,\,2 \nonumber\\
& &\quad\quad \times \Big\{-\frac{2}{N}\sum_{l=1}^{m-1}\phi_{\bold{j}_l}(a^*_{\bold{j}_l}a_{\bold{j}_l}+a^*_{-\bold{j}_l}a_{-\bold{j}_l})\Big\}\frac{\psi^{Bog}_{\bold{j}_1,\dots,\bold{j}_{m-1}}}{\|\psi^{Bog}_{\bold{j}_1,\dots,\bold{j}_{m-1}}\|}\label{end-term-1} \\
& &+W_{\bold{j}_m}\underbrace{\Big\{ R^{Bog}_{\bold{j}_1,\dots,\bold{j}_m,;\,N-2,N-2}(w)\, \prod_{l=1}^{\bar{l}}\Big[[\Gamma^{Bog\,}_{\bold{j}_1,\dots,\bold{j}_m\,;\,N-2,N-2}(w)]^{(r_l)}_{\mathcal{\tau}_h}  R^{Bog}_{\bold{j}_1,\dots,\bold{j}_m,;\,N-2,N-2}(w)\Big] \Big\}}(-)R^{Bog}_{\bold{j}_1,\dots,\bold{j}_m\,;\,i',i'}(w)\times\quad\quad \nonumber\\
& & \quad\quad\quad\quad\quad\quad\quad\quad\quad\quad\quad\quad\quad\quad Left\quad\\
& &\times \underbrace{\Big\{ \tilde{R}^{Bog}_{\bold{j}_1,\dots,\bold{j}_m,;\,N-2,N-2}(z)\,\prod_{l=1}^{\bar{l}}\Big[ [\tilde{\Gamma}^{Bog\,}_{\bold{j}_1,\dots,\bold{j}_m\,;\,N-2,N-2}(z)]^{(r_l)}_{\mathcal{\tau}_h} \tilde{R}^{Bog}_{\bold{j}_1,\dots,\bold{j}_m,;\,N-2,N-2}(z)\Big] \Big\}}\times \\
& & \quad\quad\quad\quad\quad\quad\quad\quad\quad\quad\quad\quad\quad Right\,\,\, 1\\
& &\times \tilde{R}^{Bog}_{\bold{j}_1,\dots,\bold{j}_m\,;\,i'',i''}(z)\phi_{\bold{j}_m}\frac{a_{\bold{j}_m}^{*}a^*_{-\bold{j}_m}}{N}\Big[(-4 a^*_{\bold{0}}a_{\bold{0}}-2)\sum_{l=1}^{m-1}\phi_{\bold{j}_l}a_{\bold{j}_l}a_{-\bold{j}_l}\Big]\frac{1}{\sum_{l=1}^{m-1}a^*_{\bold{j}_l}a_{-\bold{j}_l}+1}\times \label{excep-2}\\
& &\quad\times\, \underbrace{\Big\{\tilde{R}^{Bog}_{\bold{j}_1,\dots,\bold{j}_m,;\,N-2,N-2}(z)\, \prod_{l=1}^{\bar{l}}\Big[ [\tilde{\Gamma}^{Bog\,}_{\bold{j}_1,\dots,\bold{j}_m\,;\,N-2,N-2}(z)]^{(r_l)}_{\mathcal{\tau}_h} \tilde{R}^{Bog}_{\bold{j}_1,\dots,\bold{j}_m,;\,N-2,N-2}(z)\Big] \Big\}W^*_{\bold{j}_m}}\times \\
& &\quad\quad \quad\quad\quad\quad \quad\quad\quad\quad\quad\quad\quad Right\,\,2 \nonumber\\
& &\quad\quad \times \frac{\sum_{l=1}^{m-1}a^*_{\bold{j}_l}a_{-\bold{j}_l}+1}{N}\frac{\psi^{Bog}_{\bold{j}_1,\dots,\bold{j}_{m-1}}}{\|\psi^{Bog}_{\bold{j}_1,\dots,\bold{j}_{m-1}}\|}\label{end-term-2}
\end{eqnarray} 
where we have used that the operators $\{a_{\pm\bold{j}_l}\,,\,a^*_{\pm\bold{j}_l}\,;\, l=1,\dots,m-1\}$ commute with ``Right 2". An analogous structure is obtained if we consider (\ref{new-2}) replacing the exceptional RW-block,  or in the cases where the exceptional RW-block is of type $R^{Bog}_{\bold{j}_1,\dots,\bold{j}_m\,;\,i'',i''}(z)W_{\bold{j}_m}$ (see point 3. above).
\item[$\bf{III_{7}\Big)}$]
Here, we estimate the set of remainder terms of the type (\ref{remain-part1})-(\ref{remain-part2})  that have been produced in the previous step.  Due to the assumption in (\ref{ass-number-0}), the norm of the vectors of the type in (\ref{end-term-1}) and (\ref{end-term-2}) is bounded by $\mathcal{O}(\frac{1}{\sqrt{N}})$. The rest of the expression can be controlled by considering the operator norm estimate in Corollary \ref{main-lemma-H}, elementary bounds like $\sum_{\bold{j}\in \mathbb{Z}^d}a^*_{\bold{j}}a_{\bold{j}}\leq N$,  and invoking the argument in Remark \ref{estimation-proc}. 
%The estimate from Corollary  \ref{main-lemma-H} can be employed to provide an upper bound to
%\begin{equation}
%\sum_{\bar{l}=0}^{\bar{j}-1} \sum_{r_1=1}^{\bar{r}}\dots  \sum_{r_{\bar{l}}=1}^{\bar{r}}\prod_{l=1}^{\bar{l}} \Big\|\Big[(R^{Bog}_{\bold{j}_1,\dots,\bold{j}_m,;\,N-2,N-2}(w))^{\frac{1}{2}}\, [\Gamma^{Bog\,}_{\bold{j}_1,\dots,\bold{j}_m\,;\,N-2,N-2}(w)]^{(r_l)}_{\mathcal{\tau}_h}  (R^{Bog}_{\bold{j}_1,\dots,\bold{j}_m,;\,N-2,N-2}(w))^{\frac{1}{2}}\Big]\Big\| 
%\end{equation}We call this upper bound
To this end,  we recall that  due to Remark \ref{estimation-proc} for $\epsilon_{\bold{j}_m}$ small enough we can estimate
\begin{eqnarray}\label{estimation}
& &\Big\|\sum_{\bar{l}=0}^{\bar{j}-1} \sum_{r_1=1}^{\bar{r}}\dots  \sum_{r_{\bar{l}}=1}^{\bar{r}}\prod_{l=1}^{\bar{l}} \Big[(R^{Bog}_{\bold{j}_1,\dots,\bold{j}_m,;\,N-2,N-2}(w))^{\frac{1}{2}}\, [\Gamma^{Bog\,}_{\bold{j}_1,\dots,\bold{j}_m\,;\,N-2,N-2}(w)]^{(r_l)}_{\mathcal{\tau}_h}  (R^{Bog}_{\bold{j}_1,\dots,\bold{j}_m,;\,N-2,N-2}(w))^{\frac{1}{2}}\Big]\Big\| \label{subset-0}\\
&\leq &\mathcal{E}\Big(\sum_{\bar{l}=0}^{\bar{j}-1} \sum_{r_1=1}^{\bar{r}}\dots  \sum_{r_{\bar{l}}=1}^{\bar{r}}\prod_{l=1}^{\bar{l}} \Big\|\Big[(R^{Bog}_{\bold{j}_1,\dots,\bold{j}_m;\,N-2,N-2}(w))^{\frac{1}{2}}\, [\Gamma^{Bog\,}_{\bold{j}_1,\dots,\bold{j}_m\,;\,N-2,N-2}(w)]^{(r_l)}_{\mathcal{\tau}_h}  (R^{Bog}_{\bold{j}_1,\dots,\bold{j}_m;\,N-2,N-2}(w))^{\frac{1}{2}}\Big]\Big\| \Big)\quad\quad\quad\\
&=&\sum_{\bar{l}=0}^{\bar{j}-1} \sum_{r_1=1}^{\bar{r}}\dots  \sum_{r_{\bar{l}}=1}^{\bar{r}}\mathcal{E}\Big(\prod_{l=1}^{\bar{l}} \Big\|\Big[(R^{Bog}_{\bold{j}_1,\dots,\bold{j}_m;\,N-2,N-2}(w))^{\frac{1}{2}}\, [\Gamma^{Bog\,}_{\bold{j}_1,\dots,\bold{j}_m\,;\,N-2,N-2}(w)]^{(r_l)}_{\mathcal{\tau}_h}  (R^{Bog}_{\bold{j}_1,\dots,\bold{j}_m;\,N-2,N-2}(w))^{\frac{1}{2}}\Big]\Big\| \Big)\,\quad\quad\quad \\
&\leq&\mathcal{E}\Big(\Big\|\,\sum_{l_{N-2}=0}^{\bar{j}-1}\{(R^{Bog}_{\bold{j}_1,\dots,\bold{j}_m;\,N-2,N-2}(w))^{\frac{1}{2}}\Gamma^{Bog\,}_{\bold{j}_1,\dots,\bold{j}_m;\,N-2,N-2}(w)(R^{Bog}_{\bold{j}_1,\dots,\bold{j}_m;\,N-2,N-2}(w))^{\frac{1}{2}}\}^{l_{N-2}}\,\Big\|\Big)\,\label{subset-2} \\
&\leq   &\mathcal{O}(1)\label{subset-3}
\end{eqnarray}
where the symbol $\mathcal{E}$ is defined in (\ref{E-est}) and the step from (\ref{subset-2}) to (\ref{subset-3}) follows from (\ref{estimate-E-bis})-(\ref{Estimate-E-bis-bis}).

For the moment, assume that $i'<N-2$ and $i''<N-2$. Consider the expression 
\begin{eqnarray}
& &\sum_{\bar{l}=0}^{\bar{j}-1} \sum_{r_1=1}^{\bar{r}}\dots  \sum_{r_{\bar{l}}=1}^{\bar{r}}\prod_{l=1}^{\bar{l}} \Big[(R^{Bog}_{\bold{j}_1,\dots,\bold{j}_m,;\,N-2,N-2}(w))^{\frac{1}{2}}\, [\Gamma^{Bog\,}_{\bold{j}_1,\dots,\bold{j}_m\,;\,N-2,N-2}(w)]^{(r_l)}_{\mathcal{\tau}_h}  (R^{Bog}_{\bold{j}_1,\dots,\bold{j}_m,;\,N-2,N-2}(w))^{\frac{1}{2}}\Big]_{\#}\quad\quad\quad\quad  \label{diesis}
\end{eqnarray}
where the symbol $[..]_{\#}$ means that some of the resolvents  $R^{Bog}_{\bold{j}_1,\dots,\bold{j}_m\,;\,i,i}(w)$ might  have been replaced with $\tilde{R}^{Bog}_{\bold{j}_1,\dots,\bold{j}_m\,;\,i,i}(z)$, and one  separating resolvent (with $i'<N-2$) and one exceptional RW-block (with $i''<N-2$) are present, with the exceptional RW-block replaced according to the rules explained in {\bf{ STEP ${\bf{III_6}}$)}}: In the example of above, the exceptional RW-block is replaced with the operators in $(\ref{excep-1})+(\ref{excep-2})$. An analogous structure can be shown in all other cases different from the given example.
We observe that:
\begin{itemize}
\item In Corollary \ref{main-lemma-H} we can replace one or both the two resolvents $R^{Bog}_{\bold{j}_1,\dots,\bold{j}_m\,;\,i,i}(w)$, $R^{Bog}_{\bold{j}_1,\dots,\bold{j}_m\,;\,i-2,i-2}(w)$ with $\tilde{R}^{Bog}_{\bold{j}_1,\dots,\bold{j}_m\,;\,i,i}(z)$ and $\tilde{R}^{Bog}_{\bold{j}_1,\dots,\bold{j}_m\,;\,i-2,i-2}(z)$, respectively, and yet we can provide the same estimate from above for the operator norms in (\ref{main-estimate-intermediate});
\item Concerning the operator replacing the exceptional RW-block in (\ref{diesis}) we note that, in the given example,
$$\|(\tilde{R}^{Bog}_{\bold{j}_1,\dots,\bold{j}_m\,;\,i'',i''}(z))^{-\frac{1}{2}}(\ref{excep-1})(\tilde{R}^{Bog}_{\bold{j}_1,\dots,\bold{j}_m\,;\,i''+2,i''+2}(z))^{\frac{1}{2}}\|\leq C\mathcal{E}(\|(\tilde{R}^{Bog}_{\bold{j}_1,\dots,\bold{j}_m\,;\,i'',i''}(z))^{\frac{1}{2}}W^*_{\bold{j}_m}(\tilde{R}^{Bog}_{\bold{j}_1,\dots,\bold{j}_m\,;\,i''+2,i''+2}(z))^{\frac{1}{2}}\|)$$
and the same holds for the operator in (\ref{excep-2}). This type of estimate holds for the operators replacing any possible exceptional block;
\item
The norm of the separating resolvent is less than a universal constant.
\end{itemize}
Hence, we can estimate the norm of (\ref{diesis}) as we estimate (\ref{subset-0}) with the help of  $\mathcal{E}(...)$ defined in Remark \ref{estimation-proc}, i.e., for some $C'>0$
\begin{eqnarray}\label{estimation-bis}
& &\Big\|\sum_{\bar{l}=0}^{\bar{j}-1} \sum_{r_1=1}^{\bar{r}}\dots  \sum_{r_{\bar{l}}=1}^{\bar{r}}\prod_{l=1}^{\bar{l}} \Big[(R^{Bog}_{\bold{j}_1,\dots,\bold{j}_m,;\,N-2,N-2}(w))^{\frac{1}{2}}\, [\Gamma^{Bog\,}_{\bold{j}_1,\dots,\bold{j}_m\,;\,N-2,N-2}(w)]^{(r_l)}_{\mathcal{\tau}_h}  (R^{Bog}_{\bold{j}_1,\dots,\bold{j}_m,;\,N-2,N-2}(w))^{\frac{1}{2}}\Big]_{\#}\Big\| \\
&\leq &C'\,\mathcal{E}\Big(\sum_{\bar{l}=0}^{\bar{j}-1} \sum_{r_1=1}^{\bar{r}}\dots  \sum_{r_{\bar{l}}=1}^{\bar{r}}\prod_{l=1}^{\bar{l}} \Big\|\Big[(R^{Bog}_{\bold{j}_1,\dots,\bold{j}_m;\,N-2,N-2}(w))^{\frac{1}{2}}\, [\Gamma^{Bog\,}_{\bold{j}_1,\dots,\bold{j}_m\,;\,N-2,N-2}(w)]^{(r_l)}_{\mathcal{\tau}_h}  (R^{Bog}_{\bold{j}_1,\dots,\bold{j}_m;\,N-2,N-2}(w))^{\frac{1}{2}}\Big]\Big\| \Big)\quad\quad\quad\\
&=&C'\sum_{\bar{l}=0}^{\bar{j}-1} \sum_{r_1=1}^{\bar{r}}\dots  \sum_{r_{\bar{l}}=1}^{\bar{r}}\mathcal{E}\Big(\prod_{l=1}^{\bar{l}} \Big\|\Big[(R^{Bog}_{\bold{j}_1,\dots,\bold{j}_m;\,N-2,N-2}(w))^{\frac{1}{2}}\, [\Gamma^{Bog\,}_{\bold{j}_1,\dots,\bold{j}_m\,;\,N-2,N-2}(w)]^{(r_l)}_{\mathcal{\tau}_h}  (R^{Bog}_{\bold{j}_1,\dots,\bold{j}_m;\,N-2,N-2}(w))^{\frac{1}{2}}\Big]\Big\| \Big)\,.\quad\quad\quad \label{estimation-bis-1}
%&\leq&C'\mathcal{E}\Big(\Big\|\,\sum_{l_{N-2}=0}^{\bar{j}-1}\Big[(R^{Bog}_{\bold{j}_1,\dots,\bold{j}_m;\,N-2,N-2}(w))^{\frac{1}{2}}\Gamma^{Bog\,}_{\bold{j}_1,\dots,\bold{j}_m;\,N-2,N-2}(w)(R^{Bog}_{\bold{j}_1,\dots,\bold{j}_m;\,N-2,N-2}(w))^{\frac{1}{2}}\Big]^{l_{N-2}}\Big\|\Big)\,\\
%&\leq   &\mathcal{O}(1)\,\label{estimation-bis-final}
\end{eqnarray}
%where in the last step we have made use of 
%{\color{red}because the same control provided by  (\ref{complete-zero}) in Theorem \ref{Feshbach-Hbog} holds for $$\|\Big[(R^{Bog}_{\bold{j}_1,\dots,\bold{j}_m;\,N-2,N-2}(w))^{\frac{1}{2}}\Gamma^{Bog\,}_{\bold{j}_1,\dots,\bold{j}_m;\,N-2,N-2}(w)(R^{Bog}_{\bold{j}_1,\dots,\bold{j}_m;\,N-2,N-2}(w))^{\frac{1}{2}}\Big]_{\#}\|\,$$
%once we have factorized the norm of the extra type-like ``RW-block".}
%\begin{eqnarray}\label{estimation-2}
%& &\sum_{\bar{l}=0}^{\bar{j}-1} \sum_{r_1=1}^{\bar{r}}\dots  \sum_{r_{\bar{l}}=1}^{\bar{r}}\prod_{l=1}^{\bar{l}} \Big\|\Big[(\tilde{R}^{Bog}_{\bold{j}_1,\dots,\bold{j}_m,;\,N-2,N-2}(z))^{\frac{1}{2}}\, [\tilde{\Gamma}^{Bog\,}_{\bold{j}_1,\dots,\bold{j}_m\,;\,N-2,N-2}(z)]^{(r_l)}_{\mathcal{\tau}_h}  (\tilde{R}^{Bog}_{\bold{j}_1,\dots,\bold{j}_m,;\,N-2,N-2}(z))^{\frac{1}{2}}\Big]\Big\| \label{subset-0-2}\quad\quad\quad\quad\\
%&\leq&\mathcal{E}\Big\{\sum_{\bar{l}=0}^{\bar{j}-1} \sum_{r_1=1}^{\bar{r}}\dots  \sum_{r_{\bar{l}}=1}^{\bar{r}}\prod_{l=1}^{\bar{l}} \Big\|\Big[(\tilde{R}^{Bog}_{\bold{j}_1,\dots,\bold{j}_m,;\,N-2,N-2}(z))^{\frac{1}{2}}\, [\tilde{\Gamma}^{Bog\,}_{\bold{j}_1,\dots,\bold{j}_m\,;\,N-2,N-2}(z)]^{(r_l)}_{\mathcal{\tau}_h}  (\tilde{R}^{Bog}_{\bold{j}_1,\dots,\bold{j}_m,;\,N-2,N-2}(z))^{\frac{1}{2}}\Big]\Big\|\Big\}\quad\quad\quad \\
%&\leq&\mathcal{O}(1)\,.\label{estimation-2-bis}
%\end{eqnarray}

Then, we can draw the conclusion: The norm of each remainder term (where the exceptional RW-block has index $i''<N-2$ and the separating resolvent corresponds to an index $i'<N-2$) that is produced in expression (\ref{expr-406}) from the summand
\begin{eqnarray}
& &\langle  \frac{\psi^{Bog}_{\bold{j}_1,\dots,\bold{j}_{m-1}}}{\|\psi^{Bog}_{\bold{j}_1,\dots,\bold{j}_{m-1}}\|}\,, \,W_{\bold{j}_m}\,\times \label{summand}\\
& &\quad\times\Big\{ R^{Bog}_{\bold{j}_1,\dots,\bold{j}_m,;\,N-2,N-2}(w)\,\prod_{l=1}^{\bar{l}}\Big[\, [\Gamma^{Bog\,}_{\bold{j}_1,\dots,\bold{j}_m\,;\,N-2,N-2}(w)]^{(r_l)}_{\mathcal{\tau}_h} R^{Bog}_{\bold{j}_1,\dots,\bold{j}_m,;\,N-2,N-2}(w)\Big] \Big\}W_{\bold{j}_m}^* \frac{\psi^{Bog}_{\bold{j}_1,\dots,\bold{j}_{m-1}}}{\|\psi^{Bog}_{\bold{j}_1,\dots,\bold{j}_{m-1}}\|} \rangle \quad\quad\quad
%R^{Bog}_{\bold{j}_1,\dots,\bold{j}_m,;\,N-2,N-2}(w)\,\prod_{l=1}^{\bar{l}}\Big[ [\Gamma^{Bog\,}_{\bold{j}_1,\dots,\bold{j}_m\,;\,N-2,N-2}(w)]^{(r_l)}_{\mathcal{\tau}_h}  R^{Bog}_{\bold{j}_1,\dots,\bold{j}_m,;\,N-2,N-2}(w)\Big] W_{\bold{j}_m}^*
\end{eqnarray}
 is surely bounded by
\begin{eqnarray}
& &\mathcal{O}(\frac{1}{\sqrt{N}})\times
%\mathcal{E}\Big\{\prod_{l=1}^{\bar{l}} \Big\|\Big[(R^{Bog}_{\bold{j}_1,\dots,\bold{j}_m,;\,N-2,N-2}(w))^{\frac{1}{2}}\, [\Gamma^{Bog\,}_{\bold{j}_1,\dots,\bold{j}_m\,;\,N-2,N-2}(w)]^{(r_l)}_{\mathcal{\tau}_h}  (R^{Bog}_{\bold{j}_1,\dots,\bold{j}_m,;\,N-2,N-2}(w))^{\frac{1}{2}}\Big]\Big\| \Big\}\times \label{error-final} \quad\quad\quad \\
\mathcal{E}\Big(\prod_{l=1}^{\bar{l}} \Big\|\Big[(R^{Bog}_{\bold{j}_1,\dots,\bold{j}_m;\,N-2,N-2}(w))^{\frac{1}{2}}\, [\Gamma^{Bog\,}_{\bold{j}_1,\dots,\bold{j}_m\,;\,N-2,N-2}(w)]^{(r_l)}_{\mathcal{\tau}_h}  (R^{Bog}_{\bold{j}_1,\dots,\bold{j}_m;\,N-2,N-2}(w))^{\frac{1}{2}}\Big]_\#\Big\| \Big)\times \quad\quad\quad\quad \\
& &\quad\quad \times \| (\tilde{R}^{Bog}_{\bold{j}_1,\dots,\bold{j}_m,;\,N-2,N-2}(z))^{\frac{1}{2}}W_{\bold{j}_m}^* \| \| (R^{Bog}_{\bold{j}_1,\dots,\bold{j}_m,;\,N-2,N-2}(w))^{\frac{1}{2}}W_{\bold{j}_m}^* \|\, \nonumber\\
& \leq &\mathcal{O}(\frac{1}{\sqrt{N}})\times \label{error-final}
%\mathcal{E}\Big\{\prod_{l=1}^{\bar{l}} \Big\|\Big[(R^{Bog}_{\bold{j}_1,\dots,\bold{j}_m,;\,N-2,N-2}(w))^{\frac{1}{2}}\, [\Gamma^{Bog\,}_{\bold{j}_1,\dots,\bold{j}_m\,;\,N-2,N-2}(w)]^{(r_l)}_{\mathcal{\tau}_h}  (R^{Bog}_{\bold{j}_1,\dots,\bold{j}_m,;\,N-2,N-2}(w))^{\frac{1}{2}}\Big]\Big\| \Big\}\times \label{error-final} \quad\quad\quad \\
\mathcal{E}\Big(\prod_{l=1}^{\bar{l}} \Big\|(R^{Bog}_{\bold{j}_1,\dots,\bold{j}_m;\,N-2,N-2}(w))^{\frac{1}{2}}\, [\Gamma^{Bog\,}_{\bold{j}_1,\dots,\bold{j}_m\,;\,N-2,N-2}(w)]^{(r_l)}_{\mathcal{\tau}_h}  (R^{Bog}_{\bold{j}_1,\dots,\bold{j}_m;\,N-2,N-2}(w))^{\frac{1}{2}}\Big\| \Big) \quad\quad\quad\quad 
%& &\quad\quad \times \| (\tilde{R}^{Bog}_{\bold{j}_1,\dots,\bold{j}_m,;\,N-2,N-2}(z))^{\frac{1}{2}}W_{\bold{j}_m}^* \| \| (R^{Bog}_{\bold{j}_1,\dots,\bold{j}_m,;\,N-2,N-2}(w))^{\frac{1}{2}}W_{\bold{j}_m}^* \|\, \nonumber
%&=&\mathcal{O}(\frac{1}{\sqrt{N}})\nonumber
\end{eqnarray}
where the factor $\mathcal{O}(\frac{1}{\sqrt{N}})$ comes from the norm of the vectors of the type in (\ref{end-term-1}) and (\ref{end-term-2}). By a similar argument we derive the same estimate (\ref{error-final}) for each remainder term coming from the summand in (\ref{summand}) but corresponding to the cases $i'=N-2$ and/or $i''=N-2$.
Next, we make use of the estimates in (\ref{estimation})-(\ref{subset-3}) and (\ref{estimation-bis})-(\ref{estimation-bis-1}) to control the sums $\sum_{\bar{l}=0}^{\bar{j}-1} \sum_{r_1=1}^{\bar{r}}\dots  \sum_{r_{\bar{l}}=1}^{\bar{r}}$.
Since the total number of remainder terms that are produced in (\ref{expr-406}) out of each summand (\ref{summand})  is bounded by $\mathcal{O}(\bar{j}\,(2h)^{\frac{h+2}{2}}\cdot \bar{j}\,(2h)^{\frac{h+2}{2}})$,
%the estimate in (\ref{error-final}) 
%(\ref{estimation-2})-(\ref{estimation-2-bis}),
we conclude that the total contribution of the error terms produced in {\bf{ STEP III)}} is bounded by $\mathcal{O}(\frac{\bar{j}\,(2h)^{\frac{h+2}{2}}\cdot \bar{j}\,(2h)^{\frac{h+2}{2}}}{\sqrt{N}})$.
%, and on the set $\{\phi_{\bold{j}_i}, (k_{\bold{j}_i})^2;\,\, 1\leq i\leq m\}$, and the total number of remainders associated with (\ref{second-term-fp-eq-bis})  is $\mathcal{O}(\bar{j}\,h^{h}\cdot \bar{j}\,h^{h})$ at most.}
\end{itemize}

\noindent
{\bf{ STEP IV)}}

\noindent
Collecting the results in {\bf{STEP I}}, {\bf{STEP II}}, and {\bf{STEP III}}, and setting $\bar{j}\equiv h$, we conclude that
\begin{eqnarray}
& &\langle  \frac{\psi^{Bog}_{\bold{j}_1,\dots,\bold{j}_{m-1}}}{\|\psi^{Bog}_{\bold{j}_1,\dots,\bold{j}_{m-1}}\|}\,,\,\Gamma^{Bog}_{\bold{j}_1,\dots,\bold{j}_m ;N,N}(w) \frac{\psi^{Bog}_{\bold{j}_1,\dots,\bold{j}_{m-1}}}{\|\psi^{Bog}_{\bold{j}_1,\dots,\bold{j}_{m-1}}\|}\rangle\\
&=&\langle  \frac{\psi^{Bog}_{\bold{j}_1,\dots,\bold{j}_{m-1}}}{\|\psi^{Bog}_{\bold{j}_1,\dots,\bold{j}_{m-1}}\|}\,,\,W_{\bold{j}_m} \sum_{l_{N-2}=0}^{h-1}\tilde{R}^{Bog}_{\bold{j}_1,\dots,\bold{j}_m,;\,N-2,N-2}(z)\,\Big\{[\tilde{\Gamma}^{Bog\,}_{\bold{j}_1,\dots,\bold{j}_m\,;\,N-2,N-2}(z)]_{\mathcal{\tau}_h}\,\tilde{R}^{Bog}_{\bold{j}_1,\dots,\bold{j}_m\,;\,N-2,N-2}(z)\Big\}^{l_{N-2}}\times\quad\quad\quad\quad  \label{leading-step3}\\
& &\quad\quad\quad\quad\quad \times  W^*_{\bold{j}_m}\frac{\psi^{Bog}_{\bold{j}_1,\dots,\bold{j}_{m-1}}}{\|\psi^{Bog}_{\bold{j}_1,\dots,\bold{j}_{m-1}}\|}\rangle \nonumber  \\
& &+\mathcal{O}(\frac{h^2(2h)^{h+2}}{\sqrt{N}})+\mathcal{O}((\frac{4}{5})^h)+\mathcal{O}((\frac{1}{1+c\sqrt{\epsilon_{\bold{j}_m}}})^h)\,.
\end{eqnarray}
%{\color{red}where, here and the following, the multiplicative constants depend on the potential $\phi$ and the size of the box. }
%because the total number of remainders associated with (\ref{second-term-fp-eq}) produced in  {\bf{STEP III}} 
%is bounded by $\mathcal{O}(\bar{j}\,h^{h}\cdot \bar{j}\,h^{h})$. 
For convenience, we define
\begin{eqnarray}
& &[\tilde{\Gamma}^{Bog}_{\bold{j}_1,\dots,\bold{j}_m ;N,N}(z)]_{\tau_h} \label{def-tau} \\
&:=&W_{\bold{j}_m} \sum_{l_{N-2}=0}^{h-1}\tilde{R}^{Bog}_{\bold{j}_1,\dots,\bold{j}_m,;\,N-2,N-2}(z)\,\Big\{[\tilde{\Gamma}^{Bog\,}_{\bold{j}_1,\dots,\bold{j}_m\,;\,N-2,N-2}(z)]_{\mathcal{\tau}_h}\,\tilde{R}^{Bog}_{\bold{j}_1,\dots,\bold{j}_m\,;\,N-2,N-2}(z)\Big\}^{l_{N-2}} W^*_{\bold{j}_m}\,\nonumber 
\end{eqnarray}
where $[\tilde{\Gamma}^{Bog\,}_{\bold{j}_1,\dots,\bold{j}_m\,;\,N-2,N-2}(z)]_{\mathcal{\tau}_h}$ and  $\tilde{R}^{Bog}_{\bold{j}_1,\dots,\bold{j}_m\,;\,N-2,N-2}(z)$ have been defined in {\bf{STEP}} $\bf{III}$ after (\ref{tau-exp-bis}).
Now, we observe that the leading term from {\bf{STEP III}}, i.e., the term in (\ref{leading-step3}), 
%\begin{eqnarray}
%& &[\Gamma^{Bog\,}_{\bold{j}_1,\dots,\bold{j}_m\,;\,N,N}(w)]_{\mathcal{\tau}_h}\\
%&:=&W_{\bold{j}_m} \sum_{l_{N-2}=0}^{h-1}\tilde{R}^{Bog}_{\bold{j}_1,\dots,\bold{j}_m,;\,N-2,N-2}(w)\,\Big\{[\Gamma^{Bog\,}_{\bold{j}_1,\dots,\bold{j}_m\,;%\,N-2,N-2}(w)]_{\mathcal{\tau}_h}\,R^{Bog}_{\bold{j}_1,\dots,\bold{j}_m\,;\,N-2,N-2}(w)\Big\}^{l_{N-2}} W^*_{\bold{j}_m}\nonumber
%\end{eqnarray}
%want to show that 
%\begin{equation}
%\langle  \frac{\psi^{Bog}_{\bold{j}_1,\dots,\bold{j}_{m-1}}}{\|\psi^{Bog}_{\bold{j}_1,\dots,\bold{j}_{m-1}}\|}\,,\,W_{\bold{j}_m} \sum_{l_{N-2}=0}^{h-1}\tilde{R}^{Bog}_%{\bold{j}_1,\dots,\bold{j}_m,;\,N-2,N-2}(w)\,\Big\{[\Gamma^{Bog\,}_{\bold{j}_1,\dots,\bold{j}_m\,;\,N-2,N-2}(w)]_{\mathcal{\tau}_h}\,R^{Bog}_{\bold{j}_1,\dots,\bold{j}%_m\,;\,N-2,N-2}(w)\Big\}^{l_{N-2}} W^*_{\bold{j}_m}\frac{\psi^{Bog}_{\bold{j}_1,\dots,\bold{j}_{m-1}}}{\|\psi^{Bog}_{\bold{j}_1,\dots,\bold{j}_{m-1}}\|}\rangle
%\end{equation}
can be replaced with $$\langle \eta\,,\,\Gamma^{Bog}_{\bold{j}_m ;N,N}(z)\, \eta \rangle $$
up to  a small error that vanishes as $N\to \infty$. To this purpose, we
denote by $\chi_{ N-h }$ the projection onto the subspace of states with at least $N-h $ particles in the zero mode. Then, using the assumption in (\ref{ass-number-0}) we get
\begin{equation}\label{control-pro}
\|(\charf-\chi_{ N-h })\frac{\psi^{Bog}_{\bold{j}_1,\dots,\bold{j}_{m-1}}}{\|\psi^{Bog}_{\bold{j}_1,\dots,\bold{j}_{m-1}}\|}\|\leq \|(\frac{\mathcal{N}_+}{h})^{\frac{1}{2}}(\charf-\chi_{ N-h })\frac{\psi^{Bog}_{\bold{j}_1,\dots,\bold{j}_{m-1}}}{\|\psi^{Bog}_{\bold{j}_1,\dots,\bold{j}_{m-1}}\|}\|\leq \mathcal{O}(\frac{1}{\sqrt{h}})
\end{equation}
where $\mathcal{N}_+:=\sum_{\bold{j}\neq \bold{0}}a^*_{\bold{j}}a_{\bold{j}}$. Consequently, 
\begin{eqnarray}
& &\langle  \frac{\psi^{Bog}_{\bold{j}_1,\dots,\bold{j}_{m-1}}}{\|\psi^{Bog}_{\bold{j}_1,\dots,\bold{j}_{m-1}}\|}\,,\,[\tilde{\Gamma}^{Bog}_{\bold{j}_1,\dots,\bold{j}_m ;N,N}(z)]_{\tau_h} \frac{\psi^{Bog}_{\bold{j}_1,\dots,\bold{j}_{m-1}}}{\|\psi^{Bog}_{\bold{j}_1,\dots,\bold{j}_{m-1}}} \rangle \\
&= &\langle \chi_{ N-h } \frac{\psi^{Bog}_{\bold{j}_1,\dots,\bold{j}_{m-1}}}{\|\psi^{Bog}_{\bold{j}_1,\dots,\bold{j}_{m-1}}\|}\,,\,[\tilde{\Gamma}^{Bog}_{\bold{j}_1,\dots,\bold{j}_m ;N,N}(z)]_{\tau_h}\, \chi_{ N-h}\frac{\psi^{Bog}_{\bold{j}_1,\dots,\bold{j}_{m-1}}}{\|\psi^{Bog}_{\bold{j}_1,\dots,\bold{j}_{m-1}}\|} \rangle 
+\mathcal{O}(\frac{1}{\sqrt{h}})\,.
\end{eqnarray}
In the sequel, we estimate the difference between
\begin{equation}\label{deltan0expression}
\langle \chi_{ N-h } \frac{\psi^{Bog}_{\bold{j}_1,\dots,\bold{j}_{m-1}}}{\|\psi^{Bog}_{\bold{j}_1,\dots,\bold{j}_{m-1}}\|}\,,\,[\tilde{\Gamma}^{Bog}_{\bold{j}_1,\dots,\bold{j}_m ;N,N}(z)]_{\tau_h}\, \chi_{ N-h }\frac{\psi^{Bog}_{\bold{j}_1,\dots,\bold{j}_{m-1}}}{\|\psi^{Bog}_{\bold{j}_1,\dots,\bold{j}_{m-1}}\|} \rangle\,
\end{equation}
and
\begin{equation}
\langle \eta\,,\,\Gamma^{Bog}_{\bold{j}_m ;N,N}(z)\, \eta \rangle \,.
\end{equation}
In passing, it is helpful to recall the identity (see (\ref{correct-G}))
\begin{equation}
\langle \eta\,, \,\Gamma^{Bog}_{\bold{j}_{m};\,N,N}(z)\, \eta \rangle=(1-\frac{1}{N})\frac{\phi_{\bold{j}_{m}}}{2\epsilon_{\bold{j}_m}-\frac{4}{N}+2-\frac{z}{\phi_{\bold{j}_{m}}}}\check{\mathcal{G}}_{\bold{j}_{m}\,;\,N-2,N-2}(z)
\end{equation}
and to express (\ref{deltan0expression}) in a similar way. For this reason, in Lemma \ref{appendix-lemma-1} we consider  the modified expression
 $[\check{\mathcal{G}}_{\bold{j}_{*}\,;\,i,i}(z)]_{\tau_h\,;\,\Delta n_{\bold{j}_{\bold{0}}}}$ defined for $i$ even,  $N-h-2 \leq i\leq N-2$, $0\leq \Delta n_{\bold{j}_{\bold{0}}}\leq h$, $i-\Delta n_{\bold{j}_{\bold{0}}}-2\geq 0$,  by  the relation
 \begin{equation}\label{def-G-n0-1}
[\check{\mathcal{G}}_{\bold{j}_{*}\,;\,i,i}(z)]_{\tau_h\,;\,\Delta n_{\bold{j}_{\bold{0}}}}:=\sum_{l_{i}=0}^{h-1}\{[\mathcal{W}_{\bold{j}_{*}\,;i,i-2}(z)\mathcal{W}^*_{\bold{j}_*\,;i-2,i}(z)]_{\Delta n_{\bold{j}_{\bold{0}}}}\,[\check{\mathcal{G}}_{\bold{j}_{*}\,;\,i-2,i-2}(z)]_{\tau_h\,;\,\Delta n_{\bold{j}_{\bold{0}}}}\}^{l_i}
\end{equation}
where
\begin{equation}
 [\check{\mathcal{G}}_{\bold{j}_{*}\,;\,N-h-4, N-h-4}(z)]_{\tau_h\,;\,\Delta n_{\bold{j}_{\bold{0}}}}\equiv 1\,
\end{equation}
and
\begin{eqnarray}
& &[\mathcal{W}_{\bold{j}_*\,;\,i,i-2}(z)\mathcal{W}^*_{\bold{j}_*\,;\,i-2,i}(z)]_{\Delta n_{\bold{j}_{\bold{0}}}}\label{def-G-n0-med}\\
&:= & \frac{(i-\Delta n_{\bold{j}_{\bold{0}}}-1)(i-\Delta n_{\bold{j}_{\bold{0}}})}{N^2}\,\phi^2_{\bold{j}_{*}}\\
&  &\times\,\frac{(N-i+2)^2}{4\Big[(\frac{i-\Delta n_{\bold{j}_{\bold{0}}}}{N}\phi_{\bold{j}_{*}}+(k_{\bold{j}_*})^2)(N-i)-z\Big]\Big[(\frac{i-2-\Delta n_{\bold{j}_{\bold{0}}}}{N}\phi_{\bold{j}_{*}}+(k_{\bold{j}_*})^2)(N-i+2)-z\Big]}\,.\quad\quad\quad\label{def-G-n0-fin}
%&=&(\frac{(i-1)i}{N^2}+...)\,\phi^2_{\bold{j}_{*}}\\
%&  &\times\,\frac{(N-i+2)^2}{4\Big[(\frac{i}{N}\phi_{\bold{j}_{*}}+(k_{\bold{j}_*})^2)(N-i)-(z+\frac{(N-i)\Delta n_{\bold{j}_{\bold{0}}}\phi_{\bold{j}_{*}}}{N})\Big]\Big[(\frac{i-2-\Delta n_{\bold{j}_{\bold{0}}}}{N}\phi_{\bold{j}_{*}}+(k_{\bold{j}_*})^2)(N-i)+2(\frac{i-2-\Delta n_{\bold{j}_{\bold{0}}}}{N}\phi_{\bold{j}_{*}}+\epsilon_{\bold{j}_*})-z\Big]}\quad\quad\quad
\end{eqnarray}
For $N$ large enough, $(0\leq)\Delta n_{\bold{j}_{\bold{0}}}\leq h$, and 
\begin{equation}
 z\leq E^{Bog}_{\bold{j}_*}+ \sqrt{\epsilon_{\bold{j}_*}}\phi_{\bold{j}_*}\sqrt{\epsilon_{\bold{j}_*}^2+2\epsilon_{\bold{j}_*}}-\frac{(h+4)\phi_{\bold{j}_*}}{N}\,(<0)\,,
 \end{equation}
in Lemma \ref{appendix-lemma-1}  we estimate
 \begin{equation}
\Big|\frac{\partial [\check{\mathcal{G}}_{\bold{j}_{*}\,;\,i,i}(z)]_{\tau_h\,;\,\Delta n_{\bold{j}_{\bold{0}}}}}{\partial \Delta n_{\bold{j}_{\bold{0}}}}\Big|\leq K\frac{h\cdot g^{i-N+h}}{\sqrt{N}}\,
\end{equation}
where $g$ is not larger than $4$ and $K$ is a universal constant. 

\noindent
Going back to expression (\ref{deltan0expression}), we observe that the operator 
\begin{equation}
[\tilde{\Gamma}^{Bog}_{\bold{j}_1,\dots,\bold{j}_m ;N,N}(z)]_{\tau_h}
\end{equation}
 preserves the number of particles for any mode. Therefore,
\begin{eqnarray}
& &\langle \chi_{ N-h} \frac{\psi^{Bog}_{\bold{j}_1,\dots,\bold{j}_{m-1}}}{\|\psi^{Bog}_{\bold{j}_1,\dots,\bold{j}_{m-1}}\|}\,,\,[\tilde{\Gamma}^{Bog}_{\bold{j}_1,\dots,\bold{j}_m ;N,N}(z)]_{\tau_h}\, \chi_{ N-h}\frac{\psi^{Bog}_{\bold{j}_1,\dots,\bold{j}_{m-1}}}{\|\psi^{Bog}_{\bold{j}_1,\dots,\bold{j}_{m-1}}\|} \rangle \\
&= &\sum_{n_{\bold{j}_{0}}=N-h}^N \langle \Big(\chi_{ N-h} \frac{\psi^{Bog}_{\bold{j}_1,\dots,\bold{j}_{m-1}}}{\|\psi^{Bog}_{\bold{j}_1,\dots,\bold{j}_{m-1}}\|}\Big)_{n_{\bold{j}_{0}}}\,,\,[\tilde{\Gamma}^{Bog}_{\bold{j}_1,\dots,\bold{j}_m ;N,N}(z)]_{\tau_h}\, \Big(\chi_{ N-h}\frac{\psi^{Bog}_{\bold{j}_1,\dots,\bold{j}_{m-1}}}{\|\psi^{Bog}_{\bold{j}_1,\dots,\bold{j}_{m-1}}\|}\Big)_{n_{\bold{j}_{0}}} \rangle
\end{eqnarray}
where $\Big(\chi_{ N-h}\frac{\psi^{Bog}_{\bold{j}_1,\dots,\bold{j}_{m-1}}}{\|\psi^{Bog}_{\bold{j}_1,\dots,\bold{j}_{m-1}}\|}\Big)_{n_{\bold{j}_0}} $ is the component with exactly $n_{\bold{j}_{0}}$ particles in the mode $\bold{j}_{0}$. It is straightforward to check that 
\begin{eqnarray}
& &\sum_{n_{\bold{j}_{0}}=N-h}^N \langle \Big(\chi_{ N-h} \frac{\psi^{Bog}_{\bold{j}_1,\dots,\bold{j}_{m-1}}}{\|\psi^{Bog}_{\bold{j}_1,\dots,\bold{j}_{m-1}}\|}\Big)_{n_{\bold{j}_{\bold{0}}}}\,,\,[\tilde{\Gamma}^{Bog}_{\bold{j}_1,\dots,\bold{j}_m ;N,N}(z)]_{\tau_h}\, \Big(\chi_{ N-h}\frac{\psi^{Bog}_{\bold{j}_1,\dots,\bold{j}_{m-1}}}{\|\psi^{Bog}_{\bold{j}_1,\dots,\bold{j}_{m-1}}\|}\Big)_{n_{\bold{j}_{0}}} \rangle\\
&= &\sum_{n_{\bold{j}_{0}}=N-h}^N \frac{n_{\bold{j}_0}(n_{\bold{j}_0}-1)}{N^2}\,\frac{\phi_{\bold{j}_{m}}}{2\epsilon_{\bold{j}_m}+\frac{2(n_{\bold{j}_{0}}-2)}{N}-\frac{z}{\phi_{\bold{j}_{m}}}}[\check{\mathcal{G}}_{\bold{j}_{m}\,;\,N-2,N-2}(z)]_{\tau_h\,;\,N- n_{\bold{j}_{0}}}\Big\| \Big(\chi_{ N-h} \frac{\psi^{Bog}_{\bold{j}_1,\dots,\bold{j}_{m-1}}}{\|\psi^{Bog}_{\bold{j}_1,\dots,\bold{j}_{m-1}}\|}\Big)_{n_{\bold{j}_{0}}}\Big\|^2 \,.\quad\quad\quad
\end{eqnarray}
Invoking Lemma \ref{appendix-lemma-1}, we can write
\begin{eqnarray}
& &\sum_{n_{\bold{j}_{0}}=N-h}^N \frac{n_{\bold{j}_0}(n_{\bold{j}_0}-1)}{N^2}\,\frac{\phi_{\bold{j}_{m}}}{2\epsilon_{\bold{j}_m}+\frac{2(n_{\bold{j}_{0}}-2)}{N}-\frac{z}{\phi_{\bold{j}_{m}}}}[\check{\mathcal{G}}_{\bold{j}_{m}\,;\,N-2,N-2}(z)]_{\tau_h\,;\,N- n_{\bold{j}_{0}}}\Big\| \Big(\chi_{ N-h} \frac{\psi^{Bog}_{\bold{j}_1,\dots,\bold{j}_{m-1}}}{\|\psi^{Bog}_{\bold{j}_1,\dots,\bold{j}_{m-1}}\|}\Big)_{n_{\bold{j}_{0}}}\Big\|^2 \quad\quad\quad\\
%& &\langle \chi_{ N-h} \frac{\psi^{Bog}_{\bold{j}_1,\dots,\bold{j}_{m-1}}}{\|\psi^{Bog}_{\bold{j}_1,\dots,\bold{j}_{m-1}}\|}\,,\,[\tilde{\Gamma}^{Bog}_{\bold{j}_1,\dots,\bold{j}_m ;N,N}(z)]_{\tau_h}\, \chi_{ N-h }\frac{\psi^{Bog}_{\bold{j}_1,\dots,\bold{j}_{m-1}}}{\|\psi^{Bog}_{\bold{j}_1,\dots,\bold{j}_{m-1}}\|} \rangle \\
&= &(1-\frac{1}{N})\frac{\phi_{\bold{j}_{m}}}{2\epsilon_{\bold{j}_m}+2-\frac{4}{N}-\frac{z}{\phi_{\bold{j}_{m}}}}[\check{\mathcal{G}}_{\bold{j}_{m}\,;\,N-2,N-2}(z)]_{\tau_h\,;\,0}\,\Big\|\chi_{ N-h} \frac{\psi^{Bog}_{\bold{j}_1,\dots,\bold{j}_{m-1}}}{\|\psi^{Bog}_{\bold{j}_1,\dots,\bold{j}_{m-1}}\|}\Big\|^2 +\mathcal{S}(z)
\end{eqnarray}
with $|\mathcal{S}(z)|\leq \mathcal{O}(\frac{h^2g^h}{\sqrt{N}})$. In turn, we conclude that
\begin{eqnarray}
%& &\langle  \frac{\psi^{Bog}_{\bold{j}_1,\dots,\bold{j}_{m-1}}}{\|\psi^{Bog}_{\bold{j}_1,\dots,\bold{j}_{m-1}}\|}\,,\,[\tilde{\Gamma}^{Bog\,}_{\bold{j}_1,\dots,\bold{j}_m\,;\,N,N}(z_{\bold{j}_1,\dots,\bold{j}_{m-1}}+z)]_{\mathcal{\tau}_h} \frac{\psi^{Bog}_{\bold{j}_1,\dots,\bold{j}_{m-1}}}{\|\psi^{Bog}_{\bold{j}_1,\dots,\bold{j}_{m-1}}\|} \rangle \\
& &\langle  \frac{\psi^{Bog}_{\bold{j}_1,\dots,\bold{j}_{m-1}}}{\|\psi^{Bog}_{\bold{j}_1,\dots,\bold{j}_{m-1}}\|}\,,\,[\tilde{\Gamma}^{Bog}_{\bold{j}_1,\dots,\bold{j}_m ;N,N}(z)]_{\tau_h} \frac{\psi^{Bog}_{\bold{j}_1,\dots,\bold{j}_{m-1}}}{\|\psi^{Bog}_{\bold{j}_1,\dots,\bold{j}_{m-1}}\|} \rangle \\
&= &(1-\frac{1}{N})\frac{\phi_{\bold{j}_{m}}}{2\epsilon_{\bold{j}_m}+2-\frac{4}{N}-\frac{z}{\phi_{\bold{j}_{m}}}}[\check{\mathcal{G}}_{\bold{j}_{m}\,;\,N-2,N-2}(z)]_{\tau_h\,;\,0}+\mathcal{O}(\frac{1}{\sqrt{h}})+ \mathcal{O}(\frac{h^2g^h}{\sqrt{N}})\label{-1}\\
&=&(1-\frac{1}{N})\frac{\phi_{\bold{j}_{m}}}{2\epsilon_{\bold{j}_m}+2-\frac{4}{N}-\frac{z}{\phi_{\bold{j}_{m}}}}\check{\mathcal{G}}_{\bold{j}_{m}\,;\,N-2,N-2}(z)+\label{0}\\
& &+\mathcal{O}(\frac{1}{\sqrt{h}})+ \mathcal{O}(\frac{h^2 g^h}{\sqrt{N}})+\mathcal{O}((\frac{4}{5})^h)+\mathcal{O}((\frac{1}{1+c\sqrt{\epsilon_{\bold{j}_m}}})^h)\\
& =&\langle \eta \,,\,\tilde{\Gamma}^{Bog}_{\bold{j}_1,\dots,\bold{j}_m ;N,N}(z)\,\eta \rangle+\mathcal{O}(\frac{1}{\sqrt{h}})+ \mathcal{O}(\frac{h^2 g^h}{\sqrt{N}})+\mathcal{O}((\frac{4}{5})^h)+\mathcal{O}((\frac{1}{1+c\sqrt{\epsilon_{\bold{j}_m}}})^h)\quad\quad
\end{eqnarray}
where in the step from (\ref{-1}) to (\ref{0}) we essentially implement the inverse of {\bf{STEP I}} and {\bf{II}} on the quantity  $$(1-\frac{1}{N})\frac{\phi_{\bold{j}_{m}}}{2\epsilon_{\bold{j}_m}+2-\frac{4}{N}-\frac{z}{\phi_{\bold{j}_{m}}}}[\check{\mathcal{G}}_{\bold{j}_{m}\,;\,N-2,N-2}(z)]_{\tau_h\,;\,0}\equiv \langle \eta \,,\,[\tilde{\Gamma}^{Bog}_{\bold{j}_1,\dots,\bold{j}_m ;N,N}(z)]_{\tau_h}\eta \rangle.$$

As far as (\ref{primero-3-1}) is concerned, with the same arguments we arrive at
\begin{eqnarray}
& &\|\overline{\mathscr{P}_{\psi^{Bog}_{\bold{j}_1,\dots,\bold{j}_{m-1}}}}\label{laterale-est-1}\,
\Gamma^{Bog}_{\bold{j}_1,\dots,\bold{j}_m;\,N,N}(z+z^{Bog}_{\bold{j}_1,\dots,\bold{j}_{m-1}})\frac{\psi^{Bog}_{\bold{j}_1,\dots,\bold{j}_{m-1}}}{\|\psi^{Bog}_{\bold{j}_1,\dots,\bold{j}_{m-1}}\|}\| \nonumber\\
&\leq &\|\overline{\mathscr{P}_{\psi^{Bog}_{\bold{j}_1,\dots,\bold{j}_{m-1}}}}\,
[\tilde{\Gamma}^{Bog}_{\bold{j}_1,\dots,\bold{j}_m;\,N,N}(z)]_{\tau_h}\frac{\psi^{Bog}_{\bold{j}_1,\dots,\bold{j}_{m-1}}}{\|\psi^{Bog}_{\bold{j}_1,\dots,\bold{j}_{m-1}}\|}\|\\
& &+\mathcal{O}(\frac{h^2(2h)^{h+2}}{\sqrt{N}})+\mathcal{O}((\frac{4}{5})^h)+\mathcal{O}((\frac{1}{1+c\sqrt{\epsilon_{\bold{j}_m}}})^h)\nonumber\\
&\leq &\|\overline{\mathscr{P}_{\psi^{Bog}_{\bold{j}_1,\dots,\bold{j}_{m-1}}}}\,
[\tilde{\Gamma}^{Bog}_{\bold{j}_1,\dots,\bold{j}_m;\,N,N}(z)]_{\tau_h}\,\chi_{ N-h}\frac{\psi^{Bog}_{\bold{j}_1,\dots,\bold{j}_{m-1}}}{\|\psi^{Bog}_{\bold{j}_1,\dots,\bold{j}_{m-1}}\|}\| \\
& &+\mathcal{O}(\frac{1}{\sqrt{h}})+\mathcal{O}(\frac{h^2(2h)^{h+2}}{\sqrt{N}})+\mathcal{O}((\frac{4}{5})^h)+\mathcal{O}((\frac{1}{1+c\sqrt{\epsilon_{\bold{j}_m}}})^h)\nonumber\\
&=&(1-\frac{1}{N})\frac{\phi_{\bold{j}_{m}}}{2\epsilon_{\bold{j}_m}+2-\frac{4}{N}-\frac{z}{\phi_{\bold{j}_{m}}}}\,[\check{\mathcal{G}}_{\bold{j}_{m}\,;\,N-2,N-2}(z)]_{\tau_h\,;\,0}\|\overline{\mathscr{P}_{\psi^{Bog}_{\bold{j}_1,\dots,\bold{j}_{m-1}}}}
\chi_{ N-h}\frac{\psi^{Bog}_{\bold{j}_1,\dots,\bold{j}_{m-1}}}{\|\psi^{Bog}_{\bold{j}_1,\dots,\bold{j}_{m-1}}\|}\|\\
& &+\mathcal{O}(\frac{1}{\sqrt{h}})+\mathcal{O}(\frac{h^2g^h}{\sqrt{N}})+\mathcal{O}(\frac{h^2(2h)^{h+2}}{\sqrt{N}})+\mathcal{O}((\frac{4}{5})^h)+\mathcal{O}((\frac{1}{1+c\sqrt{\epsilon_{\bold{j}_m}}})^h)\nonumber\\
&= & (1-\frac{1}{N})\frac{\phi_{\bold{j}_{m}}}{2\epsilon_{\bold{j}_m}+2-\frac{4}{N}-\frac{z}{\phi_{\bold{j}_{m}}}}\,[\check{\mathcal{G}}_{\bold{j}_{m}\,;\,N-2,N-2}(z)]_{\tau_h\,;\,0}\|\overline{\mathscr{P}_{\psi^{Bog}_{\bold{j}_1,\dots,\bold{j}_{m-1}}}}
(\charf -\chi_{ N-h})\frac{\psi^{Bog}_{\bold{j}_1,\dots,\bold{j}_{m-1}}}{\|\psi^{Bog}_{\bold{j}_1,\dots,\bold{j}_{m-1}}\|}\|\quad\quad\quad\\
& &+\mathcal{O}(\frac{1}{\sqrt{h}})+\mathcal{O}(\frac{h^2g^h}{\sqrt{N}})+\mathcal{O}(\frac{h^2(2h)^{h+2}}{\sqrt{N}})+\mathcal{O}((\frac{4}{5})^h)+\mathcal{O}((\frac{1}{1+c\sqrt{\epsilon_{\bold{j}_m}}})^h)\nonumber\\
&=&\mathcal{O}(\frac{1}{\sqrt{h}})+\mathcal{O}(\frac{h^2g^h}{\sqrt{N}})+\mathcal{O}(\frac{h^2(2h)^{h+2}}{\sqrt{N}})+\mathcal{O}((\frac{4}{5})^h)+\mathcal{O}((\frac{1}{1+c\sqrt{\epsilon_{\bold{j}_m}}})^h)\,.\label{laterale-est-2}
\end{eqnarray}
}Then, we set $h\equiv \lfloor \Big(\ln N\Big)^{\frac{1}{2}} \rfloor$ (where $\lfloor \Big(\ln N\Big)^{\frac{1}{2}} \rfloor$ is assumed to be even)  and determine two positive constants $C_I$, and $C_{II}$ such that the inequalities in (\ref{primero-1-1}) and (\ref{primero-3-1}) hold true. 

\qed
\begin{lemma}\label{invertibility}
Let $2\leq m\leq M$ and assume that the Hamiltonian $H^{Bog}_{\bold{j}_1,\dots,\bold{j}_{m-1}}$ has nondegenerate ground state energy $z^{Bog}_{\bold{j}_1,\dots,\bold{j}_{m-1}}$ and ground state vector $\psi^{Bog}_{\bold{j}_1,\dots,\bold{j}_{m-1}}$ with the property in (\ref{ass-number-0}). 
Furthermore, assume that:
\begin{enumerate}
\item
There exists
 $\Delta_0\geq \Delta_{m-1}>0$ such that
\begin{equation}\label{assumption-gap}
\text{infspec}\,\Big[\Big(\hat{H}^{Bog}_{\bold{j}_1, \dots,\bold{j}_{m-1}}+\sum_{\bold{j}\in \mathbb{Z}^d \setminus \{\pm\bold{j}_1, \dots, \pm\bold{j}_{m}\}}(k_{\bold{j}})^2a^*_{\bold{j}}a_{\bold{j}}\Big)\upharpoonright_{(Q^{(>N-1)}_{\bold{j}_m}\mathcal{F}^N)\ominus \{\mathbb{C}\psi^{Bog}_{\bold{j}_1, \dots,\bold{j}_{m-1}}\}}\Big]-z^{Bog}_{\bold{j}_1,\dots,\bold{j}_{m-1}}\geq \Delta_{m-1}
\end{equation}
where $Q^{(>N-1)}_{\bold{j}_m}\mathcal{F}^N$ is the subspace of states in $\mathcal{F}^N$ with no particles in the modes $\pm \bold{j}_m$, and $\{\mathbb{C}\psi^{Bog}_{\bold{j}_1, \dots,\bold{j}_{m-1}}\}$ is the subspace generated by the vector $\psi^{Bog}_{\bold{j}_1, \dots,\bold{j}_{m-1}}$.
(Notice that $$\hat{H}^{Bog}_{\bold{j}_1, \dots,\bold{j}_{m-1}}+\sum_{\bold{j}\in \mathbb{Z}^d \setminus \{\pm\bold{j}_1, \dots, \pm\bold{j}_{m}\}}(k_{\bold{j}})^2a^*_{\bold{j}}a_{\bold{j}}=H^{Bog}_{\bold{j}_1, \dots,\bold{j}_{m-1}}-(k_{\bold{j}_m})^2(a^*_{\bold{j}_m}a_{\bold{j}_m}+a^*_{-\bold{j}_m}a_{-\bold{j}_m})\,,$$ i.e., the kinetic energy associated with the modes $\pm \bold{j}_m$ is absent.)
\item
\begin{equation}\label{lower-bound-spec}
\text{infspec}\,[\sum_{l=1}^{m-1}\hat{H}^{Bog}_{\bold{j}_l }]-z^{Bog}_{\bold{j}_1,\dots ,\bold{j}_{m-1}}\geq -\frac{m-1}{(\ln N)^{\frac{1}{8}}}\,.
\end{equation}
\end{enumerate}

\noindent
Let $\epsilon_{\bold{j}_m}$ be sufficiently small and $N$ sufficiently large such that: 

\noindent
a) for
$$z\leq E^{Bog}_{\bold{j}_m}+ \sqrt{\epsilon_{\bold{j}_m}}\phi_{\bold{j}_m}\sqrt{\epsilon_{\bold{j}_m}^2+2\epsilon_{\bold{j}_m}}\,(<0)$$the Feshbach-Schur flow associated with the couple of modes $\pm \bold{j}_m$ is well defined (see Theorem \ref{Feshbach-Hbog}); 

\noindent
b) \begin{equation}\label{def-cjm}
\frac{\ln N}{N}\ll1 \quad \text{and} \quad \frac{U_{\bold{j}_m}}{\sqrt{N}}<\frac{\Delta_{m-1}}{2}\quad \text{where}\quad U_{\bold{j}_m}:=k^2_{\bold{j}_m}+\phi_{\bold{j}_m}\,;
 \end{equation}

\noindent
c)\footnote{ This condition holds for $N$ sufficiently large  because in \emph{\underline{Corollary 4.6} of \cite{Pi1}} we show that   in the limit $N\to \infty$ the ground state energy $z_*$ of $H^{Bog}_{\bold{j}_*}$ tends to $E^{Bog}_{\bold{j}_*}$ (see Section \ref{groundstate}),  and because $\gamma\Delta_{m-1}\leq \frac{\Delta_0}{2}$.} \begin{equation}
 z_{m}+\gamma \Delta_{m-1}<E^{Bog}_{\bold{j}_m}+\frac{1}{2} \sqrt{\epsilon_{\bold{j}_m}}\phi_{\bold{j}_m}\sqrt{\epsilon_{\bold{j}_m}^2+2\epsilon_{\bold{j}_m}}\,(<0)\quad,\quad\gamma= \frac{1}{2}\,,
 \end{equation}
where  $z_{m}$ is the ground state energy of $H^{Bog}_{\bold{j}_m}$.

\noindent
Then, there exists a constant $C^{\perp}>0$ such that for  
\begin{equation}
z\leq  z_{m}-\frac{C^{\perp}}{(\ln N)^{\frac{1}{2}}}+\gamma \Delta_{m-1}  \label{bound-z-0}
\end{equation}
the following estimate holds true:
 %\begin{eqnarray}
%& &\overline{\mathscr{P}_{\psi^{Bog}_{\bold{j}_1,\dots,\bold{j}_{m-1}}}}\Gamma^{Bog}_{\bold{j}_1,\dots,\bold{j}_m;\,N,N}(z+z^{Bog}_{\bold{j}_1,\dots,\bold{j}_{m-1}})\,\overline{\mathscr{P}_{\psi^{Bog}_{\bold{j}_1,\dots,\bold{j}_{m-1}}}}\leq -z_m\overline{\mathscr{P}_{\psi^{Bog}_{\bold{j}_1,\dots,\bold{j}_{m-1}}}}\label{ineq-z_m}
%\end{eqnarray} 
%and 
\begin{equation}\label{invert-est}
\overline{\mathscr{P}_{\psi^{Bog}_{\bold{j}_1,\dots,\bold{j}_{m-1}}}}\mathscr{K}_{\bold{j}_1,\dots,\bold{j}_m}^{Bog\,(N-2)}(z+z^{Bog}_{\bold{j}_1,\dots,\bold{j}_{m-1}})\overline{\mathscr{P}_{\psi^{Bog}_{\bold{j}_1,\dots,\bold{j}_{m-1}}}}\geq (1-\gamma)\Delta_{m-1}\overline{\mathscr{P}_{\psi^{Bog}_{\bold{j}_1,\dots,\bold{j}_{m-1}}}}\,.
\end{equation}
%in the  $z$ given in (\ref{bound-z-0}).
\end{lemma}

\noindent
\emph{Proof}

As we have explained in the \emph{\bf{Outline of the proof}} (see Lemma \ref{invertibility-new}), the result is proven if the key inequality in (\ref{primero-2-new}) holds true in the interval (\ref{interval}). In order to prove  (\ref{primero-2-new}), we implement the same  truncations of {\bf{Step I}} and {\bf{Step II}} in Lemma \ref{main-relations}, and make use of (\ref{identity-Gammas}). Hereafter, we set $w=z+z^{Bog}_{\bold{j}_1,\dots,\bold{j}_{m-1}}$. For $h\geq 2$ and even, we get  
\begin{eqnarray}
& &(\ref{GammaPerp})\\
&=&\|\overline{\mathscr{P}_{\psi^{Bog}_{\bold{j}_1,\dots,\bold{j}_{m-1}}}}W_{\bold{j}_m}\,R^{Bog}_{\bold{j}_1,\dots,\bold{j}_m\,;\,N-2,N-2}(w)\times \label{express}\\
& &\quad\quad \times \sum_{l_{N-2}=0}^{\bar{j}-1}\Big[[\Gamma^{Bog}_{\bold{j}_1,\dots,\bold{j}_m\,;\,N-2,N-2}(w)]_{\tau_h}\,R^{Bog}_{\bold{j}_1,\dots,\bold{j}_m\,;\,\,N-2,N-2}(w)\Big]^{l_{N-2}}W^*_{\bold{j}_m}\,\overline{\mathscr{P}_{\psi^{Bog}_{\bold{j}_1,\dots,\bold{j}_{m-1}}}}\|\nonumber \label{step-3objiects}\\
& &+\mathcal{O}((\frac{4}{5})^{\bar{j}})+\mathcal{O}((\frac{1}{1+c\sqrt{\epsilon_{\bold{j}_m}}})^h)\,.\,\label{step-3objiects-bis}
\end{eqnarray}
We set $\bar{j}\equiv h$.
From (\ref{express})-(\ref{step-3objiects-bis}) we derive the inequality in (\ref{primero-2-new}) through steps $\bf{a)}$, $\bf{b}$, $\bf{c)}$, $\bf{d)}$, $\bf{e)}$,  and $\bf{f)}$ described below.
%\begin{remark}\label{remark-4.5}

%For the control of (\ref{express}) we implement five steps 

\noindent
{\bf{ STEP  a)}}

\noindent

In expression (\ref{express}), using the definition of $[\Gamma^{Bog}_{\bold{j}_1,\dots,\bold{j}_m\,;\,N-2,N-2}(w)]_{\tau_h}$  given in  (\ref{tau-exp})-(\ref{tau-exp-bis}) and the decomposition $$[\Gamma^{Bog\,}_{\bold{j}_1,\dots,\bold{j}_m\,;\,N-2,N-2}(w)]_{\mathcal{\tau}_h}=: \sum_{r=1}^{\bar{r}}[\Gamma^{Bog\,}_{\bold{j}_1,\dots,\bold{j}_m\,;\,N-2,N-2}(w)]^{(r)}_{\mathcal{\tau}_h}$$ in (\ref{summands-rl}),  to each (left) operator of the type 
\begin{eqnarray}\label{companion-1}
& &(R^{Bog}_{\bold{j}_1,\dots,\bold{j}_m\,;\,i,i}(w))^{\frac{1}{2}}W_{\bold{j}_m\,;\,i,i-2}\,(R^{Bog}_{\bold{j}_1,\dots,\bold{j}_m\,;\,i-2,i-2}(w))^{\frac{1}{2}}\,,\quad N-4\geq i \geq N-2-h\quad\text{and even}\,,\,\quad\quad\quad\quad%&=&(R^{Bog}_{\bold{j}_1,\dots,\bold{j}_m\,;\,i,i}(z))^{\frac{1}{2}}\phi_{\bold{j}}\frac{a^*_{\bold{0}}a^*_{\bold{0}}a_{\bold{j}}a_{-\bold{j}}}{N}\,R^{Bog}_{\bold{j}_1,\dots,\bold{j}_m\,;\,i-2,i-2}(z)\,\phi_{\bold{j}}\frac{a_{\bold{0}}a_{\bold{0}}a^*_{\bold{j}}a^*_{-\bold{j}}}{N}(R^{Bog}_{\bold{j}_1,\dots,\bold{j}_m\,;\,i,i}(z))^{\frac{1}{2}}\quad
\end{eqnarray}
that pops up from the re-expansion of $$(R^{Bog}_{\bold{j}_1,\dots,\bold{j}_m\,;\,N-2,N-2}(w))^{\frac{1}{2}}[\Gamma^{Bog}_{\bold{j}_1,\dots,\bold{j}_m\,;\,N-2,N-2}(w)]_{\tau_h}(R^{Bog}_{\bold{j}_1,\dots,\bold{j}_m\,;\,N-2,N-2}(w))^{\frac{1}{2}}$$    we can assign a (right) companion contained in the same $[\Gamma^{Bog\,}_{\bold{j}_1,\dots,\bold{j}_m\,;\,N-2,N-2}(w)]^{(r)}_{\mathcal{\tau}_h}$, precisely the closest operator
\begin{eqnarray}\label{companion-2}
& &(R^{Bog}_{\bold{j}_1,\dots,\bold{j}_m\,;\,i-2,i-2}(w))^{\frac{1}{2}}W^*_{\bold{j}_m\,;\,i-2,i}(R^{Bog}_{\bold{j}_1,\dots,\bold{j}_m\,;\,i,i}(w))^{\frac{1}{2}}\quad\quad\quad\quad
%&=&(R^{Bog}_{\bold{j}_1,\dots,\bold{j}_m\,;\,i,i}(z))^{\frac{1}{2}}\phi_{\bold{j}}\frac{a^*_{\bold{0}}a^*_{\bold{0}}a_{\bold{j}}a_{-\bold{j}}}{N}\,R^{Bog}_{\bold{j}_1,\dots,\bold{j}_m\,;\,i-2,i-2}(z)\,\phi_{\bold{j}}\frac{a_{\bold{0}}a_{\bold{0}}a^*_{\bold{j}}a^*_{-\bold{j}}}{N}(R^{Bog}_{\bold{j}_1,\dots,\bold{j}_m\,;\,i,i}(z))^{\frac{1}{2}}\quad
\end{eqnarray}
that is placed on the right of  (\ref{companion-1}) in $[\Gamma^{Bog\,}_{\bold{j}_1,\dots,\bold{j}_m\,;\,N-2,N-2}(w)]^{(r)}_{\mathcal{\tau}_h}$. The companion operator always exists because of the structure of $[\Gamma^{Bog}_{\bold{j}_1,\dots,\bold{j}_m\,;\,N-2,N-2}(w)]_{\tau_h}$ (see (\ref{tau-exp})-(\ref{tau-exp-bis})). Then, starting from the two companions 
\begin{eqnarray}\label{comp}
& &(R^{Bog}_{\bold{j}_1,\dots,\bold{j}_m\,;\, N-2-h, N-2-h}(w))^{\frac{1}{2}}[\Gamma^{Bog\,}_{\bold{j}_1,\dots,\bold{j}_m\,;\,N-2-h,N-2-h}(w)]_{\mathcal{\tau}_h}(R^{Bog}_{\bold{j}_1,\dots,\bold{j}_m\,;\, N-2-h, N-2-h}(w))^{\frac{1}{2}}\quad\quad\quad\\
&=& (R^{Bog}_{\bold{j}_1,\dots,\bold{j}_m\,;\, N-2-h, N-2-h}(w))^{\frac{1}{2}} W_{\bold{j}_m}\,R^{Bog}_{\bold{j}_1,\dots,\bold{j}_m\,;\, N-4-h, N-4-h}(w)W_{\bold{j}_m}^*(R^{Bog}_{\bold{j}_1,\dots,\bold{j}_m\,;\, N-2-h, N-2-h}(w))^{\frac{1}{2}}\nonumber
\end{eqnarray}
in (\ref{tau-exp-bis}), and using  the relation in (\ref{tau-exp}), it is not difficult to show by induction that, in expression (\ref{express}), all operators of the type (\ref{companion-1}) and (\ref{companion-2}) can be replaced with operator-valued  functions of the ``variables" $a^*_{\bold{j}_m}a_{\bold{j}_m}$ and $a^*_{-\bold{j}_m}a_{-\bold{j}_m}$,  and  the (number) operators $a^*_{\bold{j}_m}a_{\bold{j}_m}$, $a^*_{-\bold{j}_m}a_{-\bold{j}_m}$ can be replaced with c-numbers; see section 0.3 in \emph{supporting-file-Bose2.pdf}. In essence, this is due to: 1)  the projections contained in the definition of $R^{Bog}_{\bold{j}_1,\dots,\bold{j}_m\,;\,i,i}(w)$ (see (\ref{Rbog-m})); 2) the form of the interaction terms $W_{\bold{j}_m}$, $W^*_{\bold{j}_m}$; 3) the operator $\overline{\mathscr{P}_{\psi^{Bog}_{\bold{j}_1,\dots,\bold{j}_{m-1}}}}$ (see (\ref{express})) projecting onto states with no particles in the modes $\pm \bold{j}_m$.
More precisely, since 
\begin{equation}
a_{\bold{j}_m}a_{-\bold{j}_m}a^*_{\bold{j}_m}a^*_{-\bold{j}_m}=(a^*_{\bold{j}_m}a_{\bold{j}_m}+1)(a^*_{-\bold{j}_m}a_{-\bold{j}_m}+1)
\end{equation}
it turns out that (for $N-2\leq i \leq N-h-2$)
\begin{equation}
(R^{Bog}_{\bold{j}_1,\dots,\bold{j}_m\,;\,i,i}(w))^{\frac{1}{2}}\phi_{\bold{j}_m}\frac{a^*_{\bold{0}}a^*_{\bold{0}}a_{\bold{j}_m}a_{-\bold{j}_m}}{N}\,(R^{Bog}_{\bold{j}_1,\dots,\bold{j}_m\,;\,i-2,i-2}(w))^{\frac{1}{2}}\,\label{operation-1}
\end{equation}
is replaced with
\begin{eqnarray}
%& &\mathcal{W}_{\bold{j}_*\,;\,i,i-2}(z)\mathcal{W}^*_{\bold{j}_*\,;\,i-2,i}(z)\\
& & [(R^{Bog}_{\bold{j}_1,\dots,\bold{j}_m\,;\,i,i}(w))^{\frac{1}{2}}W_{\bold{j}_m\,;\,i,i-2}(R^{Bog}_{\bold{j}_1,\dots,\bold{j}_m\,;\,i-2,i-2}(w))^{\frac{1}{2}}]_{1}\label{RW*R'}\\
&:= &\phi_{\bold{j}_{m}}\,\frac{1}{\Big[\sum_{\bold{j}\notin\{\pm\bold{j}_1, \dots, \pm\bold{j}_{m}\}}(k_{\bold{j}})^2a^*_{\bold{j}}a_{\bold{j}}+\hat{H}^{Bog}_{\bold{j}_1,\dots,\bold{j}_{m-1}}-z^{Bog}_{\bold{j}_1,\dots,\bold{j}_{m-1}}+(\frac{a^*_{\bold{0}}a_{\bold{0}}}{N}\phi_{\bold{j}_{m}}+(k_{\bold{j}_{m}}^2))(n_{\bold{j}_{m}}+n_{-\bold{j}_{m}})-z\Big]^{\frac{1}{2}}}\times \quad\quad\quad\quad\label{effective-op}\\
& &\times \frac{a^*_{\bold{0}}a^*_{\bold{0}}}{N}\,\frac{ (n_{\bold{j}_{m}}+1)^{\frac{1}{2}}(n_{-\bold{j}_{m}}+1)^{\frac{1}{2}}}{\Big[\sum_{\bold{j} \{\pm\bold{j}_1, \dots, \pm\bold{j}_{m}\}}(k_{\bold{j}})^2a^*_{\bold{j}}a_{\bold{j}}+\hat{H}^{Bog}_{\bold{j}_1,\dots,\bold{j}_{m-1}}-z^{Bog}_{\bold{j}_1,\dots,\bold{j}_{m-1}}+(\frac{a^*_{\bold{0}}a_{\bold{0}}}{N}\phi_{\bold{j}_{m}}+(k_{\bold{j}_{m}}^2))(n_{\bold{j}_{m}}+n_{-\bold{j}_{m}}+2)-z\Big]^{\frac{1}{2}}}\quad\quad\quad\quad \label{effective-op-bis}
%& &\quad \times \frac{a_{\bold{0}}a_{\bold{0}}}{N}\frac{1}{\Big[\check{H}^{Bog}_{\bold{j}_1,\dots,\bold{j}_{m-1}}-z_{\bold{j}_1,\dots,\bold{j}_{m-1}}+(\frac{a^*_{\bold{j}_0}a_{\bold{j}_0}}{N}\phi_{\bold{j}_{m}}+(k_{\bold{j}_{m}}^2))(n_{\bold{j}_{m}}+n_{-\bold{j}_{m}})-z\Big]^{\frac{1}{2}}}
%&:=& \frac{(n_{\bold{j}_0}+2)(n_{\bold{j}_0}+1)}{N^2}\,\phi^2_{\bold{j}_{*}}\,\frac{ (n_{\bold{j}_{*}}+1)(n_{-\bold{j}_{*}}+1)}{\Big[(\frac{n_{\bold{j}_0}}{N}\phi_{\bold{j}_{*}}+(k_{\bold{j}_{*}}^2))(n_{\bold{j}_{*}}+n_{-\bold{j}_{*}})-z\Big]}\times\\
%& &\times \frac{1}{\Big[(\frac{(n_{\bold{j}_0}+2)}{N}\phi_{\bold{j}_{*}}+(k_{\bold{j}_{*}}^2))(n_{\bold{j}_{*}}+n_{-\bold{j}_{*}})-2(\frac{(n_{\bold{j}_0}+2)}{N}\phi_{\bold{j}_{*}}+(k_{\bold{j}_{*}}^2))-z\Big]}\quad\quad\quad
\end{eqnarray} 
where 
\begin{equation}
n_{\bold{j}_{m}}+n_{-\bold{j}_{m}}=N-i\quad,\quad n_{\bold{j}_{m}}=n_{-\bold{j}_{m}}\,.\label{operation-2}
\end{equation}
We stress that the resolvents in (\ref{effective-op})-(\ref{effective-op-bis}) act on Fock subspaces $\mathcal{F}^{M}$ with $M< N$ and even. The operator in (\ref{RW*R'}) maps $\mathcal{F}^{M}$ to $\mathcal{F}^{M+2}$ with $M<N$. Therefore, it is convenient to think of the Fock subspaces $\mathcal{F}^{M}$, $M< N$,  embedded in $\check{\mathcal{F}}^N:=\oplus_{n=0}^{N}\mathcal{F}^{n}$. 
Nevertheless, the inequality in (\ref{lower-bound-spec}) implies\footnote{It is enough to consider $\hat{H}^{Bog}_{\bold{j}_1,\dots,\bold{j}_{m-1}}-z^{Bog}_{\bold{j}_1,\dots,\bold{j}_{m-1}}$ applied to the subspace of $\mathcal{F}^N$ generated by the vectors with at least $N-M$ particles in the modes different from $\bold{0},\pm \bold{j}_1,\dots,\pm \bold{j}_{m-1}$.} that the operator 
$$\hat{H}^{Bog}_{\bold{j}_1,\dots,\bold{j}_{m-1}}-z^{Bog}_{\bold{j}_1,\dots,\bold{j}_{m-1}}$$
restricted to  $\mathcal{F}^{M}$ with $M< N$ is also bounded below by the R-H-S of (\ref{lower-bound-spec}). Hence, the resolvents in (\ref{effective-op}) and (\ref{effective-op-bis}) are well defined for $\epsilon_{\bold{j}_m}$ sufficiently small, $N$ sufficiently large,  and $z$ in the interval (\ref{interval}).

\noindent
Analogously, the operator
\begin{equation}\label{right-comp}
(R^{Bog}_{\bold{j}_1,\dots,\bold{j}_m\,;\,i-2,i-2}(w))^{\frac{1}{2}}\,\phi_{\bold{j}_m}\frac{a_{\bold{0}}a_{\bold{0}}a^*_{\bold{j}_m}a^*_{-\bold{j}_m}}{N}(R^{Bog}_{\bold{j}_1,\dots,\bold{j}_m\,;\,i,i}(w))^{\frac{1}{2}}
\end{equation}
is replaced with the hermitian conjugate of (\ref{effective-op})-(\ref{effective-op-bis}) that we denote by
\begin{equation}
[(R^{Bog}_{\bold{j}_1,\dots,\bold{j}_m\,;\,i-2,i-2}(w))^{\frac{1}{2}}W^*_{\bold{j}_m\,;\,i-2,i}(R^{Bog}_{\bold{j}_1,\dots,\bold{j}_m\,;\,i,i}(w))^{\frac{1}{2}}]_{1}\,.\label{RWR'}
\end{equation}
Similar replacements hold for the outer companion operators:
\begin{eqnarray}
& &\phi_{\bold{j}_m}\frac{a^*_{\bold{0}}a^*_{\bold{0}}a_{\bold{j}_m}a_{-\bold{j}_m}}{N}\,(R^{Bog}_{\bold{j}_1,\dots,\bold{j}_m\,;\,N-2,N-2}(w))^{\frac{1}{2}}\quad \Rightarrow\quad \phi_{\bold{j}_m}\frac{a^*_{\bold{0}}a^*_{\bold{0}}}{N}\,(R^{Bog}_{\bold{j}_1,\dots,\bold{j}_m\,;\,N-2,N-2}(w))^{\frac{1}{2}}\quad \label{outer-comp-0}\\
& &\phi_{\bold{j}_m}(R^{Bog}_{\bold{j}_1,\dots,\bold{j}_m\,;\,N-2,N-2}(w))^{\frac{1}{2}}\frac{a_{\bold{0}}a_{\bold{0}}a^*_{\bold{j}_m}a^*_{-\bold{j}_m}}{N}\quad \Rightarrow\quad \phi_{\bold{j}_m}(R^{Bog}_{\bold{j}_1,\dots,\bold{j}_m\,;\,N-2,N-2}(w))^{\frac{1}{2}}\frac{a_{\bold{0}}a_{\bold{0}}}{N}\,\quad\quad\quad\quad  \label{outer-comp}
\end{eqnarray}
where it is assumed that in $R^{Bog}_{\bold{j}_1,\dots,\bold{j}_m\,;\,N-2,N-2}(w)$ the operator $a^*_{\bold{j}_m}a_{\bold{j}_m}+a^*_{-\bold{j}_m}a_{-\bold{j}_m}$ is replaced with $2$.
\begin{remark}
For the details of the replacement  of the operators $a^*_{\pm\bold{j}_m}\,,\,a_{\pm\bold{j}_m}$ with c-numbers described in lines  (\ref{operation-1})-(\ref{outer-comp}) we refer the reader to \emph{\emph{Proposition 5.7} of \cite{Pi1}} where a similar procedure is implemented by induction.
\end{remark}

By means of the implemented operations we have shown that (more details in section 0.3 of \emph{supporting-file-Bose2.pdf})
\begin{equation}\label{1-express-0}
\overline{\mathscr{P}_{\psi^{Bog}_{\bold{j}_1,\dots,\bold{j}_{m-1}}}}W_{\bold{j}_m}\,R^{Bog}_{\bold{j}_1,\dots,\bold{j}_m\,;\,N-2,N-2}(w)\sum_{l_{N-2}=0}^{h-1}\Big[[\Gamma^{Bog}_{\bold{j}_1,\dots,\bold{j}_m\,;\,N-2,N-2}(w)]_{\tau_h}\,R^{Bog}_{\bold{j}_1,\dots,\bold{j}_m\,;\,\,N-2,N-2}(w)\Big]^{l_{N-2}}W^*_{\bold{j}_m}\,\overline{\mathscr{P}_{\psi^{Bog}_{\bold{j}_1,\dots,\bold{j}_{m-1}}}}
\end{equation}
coincides with
\begin{equation}\label{1-express}
\Big[\overline{\mathscr{P}_{\psi^{Bog}_{\bold{j}_1,\dots,\bold{j}_{m-1}}}}W_{\bold{j}_m}\,R^{Bog}_{\bold{j}_1,\dots,\bold{j}_m\,;\,N-2,N-2}(w)\sum_{l_{N-2}=0}^{h-1}\Big[[\Gamma^{Bog}_{\bold{j}_1,\dots,\bold{j}_m\,;\,N-2,N-2}(w)]_{\tau_h}\,R^{Bog}_{\bold{j}_1,\dots,\bold{j}_m\,;\,\,N-2,N-2}(w)\Big]^{l_{N-2}}W^*_{\bold{j}_m}\,\overline{\mathscr{P}_{\psi^{Bog}_{\bold{j}_1,\dots,\bold{j}_{m-1}}}}\Big]_1
\end{equation}
where the symbol $\Big[\quad\Big]_1$ means that each couple of companions
 has been transformed according to the rules described in lines (\ref{operation-1})--(\ref{outer-comp}). Concerning notation, in the following steps, we make use of the definitions
\begin{equation}
\Big[R^{Bog}_{\bold{j}_1,\dots,\bold{j}_m\,;\,i,i}(w)\Big]_1:=\frac{1}{\sum_{\bold{j}\notin\{\pm\bold{j}_1, \dots, \pm\bold{j}_{m}\}}(k_{\bold{j}})^2a^*_{\bold{j}}a_{\bold{j}}+\hat{H}^{Bog}_{\bold{j}_1,\dots,\bold{j}_{m-1}}-z^{Bog}_{\bold{j}_1,\dots,\bold{j}_{m-1}}+(\frac{a^*_{\bold{0}}a_{\bold{0}}}{N}\phi_{\bold{j}_{m}}+(k_{\bold{j}_{m}}^2))(N-i)-z}
\end{equation}
and
\begin{equation}
\Big[(R^{Bog}_{\bold{j}_1,\dots,\bold{j}_m\,;\,i,i}(w))^{\frac{1}{2}}\Big]_1:=\Big[\frac{1}{\sum_{\bold{j}\notin\{\pm\bold{j}_1, \dots, \pm\bold{j}_{m}\}}(k_{\bold{j}})^2a^*_{\bold{j}}a_{\bold{j}}+\hat{H}^{Bog}_{\bold{j}_1,\dots,\bold{j}_{m-1}}-z^{Bog}_{\bold{j}_1,\dots,\bold{j}_{m-1}}+(\frac{a^*_{\bold{0}}a_{\bold{0}}}{N}\phi_{\bold{j}_{m}}+(k_{\bold{j}_{m}}^2))(N-i)-z}\Big]^{\frac{1}{2}}\,.
\end{equation}
\\

\noindent
{\bf{ STEP b)}}

Next, we consider each couple of left and right companion operators contained in (\ref{1-express}). Up to a small remainder to be estimated, we replace  the operator $a^*_{\bold{0}}a^*_{\bold{0}}$ appearing in the numerator of   (\ref{RW*R'}) with $ a^*_{\bold{0}}a_{\bold{0}}-1$, and the operator $a_{\bold{0}}a_{\bold{0}} $ appearing in the numerator of (\ref{RWR'}) with $a^*_{\bold{0}}a_{\bold{0}}$. Roughly speaking,  we take the operator $\frac{a_{\bold{0}}a_{\bold{0}}}{N}$ (belonging to the right companion) next to the operator $\frac{a^*_{\bold{0}}a^*_{\bold{0}}}{N}$ (of the left companion),  and we write 
$$ \frac{a^*_{\bold{0}}a^*_{\bold{0}}a_{\bold{0}}a_{\bold{0}}}{N^2} =\frac{(a^*_{\bold{0}}a_{\bold{0}}-1)}{N}\frac{a^*_{\bold{0}}a_{\bold{0}}}{N}\,.$$
The operator $\frac{a^*_{\bold{0}}a_{\bold{0}}}{N}$ is then taken back to the original position (of $a_{\bold{0}}a_{\bold{0}}$) in the right companion. 
Before providing a rigorous description of this mechanism, we describe in detail the final leading expression that we obtain.

As a result of the previous operations we get as leading term  an expression identical to (\ref{1-express}) but where each operator 
$\Big[(R^{Bog}_{\bold{j}_1,\dots,\bold{j}_m\,;\,i,i}(w))^{\frac{1}{2}}W_{\bold{j}_m\,;\,i,i-2}\,(R^{Bog}_{\bold{j}_1,\dots,\bold{j}_m\,;\,i-2,i-2}(w))^{\frac{1}{2}}\Big]_{1}\,%&=&(R^{Bog}_{\bold{j}_1,\dots,\bold{j}_m\,;\,i,i}(z))^{\frac{1}{2}}\phi_{\bold{j}}\frac{a^*_{\bold{0}}a^*_{\bold{0}}a_{\bold{j}}a_{-\bold{j}}}{N}\,R^{Bog}_{\bold{j}_1,\dots,\bold{j}_m\,;\,i-2,i-2}(z)\,\phi_{\bold{j}}\frac{a_{\bold{0}}a_{\bold{0}}a^*_{\bold{j}}a^*_{-\bold{j}}}{N}(R^{Bog}_{\bold{j}_1,\dots,\bold{j}_m\,;\,i,i}(z))^{\frac{1}{2}}\quad
$
is replaced with\footnote{In (\ref{effective-op-transf})-(\ref{effective-op-bis-transf}) we use that $n_{\bold{j}_m}+n_{-\bold{j}_m}=N-i$ and $n_{\bold{j}_m}=n_{-\bold{j}_m}$.}
\begin{eqnarray}
%& &\mathcal{W}_{\bold{j}_*\,;\,i,i-2}(z)\mathcal{W}^*_{\bold{j}_*\,;\,i-2,i}(z)\\
& &\Big[(R^{Bog}_{\bold{j}_1,\dots,\bold{j}_m\,;\,i,i}(w))^{\frac{1}{2}}W_{\bold{j}_m\,;\,i,i-2}\,(R^{Bog}_{\bold{j}_1,\dots,\bold{j}_m\,;\,i-2,i-2}(w))^{\frac{1}{2}}\Big]_{2}\\
&:= &\phi_{\bold{j}_{m}}\,\frac{1}{\Big[\sum_{\bold{j}\notin\{\pm\bold{j}_1, \dots, \pm\bold{j}_{m}\}}(k_{\bold{j}})^2a^*_{\bold{j}}a_{\bold{j}}+\hat{H}^{Bog}_{\bold{j}_1,\dots,\bold{j}_{m-1}}-z^{Bog}_{\bold{j}_1,\dots,\bold{j}_{m-1}}+(\frac{a^*_{\bold{0}}a_{\bold{0}}}{N}\phi_{\bold{j}_{m}}+(k_{\bold{j}_{m}}^2))(N-i)-z\Big]^{\frac{1}{2}}}\times \quad\quad\quad\quad\label{effective-op-transf}\\
& &\times \frac{(a^*_{\bold{0}}a_{\bold{0}}-1)}{N}\,\frac{ (N-i+2)^{\frac{1}{2}}(N-i+2)^{\frac{1}{2}}}{2\Big[\sum_{\bold{j}\notin \{\pm\bold{j}_1, \dots, \pm\bold{j}_{m}\}}(k_{\bold{j}})^2a^*_{\bold{j}}a_{\bold{j}}+\hat{H}^{Bog}_{\bold{j}_1,\dots,\bold{j}_{m-1}}-z^{Bog}_{\bold{j}_1,\dots,\bold{j}_{m-1}}+(\frac{a^*_{\bold{0}}a_{\bold{0}}}{N}\phi_{\bold{j}_{m}}+(k_{\bold{j}_{m}}^2))(N-i+2)-z\Big]^{\frac{1}{2}}}\,,\quad\quad\quad\,\label{effective-op-bis-transf}
%& &\quad \times \frac{a_{\bold{0}}a_{\bold{0}}}{N}\frac{1}{\Big[\check{H}^{Bog}_{\bold{j}_1,\dots,\bold{j}_{m-1}}-z_{\bold{j}_1,\dots,\bold{j}_{m-1}}+(\frac{a^*_{\bold{j}_0}a_{\bold{j}_0}}{N}\phi_{\bold{j}_{m}}+(k_{\bold{j}_{m}}^2))(n_{\bold{j}_{m}}+n_{-\bold{j}_{m}})-z\Big]^{\frac{1}{2}}}
%&:=& \frac{(n_{\bold{j}_0}+2)(n_{\bold{j}_0}+1)}{N^2}\,\phi^2_{\bold{j}_{*}}\,\frac{ (n_{\bold{j}_{*}}+1)(n_{-\bold{j}_{*}}+1)}{\Big[(\frac{n_{\bold{j}_0}}{N}\phi_{\bold{j}_{*}}+(k_{\bold{j}_{*}}^2))(n_{\bold{j}_{*}}+n_{-\bold{j}_{*}})-z\Big]}\times\\
%& &\times \frac{1}{\Big[(\frac{(n_{\bold{j}_0}+2)}{N}\phi_{\bold{j}_{*}}+(k_{\bold{j}_{*}}^2))(n_{\bold{j}_{*}}+n_{-\bold{j}_{*}})-2(\frac{(n_{\bold{j}_0}+2)}{N}\phi_{\bold{j}_{*}}+(k_{\bold{j}_{*}}^2))-z\Big]}\quad\quad\quad
\end{eqnarray} 
and its right companion $\Big[R^{Bog}_{\bold{j}_1,\dots,\bold{j}_m\,;\,i-2,i-2}(w))^{\frac{1}{2}}W^*_{\bold{j}_m\,;\,i-2,i}\,(R^{Bog}_{\bold{j}_1,\dots,\bold{j}_m\,;\,i,i}(w))^{\frac{1}{2}}\Big]_{1}$
with the operator
\begin{eqnarray}
%& &\mathcal{W}_{\bold{j}_*\,;\,i,i-2}(z)\mathcal{W}^*_{\bold{j}_*\,;\,i-2,i}(z)\\
& &\Big[R^{Bog}_{\bold{j}_1,\dots,\bold{j}_m\,;\,i-2,i-2}(w))^{\frac{1}{2}}W^*_{\bold{j}_m\,;\,i-2,i}\,(R^{Bog}_{\bold{j}_1,\dots,\bold{j}_m\,;\,i,i}(w))^{\frac{1}{2}}\Big]_{2}\\
&:= &\phi_{\bold{j}_{m}}\,\frac{1}{\Big[\sum_{\bold{j}\notin\{\pm\bold{j}_1, \dots, \pm\bold{j}_{m}\}}(k_{\bold{j}})^2a^*_{\bold{j}}a_{\bold{j}}+\hat{H}^{Bog}_{\bold{j}_1,\dots,\bold{j}_{m-1}}-z^{Bog}_{\bold{j}_1,\dots,\bold{j}_{m-1}}+(\frac{a^*_{\bold{0}}a_{\bold{0}}}{N}\phi_{\bold{j}_{m}}+(k_{\bold{j}_{m}}^2))(N-i+2)-z\Big]^{\frac{1}{2}}}\times \quad\quad\quad\quad\label{effective-op-transf-comp}\\
& &\times \,\frac{a^*_{\bold{0}}a_{\bold{0}}}{N}\frac{ (N-i+2)^{\frac{1}{2}}(N-i+2)^{\frac{1}{2}}}{2\Big[\sum_{\bold{j}\notin \{\pm\bold{j}_1, \dots, \pm\bold{j}_{m}\}}(k_{\bold{j}})^2a^*_{\bold{j}}a_{\bold{j}}+\hat{H}^{Bog}_{\bold{j}_1,\dots,\bold{j}_{m-1}}-z^{Bog}_{\bold{j}_1,\dots,\bold{j}_{m-1}}+(\frac{a^*_{\bold{0}}a_{\bold{0}}}{N}\phi_{\bold{j}_{m}}+(k_{\bold{j}_{m}}^2))(N-i)-z\Big]^{\frac{1}{2}}}\,.\label{effective-op-bis-transf-comp}
%& &\quad \times \frac{a_{\bold{0}}a_{\bold{0}}}{N}\frac{1}{\Big[\check{H}^{Bog}_{\bold{j}_1,\dots,\bold{j}_{m-1}}-z_{\bold{j}_1,\dots,\bold{j}_{m-1}}+(\frac{a^*_{\bold{j}_0}a_{\bold{j}_0}}{N}\phi_{\bold{j}_{m}}+(k_{\bold{j}_{m}}^2))(n_{\bold{j}_{m}}+n_{-\bold{j}_{m}})-z\Big]^{\frac{1}{2}}}
\end{eqnarray}
We point out that the resolvents in (\ref{effective-op-transf})-(\ref{effective-op-bis-transf}) and in (\ref{effective-op-transf-comp})-(\ref{effective-op-bis-transf-comp}) are now operators from $\mathcal{F}^N$ to $\mathcal{F}^N$. Analogous replacements are implemented on the outer companions in (\ref{outer-comp-0})-(\ref{outer-comp}), and we denote them by $\Big[W_{\bold{j}_m} (R^{Bog}_{\bold{j}_1,\dots,\bold{j}_m\,;\,N-2,N-2})^{\frac{1}{2}}\Big]_2$ and $\Big[ (R^{Bog}_{\bold{j}_1,\dots,\bold{j}_m\,;\,N-2,N-2})^{\frac{1}{2}}W^*_{\bold{j}_m}\Big]_2$, respectively.

\noindent
We denote this (leading) collection of terms as
\begin{equation}\label{2-express}
\Big[\overline{\mathscr{P}_{\psi^{Bog}_{\bold{j}_1,\dots,\bold{j}_{m-1}}}}W_{\bold{j}_m}\,R^{Bog}_{\bold{j}_1,\dots,\bold{j}_m\,;\,N-2,N-2}(w)\sum_{l_{N-2}=0}^{h-1}\Big[[\Gamma^{Bog}_{\bold{j}_1,\dots,\bold{j}_m\,;\,N-2,N-2}(w)]_{\tau_h}\,R^{Bog}_{\bold{j}_1,\dots,\bold{j}_m\,;\,\,N-2,N-2}(w)\Big]^{l_{N-2}}W^*_{\bold{j}_m}\,\overline{\mathscr{P}_{\psi^{Bog}_{\bold{j}_1,\dots,\bold{j}_{m-1}}}}\Big]_2\,.
\end{equation}
%\begin{equation}
%\Big[\overline{\mathscr{P}_{\psi^{Bog}_{\bold{j}_1,\dots,\bold{j}_{m-1}}}}\,\phi^2_{\bold{j}_{m}}\frac{(a^*_{\bold{0}}a_{\bold{0}}-1)}{N}\Big\{R^{Bog}_{\bold{j}_1,\dots,\bold{j}_m,;\,N-2,N-2}(w)\,\prod_{l=1}^{\bar{l}}\Big[ [\Gamma^{Bog\,}_{\bold{j}_1,\dots,\bold{j}_m\,;\,N,N}(w)]^{(r_l)}_{\mathcal{\tau}_h}  R^{Bog}_{\bold{j}_1,\dots,\bold{j}_m,;\,N-2,N-2}(w)\Big] \Big\}\frac{a^*_{\bold{0}}a_{\bold{0}}}{N}\,.\overline{\mathscr{P}_{\psi^{Bog}_{\bold{j}_1,\dots,\bold{j}_{m-1}}}}\Big]_{2}\,.\label{l-block-inv-lead}
%\end{equation} 

Now, we describe the procedure that yields (\ref{2-express}) and  the way we organize the remainder terms corresponding to $(\ref{1-express})-(\ref{2-express})$.  Starting from the definition in (\ref{tau-exp}), for $N-2\geq i \geq N-h$ we can write
\begin{eqnarray}
& &\Big[(R^{Bog}_{\bold{j}_1,\dots,\bold{j}_m;\,i,i}(w))^{\frac{1}{2}}[\Gamma^{Bog\,}_{\bold{j}_1,\dots,\bold{j}_m\,;\,i,i}(w)]_{\mathcal{\tau}_h}(R^{Bog}_{\bold{j}_1,\dots,\bold{j}_m;\,i,i}(w))^{\frac{1}{2}}\Big]_1\\
&=&\Big[(R^{Bog}_{\bold{j}_1,\dots,\bold{j}_m;\,i,i}(w))^{\frac{1}{2}}W_{\bold{j}_m} (R^{Bog}_{\bold{j}_1,\dots,\bold{j}_m;\,i-2,i-2}(w))^{\frac{1}{2}}\,\times\\
& &\quad\quad \times \,\sum_{l_{i-2}=0}^{h-1}\Big\{(R^{Bog}_{\bold{j}_1,\dots,\bold{j}_m;\,i-2,i-2}(w))^{\frac{1}{2}}[\Gamma^{Bog\,}_{\bold{j}_1,\dots,\bold{j}_m\,;\,i-2,i-2}(w)]_{\mathcal{\tau}_h}\,(R^{Bog}_{\bold{j}_1,\dots,\bold{j}_m\,;\,i-2,i-2}(w))^{\frac{1}{2}}\Big\}^{l_{i-2}} \times\quad\quad\quad\quad\quad \\
& &\quad\quad \times (R^{Bog}_{\bold{j}_1,\dots,\bold{j}_m;\,i-2,i-2}(w))^{\frac{1}{2}}W^*_{\bold{j}_m}(R^{Bog}_{\bold{j}_1,\dots,\bold{j}_m;\,i,i}(w))^{\frac{1}{2}}\Big]_1\nonumber
\end{eqnarray}
where
\begin{equation}
[\Gamma^{Bog\,}_{\bold{j}_1,\dots,\bold{j}_m\,;\,N-2-h,N-2-h}(w)]_{\mathcal{\tau}_h}
:=  W_{\bold{j}_m}\,R^{Bog}_{\bold{j}_1,\dots,\bold{j}_m\,;\, N-4-h, N-4-h}(w)W_{\bold{j}_m}^*\,.\quad\quad
\end{equation}
\emph{{\bf{Warning.}} As no confusion can arise, to make our formulae shorter in the remaining part  of {\bf{STEP b)}} we omit the label $\bold{j}_1,\dots,\bold{j}_m$ and the argument $w$ in $R^{Bog}_{\bold{j}_1,\dots,\bold{j}_m;\,i,i}(w)$, and the label $\psi^{Bog}_{\bold{j}_1,\dots,\bold{j}_{m-1}}$ in $\overline{\mathscr{P}_{\psi^{Bog}_{\bold{j}_1,\dots,\bold{j}_{m-1}}}}$.} 

\noindent
Using the formulae of above, we can write (in the new notation)
\begin{eqnarray}
& &\Big[\overline{\mathscr{P}}W\,R^{Bog}_{N-2,N-2}\sum_{l_{N-2}=0}^{h-1}\Big\{[\Gamma^{Bog}_{N-2,N-2}]_{\tau_h}\,R^{Bog}_{N-2,N-2}\Big\}^{l_{N-2}}W^*\,\overline{\mathscr{P}}\Big]_1\nonumber\\
&=&\overline{\mathscr{P}}\Big[W\,(R^{Bog}_{N-2,N-2})^{\frac{1}{2}}\sum_{l_{N-2}=0}^{h-1}\Big\{(R^{Bog}_{N-2,N-2})^{\frac{1}{2}}[\Gamma^{Bog}_{N-2,N-2}]_{\tau_h}\,(R^{Bog}_{N-2,N-2})^{\frac{1}{2}}\Big\}^{l_{N-2}}(R^{Bog}_{N-2,N-2})^{\frac{1}{2}}W^*\,\Big]_1\overline{\mathscr{P}}\nonumber \\
%&=&\Big[\overline{\mathscr{P}_{\psi^{Bog}_{\bold{j}_1,\dots,\bold{j}_{m-1}}}}W_{\bold{j}_m}\,R^{Bog}_{\bold{j}_1,\dots,\bold{j}_m\,;\,N-2,N-2}(w)\sum_{l_{N-2}=0}^{h-1}\Big[[\Gamma^{Bog}_{\bold{j}_1,\dots,\bold{j}_m\,;\,N-2,N-2}(w)]_{\tau_h}\,R^{Bog}_{\bold{j}_1,\dots,\bold{j}_m\,;\,\,N-2,N-2}(w)\Big]^{l_{N-2}}W^*_{\bold{j}_m}\,\overline{\mathscr{P}_{\psi^{Bog}_{\bold{j}_1,\dots,\bold{j}_{m-1}}}}\Big]_2 \nonumber\\
&= &\overline{\mathscr{P}}\Big[W (R^{Bog}_{N-2,N-2})^{\frac{1}{2}}\Big]_2\,\sum_{l_{N-2}=0}^{h-1}\Big\{\Big[(R^{Bog}_{N-2,N-2})^{\frac{1}{2}}[\Gamma^{Bog\,}_{N-2,N-2}]_{\mathcal{\tau}_h}\,(R^{Bog}_{N-2,N-2})^{\frac{1}{2}}\Big]_1\Big\}^{l_{N-2}}  \Big[(R^{Bog}_{N-2,N-2})^{\frac{1}{2}}W^*\Big]_2\overline{\mathscr{P}}\quad\quad\quad\quad \label{lead-1}\\
& &+\overline{\mathscr{P}}\Big[\underbrace{W (R^{Bog}_{N-2,N-2})^{\frac{1}{2}}\Big]_1\,\sum_{l_{N-2}=0}^{h-1}\Big\{\Big[(R^{Bog}_{N-2,N-2})^{\frac{1}{2}}[\Gamma^{Bog\,}_{N-2,N-2}]_{\mathcal{\tau}_h}\,(R^{Bog}_{N-2,N-2})^{\frac{1}{2}}\Big]_1\Big\}^{l_{N-2}}  \Big[(R^{Bog}_{N-2,N-2})^{\frac{1}{2}}W^*}\Big]_1\overline{\mathscr{P}}\quad \label{rem-1}
%&&+\overline{\mathscr{P}}\Big[W (R^{Bog}_{N-2,N-2})^{\frac{1}{2}}\Big]_1\,\widetilde{\sum}_{l_{N-2}=0}^{h-1}\Big\{\Big[(R^{Bog}_{N-2,N-2})^{\frac{1}{2}}[\Gamma^{Bog\,}_{N-2,N-2}]_{\mathcal{\tau}_h}\,(R^{Bog}_{N-2,N-2})^{\frac{1}{2}}\Big]^{rem}_1\Big\}^{l_{N-2}}  \Big[(R^{Bog}_{N-2,N-2})^{\frac{1}{2}}W^*\Big]_1\overline{\mathscr{P}}\quad \quad\quad
\end{eqnarray}
where the symbol $\,\underbrace{\quad}\,$ means that the embraced expression  (including the outer $W$ and $W^*$) is replaced with 
%the sum of two commutators: namely the commutators of $$\frac{a_{\bold{0}}a_{\bold{0}}}{N}\frac{N-i+2}{2}|_{i=N} \quad\text{and}\quad -\frac{a^*_{\bold{0}}a_{\bold{0}}}{N}\frac{N-i+2}{2}|_{i=N},$$ with the operator in between the left and right edge, i.e.,}
\begin{eqnarray}
& &\frac{a^*_{\bold{0}}a^*_{\bold{0}}}{N}\Big[\Big[(R^{Bog}_{N-2,N-2})^{\frac{1}{2}}\,\sum_{l_{N-2}=0}^{h-1}\Big\{(R^{Bog}_{N-2,N-2})^{\frac{1}{2}}[\Gamma^{Bog\,}_{N-2,N-2}]_{\mathcal{\tau}_h}\,(R^{Bog}_{N-2,N-2})^{\frac{1}{2}}\Big\}^{l_{N-2}} (R^{Bog}_{N-2,N-2})^{\frac{1}{2}}\Big]_1\,,\,\frac{a_{\bold{0}}a_{\bold{0}}}{N}\frac{N-i+2}{2}|_{i=N} \Big] \nonumber \\
&+&\frac{(a^*_{\bold{0}}a_{\bold{0}}-1)}{N}\times \nonumber\\
& &\quad \times \Big[\frac{a^*_{\bold{0}}a_{\bold{0}}}{N}(\frac{N-i+2}{2}|_{i=N}) \,,\,\Big[(R^{Bog}_{N-2,N-2})^{\frac{1}{2}}\,\sum_{l_{N-2}=0}^{h-1}\Big\{\Big[(R^{Bog}_{N-2,N-2})^{\frac{1}{2}}[\Gamma^{Bog\,}_{N-2,N-2}]_{\mathcal{\tau}_h}\,(R^{Bog}_{N-2,N-2})^{\frac{1}{2}}\Big\}^{l_{N-2}} (R^{Bog}_{N-2,N-2})^{\frac{1}{2}}\Big]_1\Big]\,.\nonumber
%&=:&\mathcal{D}_i\Big(\Big[(R^{Bog}_{i,i})^{\frac{1}{2}}[\Gamma^{Bog\,}_{i,i}]_{\mathcal{\tau}_h}\,(R^{Bog}_{i,i})^{\frac{1}{2}}\Big]_1\Big)
\end{eqnarray}

\noindent
Next, in a similar way, for $N-h-2\leq i\leq N-2$, we define\footnote{If $i=N-h-2$ the operator $\sum_{l_{i-2}=0}^{h-1}\Big\{\Big[(R^{Bog}_{i-2,i-2})^{\frac{1}{2}}[\Gamma^{Bog\,}_{i-2,i-2}]_{\mathcal{\tau}_h}\,(R^{Bog}_{i-2,i-2})^{\frac{1}{2}}\Big]_1\Big\}^{l_{i-2}}$ is absent.} the operation $\mathcal{D}_i$ where $i$ stands for level $i$ (recall that $i$ is an even number):
\begin{eqnarray}
& &\mathcal{D}_i\Big(\Big[(R^{Bog}_{i,i})^{\frac{1}{2}}[\Gamma^{Bog\,}_{i,i}]_{\mathcal{\tau}_h}\,(R^{Bog}_{i,i})^{\frac{1}{2}}\Big]_1\Big)\label{def-D-0}\\
%&:=&\Big[(R^{Bog}_{N-2,N-2})^{\frac{1}{2}}[\Gamma^{Bog\,}_{N-2,N-2}]_{\mathcal{\tau}_h}\,(R^{Bog}_{N-2,N-2})^{\frac{1}{2}}\Big]_1\\
&:= &\Big[(R^{Bog}_{i,i})^{\frac{1}{2}}\underbrace{W (R^{Bog}_{i-2,i-2})^{\frac{1}{2}}\Big]_1\,\sum_{l_{i-2}=0}^{h-1}\Big\{\Big[(R^{Bog}_{i-2,i-2})^{\frac{1}{2}}[\Gamma^{Bog\,}_{i-2,i-2}]_{\mathcal{\tau}_h}\,(R^{Bog}_{i-2,i-2})^{\frac{1}{2}}\Big]_1\Big\}^{l_{i-2}}  \Big[(R^{Bog}_{i-2,i-2})^{\frac{1}{2}}W^*}(R^{Bog}_{i,i})^{\frac{1}{2}}\Big]_1\,,\nonumber
\end{eqnarray}
where the symbol $\,\underbrace{\quad}\,$ means that the embraced expression is replaced with 
%the sum of two commutators: namely the commutators of $$\frac{a_{\bold{0}}a_{\bold{0}}}{N}\frac{N-i+2}{2}|_{i=N} \quad\text{and}\quad -\frac{a^*_{\bold{0}}a_{\bold{0}}}{N}\frac{N-i+2}{2}|_{i=N},$$ with the operator in between the left and right edge, i.e.,}
\begin{eqnarray}
& &\frac{N-i+2}{2}\frac{a^*_{\bold{0}}a^*_{\bold{0}}}{N}\times \label{embraced-comm-1}\\
& &\quad \times \Big[\Big[(R^{Bog}_{i-2,i-2})^{\frac{1}{2}}\,\sum_{l_{i-2}=0}^{h-1}\Big\{(R^{Bog}_{i-2,i-2})^{\frac{1}{2}}[\Gamma^{Bog\,}_{i-2,i-2}]_{\mathcal{\tau}_h}\,(R^{Bog}_{i-2,i-2})^{\frac{1}{2}}\Big\}^{l_{i-2}} (R^{Bog}_{i-2,i-2})^{\frac{1}{2}}\Big]_1\,,\,\frac{a_{\bold{0}}a_{\bold{0}}}{N}\frac{N-i+2}{2} \Big] \quad\quad\quad  \nonumber\\
&+&\frac{N-i+2}{2}\frac{(a^*_{\bold{0}}a_{\bold{0}}-1)}{N}\times \label{embraced-comm-2} \\
& &\quad \times \Big[\frac{a^*_{\bold{0}}a_{\bold{0}}}{N}(\frac{N-i+2}{2}) \,,\,\Big[(R^{Bog}_{i-2,i-2})^{\frac{1}{2}}\,\sum_{l_{i-2}=0}^{h-1}\Big\{(R^{Bog}_{i-2,i-2})^{\frac{1}{2}}[\Gamma^{Bog\,}_{i-2,i-2}]_{\mathcal{\tau}_h}\,(R^{Bog}_{i-2,i-2})^{\frac{1}{2}}\Big\}^{l_{i-2}} (R^{Bog}_{i-2,i-2})^{\frac{1}{2}}\Big]_1\Big]\,.\quad\quad\quad \nonumber
%&=:&\mathcal{D}_i\Big(\Big[(R^{Bog}_{i,i})^{\frac{1}{2}}[\Gamma^{Bog\,}_{i,i}]_{\mathcal{\tau}_h}\,(R^{Bog}_{i,i})^{\frac{1}{2}}\Big]_1\Big)
\end{eqnarray}

\noindent
On a product of $n$ operators $\Big[(R^{Bog}_{i,i})^{\frac{1}{2}}[\Gamma^{Bog\,}_{i,i}]_{\mathcal{\tau}_h}\,(R^{Bog}_{i,i})^{\frac{1}{2}}\Big]_1$,  by definition $\mathcal{D}_i$  acts as a ``derivative" according to the Leibinz rule:
\begin{eqnarray}
& &\mathcal{D}_i\Big(\Big\{\Big[(R^{Bog}_{i,i})^{\frac{1}{2}}[\Gamma^{Bog\,}_{i,i}]_{\mathcal{\tau}_h}\,(R^{Bog}_{i,i})^{\frac{1}{2}}\Big]_1\Big\}^n)\label{def-D_i}\\
&:=&\sum_{j=0}^{n-1}\Big\{\mathcal{L}_i\Big(\Big[(R^{Bog}_{i,i})^{\frac{1}{2}}[\Gamma^{Bog\,}_{i,i}]_{\mathcal{\tau}_h}\,(R^{Bog}_{i,i})^{\frac{1}{2}}\Big]_1\Big)\Big\}^j\mathcal{D}_i\Big(\Big[(R^{Bog}_{i,i})^{\frac{1}{2}}[\Gamma^{Bog\,}_{i,i}]_{\mathcal{\tau}_h}\,(R^{Bog}_{i,i})^{\frac{1}{2}}\Big]_1\Big)\,\Big\{\Big[(R^{Bog}_{i,i})^{\frac{1}{2}}[\Gamma^{Bog\,}_{i,i}]_{\mathcal{\tau}_h}\,(R^{Bog}_{i,i})^{\frac{1}{2}}\Big]_1\Big\}^{n-j-1}\nonumber
\end{eqnarray}
where
\begin{eqnarray}
& &\mathcal{L}_i\Big(\Big[(R^{Bog}_{i,i})^{\frac{1}{2}}[\Gamma^{Bog\,}_{i,i}]_{\mathcal{\tau}_h}\,(R^{Bog}_{i,i})^{\frac{1}{2}}\Big]_1\Big)\label{def-Li}\\
%&:=&\Big[(R^{Bog}_{N-2,N-2})^{\frac{1}{2}}[\Gamma^{Bog\,}_{N-2,N-2}]_{\mathcal{\tau}_h}\,(R^{Bog}_{N-2,N-2})^{\frac{1}{2}}\Big]_1\\
&:= &\Big[(R^{Bog}_{i,i})^{\frac{1}{2}}W (R^{Bog}_{i-2,i-2})^{\frac{1}{2}}\Big]_2\,\sum_{l_{i-2}=0}^{h-1}\Big\{\Big[(R^{Bog}_{i-2,i-2})^{\frac{1}{2}}[\Gamma^{Bog\,}_{i-2,i-2}]_{\mathcal{\tau}_h}\,(R^{Bog}_{i-2,i-2})^{\frac{1}{2}}\Big]_1\Big\}^{l_{i-2}}  \Big[(R^{Bog}_{i-2,i-2})^{\frac{1}{2}}W^*(R^{Bog}_{i,i})^{\frac{1}{2}}\Big]_2\,.\nonumber
\end{eqnarray}
On a product of $n$ operators $\Big[(R^{Bog}_{i,i})^{\frac{1}{2}}[\Gamma^{Bog\,}_{i,i}]_{\mathcal{\tau}_h}\,(R^{Bog}_{i,i})^{\frac{1}{2}}\Big]_1$ not necessarely contiguous (i.e., the operators  $\Big[(R^{Bog}_{i,i})^{\frac{1}{2}}[\Gamma^{Bog\,}_{i,i}]_{\mathcal{\tau}_h}\,(R^{Bog}_{i,i})^{\frac{1}{2}}\Big]_1$ can be factors of a larger product of operators, and as factors of this product they may be not contiguous) we define the action of $\mathcal{D}_i$ as in (\ref{def-D_i}) with $\mathcal{D}_i(A)=0$ and $\mathcal{L}_i(A)=A$ if $A\neq \Big[(R^{Bog}_{i,i})^{\frac{1}{2}}[\Gamma^{Bog\,}_{i,i}]_{\mathcal{\tau}_h}\,(R^{Bog}_{i,i})^{\frac{1}{2}}\Big]_1$
%If the factors are not contiguous the ordering of the factors is respected by the Leibniz rule of the derivative $\mathcal{D}_i$
, e.g.,
\begin{eqnarray}
& &\mathcal{D}_i\Big(\Big[(R^{Bog}_{i,i})^{\frac{1}{2}}[\Gamma^{Bog\,}_{i,i}]_{\mathcal{\tau}_h}\,(R^{Bog}_{i,i})^{\frac{1}{2}}\Big]_1\, A\,\Big[(R^{Bog}_{i,i})^{\frac{1}{2}}[\Gamma^{Bog\,}_{i,i}]_{\mathcal{\tau}_h}\,(R^{Bog}_{i,i})^{\frac{1}{2}}\Big]_1\Big)\\
&=&\mathcal{D}_i\Big(\Big[(R^{Bog}_{i,i})^{\frac{1}{2}}[\Gamma^{Bog\,}_{i,i}]_{\mathcal{\tau}_h}\,(R^{Bog}_{i,i})^{\frac{1}{2}}\Big]_1\Big)\,A\,\Big[(R^{Bog}_{i,i})^{\frac{1}{2}}[\Gamma^{Bog\,}_{i,i}]_{\mathcal{\tau}_h}\,(R^{Bog}_{i,i})^{\frac{1}{2}}\Big]_1\\
& &+\mathcal{L}_i\Big(\Big[(R^{Bog}_{i,i})^{\frac{1}{2}}[\Gamma^{Bog\,}_{i,i}]_{\mathcal{\tau}_h}\Big)\,(R^{Bog}_{i,i})^{\frac{1}{2}}\Big]_1\Big)\,A\,\mathcal{D}_i\Big(\Big[(R^{Bog}_{i,i})^{\frac{1}{2}}[\Gamma^{Bog\,}_{i,i}]_{\mathcal{\tau}_h}\,(R^{Bog}_{i,i})^{\frac{1}{2}}\Big]_1\Big)\,.
\end{eqnarray}
By definition, $\mathcal{D}_i$ acts linearly with respect to the sum.

\noindent
Using the definitions of above (see (\ref{def-D-0}), (\ref{embraced-comm-1}), (\ref{embraced-comm-2}), (\ref{def-D_i}), and (\ref{def-Li})), it is straightforward to check that for any product of $n$ operators $\Big[(R^{Bog}_{i,i})^{\frac{1}{2}}[\Gamma^{Bog\,}_{i,i}]_{\mathcal{\tau}_h}\,(R^{Bog}_{i,i})^{\frac{1}{2}}\Big]_1$ the following identity holds (for the details see section 0.4 in \emph{supporting-file-Bose2.pdf})
\begin{eqnarray}
& &\Big\{\Big[(R^{Bog}_{i,i})^{\frac{1}{2}}[\Gamma^{Bog\,}_{i,i}]_{\mathcal{\tau}_h}\,(R^{Bog}_{i,i})^{\frac{1}{2}}\Big]_1\Big\}^{n}  \label{D-product}\quad\quad\quad\quad\quad\quad \\
&= &\Big\{\mathcal{L}_i\Big(\Big[(R^{Bog}_{i,i})^{\frac{1}{2}}[\Gamma^{Bog\,}_{i,i}]_{\mathcal{\tau}_h}\,(R^{Bog}_{i,i})^{\frac{1}{2}}\Big]_1\Big)\Big\}^{n}  \\
& &+\mathcal{D}_i\Big(\Big\{\Big[(R^{Bog}_{i,i})^{\frac{1}{2}}[\Gamma^{Bog\,}_{i,i}]_{\mathcal{\tau}_h}\,(R^{Bog}_{i,i})^{\frac{1}{2}}\Big]_1\Big\}^{n}\Big) \label{D-product-fin}\,.
\end{eqnarray}
The same property is true for  finite products where the factors $\Big[(R^{Bog}_{i,i})^{\frac{1}{2}}[\Gamma^{Bog\,}_{i,i}]_{\mathcal{\tau}_h}\,(R^{Bog}_{i,i})^{\frac{1}{2}}\Big]_1$ are not contiguous, e.g.,
\begin{eqnarray}
& &\Big[(R^{Bog}_{i,i})^{\frac{1}{2}}[\Gamma^{Bog\,}_{i,i}]_{\mathcal{\tau}_h}\,(R^{Bog}_{i,i})^{\frac{1}{2}}\Big]_1\, A\,\Big[(R^{Bog}_{i,i})^{\frac{1}{2}}[\Gamma^{Bog\,}_{i,i}]_{\mathcal{\tau}_h}\,(R^{Bog}_{i,i})^{\frac{1}{2}}\Big]_1\\
&= &\mathcal{L}_i\Big(\Big[(R^{Bog}_{i,i})^{\frac{1}{2}}[\Gamma^{Bog\,}_{i,i}]_{\mathcal{\tau}_h}\,(R^{Bog}_{i,i})^{\frac{1}{2}}\Big]_1\Big)\,A\,\mathcal{L}_i\Big(\Big[(R^{Bog}_{i,i})^{\frac{1}{2}}[\Gamma^{Bog\,}_{i,i}]_{\mathcal{\tau}_h}\,(R^{Bog}_{i,i})^{\frac{1}{2}}\Big]_1\Big)\,\\
& &+\mathcal{D}_i\Big(\Big[(R^{Bog}_{i,i})^{\frac{1}{2}}[\Gamma^{Bog\,}_{i,i}]_{\mathcal{\tau}_h}\,(R^{Bog}_{i,i})^{\frac{1}{2}}\Big]_1\,A\,\Big[(R^{Bog}_{i,i})^{\frac{1}{2}}[\Gamma^{Bog\,}_{i,i}]_{\mathcal{\tau}_h}\,(R^{Bog}_{i,i})^{\frac{1}{2}}\Big]_1\Big)\,.
\end{eqnarray}

\noindent
Therefore, we can re-write 
\begin{eqnarray}
& &(\ref{lead-1})\quad\quad\quad\quad\quad\quad \\
&= &\overline{\mathscr{P}}\Big[W (R^{Bog}_{N-2,N-2})^{\frac{1}{2}}\Big]_2\,\sum_{l_{N-2}=0}^{h-1}\Big\{\mathcal{L}_{N-2}\Big(\Big[(R^{Bog}_{N-2,N-2})^{\frac{1}{2}}[\Gamma^{Bog\,}_{N-2,N-2}]_{\mathcal{\tau}_h}\,(R^{Bog}_{N-2,N-2})^{\frac{1}{2}}\Big]_1\Big)\Big\}^{l_{N-2}}  \Big[(R^{Bog}_{N-2,N-2})^{\frac{1}{2}}W^*\Big]_2\overline{\mathscr{P}}\label{lead-1-0}\\
& &+\overline{\mathscr{P}}\Big[W (R^{Bog}_{N-2,N-2})^{\frac{1}{2}}\Big]_2\,\sum_{l_{N-2}=0}^{h-1}\mathcal{D}_{N-2}\Big(\Big\{\Big[(R^{Bog}_{N-2,N-2})^{\frac{1}{2}}[\Gamma^{Bog\,}_{N-2,N-2}]_{\mathcal{\tau}_h}\,(R^{Bog}_{N-2,N-2})^{\frac{1}{2}}\Big]_1\Big\}^{l_{N-2}}\Big)  \Big[(R^{Bog}_{N-2,N-2})^{\frac{1}{2}}W^*\Big]_2\overline{\mathscr{P}}\,.\quad\quad\quad\label{rem-2}
\end{eqnarray}
We observe that the expression in  (\ref{lead-1-0}) coincides with the original one in (\ref{1-express}) with the only difference that all the companion operators $[...]_1$on level $N-2$ (i.e., the highest index amongst the two resolvents of a companion operator) are replaced with the corresponding $[...]_2$. Therefore, we re-write (\ref{lead-1-0}) as follows
\begin{equation}
\overline{\mathscr{P}}\Big[W (R^{Bog}_{N-2,N-2})^{\frac{1}{2}}\Big]_2\,\sum_{l_{N-2}=0}^{h-1}\Big\{\Big[(R^{Bog}_{N-2,N-2})^{\frac{1}{2}}[\Gamma^{Bog\,}_{N-2,N-2}]_{\mathcal{\tau}_h}\,(R^{Bog}_{N-2,N-2})^{\frac{1}{2}}\Big]_{2\,;\,N-2}\Big\}^{l_{N-2}}  \Big[(R^{Bog}_{N-2,N-2})^{\frac{1}{2}}W^*\Big]_2\overline{\mathscr{P}}
\end{equation}
where, in general, we denote by
\begin{equation}
\Big[(R^{Bog}_{N-2,N-2})^{\frac{1}{2}}[\Gamma^{Bog\,}_{N-2,N-2}]_{\mathcal{\tau}_h}\,(R^{Bog}_{N-2,N-2})^{\frac{1}{2}}\Big]_{2\,;\,i}
\end{equation}
the expression $\Big[(R^{Bog}_{N-2,N-2})^{\frac{1}{2}}[\Gamma^{Bog\,}_{N-2,N-2}]_{\mathcal{\tau}_h}\,(R^{Bog}_{N-2,N-2})^{\frac{1}{2}}\Big]_{1}$ where the companion operators $[...]_1$ on all levels from $N-2$ down to $i$  are replaced with the corresponding operators $[...]_2$.
We keep (\ref{rem-1}) and (\ref{rem-2}) aside and implement  the decomposition  (\ref{D-product})-(\ref{D-product-fin}) on level $N-4$ for the expression in   (\ref{lead-1-0}). We iterate this scheme down to level $i=N-2-h$.
By induction, it is immediate to verify that
\begin{eqnarray}
& &\overline{\mathscr{P}}\Big[W (R^{Bog}_{N-2,N-2})^{\frac{1}{2}}\,\sum_{l_{N-2}=0}^{h-1}\Big\{(R^{Bog}_{N-2,N-2})^{\frac{1}{2}}[\Gamma^{Bog\,}_{N-2,N-2}]_{\mathcal{\tau}_h}\,(R^{Bog}_{N-2,N-2})^{\frac{1}{2}}\Big\}^{l_{N-2}}  \Big[(R^{Bog}_{N-2,N-2})^{\frac{1}{2}}W^*\Big]_1\overline{\mathscr{P}}\quad\quad\quad\quad\quad\quad \\
&= &\overline{\mathscr{P}}\Big[W (R^{Bog}_{N-2,N-2})^{\frac{1}{2}}\Big]_2\,\sum_{l_{N-2}=0}^{h-1}\Big\{\Big[(R^{Bog}_{N-2,N-2})^{\frac{1}{2}}[\Gamma^{Bog\,}_{N-2,N-2}]_{\mathcal{\tau}_h}\,(R^{Bog}_{N-2,N-2})^{\frac{1}{2}}\Big]_{2}\Big\}^{l_{N-2}}  \Big[(R^{Bog}_{N-2,N-2})^{\frac{1}{2}}W^*\Big]_2\overline{\mathscr{P}}\label{lead-1-bis}\\
& &+\sum_{j=N-4\,,\,even}^{N-2-h}\overline{\mathscr{P}}\Big[W (R^{Bog}_{N-2,N-2})^{\frac{1}{2}}\Big]_2\,\Big\{\sum_{l_{N-2}=0}^{h-1}\mathcal{D}_{j}\Big(\Big\{\Big[(R^{Bog}_{N-2,N-2})^{\frac{1}{2}}[\Gamma^{Bog\,}_{N-2,N-2}]_{\mathcal{\tau}_h}\,(R^{Bog}_{N-2,N-2})^{\frac{1}{2}}\Big]_{2\,;\,j+2}\Big\}^{l_{N-2}}\Big)\Big\}\label{rem-in} \\
& &\quad\quad\quad\quad \times  \Big[(R^{Bog}_{N-2,N-2})^{\frac{1}{2}}W^*\Big]_2\overline{\mathscr{P}}\nonumber\\
& &+\overline{\mathscr{P}}\Big[W (R^{Bog}_{N-2,N-2})^{\frac{1}{2}}\Big]_2\,\Big\{\sum_{l_{N-2}=0}^{h-1}\mathcal{D}_{N-2}\Big(\Big\{\Big[(R^{Bog}_{N-2,N-2})^{\frac{1}{2}}[\Gamma^{Bog\,}_{N-2,N-2}]_{\mathcal{\tau}_h}\,(R^{Bog}_{N-2,N-2})^{\frac{1}{2}}\Big]_1\Big\}^{l_{N-2}}\Big)\Big\} \times \label{rem-fin-1}\\
& &\quad\quad\quad\quad\times  \Big[(R^{Bog}_{N-2,N-2})^{\frac{1}{2}}W^*\Big]_2\overline{\mathscr{P}}\nonumber\\
& &+(\ref{rem-1})\,.\label{rem-fin}
\end{eqnarray}
The structure of the remainder terms displayed in (\ref{rem-in})-(\ref{rem-fin}) is important to estimate them in {\bf{STEP} e)}.
\\
% {\color{red}the variables $a_{\bold{j}_m},a^*_{\bold{j}_m}$ have been contracted  as  described in lines (\ref{operation-1})-(\ref{outer-comp}).}

%As a byproduct of this  operation we get two remainders $\mathcal{X}$ and $\mathcal{Y}$ defined below: 

%\noindent
%The operators  $W_{\bold{j}_m}$ and $W^*_{\bold{j}_m}$ of the left and the right companion, respectively, are separated by a product of ``RW" blocks and a resolvent, that we denote in a compact way by $\prod_{j} (RW)_j\,R$
% then the remainder $\mathcal{X}$ corresponding to the coincides with the original product of operators that defines (\ref{l-block-inv}) except that the two considered companions $\mathcal{X}:=[\prod_{j} (RW)_j\,R\,,\,\frac{a_{\bold{0}}a_{\bold{0}}}{N}]$ and $\mathcal{Y}:=-[\prod_j (RW)_j\,R\,,\,\frac{a^*_{\bold{0}}a_{\bold{0}}}{N}]$.

\noindent
{\bf{ STEP c)}}

\noindent

We proceed by implementing Steps $\bf{III_{1}}$ and $\bf{III_2}$ of Lemma \ref{main-relations} on (\ref{2-express}). Hence, we express (\ref{2-express}) as a sum of expressions of the type
\begin{equation}
\Big[\overline{\mathscr{P}_{\psi^{Bog}_{\bold{j}_1,\dots,\bold{j}_{m-1}}}}\,\phi^2_{\bold{j}_{m}}\frac{a^*_{\bold{0}}a^*_{\bold{0}}}{N}\Big\{R^{Bog}_{\bold{j}_1,\dots,\bold{j}_m,;\,N-2,N-2}(w)\,\prod_{l=1}^{\bar{l}}\Big[ [\Gamma^{Bog\,}_{\bold{j}_1,\dots,\bold{j}_m\,;\,N,N}(w)]^{(r_l)}_{\mathcal{\tau}_h}  R^{Bog}_{\bold{j}_1,\dots,\bold{j}_m,;\,N-2,N-2}(w)\Big] \Big\}\frac{a_{\bold{0}}a_{\bold{0}}}{N}\overline{\mathscr{P}_{\psi^{Bog}_{\bold{j}_1,\dots,\bold{j}_{m-1}}}}\Big]_2\label{l-block-inv}
\end{equation} 
where the symbol $\Big[\quad\Big]_2$ means that each couple of companion operators has been replaced according to the rules of above.
Next, starting from the very right of (\ref{l-block-inv}),  in front (i.e., on the right) of each resolvent we split the identity (in $Q^{(>N-1)}_{\bold{j}_m}\mathcal{F}^N$) into
\begin{equation}
\charf_{Q^{(>N-1)}_{\bold{j}_m}\mathcal{F}^N}=\mathscr{P}_{\psi^{Bog}_{\bold{j}_1,\dots,\bold{j}_{m-1}}}+\overline{\mathscr{P}_{\psi^{Bog}_{\bold{j}_1,\dots,\bold{j}_{m-1}}}}\,.
\end{equation}
%Since on the very right of (\ref{l-block-inv}) there is 
Due to the operator $\overline{\mathscr{P}_{\psi^{Bog}_{\bold{j}_1,\dots,\bold{j}_{m-1}}}}$, for the very first resolvent from the right we have
\begin{equation}
\frac{1}{T_{\bold{j}\neq \{\pm\bold{j}_1, \dots, \pm\bold{j}_{m}\}} +\hat{H}^{Bog}_{\bold{j}_1,\dots,\bold{j}_{m-1}}-z^{Bog}_{\bold{j}_1,\dots,\bold{j}_{m-1}}+(\frac{a^*_{\bold{0}}a_{\bold{0}}}{N}\phi_{\bold{j}_{m}}+k_{\bold{j}_{m}}^2)(N-i)|_{i=N-2}-z}(\mathscr{P}_{\psi^{Bog}_{\bold{j}_1,\dots,\bold{j}_{m-1}}}+\overline{\mathscr{P}_{\psi^{Bog}_{\bold{j}_1,\dots,\bold{j}_{m-1}}}})\frac{a^*_{\bold{0}}a_{\bold{0}}}{N}\overline{\mathscr{P}_{\psi^{Bog}_{\bold{j}_1,\dots,\bold{j}_{m-1}}}}\label{res+id}
\end{equation}
where $T_{\bold{j}\neq \{\pm\bold{j}_1, \dots, \pm\bold{j}_{m}\}}:=\sum_{\bold{j}\neq \{\pm\bold{j}_1, \dots, \pm\bold{j}_{m}\}}k_{\bold{j}}^2a^*_{\bold{j}}a_{\bold{j}}$.
Note that the property in  (\ref{ass-number-0}) implies
\begin{equation}
\|\mathscr{P}_{\psi^{Bog}_{\bold{j}_1,\dots,\bold{j}_{m-1}}}\sum_{\bold{j}\neq \bold{0}}a^*_{\bold{j}}a_{\bold{j}}\|\leq \mathcal{O}(\sqrt{N})\,.
\end{equation}
Consequently, the norm of the operator that in (\ref{res+id}) is proportional to $\mathscr{P}_{\psi^{Bog}_{\bold{j}_1,\dots,\bold{j}_{m-1}}}\frac{a^*_{\bold{0}}a_{\bold{0}}}{N}\overline{\mathscr{P}_{\psi^{Bog}_{\bold{j}_1,\dots,\bold{j}_{m-1}}}}$ is small, given that 
\begin{equation}\label{a0sand}
\|\mathscr{P}_{\psi^{Bog}_{\bold{j}_1,\dots,\bold{j}_{m-1}}}\frac{a^*_{\bold{0}}a_{\bold{0}}}{N}\,\overline{\mathscr{P}_{\psi^{Bog}_{\bold{j}_1,\dots,\bold{j}_{m-1}}}}\|=\|\mathscr{P}_{\psi^{Bog}_{\bold{j}_1,\dots,\bold{j}_{m-1}}}\frac{N-\sum_{\bold{j}\neq \bold{0}}a^*_{\bold{j}}a_{\bold{j}}}{N}\,\overline{\mathscr{P}_{\psi^{Bog}_{\bold{j}_1,\dots,\bold{j}_{m-1}}}}\| \leq \mathcal{O}(\frac{1}{\sqrt{N}})\,.
\end{equation}

\noindent
Using the property in  (\ref{ass-number-0}) once more we can estimate\footnote{Recall that $\mathscr{P}_{\psi^{Bog}_{\bold{j}_1,\dots,\bold{j}_{m-1}}}\Big[T_{\bold{j}\neq \{\pm\bold{j}_1, \dots, \pm\bold{j}_{m}\}}+\hat{H}^{Bog}_{\bold{j}_1,\dots,\bold{j}_{m-1}}-z^{Bog}_{\bold{j}_1,\dots,\bold{j}_{m-1}}\Big]=0$.}
\begin{equation}\label{a0sand-bis}
\|\mathscr{P}_{\psi^{Bog}_{\bold{j}_1,\dots,\bold{j}_{m-1}}}\Big[T_{\bold{j}\neq \{\pm\bold{j}_1, \dots, \pm\bold{j}_{m}\}}+\hat{H}^{Bog}_{\bold{j}_1,\dots,\bold{j}_{m-1}}-z^{Bog}_{\bold{j}_1,\dots,\bold{j}_{m-1}}+(\frac{a^*_{\bold{0}}a_{\bold{0}}}{N}\phi_{\bold{j}_{m}}+k_{\bold{j}_{m}}^2)(N-i)-z\Big]\overline{\mathscr{P}_{\psi^{Bog}_{\bold{j}_1,\dots,\bold{j}_{m-1}}}}\|\leq \mathcal{O}(\frac{N-i}{\sqrt{N}})\,.
\end{equation}
With regard to the operator that in (\ref{res+id}) is proportional to $\overline{\mathscr{P}_{\psi^{Bog}_{\bold{j}_1,\dots,\bold{j}_{m-1}}}}\frac{a^*_{\bold{0}}a_{\bold{0}}}{N}\overline{\mathscr{P}_{\psi^{Bog}_{\bold{j}_1,\dots,\bold{j}_{m-1}}}}$, combining (\ref{a0sand-bis}) with the (formal) resolvent identity
\begin{equation}\label{res-ide}
\frac{1}{A}=\frac{1}{B}+\frac{1}{A}(B-A)\frac{1}{B}
\end{equation}
where $$A\equiv T_{\bold{j}\neq \{\pm\bold{j}_1, \dots, \pm\bold{j}_{m}\}}+\hat{H}^{Bog}_{\bold{j}_1,\dots,\bold{j}_{m-1}}-z^{Bog}_{\bold{j}_1,\dots,\bold{j}_{m-1}}+(\frac{a^*_{\bold{0}}a_{\bold{0}}}{N}\phi_{\bold{j}_{m}}+k_{\bold{j}_{m}}^2)(N-i)|_{i=N-2}-z$$
and $B\equiv \overline{\mathscr{P}_{\psi^{Bog}_{\bold{j}_1,\dots,\bold{j}_{m-1}}}}A\overline{\mathscr{P}_{\psi^{Bog}_{\bold{j}_1,\dots,\bold{j}_{m-1}}}}$, up to an operator whose norm is bounded by $\mathcal{O}(\frac{N-i}{\sqrt{N}})|_{i=N-2}$
we can replace
\begin{eqnarray}
& &\,\frac{1}{T_{\bold{j}\neq \{\pm\bold{j}_1, \dots, \pm\bold{j}_{m}\}}+\hat{H}^{Bog}_{\bold{j}_1,\dots,\bold{j}_{m-1}}-z^{Bog}_{\bold{j}_1,\dots,\bold{j}_{m-1}}+(\frac{a^*_{\bold{0}}a_{\bold{0}}}{N}\phi_{\bold{j}_{m}}+k_{\bold{j}_{m}}^2)(N-i)|_{i=N-2}-z}\overline{\mathscr{P}_{\psi^{Bog}_{\bold{j}_1,\dots,\bold{j}_{m-1}}}}\quad\quad\quad\quad
\end{eqnarray}
with
\begin{equation}
\,\frac{1}{\overline{\mathscr{P}_{\psi^{Bog}_{\bold{j}_1,\dots,\bold{j}_{m-1}}}}\Big[T_{\bold{j}\neq \{\pm\bold{j}_1, \dots, \pm\bold{j}_{m}\}}+\hat{H}^{Bog}_{\bold{j}_1,\dots,\bold{j}_{m-1}}-z^{Bog}_{\bold{j}_1,\dots,\bold{j}_{m-1}}+(\frac{a^*_{\bold{0}}a_{\bold{0}}}{N}\phi_{\bold{j}_{m}}+k_{\bold{j}_{m}}^2)(N-i)|_{i=N-2}-z\Big]\overline{\mathscr{P}_{\psi^{Bog}_{\bold{j}_1,\dots,\bold{j}_{m-1}}}}}\overline{\mathscr{P}_{\psi^{Bog}_{\bold{j}_1,\dots,\bold{j}_{m-1}}}}\,.
\end{equation}

%(Similarly for the same expression with $N-i+2$ instead of $N-i$.) 
Proceeding from the right to the left, an analogous argument can be repeated for all the resolvents in (\ref{l-block-inv}). Hence, up to an operator whose norm is bounded by $\mathcal{O}(\frac{N-i}{\sqrt{N}})$, each operator of the type   
\begin{equation}
\Big[(R^{Bog}_{\bold{j}_1,\dots,\bold{j}_m\,;\,i,i}(w))^{\frac{1}{2}}W_{\bold{j}_m\,;\,i,i-2}(R^{Bog}_{\bold{j}_1,\dots,\bold{j}_m\,;\,i-2,i-2}(w))^{\frac{1}{2}}\Big]_2 \label{comp-2}
\end{equation} is replaced with 
%\begin{eqnarray}
%& &\,\frac{1}{\overline{\mathscr{P}_{\psi^{Bog}_{\bold{j}_1,\dots,\bold{j}_{m-1}}}}\Big[\sum_{\bold{j}\neq \{\pm\bold{j}_1, \dots, \pm\bold{j}_{m}\}}k_{\bold{j}}^2a^*_{\bold{j}}a_{\bold{j}}+\hat{H}^{Bog}_{\bold{j}_1,\dots,\bold{j}_{m-1}}-z^{Bog}_{\bold{j}_1,\dots,\bold{j}_{m-1}}+(\frac{a^*_{\bold{0}}a_{\bold{0}}}{N}\phi_{\bold{j}_{m}}+k_{\bold{j}_{m}}^2)(n_{\bold{j}_{m}}+n_{-\bold{j}_{m}})-z\Big]\overline{\mathscr{P}_{\psi^{Bog}_{\bold{j}_1,\dots,\bold{j}_{m-1}}}}}\times \nonumber\\
%& &\times \overline{\mathscr{P}_{\psi^{Bog}_{\bold{j}_1,\dots,\bold{j}_{m-1}}}} \frac{(a^*_{\bold{0}}a_{\bold{0}}-1)}{N}\mathscr{P}'_{\psi^{Bog}_{\bold{j}_1,\dots,\bold{j}_{m-1}}}(n_{\bold{j}_{m}}+1)^{\frac{1}{2}}(n_{-\bold{j}_{m}}+1)^{\frac{1}{2}}\,.\label{RW}
%\end{eqnarray}
%An analogous operation must be implement on the RW-blocks of the type $R^{Bog}_{\bold{j}_1,\dots,\bold{j}_m\,;\,i,i}(w)W^*_{\bold{j}_m\,;\,i,i+2}$.
%After transforming each RW-block in (\ref{l-block}) according to the procedure just explained, we make use of (\ref{a0sand}) one more time and get for (\ref{RW})
\begin{eqnarray}
& &\Big[(R^{Bog}_{\bold{j}_1,\dots,\bold{j}_m\,;\,i,i}(w))^{\frac{1}{2}}W_{\bold{j}_m\,;\,i,i-2}(R^{Bog}_{\bold{j}_1,\dots,\bold{j}_m\,;\,i-2,i-2}(w))^{\frac{1}{2}}\Big]_{\overline{\mathscr{P}}}\\
&:= &\overline{\mathscr{P}_{\psi^{Bog}_{\bold{j}_1,\dots,\bold{j}_{m-1}}}}\Big(\frac{1}{\overline{\mathscr{P}_{\psi^{Bog}_{\bold{j}_1,\dots,\bold{j}_{m-1}}}}\Big[T_{\bold{j}\neq \{\pm\bold{j}_1, \dots, \pm\bold{j}_{m}\}}+\hat{H}^{Bog}_{\bold{j}_1,\dots,\bold{j}_{m-1}}-z^{Bog}_{\bold{j}_1,\dots,\bold{j}_{m-1}}+(\frac{a^*_{\bold{0}}a_{\bold{0}}}{N}\phi_{\bold{j}_{m}}+k_{\bold{j}_{m}}^2)(N-i)-z\Big]\overline{\mathscr{P}_{\psi^{Bog}_{\bold{j}_1,\dots,\bold{j}_{m-1}}}}}\Big)^{\frac{1}{2}}\times \nonumber\\
& &\times\, \overline{\mathscr{P}_{\psi^{Bog}_{\bold{j}_1,\dots,\bold{j}_{m-1}}}}\,\phi_{\bold{j}_{m}}\frac{(a^*_{\bold{0}}a_{\bold{0}}-1)}{2N}(N-i+2)\overline{\mathscr{P}_{\psi^{Bog}_{\bold{j}_1,\dots,\bold{j}_{m-1}}}}\times\,\\
& &\times\, \Big(\frac{1}{\overline{\mathscr{P}_{\psi^{Bog}_{\bold{j}_1,\dots,\bold{j}_{m-1}}}}\Big[T_{\bold{j}\neq \{\pm\bold{j}_1, \dots, \pm\bold{j}_{m}\}}+\hat{H}^{Bog}_{\bold{j}_1,\dots,\bold{j}_{m-1}}-z^{Bog}_{\bold{j}_1,\dots,\bold{j}_{m-1}}+(\frac{a^*_{\bold{0}}a_{\bold{0}}}{N}\phi_{\bold{j}_{m}}+k_{\bold{j}_{m}}^2)(N-i+2)-z\Big]\overline{\mathscr{P}_{\psi^{Bog}_{\bold{j}_1,\dots,\bold{j}_{m-1}}}}}\Big)^{\frac{1}{2}}\overline{\mathscr{P}_{\psi^{Bog}_{\bold{j}_1,\dots,\bold{j}_{m-1}}}}\,.\quad\quad\quad\quad\quad\quad\nonumber
\end{eqnarray}

The companion operator of (\ref{comp-2}) is replaced with
\begin{eqnarray}
& &\Big[(R^{Bog}_{\bold{j}_1,\dots,\bold{j}_m\,;\,i-2,i-2}(w))^{\frac{1}{2}}W^*_{\bold{j}_m\,;\,i-2,i}(R^{Bog}_{\bold{j}_1,\dots,\bold{j}_m\,;\,i,i}(w))^{\frac{1}{2}}\Big]_{\overline{\mathscr{P}}}\\
&:= &\overline{\mathscr{P}_{\psi^{Bog}_{\bold{j}_1,\dots,\bold{j}_{m-1}}}}\Big(\frac{1}{\overline{\mathscr{P}_{\psi^{Bog}_{\bold{j}_1,\dots,\bold{j}_{m-1}}}}\Big[T_{\bold{j}\neq \{\pm\bold{j}_1, \dots, \pm\bold{j}_{m}\}}+\hat{H}^{Bog}_{\bold{j}_1,\dots,\bold{j}_{m-1}}-z^{Bog}_{\bold{j}_1,\dots,\bold{j}_{m-1}}+(\frac{a^*_{\bold{0}}a_{\bold{0}}}{N}\phi_{\bold{j}_{m}}+k_{\bold{j}_{m}}^2)(N-i+2)-z\Big]\overline{\mathscr{P}_{\psi^{Bog}_{\bold{j}_1,\dots,\bold{j}_{m-1}}}}}\Big)^{\frac{1}{2}}\times \nonumber\\
& &\times \overline{\mathscr{P}_{\psi^{Bog}_{\bold{j}_1,\dots,\bold{j}_{m-1}}}}\,\phi_{\bold{j}_{m}}\frac{a^*_{\bold{0}}a_{\bold{0}}}{2N}(N-i+2)\overline{\mathscr{P}_{\psi^{Bog}_{\bold{j}_1,\dots,\bold{j}_{m-1}}}}\times\,\\
& &\times \Big(\frac{1}{\overline{\mathscr{P}_{\psi^{Bog}_{\bold{j}_1,\dots,\bold{j}_{m-1}}}}\Big[T_{\bold{j}\neq \{\pm\bold{j}_1, \dots, \pm\bold{j}_{m}\}}+\hat{H}^{Bog}_{\bold{j}_1,\dots,\bold{j}_{m-1}}-z^{Bog}_{\bold{j}_1,\dots,\bold{j}_{m-1}}+(\frac{a^*_{\bold{0}}a_{\bold{0}}}{N}\phi_{\bold{j}_{m}}+k_{\bold{j}_{m}}^2)(N-i)-z\Big]\overline{\mathscr{P}_{\psi^{Bog}_{\bold{j}_1,\dots,\bold{j}_{m-1}}}}}\Big)^{\frac{1}{2}}\overline{\mathscr{P}_{\psi^{Bog}_{\bold{j}_1,\dots,\bold{j}_{m-1}}}}\,.\nonumber
\end{eqnarray}
Analogous replacements hold for the outer companions (i.e., the operators in (\ref{outer-comp-0}) and (\ref{outer-comp})).

\noindent
{\bf{ STEP d)}}

\noindent
Here, we estimate from above the norm of the leading expression resulting from {\bf{ STEPS a), b), c)}}. We observe that, under the assumptions on $\epsilon_{\bold{j}_m}$, $N$, and $z$,
\begin{eqnarray}
& &\|\Big(\frac{1}{\overline{\mathscr{P}_{\psi^{Bog}_{\bold{j}_1,\dots,\bold{j}_{m-1}}}}\Big[\sum_{\bold{j}\neq \{\pm\bold{j}_1, \dots, \pm\bold{j}_{m}\}}k_{\bold{j}}^2a^*_{\bold{j}}a_{\bold{j}}+\hat{H}^{Bog}_{\bold{j}_1,\dots,\bold{j}_{m-1}}-z^{Bog}_{\bold{j}_1,\dots,\bold{j}_{m-1}}+(\frac{a^*_{\bold{0}}a_{\bold{0}}}{N}\phi_{\bold{j}_{m}}+k_{\bold{j}_{m}}^2)(N-i)-z\Big]\overline{\mathscr{P}_{\psi^{Bog}_{\bold{j}_1,\dots,\bold{j}_{m-1}}}}}\Big)^{\frac{1}{2}}\nonumber \\
& &\quad\quad\times \Big(\overline{\mathscr{P}_{\psi^{Bog}_{\bold{j}_1,\dots,\bold{j}_{m-1}}}}\Big[\Delta_{m-1}+(\frac{a^*_{\bold{0}}a_{\bold{0}}}{N}\phi_{\bold{j}_{m}}+k_{\bold{j}_{m}}^2)(N-i)-z\Big]\overline{\mathscr{P}_{\psi^{Bog}_{\bold{j}_1,\dots,\bold{j}_{m-1}}}}\Big)^{\frac{1}{2}}\| \nonumber\\
& \leq &1
\end{eqnarray}
thanks to (\ref{assumption-gap}).  (See a similar argument in Corollary \ref{main-lemma-H}.)  Hence, we can estimate
\begin{eqnarray}
& &\|[(R^{Bog}_{\bold{j}_1,\dots,\bold{j}_m\,;\,i,i}(w))^{\frac{1}{2}}W_{\bold{j}_m\,;\,i,i-2}(R^{Bog}_{\bold{j}_1,\dots,\bold{j}_m\,;\,i-2,i-2}(w))^{\frac{1}{2}}]_{\overline{\mathscr{P}}}\|\\
&\leq &\|\Big(\frac{1}{\overline{\mathscr{P}_{\psi^{Bog}_{\bold{j}_1,\dots,\bold{j}_{m-1}}}}\Big[\Delta_{m-1}+(\frac{a^*_{\bold{0}}a_{\bold{0}}}{N}\phi_{\bold{j}_{m}}+k_{\bold{j}_{m}}^2)(N-i)-z\Big]\overline{\mathscr{P}_{\psi^{Bog}_{\bold{j}_1,\dots,\bold{j}_{m-1}}}}}\Big)^{\frac{1}{2}}\times\\
& &\quad\times \overline{\mathscr{P}_{\psi^{Bog}_{\bold{j}_1,\dots,\bold{j}_{m-1}}}}\,\phi_{\bold{j}_{m}}\frac{(a^*_{\bold{0}}a_{\bold{0}}-1)}{2N}(N-i+2)\overline{\mathscr{P}_{\psi^{Bog}_{\bold{j}_1,\dots,\bold{j}_{m-1}}}}\times\,\nonumber\\
& &\quad \times \Big(\frac{1}{\overline{\mathscr{P}_{\psi^{Bog}_{\bold{j}_1,\dots,\bold{j}_{m-1}}}}\Big[\Delta_{m-1}+(\frac{a^*_{\bold{0}}a_{\bold{0}}}{N}\phi_{\bold{j}_{m}}+k_{\bold{j}_{m}}^2)(N-i+2)-z\Big]\overline{\mathscr{P}_{\psi^{Bog}_{\bold{j}_1,\dots,\bold{j}_{m-1}}}}}\Big)^{\frac{1}{2}}\overline{\mathscr{P}_{\psi^{Bog}_{\bold{j}_1,\dots,\bold{j}_{m-1}}}}\|\,.\nonumber
\end{eqnarray}
We set $h=\lfloor (\ln N)^{\frac{1}{2}} \rfloor$ with $\ln N\ll N$. The (even) index $i$ ranges from $N-2-\lfloor (\ln N)^{\frac{1}{2}} \rfloor $ to $N-2$ where, for simplicity of the notation,  we have assumed that $\lfloor (\ln N)^{\frac{1}{2}} \rfloor $ is even. 

\noindent
With the help of the spectral theorem for commuting self-adjoint operators (i.e., $\overline{\mathscr{P}_{\psi^{Bog}_{\bold{j}_1,\dots,\bold{j}_{m-1}}}}$ and $\overline{\mathscr{P}_{\psi^{Bog}_{\bold{j}_1,\dots,\bold{j}_{m-1}}}}a^*_{\bold{0}}a_{\bold{0}}\overline{\mathscr{P}_{\psi^{Bog}_{\bold{j}_1,\dots,\bold{j}_{m-1}}}}$) and the bound $a^*_{\bold{0}}a_{\bold{0}}\leq N$, we deduce (see section 0.5 of the file \emph{supporting-file-Bose2.pdf}) that
\begin{eqnarray}
& &\|\Big[(R^{Bog}_{\bold{j}_1,\dots,\bold{j}_m\,;\,i,i}(w))^{\frac{1}{2}}W_{\bold{j}_m\,;\,i,i-2}(R^{Bog}_{\bold{j}_1,\dots,\bold{j}_m\,;\,i-2,i-2}(w))^{\frac{1}{2}}\Big]_{\overline{\mathscr{P}}}\|\\
&\leq &\frac{N-i+2}{2\Big[\Delta_{m-1}+(\phi_{\bold{j}_{m}}+k_{\bold{j}_{m}}^2)(N-i+2)-z\Big]^{\frac{1}{2}} }\,\frac{\phi_{\bold{j}_{m}}}{\Big[\Delta_{m-1}+(\phi_{\bold{j}_{m}}+k_{\bold{j}_{m}}^2)(N-i)-z\Big]^{\frac{1}{2}} }\,\quad\quad \quad\quad \label{spectral}
\end{eqnarray}
for $N$ large enough and $z$ in the interval (\ref{interval}). The same bound holds for 
$$\|\Big[(R^{Bog}_{\bold{j}_1,\dots,\bold{j}_m\,;\,i-2,i-2}(w))^{\frac{1}{2}}W^*_{\bold{j}_m\,;\,i-2,i}(R^{Bog}_{\bold{j}_1,\dots,\bold{j}_m\,;\,i,i}(w))^{\frac{1}{2}}\Big]_{\overline{\mathscr{P}}}\|\,.$$

\noindent
Finally, we collect the leading terms obtained for each expression (\ref{l-block-inv}) after the implementation of {\bf{STEP c)}}, and recall that, by construction, the sum of expressions (\ref{l-block-inv}) coincides with 
(\ref{2-express}).
%\begin{equation}
%\overline{\mathscr{P}_{\psi^{Bog}_{\bold{j}_1,\dots,\bold{j}_{m-1}}}}W_{\bold{j}_m}\,R^{Bog}_{\bold{j}_1,\dots,\bold{j}_m\,;\,N-2,N-2}(w)\, \sum_{l_{N-2}=0}^{h-1}\Big[[\Gamma^{Bog}_{\bold{j}_1,\dots,\bold{j}_m\,;\,N-2,N-2}(w)]_{\tau_h}\,R^{Bog}_{\bold{j}_1,\dots,\bold{j}_m\,;\,\,N-2,N-2}(w)\Big]^{l_{N-2}}W^*_{\bold{j}_m}\,\overline{\mathscr{P}_{\psi^{Bog}_{\bold{j}_1,\dots,\bold{j}_{m-1}}}}
%\end{equation}
%where $[\Gamma^{Bog}_{\bold{j}_1,\dots,\bold{j}_m\,;\,N-2,N-2}(w)]_{\tau_h}$ is defined in (\ref{def-tau}).
Therefore, due to (\ref{spectral}) and to an analogous operator norm estimate for the outer companions,  the norm of the sum of the collected leading terms associated with (\ref{2-express}) is bounded by 
\begin{equation}\label{Ktrunc}
\frac{\phi^2_{\bold{j}_{m}}}{\Big[\Delta_{m-1}+2(\phi_{\bold{j}_{m}}+k_{\bold{j}_{m}}^2)-z\Big] }\check{\mathcal{K}}^{(h)}_{\bold{j}_{m}\,;\,N-2,N-2}(z)
\end{equation}
 where $\check{\mathcal{K}}^{(h)}_{\bold{j}_{m}\,;\,N-2,N-2}(z)$ is defined by recursion:
\begin{equation}
\check{\mathcal{K}}^{(h)}_{\bold{j}_{m}\,;\,N-4-h,N-4-h}(z)\equiv 1
\end{equation}
and for $N-2\geq i\geq N-2-h$ (and even)
\begin{equation}\label{def-Kcheck}
\check{\mathcal{K}}^{(h)}_{\bold{j}_{m}\,;\,i,i}(z):=\sum_{l_{i}=0}^{h-1}\Big[\frac{(N-i+2)^{2}}{4\Big[\Delta_{m-1}+(\phi_{\bold{j}_{m}}+k_{\bold{j}_{m}}^2)(N-i+2)-z\Big] }\,\frac{\phi^2_{\bold{j}_{m}}}{\Big[\Delta_{m-1}+(\phi_{\bold{j}_{m}}+k_{\bold{j}_{m}}^2)(N-i)-z\Big]}\check{\mathcal{K}}^{(h)}_{\bold{j}_{m}\,;\,i-2,i-2}(z)\Big]^{l_{i}}\,.\\
\end{equation}
Eventually, we want to show that, for $z$ in the  interval in  (\ref{interval}) and assuming the condition in (\ref{def-cjm}),  for the chosen $h$ the quantity in (\ref{Ktrunc}) is bounded by
\begin{equation}\label{bound-lemma4.5}
(1-\frac{1}{N})\frac{\phi_{\bold{j}_{m}}}{2\epsilon_{\bold{j}_m}+2-\frac{4}{N}-\frac{z-\Delta_{m-1}+\frac{U_{\bold{j}_m}}{\sqrt{N}}}{\phi_{\bold{j}_{m}}}}\,\check{\mathcal{G}}_{\bold{j}_{m}\,;\,N-2,N-2}(z-\Delta_{m-1}+\frac{U_{\bold{j}_m}}{\sqrt{N}})\,.
\end{equation}
where $U_{\bold{j_m}}=k^2_{\bold{j}_m}+\phi_{\bold{j}_m}$.
%{\color{red}where $\check{\mathcal{G}}_{\bold{j}_{m}\,;\,N-2,N-2}(z)$ is introduced in (\ref{in-formula-G})-(\ref{fin-formula-G}) and $C_{\bold{j_m}}$ is a positive constant depending on $k_{\bold{j}_m}^2$ and $\phi_{\bold{j}_m}$.} Notice that the quantity in ...
%coincides with
%%f_{\bold{j}_m}(z-\Delta_{m-1}+\frac{C_{\bold{j}_m}}{\sqrt{N}})+z-\Delta_{m-1}+\frac{C_{\bold{j}_m}}{\sqrt{N}}\,.
%\end{equation}
(Recall that we have set $h=\lfloor (\ln N)^{\frac{1}{2}} \rfloor$ with $\ln N\ll N$. The (even) index $i$ ranges from $N-2-\lfloor (\ln N)^{\frac{1}{2}} \rfloor $ to $N-2$ where, for simplicity of the notation,  we have assumed that $\lfloor (\ln N)^{\frac{1}{2}} \rfloor $ is even.) Then, for  $z$ in the interval in (\ref{interval}), and assuming the condition in (\ref{def-cjm}), a direct computation (see section 0.6 in \emph{supporting-file-Bose2.pdf}) shows that 
the following inequality holds true (for $N-2\geq i\geq N-2-h$)
\begin{eqnarray}
& &\frac{(N-i+2)^{2}}{4\Big[\Delta_{m-1}+(\phi_{\bold{j}_{m}}+k_{\bold{j}_{m}}^2)(N-i+2)-z\Big] }\,\frac{\phi^2_{\bold{j}_{m}}}{\Big[\Delta_{m-1}+(\phi_{\bold{j}_{m}}+k_{\bold{j}_{m}}^2)(N-i)-z\Big]}\quad\quad\quad \label{Delta-W}\\
&\leq &\mathcal{W}_{\bold{j}_{m}\,;\,i,i-2}(z-\Delta_{m-1}+\frac{U_{\bold{j}_m}}{\sqrt{N}})\mathcal{W}^*_{\bold{j}_{m}\,;\,i-2,i}(z-\Delta_{m-1}+\frac{U_{\bold{j}_m}}{\sqrt{N}})\nonumber
\end{eqnarray}
{\color{red}}where $\mathcal{W}_{\bold{j}_{m}\,;\,i,i-2}(z)\mathcal{W}^*_{\bold{j}_{m}\,;\,i-2,i}(z)$ enters the definition of $\check{\mathcal{G}}_{\bold{j}_{m}\,;\,N-2,N-2}(z)$  (see (\ref{in-formula-G})-(\ref{fin-formula-G})). Now, we define $\check{\mathcal{G}}^{(h)}_{\bold{j}_{m}\,;\,i,i}(z)$ similarly to $\check{\mathcal{G}}_{\bold{j}_{m}\,;\,i,i}(z)$  but replacing $\sum_{l_{i}=0}^{\infty}$ with $\sum_{l_{i}=0}^{h-1}$ in the recursive definition (see (\ref{in-formula-G})). Then, for $N$ sufficiently large we readily obtain
\begin{equation}
\check{\mathcal{K}}^{(h)}_{\bold{j}_{m}\,;\,N-2,N-2}(z)\leq \check{\mathcal{G}}^{(h)}_{\bold{j}_{m}\,;\,N-2,N-2}(z-\Delta_{m-1}+\frac{U_{\bold{j}_m}}{\sqrt{N}})\leq \check{\mathcal{G}}_{\bold{j}_{m}\,;\,N-2,N-2}(z-\Delta_{m-1}+\frac{U_{\bold{j}_m}}{\sqrt{N}})\,
\end{equation}
and
\begin{equation}
\frac{\phi^2_{\bold{j}_{m}}}{\Big[\Delta_{m-1}+2(\phi_{\bold{j}_{m}}+k_{\bold{j}_{m}}^2)-z\Big] }\check{\mathcal{K}}^{(h)}_{\bold{j}_{m}\,;\,N-2,N-2}(z)\leq (\ref{bound-lemma4.5})\,.
\end{equation}
%(\ref{l-block}) by replacing the identity in front of each resolvent with
%\begin{equation}
%(\charf_{\mathcal{F}^N\ominus \mathcal{F}^N_{\pm \bold{j}_m}}-
%\mathscr{P}_{\psi^{Bog}_{\bold{j}_1,\dots,\bold{j}_{m-1}}})
%\end{equation}

\noindent
{\bf{ STEP e)}}

\noindent
Now, we estimate (all) the remainder terms produced implementing {\bf{ STEP b)}} on the expression in (\ref{1-express}).
%that remainder term associated with $\mathcal{X}$  is surely bounded $$\mathcal{O}(\frac{1}{\sqrt{N}})\times  [ \text{maximum number of "RW" blocks in expression of the type}\, (\ref{l-block-inv})]^2$$ which in STEP ... of Lemma is estimated $\mathcal{O}(...)$ at most. The number of commutators $\mathcal{X}, \mathcal{Y}, \mathcal{Z}$ that is produced is proportional to the number of blocks that is at most $\mathcal{O}(...)$ (see Step ... in  Lemma...)
We claim that each of the $h/2+2$ remainder terms in (\ref{rem-in})-(\ref{rem-fin}) is bounded in norm by
$$\mathcal{O}(h\frac{h\,(2h)^{\frac{h+2}{2}}\cdot h\,(2h)^{\frac{h+2}{2}}}{N})\,.$$
Concerning notation, we remind that in {\bf{ STEP b)}} we omit the label $\bold{j}_1,\dots,\bold{j}_m$ and the argument $w$ in $R^{Bog}_{\bold{j}_1,\dots,\bold{j}_m;\,i,i}(w)$, and the label $\psi^{Bog}_{\bold{j}_1,\dots,\bold{j}_{m-1}}$ in $\overline{\mathscr{P}_{\psi^{Bog}_{\bold{j}_1,\dots,\bold{j}_{m-1}}}}$. In the present step where we estimate the norm of the (remainder) terms produced in {\bf{ STEP b)}} the notation is consistent with that choice, except for points 4. and 5. where it is useful to re-introduce the complete notation.

To estimate the norm of the $j-th$ summand in (\ref{rem-in}), i.e.,
\begin{equation}
\overline{\mathscr{P}}\Big[W (R^{Bog}_{N-2,N-2})^{\frac{1}{2}}\Big]_2\,\Big\{\sum_{l_{N-2}=0}^{h-1}\mathcal{D}_{j}\Big(\Big\{\Big[(R^{Bog}_{N-2,N-2})^{\frac{1}{2}}[\Gamma^{Bog\,}_{N-2,N-2}]_{\mathcal{\tau}_h}\,(R^{Bog}_{N-2,N-2})^{\frac{1}{2}}\Big]_{2\,;\,j+2}\Big\}^{l_{N-2}}\Big)\Big\} \Big[(R^{Bog}_{N-2,N-2})^{\frac{1}{2}}W^*\Big]_2\overline{\mathscr{P}}\,,\label{j-summand}
\end{equation}
at first we express
\begin{equation}
\sum_{l_{N-2}=0}^{h-1}\Big\{\Big[(R^{Bog}_{N-2,N-2})^{\frac{1}{2}}[\Gamma^{Bog\,}_{N-2,N-2}]_{\mathcal{\tau}_h}\,(R^{Bog}_{N-2,N-2})^{\frac{1}{2}}\Big]_{2\,;\,j+2}\Big\}^{l_{N-2}}
\end{equation} 
as a sum of ``monomials" similarly to the procedure in {\bf{STEP ${\bf{III_1}}$}} of Lemma \ref{main-relations}.   Next,
\begin{itemize}
\item for each of these monomials  we evaluate how many new monomials are created due to the action of $\mathcal{D}_j$;
\item we estimate the norm of each resulting term;
% by $\mathcal{O}(\frac{N-j}{N})$ times the norm of the monomial it is derived from};
\item we apply the argument of Remark \ref{estimation-proc} and conclude that the norm of the whole expression in (\ref{j-summand}) is bounded by $\mathcal{O}(\frac{N-j}{N})$ times the maximum number of new terms created by the action of $\mathcal{D}_j$ on a single monomial.
\end{itemize}

Below we explain the procedure.
\begin{enumerate}
\item
Like in {\bf{STEP ${\bf{III_1}}$}} of Lemma \ref{main-relations}, we use  formula (\ref{tau-exp})  iteratively in order to re-express 
\begin{equation}\label{point1}
\sum_{l_{N-2}=0}^{h-1}\Big\{\Big[(R^{Bog}_{N-2,N-2})^{\frac{1}{2}}[\Gamma^{Bog\,}_{N-2,N-2}]_{\mathcal{\tau}_h}\,(R^{Bog}_{N-2,N-2})^{\frac{1}{2}}\Big]_{2\,;\,j+2}\Big\}^{l_{N-2}}\,.
\end{equation}
In detail, similarly to (\ref{summands-rl}),  we get (for some $j-$dependent $\bar{r}$)
 \begin{equation}\label{summands-rl-bis-bis}
\Big[(R^{Bog}_{N-2,N-2})^{\frac{1}{2}}[\Gamma^{Bog\,}_{N-2,N-2}]_{\mathcal{\tau}_h}(R^{Bog}_{N-2,N-2})^{\frac{1}{2}}\Big]_{2\,;\,j+2}=: \sum_{r=1}^{\bar{r}}\Big[(R^{Bog}_{N-2,N-2})^{\frac{1}{2}}[\Gamma^{Bog\,}_{N-2,N-2}]^{(r)}_{\mathcal{\tau}_h}(R^{Bog}_{N-2,N-2})^{\frac{1}{2}}\Big]_{2\,;\,j+2}
 \end{equation}
where each $[(R^{Bog}_{N-2,N-2})^{\frac{1}{2}}[\Gamma^{Bog\,}_{N-2,N-2}]^{(r)}_{\mathcal{\tau}_h}(R^{Bog}_{N-2,N-2})^{\frac{1}{2}}]_{2\,;\,j+2}$ is a ``monomial" that corresponds to $[(R^{Bog}_{N-2,N-2})^{\frac{1}{2}}W(R^{Bog}_{N-4,N-4})^{\frac{1}{2}}]_2$ multiplying (on the right) factors of the type
\begin{equation}
[(R^{Bog}_{s-2,s-2})^{\frac{1}{2}}W^*(R^{Bog}_{s,s})^{\frac{1}{2}}]_2\quad,\quad [(R^{Bog}_{s,s})^{\frac{1}{2}}W(R^{Bog}_{s-2,s-2})^{\frac{1}{2}}]_2\quad,\quad \Big[(R^{Bog}_{j,j})^{\frac{1}{2}}[\Gamma^{Bog\,}_{j,j}]_{\mathcal{\tau}_h}\,(R^{Bog}_{j,j})^{\frac{1}{2}}\Big]_1\,
\end{equation} 
with $j+2\leq s \leq N-2$ and even. Therefore, (\ref{point1}) can be written
\begin{equation}\label{sum-mon-0}
\sum_{\bar{l}=0}^{h-1} \sum_{r_1=1}^{\bar{r}}\dots  \sum_{r_{\,\bar{l}\,}=1}^{\bar{r}}\,\prod_{l=1}^{\bar{l}}\Big\{\,\Big[(R^{Bog}_{N-2,N-2})^{\frac{1}{2}}[\Gamma^{Bog\,}_{N-2,N-2}]^{(r_l)}_{\mathcal{\tau}_h}(R^{Bog}_{N-2,N-2})^{\frac{1}{2}}\Big]_{2\,;\,j+2}\Big\}
\end{equation}
where each $\Big[(R^{Bog}_{N-2,N-2})^{\frac{1}{2}}[\Gamma^{Bog\,}_{N-2,N-2}]^{(r_l)}_{\mathcal{\tau}_h}(R^{Bog}_{N-2,N-2})^{\frac{1}{2}}\Big]_{2\,;\,j+2}$ is a summand in (\ref{summands-rl-bis-bis}) and $\prod_{l=1}^{\bar{l}=0}[\dots]\equiv \charf$.

We define
\begin{eqnarray}
\#_{R\Gamma R}&:=&\text{the maximum number of factors}\, \Big[(R^{Bog}_{j,j})^{\frac{1}{2}}[\Gamma^{Bog\,}_{j,j}]_{\mathcal{\tau}_h}\,(R^{Bog}_{j,j})^{\frac{1}{2}}\Big]_1 \,\,\text{contained in each}\nonumber\\
& &\text{product}\,\,\prod_{l=1}^{\bar{l}}\Big\{\Big[(R^{Bog}_{N-2,N-2})^{\frac{1}{2}}[\Gamma^{Bog\,}_{N-2,N-2}]^{(r_l)}_{\mathcal{\tau}_h}(R^{Bog}_{N-2,N-2})^{\frac{1}{2}}\Big]_{2\,;\,j+2}\Big\}\,.\nonumber
\end{eqnarray}
\item Consider the action of $\mathcal{D}_i$ on a product like the one in (\ref{def-D_i})  but possibly with extra-factors $ A_{r}\dots B_{r}$, $r=1,\dots l_{i-1}$, namely 
\begin{equation}
\Big[(R^{Bog}_{i,i})^{\frac{1}{2}}[\Gamma^{Bog\,}_{i,i}]_{\mathcal{\tau}_h}\,(R^{Bog}_{i,i})^{\frac{1}{2}}\Big]_1 A_1\dots B_1 \Big[(R^{Bog}_{i,i})^{\frac{1}{2}}[\Gamma^{Bog\,}_{i,i}]_{\mathcal{\tau}_h}\,(R^{Bog}_{i,i})^{\frac{1}{2}}\Big]_1A_{l_{i-1}}\dots B_{l_{i-1}} \Big[(R^{Bog}_{i,i})^{\frac{1}{2}}[\Gamma^{Bog\,}_{i,i}]_{\mathcal{\tau}_h}\,(R^{Bog}_{i,i})^{\frac{1}{2}}\Big]_1  \label{D-product-bis}
\end{equation} 
Out of the product in (\ref{D-product-bis}), the operation $\mathcal{D}_i$ yields a sum of products (see (\ref{def-D_i})) where: 

\noindent
a)  the number of summands equals the number, $l_i$,  of factors $\Big[(R^{Bog}_{i,i})^{\frac{1}{2}}[\Gamma^{Bog\,}_{i,i}]_{\mathcal{\tau}_h}\,(R^{Bog}_{i,i})^{\frac{1}{2}}\Big]_1$ in the product (\ref{D-product-bis}); 

\noindent
b)  each summand is a product like the one in (\ref{D-product-bis}) where some of the factors $\Big[(R^{Bog}_{i,i})^{\frac{1}{2}}[\Gamma^{Bog\,}_{i,i}]_{\mathcal{\tau}_h}\,(R^{Bog}_{i,i})^{\frac{1}{2}}\Big]_1$ may have been replaced with $\mathcal{L}_i(\Big[(R^{Bog}_{i,i})^{\frac{1}{2}}[\Gamma^{Bog\,}_{i,i}]_{\mathcal{\tau}_h}\,(R^{Bog}_{i,i})^{\frac{1}{2}}\Big]_1)$, and one single factor\\ $\Big[(R^{Bog}_{i,i})^{\frac{1}{2}}[\Gamma^{Bog\,}_{i,i}]_{\mathcal{\tau}_h}\,(R^{Bog}_{i,i})^{\frac{1}{2}}\Big]_1$ has been replaced with   $\mathcal{D}_i\Big(\Big[(R^{Bog}_{i,i})^{\frac{1}{2}}[\Gamma^{Bog\,}_{i,i}]_{\mathcal{\tau}_h}\,(R^{Bog}_{i,i})^{\frac{1}{2}}\Big]_1\Big)$.
%that contains the commutators associated with the symbol $\underbrace{\quad}$ described after line (\ref{rem-1}).

\noindent
Therefore, the maximum number of new summands that are produced (due to  (\ref{def-D_i})) from a monomial of (\ref{sum-mon-0})   is  $\#_{R\Gamma R}$ (defined in point 1.). 
\item As specified in point 2., each summand produced in point 2. is a product containing only one 
\begin{eqnarray}
& &\mathcal{D}_j\Big(\Big[(R^{Bog}_{j,j})^{\frac{1}{2}}[\Gamma^{Bog\,}_{j,j}]_{\mathcal{\tau}_h}\,(R^{Bog}_{j,j})^{\frac{1}{2}}\Big]_1\Big)\label{def-D}\\
%&:=&\Big[(R^{Bog}_{N-2,N-2})^{\frac{1}{2}}[\Gamma^{Bog\,}_{N-2,N-2}]_{\mathcal{\tau}_h}\,(R^{Bog}_{N-2,N-2})^{\frac{1}{2}}\Big]_1\\
&:= &\Big[(R^{Bog}_{j,j})^{\frac{1}{2}}\underbrace{W (R^{Bog}_{j-2,j-2})^{\frac{1}{2}}\Big]_1\,\sum_{l_{j-2}=0}^{h-1}\Big\{\Big[(R^{Bog}_{j-2,j-2})^{\frac{1}{2}}[\Gamma^{Bog\,}_{j-2,j-2}]_{\mathcal{\tau}_h}\,(R^{Bog}_{j-2,j-2})^{\frac{1}{2}}\Big]_1\Big\}^{l_{j-2}}  \Big[(R^{Bog}_{j-2,j-2})^{\frac{1}{2}}W^*}(R^{Bog}_{j,j})^{\frac{1}{2}}\Big]_1\,.\nonumber
\end{eqnarray}
The symbol $\underbrace{\quad}$ stands for two commutators (see (\ref{embraced-comm-1})-(\ref{embraced-comm-2})).  After computing the commutators,  the R-H-S of (\ref{def-D}) can be written as a sum of product of operators. We want to estimate the number of the new summands due to the commutators as specified below after (\ref{sum-mon}). To this purpose, using formula (\ref{tau-exp}),  we re-write $[[\Gamma^{Bog\,}_{j-2,j-2}]_{\mathcal{\tau}_h}]_1$ as
 \begin{equation}\label{summands-rl-bis}
[[\Gamma^{Bog\,}_{j-2,j-2}]_{\mathcal{\tau}_h}]_1=: \sum_{r=1}^{\bar{r}}[[\Gamma^{Bog\,}_{j-2,j-2}]^{(r)}_{\mathcal{\tau}_h}]_1
 \end{equation}
(for some $h-$dependent $\bar{r}$)  where   each summand $[[\Gamma^{Bog\,}_{j-2,j-2}]^{(r)}_{\mathcal{\tau}_h}]_1$ corresponds to $\phi_{\bold{j}_m}\frac{a^*_{\bold{0}}a^*_{\bold{0}}}{N}\frac{N-i+2}{2}|_{i=j-2}$ multiplying on the right a finite product of modified ``RW-blocks", i.e., operators of the type
 \begin{equation}\label{blocks-bis}
[R^{Bog}_{i,i}W]_1:=[R^{Bog}_{i,i}]_1\phi_{\bold{j}_m}\frac{a^*_{\bold{0}}a^*_{\bold{0}}}{N}\frac{N-i+2}{2}\quad,\quad [R^{Bog}_{i,i}W^*]_1:= [R^{Bog}_{i,i}]_1\phi_{\bold{j}_m}\frac{a_{\bold{0}}a_{\bold{0}}}{N}\frac{N-i}{2}\,,
\end{equation}
where $i$ is even and ranges\footnote{For the expression $[R^{Bog}_{i,i}W]_1$, the index $i $ ranges from $j-2-h$ to $j-4$.} from $j-4-h$ to $j-4$.  The number of modified RW-blocks for each $\Big[[\Gamma^{Bog\,}_{j-2,j-2}]^{(r)}_{\mathcal{\tau}_h}\Big]_1$ is bounded by $\mathcal{O}((2h)^{\frac{h+2}{2}})$. Then, we re-express 
\begin{equation}\label{sumtobeexp}
\sum_{l_{j-2}=0}^{h-1}\Big\{\Big[(R^{Bog}_{j-2,j-2})^{\frac{1}{2}}[\Gamma^{Bog\,}_{j-2,j-2}]_{\mathcal{\tau}_h}\,(R^{Bog}_{j-2,j-2})^{\frac{1}{2}}\Big]_1\Big\}^{l_{j-2}}
\end{equation}
as a sum of ``monomials"
\begin{equation}\label{sum-mon}
\sum_{\bar{l}=0}^{h-1} \sum_{r_1=1}^{\bar{r}}\dots  \sum_{r_{\,\bar{l}\,}=1}^{\bar{r}}\,\prod_{l=1}^{\bar{l}}\Big\{\Big[\, (R^{Bog}_{j-2,j-2})^{\frac{1}{2}}\Big]_1\Big[[\Gamma^{Bog\,}_{j-2,j-2}]^{(r_l)}_{\mathcal{\tau}_h}\Big]_1\, \Big[(R^{Bog}_{j-2,j-2})^{\frac{1}{2}}\Big]_1 \Big\}
\end{equation}
where each $\Big[[\Gamma^{Bog\,}_{j-2,j-2}]^{(r_l)}_{\mathcal{\tau}_h}\Big]_1$ is a summand in (\ref{summands-rl-bis}) and $\prod_{l=1}^{\bar{l}=0}[\dots]\equiv \charf$.
Out of each monomial  in expression (\ref{sum-mon}) the commutators on the R-H-S of (\ref{def-D})  create new monomials the number of which can be estimated less than $\mathcal{O}(\#_{RW})$ where 
\begin{eqnarray}
\#_{RW}&:=&\text{the maximum number of modified "RW" blocks contained in each}\quad\quad \nonumber\\
& &\text{product}\,\,\prod_{l=1}^{\bar{l}}\Big\{\Big[\, (R^{Bog}_{j-2,j-2})^{\frac{1}{2}}\Big]_1\Big[[\Gamma^{Bog\,}_{j-2,j-2}]^{(r_l)}_{\mathcal{\tau}_h}\Big]_1\, \Big[(R^{Bog}_{j-2,j-2})^{\frac{1}{2}}\Big]_1 \Big\}\,.  \nonumber
\end{eqnarray}
\item
%{\color{red}For the estimate of the operator norm of the product of operators corresponding to each summand resulting from the operations in point 2. and 3., we set $h=\lfloor (\ln N)^{\frac{1}{2}} \rfloor$ (or  $h=\lfloor (\ln N)^{\frac{1}{2}} \rfloor+1$ in case $\lfloor (\ln N)^{\frac{1}{2}} \rfloor$ is odd)}. 
Starting from the assumption in (\ref{lower-bound-spec}), we derive the inequality
\begin{eqnarray}
& &\|\frac{1}{\Big[\sum_{\bold{j}\notin\{\pm\bold{j}_1, \dots, \pm\bold{j}_{m}\}}(k_{\bold{j}})^2a^*_{\bold{j}}a_{\bold{j}}+\hat{H}^{Bog}_{\bold{j}_1,\dots,\bold{j}_{m-1}}-z^{Bog}_{\bold{j}_1,\dots,\bold{j}_{m-1}}+(\frac{a^*_{\bold{0}}a_{\bold{0}}}{N}\phi_{\bold{j}_{m}}+(k_{\bold{j}_{m}}^2))(N-i)-z\Big]^{\frac{1}{2}}}\times\quad\quad\quad\quad\quad  \\
& &\quad \times\Big[(\frac{a^*_{\bold{0}}a_{\bold{0}}}{N}\phi_{\bold{j}_{m}}+(k_{\bold{j}_{m}}^2))(N-i)-z-\frac{m-1}{(\ln N)^{\frac{1}{8}}}\Big]^{\frac{1}{2}}\|\\
&\leq &1\,,
\end{eqnarray}
where the operator norm is referred to the restriction of the operator  to  $\check{\mathcal{F}}^N$. (Likewise, this restriction is assumed below  whenever we consider operators of the type $[\dots]_1$, otherwise the operator acts from $\mathcal{F}^N$ to $\mathcal{F}^N$.) Then, we observe that for $w=z^{Bog}_{\bold{j}_1,\dots,\bold{j}_{m-1}}+z$
\begin{eqnarray}
%& &\mathcal{W}_{\bold{j}_*\,;\,i,i-2}(z)\mathcal{W}^*_{\bold{j}_*\,;\,i-2,i}(z)\\
& &\| [(R^{Bog}_{\bold{j}_1,\dots,\bold{j}_m\,;\,i-2,i-2}(w))^{\frac{1}{2}}W^*_{\bold{j}_m\,;\,i-2,i}(R^{Bog}_{\bold{j}_1,\dots,\bold{j}_m\,;\,i,i}(w))^{\frac{1}{2}}]_{1}\| \label{esti-1}\\
&\leq &\|\phi_{\bold{j}_{m}}\,\frac{1}{\Big[(\frac{a^*_{\bold{0}}a_{\bold{0}}}{N}\phi_{\bold{j}_{m}}+(k_{\bold{j}_{m}}^2))(N-i)-z-\frac{m-1}{(\ln N)^{\frac{1}{8}}}\Big]^{\frac{1}{2}}}\times \label{S-in} \quad\quad\quad\quad \\
& &\quad\quad\times \frac{a_{\bold{0}}a_{\bold{0}}}{N}\,\frac{ N-i+2}{2\Big[(\frac{a^*_{\bold{0}}a_{\bold{0}}}{N}\phi_{\bold{j}_{m}}+(k_{\bold{j}_{m}}^2))(N-i+2)-z-\frac{m-1}{(\ln N)^{\frac{1}{8}}}\Big]^{\frac{1}{2}}}\|\quad\quad\quad\quad  \label{S-in2} \\
&\leq & \mathcal{E}(\|(R^{Bog}_{\bold{j}_1,\dots,\bold{j}_m\,;\,i-2,i-2}(w)^{\frac{1}{2}}W^*_{\bold{j}_m\,;\,i-2,i}(R^{Bog}_{\bold{j}_1,\dots,\bold{j}_m\,;\,i,i}(w))^{\frac{1}{2}}\|)\,\label{esti-1}
%& &\quad \times \frac{a_{\bold{0}}a_{\bold{0}}}{N}\frac{1}{\Big[\check{H}^{Bog}_{\bold{j}_1,\dots,\bold{j}_{m-1}}-z_{\bold{j}_1,\dots,\bold{j}_{m-1}}+(\frac{a^*_{\bold{j}_0}a_{\bold{j}_0}}{N}\phi_{\bold{j}_{m}}+(k_{\bold{j}_{m}}^2))(n_{\bold{j}_{m}}+n_{-\bold{j}_{m}})-z\Big]^{\frac{1}{2}}}
%&:=& \frac{(n_{\bold{j}_0}+2)(n_{\bold{j}_0}+1)}{N^2}\,\phi^2_{\bold{j}_{*}}\,\frac{ (n_{\bold{j}_{*}}+1)(n_{-\bold{j}_{*}}+1)}{\Big[(\frac{n_{\bold{j}_0}}{N}\phi_{\bold{j}_{*}}+(k_{\bold{j}_{*}}^2))(n_{\bold{j}_{*}}+n_{-\bold{j}_{*}})-z\Big]}\times\\
%& &\times \frac{1}{\Big[(\frac{(n_{\bold{j}_0}+2)}{N}\phi_{\bold{j}_{*}}+(k_{\bold{j}_{*}}^2))(n_{\bold{j}_{*}}+n_{-\bold{j}_{*}})-2(\frac{(n_{\bold{j}_0}+2)}{N}\phi_{\bold{j}_{*}}+(k_{\bold{j}_{*}}^2))-z\Big]}\quad\quad\quad
\end{eqnarray}
where for the step from (\ref{S-in}) to (\ref{esti-1}) we  bound (\ref{S-in})-(\ref{S-in2}) by $\sqrt{(\ref{bound-S-1})}$  and proceed as in Corollary \ref{main-lemma-H}; recall that, by definition, $\mathcal{E}(\|(R^{Bog}_{\bold{j}_1,\dots,\bold{j}_m\,;\,i-2,i-2}(w)^{\frac{1}{2}}W^*_{\bold{j}_m\,;\,i-2,i}(R^{Bog}_{\bold{j}_1,\dots,\bold{j}_m\,;\,i,i}(w))^{\frac{1}{2}}\|)$ coincides with the R-H-S of (\ref{main-estimate-intermediate}).
%where
%$$w'=w+\frac{2\phi_{\bold{j}_m}(N-i+2)}{N}=z^{Bog}_{\bold{j}_1,\dots,\bold{j}_m}+z+\frac{2\phi_{\bold{j}_m}(N-i+2)}{N}\leq z^{Bog}_{\bold{j}_1,\dots,\bold{j}_m}+E^{Bog}_{\bold{j}_m}+\sqrt{\epsilon_{\bold{j}_m}}\phi_{\bold{j}_m}\sqrt{\epsilon_{\bold{j}_m}^2+2\epsilon_{\bold{j}_m}}$$ for $N$ sufficiently large and $h=\mathcal{O}(\lfloor (\ln N)^{\frac{1}{2}} \rfloor)$, because $N-i+2\leq h+2$ and
% $z<z_{m}+\gamma \Delta_{m-1}<E^{Bog}_{\bold{j}_m}+\frac{1}{2} \sqrt{\epsilon_{\bold{j}_m}}\phi_{\bold{j}_m}\sqrt{\epsilon_{\bold{j}_m}^2+2\epsilon_{\bold{j}_m}}$ by assumption (see (\ref{bound-z-0})).
In the same way we bound $\| [(R^{Bog}_{\bold{j}_1,\dots,\bold{j}_m\,;\,i,i}(w))^{\frac{1}{2}}W_{\bold{j}_m\,;\,i,i-2}(R^{Bog}_{\bold{j}_1,\dots,\bold{j}_m\,;\,i-2,i-2}(w))^{\frac{1}{2}}]_{1}\|$. Analogously, we can state 
%\begin{eqnarray}
%& &\| [(R^{Bog}_{\bold{j}_1,\dots,\bold{j}_m\,;\,i-2,i-2}(w))^{\frac{1}{2}}W^*_{\bold{j}_m\,;\,i-2,i}(R^{Bog}_{\bold{j}_1,\dots,\bold{j}_m\,;\,i,i}(w))^{\frac{1}{2}}]_{2}\| \label{esti-2}\\
%&\leq &  \mathcal{E}(\|(R^{Bog}_{\bold{j}_1,\dots,\bold{j}_m\,;\,i-2,i-2}(w)^{\frac{1}{2}}W^*_{\bold{j}_m\,;\,i-2,i}(R^{Bog}_{\bold{j}_1,\dots,\bold{j}_m\,;\,i,i}(w))^{\frac{1}{2}}\|)\,.\nonumber
%\mathcal{E}((R^{Bog}_{\bold{j}_1,\dots,\bold{j}_m\,;\,i,i}(w))^{\frac{1}{2}}W_{\bold{j}_m\,;\,i,i-2}(R^{Bog}_{\bold{j}_1,\dots,\bold{j}_m\,;\,i-2,i-2}(w))^{\frac{1}{2}})
%& &\quad \times \frac{a_{\bold{0}}a_{\bold{0}}}{N}\frac{1}{\Big[\check{H}^{Bog}_{\bold{j}_1,\dots,\bold{j}_{m-1}}-z_{\bold{j}_1,\dots,\bold{j}_{m-1}}+(\frac{a^*_{\bold{j}_0}a_{\bold{j}_0}}{N}\phi_{\bold{j}_{m}}+(k_{\bold{j}_{m}}^2))(n_{\bold{j}_{m}}+n_{-\bold{j}_{m}})-z\Big]^{\frac{1}{2}}}
%&:=& \frac{(n_{\bold{j}_0}+2)(n_{\bold{j}_0}+1)}{N^2}\,\phi^2_{\bold{j}_{*}}\,\frac{ (n_{\bold{j}_{*}}+1)(n_{-\bold{j}_{*}}+1)}{\Big[(\frac{n_{\bold{j}_0}}{N}\phi_{\bold{j}_{*}}+(k_{\bold{j}_{*}}^2))(n_{\bold{j}_{*}}+n_{-\bold{j}_{*}})-z\Big]}\times\\
%& &\times \frac{1}{\Big[(\frac{(n_{\bold{j}_0}+2)}{N}\phi_{\bold{j}_{*}}+(k_{\bold{j}_{*}}^2))(n_{\bold{j}_{*}}+n_{-\bold{j}_{*}})-2(\frac{(n_{\bold{j}_0}+2)}{N}\phi_{\bold{j}_{*}}+(k_{\bold{j}_{*}}^2))-z\Big]}\quad\quad\quad
%\end{eqnarray} 
%and
\begin{eqnarray}
& &\| [(R^{Bog}_{\bold{j}_1,\dots,\bold{j}_m\,;\,i-2,i-2}(w))^{\frac{1}{2}}W^*_{\bold{j}_m\,;\,i-2,i}(R^{Bog}_{\bold{j}_1,\dots,\bold{j}_m\,;\,i,i}(w))^{\frac{1}{2}}]_{2}\|\times \\
& &\quad\quad\quad \times \| [(R^{Bog}_{\bold{j}_1,\dots,\bold{j}_m\,;\,i,i}(w))^{\frac{1}{2}}W_{\bold{j}_m\,;\,i,i-2}(R^{Bog}_{\bold{j}_1,\dots,\bold{j}_m\,;\,i-2,i-2}(w))^{\frac{1}{2}}]_{2}\| \nonumber \\
&\leq &  \mathcal{E}(\|(R^{Bog}_{\bold{j}_1,\dots,\bold{j}_m\,;\,i,i}(w)^{\frac{1}{2}}W_{\bold{j}_m\,;\,i,i-2}(R^{Bog}_{\bold{j}_1,\dots,\bold{j}_m\,;\,i-2,i-2}(w))^{\frac{1}{2}}\|^2)\,.\label{esti-2}
\end{eqnarray}
Regarding the commutators corresponding to the symbol $\underbrace{\quad}$ in the R-H-S of (\ref{def-D}), we estimate
\begin{eqnarray}
%& &\|[(R^{Bog}_{i,i})^{\frac{1}{2}}\frac{a^*_{\bold{0}}a^*_{\bold{0}}}{N}\frac{N-i}{2}(R^{Bog}_{i,i})^{\frac{1}{2}}\,\, \frac{a_{\bold{0}}a_{\bold{0}}}{N}\frac{N-j}{2}]\|\\
& &\|[(R^{Bog}_{\bold{j}_1,\dots,\bold{j}_m\,;\,i,i}(w))^{-\frac{1}{2}}]_1\,\Big[\, [R^{Bog}_{\bold{j}_1,\dots,\bold{j}_m\,;\,i,i}(w)]_1\frac{a^*_{\bold{0}}a^*_{\bold{0}}}{N}\frac{N-i+2}{2}\,,\,\frac{a_{\bold{0}}a_{\bold{0}}}{N}\frac{N-j+2}{2}\,\Big]\times \label{comm}\\
& &\quad\quad\quad\quad \quad \times [(R^{Bog}_{\bold{j}_1,\dots,\bold{j}_m\,;\,i-2,i-2}(w))^{\frac{1}{2}}]_1\|\quad\quad\quad \nonumber  \\
&\leq &\frac{N-i+2}{2}\frac{N-j+2}{2}\times\\
& &\quad\quad\quad \times \|[(R^{Bog}_{\bold{j}_1,\dots,\bold{j}_m\,;\,i,i}(w))^{-\frac{1}{2}}]_1\,\Big[\,[R^{Bog}_{\bold{j}_1,\dots,\bold{j}_m\,;\,i,i}(w)]_1\,,\,\frac{a_{\bold{0}}a_{\bold{0}}}{N}\,\Big]\frac{a^*_{\bold{0}}a^*_{\bold{0}}}{N}[(R^{Bog}_{\bold{j}_1,\dots,\bold{j}_m\,;\,i-2,i-2}(w))^{\frac{1}{2}}]_1\| \nonumber\\
& &+\frac{N-i+2}{2}\frac{N-j+2}{2}\|[(R^{Bog}_{\bold{j}_1,\dots,\bold{j}_m\,;\,i,i}(w))^{\frac{1}{2}}]_1\,\Big[\,\frac{a^*_{\bold{0}}a^*_{\bold{0}}}{N}\,,\,\frac{a_{\bold{0}}a_{\bold{0}}}{N}\,\Big]\, [(R^{Bog}_{\bold{j}_1,\dots,\bold{j}_m\,;\,i-2,i-2}(w))^{\frac{1}{2}}]_1\|\quad\quad\quad\quad\\
&\leq&\mathcal{O}\Big(\,\frac{N-j}{N}\|[(R^{Bog}_{\bold{j}_1,\dots,\bold{j}_m\,;\,i,i}(w))^{\frac{1}{2}}]_1\,\frac{a^*_{\bold{0}}a^*_{\bold{0}}}{N}\frac{N-i+2}{2}\,[(R^{Bog}_{\bold{j}_1,\dots,\bold{j}_m\,;\,i-2,i-2}(w))^{\frac{1}{2}}]_1\|\,\Big)\\
& &+\mathcal{O}\Big(\,\frac{N-j}{N}\|[(R^{Bog}_{\bold{j}_1,\dots,\bold{j}_m\,;\,i,i}(w))^{\frac{1}{2}}]_1\frac{a^*_{\bold{0}}a_{\bold{0}}}{N}\frac{N-i+2}{2}[(R^{Bog}_{\bold{j}_1,\dots,\bold{j}_m\,;\,i-2,i-2}(w))^{\frac{1}{2}}]_1\|\,\Big)\\
&\leq &\mathcal{O}(\frac{N-j}{N}) \mathcal{E}(\|(R^{Bog}_{\bold{j}_1,\dots,\bold{j}_m\,;\,i,i}(w)^{\frac{1}{2}}W_{\bold{j}_m\,;\,i,i-2}(R^{Bog}_{\bold{j}_1,\dots,\bold{j}_m\,;\,i-2,i-2}(w))^{\frac{1}{2}}\|)\,.
\end{eqnarray}
Here,  we have used
\begin{equation}
\|[(R^{Bog}_{\bold{j}_1,\dots,\bold{j}_m\,;\,i,i}(w))^{-\frac{1}{2}}]_1\Big[\,[R^{Bog}_{\bold{j}_1,\dots,\bold{j}_m\,;\,i,i}(w)]_1\,,\,\frac{a_{\bold{0}}a_{\bold{0}}}{N}\,\Big][(R^{Bog}_{\bold{j}_1,\dots,\bold{j}_m\,;\,i,i}(w))^{-\frac{1}{2}}]_1\|\leq \mathcal{O}(\frac{1}{N})
\end{equation}
that follows from a standard computation and the bound $$\hat{H}^{Bog}_{\bold{j}_1,\dots,\bold{j}_{m-1}}\geq \Delta_0\sum_{\bold{j}\in \{\pm \bold{j}_1,\dots,\pm \bold{j}_{m-1}\}}a^*_{\bold{j}}a_{\bold{j}} -\sum_{l=1}^{m-1}\phi_{\bold{j}_l} $$ (see (\ref{id-1}), (\ref{energy-est-in}), and (\ref{energy-est-in-0}) and extend the same inequalities to the case of  $m-1$ couples of modes).
With the same argument, we get
\begin{eqnarray}
& &\|[(R^{Bog}_{\bold{j}_1,\dots,\bold{j}_m\,;\,i,i}(w))^{-\frac{1}{2}}]_1\,\Big[\frac{a^*_{\bold{0}}a_{\bold{0}}-1}{N}\frac{N-j+2}{2}\,,\,[R^{Bog}_{\bold{j}_1,\dots,\bold{j}_m\,;\,i,i}(w)]_1\frac{a_{\bold{0}}a_{\bold{0}}}{N}\frac{N-i}{2}\Big]\,\times \quad\quad\quad\quad\quad\quad \label{comm-2}\\
& &\quad\quad\quad\quad [(R^{Bog}_{\bold{j}_1,\dots,\bold{j}_m\,;\,i-2,i-2}(w))^{\frac{1}{2}}]_1\|\quad\quad\quad \nonumber\\
%&\leq &\|[(R^{Bog}_{i,i})^{\frac{1}{2}}]_1\,\Big[\frac{a^*_{\bold{0}}a_{\bold{0}}-1}{N}\frac{N-j}{2}\,,\,\frac{(a_{\bold{0}}a_{\bold{0}})^{\#}}{N}\frac{N-i}{2}\Big]\,[(R^{Bog}_{i,i})^{\frac{1}{2}}]_1\|\\
%& &+\|[(R^{Bog}_{i,i})^{-\frac{1}{2}}]_1\,\Big[\frac{a^*_{\bold{0}}a_{\bold{0}}-1}{N}\frac{N-j}{2}\,,\,[R^{Bog}_{i,i}]_1\Big]\,\frac{(a_{\bold{0}}a_{\bold{0}})^{\#}}{N}\frac{N-i}{2}[(R^{Bog}_{i,i})^{\frac{1}{2}}]_1\|\quad\quad\quad\quad \\
%&\leq &\|(R^{Bog}_{i,i})^{-\frac{1}{2}}[R^{Bog}_{i,i}\,,\,\frac{a_{\bold{0}}a_{\bold{0}}}{N}]\frac{a^*_{\bold{0}}a^*_{\bold{0}}}{N}\frac{N-i}{2}(R^{Bog}_{i,i})^{\frac{1}{2}}\|\\
%& &+\|(R^{Bog}_{i,i})^{-\frac{1}{2}}R^{Bog}_{i,i}[ \frac{a^*_{\bold{0}}a^*_{\bold{0}}}{N}\frac{N-i}{2}\,,\,\frac{a_{\bold{0}}a_{\bold{0}}}{N}](R^{Bog}_{i,i})^{\frac{1}{2}}\|\quad\quad\\
&\leq&\mathcal{O}\Big(\,\frac{N-j+2}{N}\|[(R^{Bog}_{\bold{j}_1,\dots,\bold{j}_m\,;\,i,i})^{\frac{1}{2}}(w)]_1\,\frac{a_{\bold{0}}a_{\bold{0}}}{N}\frac{N-i+2}{2}\,[(R^{Bog}_{\bold{j}_1,\dots,\bold{j}_m\,;\,i-2,i-2}(w))^{\frac{1}{2}}]_1\|\,\Big)\,.
\end{eqnarray}
We get analogous estimates for the commutators with an operator of the type $[R^{Bog}_{i,i}W^*]_1$ in the first case (see (\ref{comm})), and of type $[R^{Bog}_{i,i}W]_1$ in the second case (see (\ref{comm-2})).
%where $(a_{\bold{0}}a_{\bold{0}})^{\#}$ is either $a_{\bold{0}}a_{\bold{0}}$ or $a^{*}_{\bold{0}}a^*_{\bold{0}}$.

\item 
We follow the fate  through the operations described in points 2. and 3. of each monomial $$\prod_{l=1}^{\bar{l}}\Big\{\Big[(R^{Bog}_{N-2,N-2})^{\frac{1}{2}}[\Gamma^{Bog\,}_{N-2,N-2}]^{(r_l)}_{\mathcal{\tau}_h}(R^{Bog}_{N-2,N-2})^{\frac{1}{2}}(R^{Bog}_{N-2,N-2})^{\frac{1}{2}}\Big]_{2\,;\,j+2}\Big\}$$
%$$\Big[\, (R^{Bog}_{N-2,N-2})^{\frac{1}{2}}[\Gamma^{Bog\,}_{N-2,N-2}]^{(r_l)}_{\mathcal{\tau}_h\,;\,j+2} (R^{Bog}_{N-2,N-2})^{\frac{1}{2}}\Big]_1$$ 
that appears in (\ref{sum-mon-0}) (see point 1.), and estimate the ratio between the total number of resulting monomials and the number (of monomials) that is obtained if: 

\noindent
a) we just expand one single factor $$\Big[(R^{Bog}_{j,j})^{\frac{1}{2}}[\Gamma^{Bog\,}_{j,j}]_{\mathcal{\tau}_h}\,(R^{Bog}_{j,j})^{\frac{1}{2}}\Big]_1$$
(contained in $\Big[ (R^{Bog}_{N-2,N-2})^{\frac{1}{2}}[\Gamma^{Bog\,}_{N-2,N-2}]^{(r_l)}_{\mathcal{\tau}_h} (R^{Bog}_{N-2,N-2})^{\frac{1}{2}}\Big]_{2\,;\,j+2}$) according to the identity
\begin{eqnarray}
& &\Big[(R^{Bog}_{j,j})^{\frac{1}{2}}[\Gamma^{Bog\,}_{j,j}]_{\mathcal{\tau}_h}\,(R^{Bog}_{j,j})^{\frac{1}{2}}\Big]_1\label{5.233}\\
&=&\Big[(R^{Bog}_{j,j})^{\frac{1}{2}}W (R^{Bog}_{j-2,j-2})^{\frac{1}{2}}\Big]_1\,\sum_{l_{j-2}=0}^{h-1}\Big\{\Big[(R^{Bog}_{j-2,j-2})^{\frac{1}{2}}[\Gamma^{Bog\,}_{j-2,j-2}]_{\mathcal{\tau}_h}\,(R^{Bog}_{j-2,j-2})^{\frac{1}{2}}\Big]_1\Big\}^{l_{j-2}}  \Big[(R^{Bog}_{j-2,j-2})^{\frac{1}{2}}W^*(R^{Bog}_{j,j})^{\frac{1}{2}}\Big]_1\,;\nonumber
\end{eqnarray}
b) we replace $\sum_{l_{j-2}=0}^{h-1}\Big\{\Big[(R^{Bog}_{j-2,j-2})^{\frac{1}{2}}[\Gamma^{Bog\,}_{j-2,j-2}]_{\mathcal{\tau}_h}\,(R^{Bog}_{j-2,j-2})^{\frac{1}{2}}\Big]_1\Big\}^{l_{j-2}}$ contained in (\ref{5.233}) with (\ref{sum-mon}).

\noindent
This ratio is bounded by $\mathcal{O}(\#_{RW})\times  \mathcal{O}(\#_{R \Gamma R})$ because:
\begin{itemize}
\item  In point 2.,  out of $\prod_{l=1}^{\bar{l}}\Big\{\Big[(R^{Bog}_{N-2,N-2})^{\frac{1}{2}}[\Gamma^{Bog\,}_{N-2,N-2}]^{(r_l)}_{\mathcal{\tau}_h}(R^{Bog}_{N-2,N-2})^{\frac{1}{2}}\Big]_{2\,;\,j+2}\Big\}$ we produce at most $ \mathcal{O}(\#_{R \Gamma R})$ new monomials of the same type except that some factors  $\Big[(R^{Bog}_{j,j})^{\frac{1}{2}}[\Gamma^{Bog\,}_{j,j}]_{\mathcal{\tau}_h}\,(R^{Bog}_{j,j})^{\frac{1}{2}}\Big]_1$ have been replaced with  $\mathcal{L}_j(\Big[(R^{Bog}_{j,j})^{\frac{1}{2}}[\Gamma^{Bog\,}_{j,j}]_{\mathcal{\tau}_h}\,(R^{Bog}_{j,j})^{\frac{1}{2}}\Big]_1)$ and one single factor\\ $\Big[(R^{Bog}_{j,j})^{\frac{1}{2}}[\Gamma^{Bog\,}_{j,j}]_{\mathcal{\tau}_h}\,(R^{Bog}_{j,j})^{\frac{1}{2}}\Big]_1$ with $\mathcal{D}_j( \Big[(R^{Bog}_{j,j})^{\frac{1}{2}}[\Gamma^{Bog\,}_{j,j}]_{\mathcal{\tau}_h}\,(R^{Bog}_{j,j})^{\frac{1}{2}}\Big]_1)$. 
\item In point 3., for each monomial (see (\ref{sum-mon})) of the type 
%\Big[\, (R^{Bog}_{j-2,j-2})^{\frac{1}{2}}[\Gamma^{Bog\,}_{j-2,j-2}]^{(r_l)}_{\mathcal{\tau}_h} (R^{Bog}_{j-2,j-2})^{\frac{1}{2}}\Big]_1
$$\prod_{l=1}^{\bar{l}}\Big\{\Big[\, (R^{Bog}_{j-2,j-2})^{\frac{1}{2}}\Big]_1\Big[[\Gamma^{Bog\,}_{j-2,j-2}]^{(r_l)}_{\mathcal{\tau}_h}\Big]_1\, \Big[(R^{Bog}_{j-2,j-2})^{\frac{1}{2}}\Big]_1 \Big\}$$
 the commutators create at most  $\mathcal{O}(\#_{RW})$ more terms of the same type except  for the  factor coming from a commutator; see (\ref{comm})) and (\ref{comm-2})) in point 4..
 \end{itemize}

We intend to make use of the same argument exploited at {\bf{STEP ${\bf{III_7}}$}} of Lemma \ref{main-relations}. When implementing this argument, due to the estimates in (\ref{esti-1})-(\ref{esti-2}) combined with the constraint on $z$ assumed in (\ref{bound-z-0}),  we can ignore that in
$$\Big\{\Big[(R^{Bog}_{N-2,N-2})^{\frac{1}{2}}[\Gamma^{Bog\,}_{N-2,N-2}]_{\mathcal{\tau}_h}\,(R^{Bog}_{N-2,N-2})^{\frac{1}{2}}\Big]_{2\,;\,j+2}\Big\}^{l_{N-2}}$$
the original companion operators have been replaced with $[...]_2$ from level $N-2$ down to level $j+2$ and with  $[...]_1$ from level $j-4$ down to $N-2-h$. Likewise, as far as the operator norm is concerned, we can ignore the replacement of factors $\Big[(R^{Bog}_{j,j})^{\frac{1}{2}}[\Gamma^{Bog\,}_{j,j}]_{\mathcal{\tau}_h}\,(R^{Bog}_{j,j})^{\frac{1}{2}}\Big]_1$ with  $\mathcal{L}_j(\Big[(R^{Bog}_{j,j})^{\frac{1}{2}}[\Gamma^{Bog\,}_{j,j}]_{\mathcal{\tau}_h}\,(R^{Bog}_{j,j})^{\frac{1}{2}}\Big]_1)$ which is due to the action of $\mathcal{D}_j$. Furthermore, due to the result in point 4. the estimate of the norm of each resulting monomial carries an extra factor bounded by $\mathcal{O}(\frac{N-j}{N})$. Using the same rationale of  {\bf{STEP ${\bf{III_7}}$}} of Lemma \ref{main-relations}, after the previous analysis concerning the structure of the monomials we can conclude that the norm of the  sum of monomials resulting from the operations in point 2. and 3. applied to
$$\prod_{l=1}^{\bar{l}}\Big\{\Big[(R^{Bog}_{\bold{j}_1,\dots,\bold{j}_m;\,N-2,N-2}(w))^{\frac{1}{2}}[\Gamma^{Bog\,}_{\bold{j}_1,\dots,\bold{j}_m;\,N-2,N-2}(w)]^{(r_l)}_{\mathcal{\tau}_h}(R^{Bog}_{\bold{j}_1,\dots,\bold{j}_m;\,N-2,N-2}(w))^{\frac{1}{2}}\Big]_{2\,;\,j+2} \Big\}$$
%$$\Big[\, (R^{Bog}_{\bold{j}_1,\dots,\bold{j}_m;\,N-2,N-2}(w))^{\frac{1}{2}}[\Gamma^{Bog\,}_{\bold{j}_1,\dots,\bold{j}_m;\,N-2,N-2}(w)]^{(r_l)}_{\mathcal{\tau}_h\,;\,j+2} (R^{Bog}_{\bold{j}_1,\dots,\bold{j}_m;\,N-2,N-2}(w))^{\frac{1}{2}}\Big]_1$$ 
can be estimated less than
\begin{eqnarray}
& &\mathcal{O}(\frac{N-j}{N})\times  \mathcal{O}(\#_{RW})\times  \mathcal{O}(\#_{R \Gamma R})\times \\
& &\quad \times \mathcal{E}\Big(\Big\|\,\prod_{l=1}^{\bar{l}}\Big\{\Big[(R^{Bog}_{\bold{j}_1,\dots,\bold{j}_m;\,N-2,N-2}(w))^{\frac{1}{2}}[\Gamma^{Bog\,}_{\bold{j}_1,\dots,\bold{j}_m;\,N-2,N-2}(w)]^{(r_l)}_{\mathcal{\tau}_h}(R^{Bog}_{\bold{j}_1,\dots,\bold{j}_m;\,N-2,N-2}(w))^{\frac{1}{2}}\Big]_{2\,;\,j+2} \Big\|\Big)\,.\nonumber
\end{eqnarray}

\noindent
By  invoking Remark \ref{estimation-proc}, and making use of  (\ref{estimate-E-bis})-(\ref{Estimate-E-bis-bis}), we get
\begin{eqnarray}
& &\sum_{\bar{l}=0}^{h-1} \sum_{r_1=1}^{\bar{r}}\dots  \\
& &\quad \dots \sum_{r_{\,\bar{l}\,}=1}^{\bar{r}}\,\mathcal{E}\Big(\Big\|\prod_{l=1}^{\bar{l}}\Big\{\Big[(R^{Bog}_{\bold{j}_1,\dots,\bold{j}_m;\,N-2,N-2}(w))^{\frac{1}{2}}[\Gamma^{Bog\,}_{\bold{j}_1,\dots,\bold{j}_m;\,N-2,N-2}(w)]^{(r_l)}_{\mathcal{\tau}_h}(R^{Bog}_{\bold{j}_1,\dots,\bold{j}_m;\,N-2,N-2}(w))^{\frac{1}{2}}\Big]_{2\,;\,j+2} \Big\}\Big\|\Big)\nonumber \\
&\leq  &\mathcal{E}\Big(\Big\|\sum_{l_{N-2}=0}^{h-1}\{(R^{Bog}_{\bold{j}_1,\dots,\bold{j}_m;\,N-2,N-2}(w))^{\frac{1}{2}}\Gamma^{Bog\,}_{\bold{j}_1,\dots,\bold{j}_m;\,N-2,N-2}(w)(R^{Bog}_{\bold{j}_1,\dots,\bold{j}_m;\,N-2,N-2}(w))^{\frac{1}{2}}\}^{l_{N-2}}\Big\|\Big)\,\nonumber\\
&\leq & \mathcal{O}(1) \nonumber
\end{eqnarray}
Finally, we conclude that  the norm of the $j-th$ summand in (\ref{rem-in}) is bounded by
\begin{equation}
\mathcal{O}(\frac{N-j}{N})\times  \mathcal{O}(\#_{RW})\times  \mathcal{O}(\#_{R \Gamma R})\,
\end{equation}
where $N-j\leq \mathcal{O}(h)$.  Likewise, we can bound (\ref{rem-fin-1}) and (\ref{rem-fin}).
\item
The numbers $\#_{RW}$  and $\#_{R\Gamma R}$ are less than $\mathcal{O}(h\,(2h)^{\frac{h+2}{2}})$ (see {\bf{STEP ${\bf{III_1}}$}} of Lemma \ref{main-relations}). 
%The index $j$ in (\ref{rem-in}) ranges from $N-2$ to $N-h-2$.  
%the total contribution of the error terms produced in {\bf{ STEP b)}} is bounded by ....
%\item
% Finally, using the argument of Remark ... thanks to the estimates in (\ref{subset-0})-(\ref{subset-3})
%(\ref{estimation-2})-(\ref{estimation-2-bis}),
%and in (\ref{error-final-bis}), the total contribution of the error terms produced in {\bf{ STEP b)}} is bounded by.
\end{enumerate}

\noindent
{\bf{ STEP f)}}

\noindent
Due to the inequalities in (\ref{spectral}), (\ref{Delta-W}), and the estimates in {\bf{ STEP c)}}, we can argue as in {\bf{ STEP e)}} and bound the norm of the sum of the remainder terms produced in {\bf{ STEP c)}}} by
\begin{equation}
\mathcal{O}(\frac{h}{\sqrt{N}})\times  \mathcal{O}(h\,(2h)^{\frac{h+2}{2}})\,.
\end{equation}

%, and on the set $\{\phi_{\bold{j}_i}, (k_{\bold{j}_i})^2;\,\, 1\leq i\leq m\}$, and the total number of remainders associated with (\ref{second-term-fp-eq-bis})  is $\mathcal{O}(\bar{j}\,h^{h}\cdot \bar{j}\,h^{h})$ at most.}
%and on the set $\{\phi_{\bold{j}_i}, (k_{\bold{j}_i})^2;\,\, 1\leq i\leq m\}$.
%Finally, ,  the assumption in (\ref{assumption-gap}) and an argument (partially) similar to \emph{\underline{Corollary 4.13} of \cite{Pi1}} imply that the leading term of
%\begin{eqnarray}
%& &(\ref{GammaPerp})\\
%&&\|\overline{\mathscr{P}_{\psi^{Bog}_{\bold{j}_1,\dots,\bold{j}_{m-1}}}}W_{\bold{j}_m}\,R^{Bog}_{\bold{j}_1,\dots,\bold{j}_m\,;\,N-2,N-2}(w)\times \\
%& &\quad\quad \times \sum_{l_{N-2}=0}^{h-1}\Big[[\Gamma^{Bog}_{\bold{j}_1,\dots,\bold{j}_m\,;\,N-2,N-2}(w)]_{\tau_h}\,R^{Bog}_{\bold{j}_1,\dots,\bold{j}_m\,;\,\,N-2,N-2}(w)\Big]^{l_{N-2}}W^*_{\bold{j}%_m}\,\overline{\mathscr{P}_{\psi^{Bog}_{\bold{j}_1,\dots,\bold{j}_{m-1}}}}\|\nonumber \\
%\end{eqnarray}
%that has been isolated is bounded by
%\begin{equation}\label{invertibility-G}
%\frac{\phi_{\bold{j}_{m}}}{2\epsilon_{\bold{j}_m}+2-\frac{z-\Delta_{m-1}(1-\frac{\phi_{\bold{j}_m}\lfloor (\ln N)^{\frac{1}{2}} \rfloor}{N\Delta_{m-1}})}{\phi_{\bold{j}_{m}}}}[\check{\mathcal{G}}_{\bold{j}_{m}%\,;\,N-2,N-2}(z-\Delta_{m-1}(1-\frac{\phi_{\bold{j}_m}\lfloor (\ln N)^{\frac{1}{2}} \rfloor}{N\Delta_{m-1}}))]_{\tau_{h\equiv \lfloor (\ln N)^{\frac{1}{2}} \rfloor-2}\,;\,0}\,
%\end{equation}

As a final result of {\bf{STEPS} a), b), c), d), e), f)},  we have derived 
\begin{eqnarray}
& &\|\overline{\mathscr{P}_{\psi^{Bog}_{\bold{j}_1,\dots,\bold{j}_{m-1}}}}\Gamma^{Bog}_{\bold{j}_1,\dots,\bold{j}_m;\,N,N}(z+z^{Bog}_{\bold{j}_1,\dots,\bold{j}_{m-1}})\,\overline{\mathscr{P}_{\psi^{Bog}_{\bold{j}_1,\dots,\bold{j}_{m-1}}}}\|\\
&\leq &(\ref{bound-lemma4.5})+\mathcal{O}((\frac{4}{5})^{h})+\mathcal{O}((\frac{1}{1+c\sqrt{\epsilon_{\bold{j}_m}}})^h)+\mathcal{O}(\frac{h^4(2h)^{h+2}}{\sqrt{N}})\\
&\leq &(\ref{bound-lemma4.5})+\mathcal{O}(\frac{1}{\sqrt{(\ln N)}})\,
\end{eqnarray}
where in the last step we have used that $h=\mathcal{O}((\ln N)^{\frac{1}{2}} $).
\qed

\begin{remark}\label{invert}
For the estimate of the norm of (\ref{invert-G}) (see Property 4 in Theorem \ref{induction-many-modes}), we can adapt the steps of Lemma \ref{invertibility} because the operator $\hat{Q}_{\bold{j}_m}^{(>N-1)}$ projects onto a subspace of states with fixed number, say $j$, of particles in the modes different from $\pm\bold{j}_1\,,\dots \,,\pm\bold{j}_m$ and $\bold{0}$, and without particles in the modes $\pm \bold{j}_{m}$. Hence, we can implement {\bf{Steps a)}} and {\bf{b)}} but the $\bar{j}$ in formula (\ref{express}) in general does not not coincide with $h$ because, depending on the number of particles, $j$,  in the modes different from $\pm\bold{j}_1\,,\dots \,,\pm\bold{j}_m$ and $\bold{0}$, the number $h$ to be chosen  may be strictly  smaller than $\lfloor (\ln N)^{\frac{1}{2}} \rfloor$. Analogously, the sums over $l_{N-2}$ are up to $\bar{j}-1$. $\bar{j}$ is set equal to $\lfloor (\ln N)^{\frac{1}{2}} \rfloor$.  We skip {\bf{Step c)}} because we do not split $\hat{Q}_{\bold{j}_m}^{(>N-1)}$ into the sum of two projections. For the estimate of the leading terms resulting from {\bf{Steps a)}} and {\bf{b)}} we proceed like in {\bf{Step d)}} but replacing $\Delta_{m-1}$ with $-\frac{m-1}{(\ln N)^{\frac{1}{8}}}$ due to the inequality in (\ref{lower}) that is assumed to hold at the inductive step $m-1$. Furthermore, the projection $\overline{\mathscr{P}_{\psi^{Bog}_{\bold{j}_1,\dots,\bold{j}_{m-1}}}}$ is replaced with $\hat{Q}_{\bold{j}_m}^{(>N-1)}$ and the sum over $l_i$ in formula (\ref{def-Kcheck}) is up to $\bar{j}-1$.  {\bf{Step e)}} is essentially the same. {\bf{Step f)}} is absent because {\bf{Step c)}} is absent.
\end{remark}

\begin{lemma}\label{truncation}
On the basis of Definition \ref{def-sums}, for $N-4\geq h\geq 2$ and even, the identity below holds true
\begin{equation}\label{id-gammatau}
\sum_{l=N-2-h,\,l\, even}^{N-4}[\Gamma^{Bog\,}_{\bold{j}_1,\dots,\bold{j}_m\,;\,N-2,N-2}(w)]_{(l,h_-;l+2,h_-;\dots ; N-4,h_-)} = [\Gamma^{Bog\,}_{\bold{j}_1,\dots,\bold{j}_m\,;\,N-2,N-2}(w)]_{\mathcal{\tau}_h}
\end{equation} 
where the R-H-S is defined by 
\begin{eqnarray}
& &[\Gamma^{Bog\,}_{\bold{j}_1,\dots,\bold{j}_m\,;\,i,i}(w)]_{\mathcal{\tau}_h} \label{gammatau}\\
&:=&W_{\bold{j}_m} \sum_{l_{i-2}=0}^{h-1}R^{Bog}_{\bold{j}_1,\dots,\bold{j}_m,;\,i-2,i-2}(w)\,\Big\{[\Gamma^{Bog\,}_{\bold{j}_1,\dots,\bold{j}_m\,;\,i-2,i-2}(w)]_{\mathcal{\tau}_h}\,R^{Bog}_{\bold{j}_1,\dots,\bold{j}_m\,;\,i-2,i-2}(w)\Big\}^{l_{i-2}} W^*_{\bold{j}_m}\nonumber
\end{eqnarray}
for $N-2\geq i \geq N-h$ with
\begin{equation}
[\Gamma^{Bog\,}_{\bold{j}_1,\dots,\bold{j}_m\,;\,N-2-h,N-2-h}(w)]_{\mathcal{\tau}_h} \label{def-gammatau-initial}
:=  W_{\bold{j}_m}\,R^{Bog}_{\bold{j}_1,\dots,\bold{j}_m\,;\, N-4-h, N-4-h}(w)W_{\bold{j}_m}^*\,.\quad\quad
\end{equation}

\end{lemma}

\noindent
\emph{Proof}

If $h=2$ the identity in (\ref{id-gammatau})  follows directly from the definitions in (\ref{initial-gamma-1}), (\ref{def-0}), (\ref{gammatau}),  and  (\ref{def-gammatau-initial}). Therefore, from now on we assume $h\geq 4$. 

\noindent
For $r=0$ we define (see (\ref{def-0}))
\begin{equation}\label{def-gammatau-0}
\{[\Gamma^{Bog\,}_{\bold{j}_1,\dots,\bold{j}_m\,;\,N-2-h,N-2-h}(w)]_{\mathcal{\tau}_h}\}_0:=[\Gamma^{Bog\,}_{\bold{j}_1,\dots,\bold{j}_m\,,;\,N-2-h,N-2-h}(w)]^{(0)}_{( N-4-h,h_-)}\,,
\end{equation}
%\begin{eqnarray}
%& &[\Gamma^{Bog\,}_{\bold{j}_1,\dots,\bold{j}_m\,;\,N-h,N-h}(w)]_{\mathcal{\tau}_h, 2}\\
%&:= &[\Gamma^{Bog\,}_{\bold{j}_1,\dots,\bold{j}_m\,,;\,N-h,N-h}(w)]_{( N-4-h,h_-;N-h-2, h_-)}\\
%&&+[\Gamma^{Bog\,}_{\bold{j}_1,\dots,\bold{j}_m\,,;\,N-h,N-h}(w)]_{(N-h-2, h_-)}\\
%&=&[\Gamma^{Bog\,}_{\bold{j}_1,\dots,\bold{j}_m\,,;\,N-h,N-h}(w)]_{( N-4-h,h_-;N-h-2, h_-)}\\
%&&+[\Gamma^{Bog\,}_{\bold{j}_1,\dots,\bold{j}_m\,;\,N-h,N-h}(w)]_{\mathcal{\tau}_h, 0}
%\end{eqnarray}
and, for  $2\leq r \leq h$ with $r$ even, 
\begin{eqnarray}
& &\{[\Gamma^{Bog\,}_{\bold{j}_1,\dots,\bold{j}_m\,;\,N-h-2+r,N-h-2+r}(w)]_{\mathcal{\tau}_h}\}_r\label{general-term-taur}\\
&:= &[\Gamma^{Bog\,}_{\bold{j}_1,\dots,\bold{j}_m\,,;\,N-h-2+r,N-h-2+r}(w)]_{( N-2-h,\,h_-;\dots;N-h-6+r, h_-;N-h-4+r, h_-)}\\
& &+[\Gamma^{Bog\,}_{\bold{j}_1,\dots,\bold{j}_m\,,;\,N-h-2+r,N-h-2+r}(w)]_{( N-2-h+2,\,h_-;\dots;N-h-6+r, h_-;N-h-4+r, h_-)}\\
& &+\dots\\
&&+[\Gamma^{Bog\,}_{\bold{j}_1,\dots,\bold{j}_m\,,;\,N-h-2+r,\,N-h-2+r}(w)]_{(N-h-4+r, h_-)}\,.
%&=&[\Gamma^{B og\,}_{\bold{j}_1,\dots,\bold{j}_m\,,;\,N-h,N-h}(w)]_{( N-4-h,h_-;N-h-2, h_-)}\\
%&&+[\Gamma^{Bog\,}_{\bold{j}_1,\dots,\bold{j}_m\,;\,N-h,N-h}(w)]_{\mathcal{\tau}_h, 2}
\end{eqnarray}
We shall show by induction that for $N-2\geq i\geq N-h-2$
\begin{equation}\label{identity-ind}
[\Gamma^{Bog\,}_{\bold{j}_1,\dots,\bold{j}_m\,;\,i,i}(w)]_{\mathcal{\tau}_h}=\{[\Gamma^{Bog\,}_{\bold{j}_1,\dots,\bold{j}_m\,;\,N-h-2+r_i,N-h-2+r_i}(w)]_{\mathcal{\tau}_h}\}_{r_i}\Big|_{r_i\equiv i-N+h+2}\,.
\end{equation}
This readily implies
\begin{eqnarray}
& &[\Gamma^{Bog\,}_{\bold{j}_1,\dots,\bold{j}_m\,;\,N-2,N-2}(w)]_{\mathcal{\tau}_h}\\
&=&\{[\Gamma^{Bog\,}_{\bold{j}_1,\dots,\bold{j}_m\,;\,N-h-2+r,N-h-2+r}(w)]_{\mathcal{\tau}_h}\}_r\Big|_{r\equiv h}\\
&=&\sum_{l=N-2-h,\,l\, even}^{N-4}[\Gamma^{Bog\,}_{\bold{j}_1,\dots,\bold{j}_m\,,;\,N-2,N-2}(w)]_{(l,h_-;l+2,h_-;\dots ; N-4,h_-)}\,.
\end{eqnarray}

For $i=N-h-2$ the statement in (\ref{identity-ind}) is true because the R-H-S of (\ref{def-gammatau-initial}) is equal to the R-H-S of (\ref{def-gammatau-0}).
Next, we assume that it holds for $i-2\geq N-h-2$ and show that it is also true for $i$. To this end, for $N-2\geq i \geq N-h$ and $r_i:=i-N+h+2$ ($\Rightarrow\,\,h\geq r_i \geq 2$), we make use of the definition in (\ref{gammatau}) and assume (\ref{identity-ind}) for $i-2$. We get:
%that  $\{[\Gamma^{Bog\,}_{\bold{j}_1,\dots,\bold{j}_m\,;\,i-2,i-2}(w)]_{\mathcal{\tau}_h}\}_{r_i-2}$ corresponds to (\ref{general-term-taur})  with $r\equiv r_i-2$. We get:
\begin{eqnarray}
& &[\Gamma^{Bog\,}_{\bold{j}_1,\dots,\bold{j}_m\,;\,i,i}(w)]_{\mathcal{\tau}_h}\\
&=&W_{\bold{j}_m} \sum_{l_{i-2}=0}^{h-1}R^{Bog}_{\bold{j}_1,\dots,\bold{j}_m,;\,i-2,i-2}(w)\,\Big\{\{[\Gamma^{Bog\,}_{\bold{j}_1,\dots,\bold{j}_m\,;\,i-2,i-2}(w)]_{\mathcal{\tau}_h}\}_{r_{i-2}}\,R^{Bog}_{\bold{j}_1,\dots,\bold{j}_m\,;\,i-2,i-2}(w)\Big\}^{l_{i-2}} W^*_{\bold{j}_m}\nonumber\\
& =&W_{\bold{j}_m} R^{Bog}_{\bold{j}_1,\dots,\bold{j}_m,;\,i-2,i-2}(w)\, W^*_{\bold{j}_m}\\
& &+W_{\bold{j}_m} \sum_{l_{i-2}=1}^{h-1}R^{Bog}_{\bold{j}_1,\dots,\bold{j}_m,;\,i-2,i-2}(w)\,\Big\{\{[\Gamma^{Bog\,}_{\bold{j}_1,\dots,\bold{j}_m\,;\,i-2,i-2}(w)]_{\mathcal{\tau}_h}\}_{r_{i-2}}\,R^{Bog}_{\bold{j}_1,\dots,\bold{j}_m\,;\,i-2,i-2}(w)\Big\}^{l_{i-2}} W^*_{\bold{j}_m}\,\quad\quad\quad \label{sum-gamma-r}\\
&=&\{[\Gamma^{Bog\,}_{\bold{j}_1,\dots,\bold{j}_m\,;\,i,i}(w)]_{\mathcal{\tau}_h}\}_{r_i}\label{gammari}
\end{eqnarray}
where  $\{[\Gamma^{Bog\,}_{\bold{j}_1,\dots,\bold{j}_m\,;\,i,i}(w)]_{\mathcal{\tau}_h}\}_{r_i}$ is given in (\ref{general-term-taur}) with  $r\equiv r_i$. 
The latter identity is evident if we take into account that:
\begin{itemize}
\item by definition (see (\ref{initial-gamma-1}), (\ref{initial-gamma-2}), and (\ref{initial-gamma-3}))
\begin{eqnarray}
& &[\Gamma^{Bog\,}_{\bold{j}_1,\dots,\bold{j}_m\,;\,N-h-2+r_i,N-h-2+r_i}(w)]_{(N-h-4+r_i, h_-)}\\
%& =&W_{\bold{j}_m} \sum_{l_{i-2}=0}^{h-1}R^{Bog}_{\bold{j}_1,\dots,\bold{j}_m,;\,i-2,i-2}(w)\,\Big\{[\Gamma^{Bog\,}_{\bold{j}_1,\dots,\bold{j}_m\,;\,i-2,i-2}(w)]_{\mathcal{\tau}_h, r_i-2}\,R^{Bog}_{\bold{j}_1,\dots,\bold{j}_m\,;\,i-2,i-2}(w)\Big\}^{l_{i-2}} W^*_{\bold{j}_m}\nonumber\\
& =&W_{\bold{j}_m} R^{Bog}_{\bold{j}_1,\dots,\bold{j}_m,;\,i-2,i-2}(w)\, W^*_{\bold{j}_m}\\
& &+W_{\bold{j}_m} \sum_{l_{i-2}=1}^{h-1}R^{Bog}_{\bold{j}_1,\dots,\bold{j}_m,;\,i-2,i-2}(w)\,\times\\
& &\quad\quad\quad\quad\times \Big\{[\Gamma^{Bog\,}_{\bold{j}_1,\dots,\bold{j}_m\,,;\,N-h-2+r_i-2,N-h-2+r_i-2}(w)]^{(0)}_{(N-h-4+r_i-4, h_-)}\,R^{Bog}_{\bold{j}_1,\dots,\bold{j}_m\,;\,i-2,i-2}(w)\Big\}^{l_{i-2}} W^*_{\bold{j}_m}\quad\quad\quad\quad
\end{eqnarray}
\item each other term in (\ref{gammari}) of the type
\begin{equation}
[\Gamma^{Bog\,}_{\bold{j}_1,\dots,\bold{j}_m\,,;\,N-h-2+r_i,\,N-h-2+r_i}(w)]_{( N-4-h+q,h_-;\dots;N-h-4+r_i, h_-)}
\end{equation}
with $0\leq q \leq r_i-2$ (and even) is obtained from the sum in (\ref{sum-gamma-r}) according to the definition in  (\ref{def-gamma-relation}).
\end{itemize}
\qed

\begin{lemma} \label{appendix-lemma-1}
Assume that $\epsilon_{\bold{j}_*}$ is small enough and $N$ large enough such that for $z\leq E^{Bog}_{\bold{j}_*}+ \sqrt{\epsilon_{\bold{j}_*}}\phi_{\bold{j}_*}\sqrt{\epsilon_{\bold{j}_*}^2+2\epsilon_{\bold{j}_*}}(<0)$ the functions $\check{\mathcal{G}}_{\bold{j}_{*}\,;\,i,i}(z)$ (see (\ref{in-formula-G})) are well defined and fulfill the bound $\check{\mathcal{G}}_{\bold{j}_{*}\,;\,i,i}(z)\leq \frac{1}{X_i}$ where $X_i$ is defined in (\ref{def-X}). Then, for $N$ large enough, $(0\leq)\Delta n_{\bold{j}_{\bold{0}}}\leq h$, and
\begin{equation}
 z\leq E^{Bog}_{\bold{j}_*}+ \sqrt{\epsilon_{\bold{j}_*}}\phi_{\bold{j}_*}\sqrt{\epsilon_{\bold{j}_*}^2+2\epsilon_{\bold{j}_*}}-\frac{(h+4)\phi_{\bold{j}_*}}{N}\label{range-z-G}
\end{equation}
the estimate below holds true for some $g>0$ not larger than $4$
 \begin{equation}
\Big|\frac{\partial [\check{\mathcal{G}}_{\bold{j}_{*}\,;\,i,i}(z)]_{\tau_h\,;\,\Delta n_{\bold{j}_{\bold{0}}}}}{\partial \Delta n_{\bold{j}_{\bold{0}}}}\Big|\leq K\frac{h\cdot g^{i-N+h}}{N^{\frac{1}{2}}}\,
\end{equation}
where $[\check{\mathcal{G}}_{\bold{j}_{*}\,;\,i,i}(z)]_{\tau_h\,;\,\Delta n_{\bold{j}_{\bold{0}}}}$ is defined in (\ref{def-G-n0-1})-(\ref{def-G-n0-fin}) for $i$ even, $N-h-2\leq i \leq N-2$,  and $i-\Delta n_{\bold{j}_{\bold{0}}}-2\geq 0$. $K$ is a universal constant.
\end{lemma}

\noindent
\emph{Proof}

First, we observe that we can apply \emph{\underline{Theorem 3.1 of \cite{Pi1}}} (see Section \ref{Feshbach}) and the following inequality holds (see (\ref{ineq-Feshbach}))
\begin{equation}\label{ineq-G}
\check{\mathcal{G}}_{\bold{j}_{*}\,;\,i,i}(z)\leq  \|\sum_{l_{i}=0}^{\infty}\Big[(R^{Bog}_{\bold{j}_*\,;\,i,i}(z))^{\frac{1}{2}}\Gamma^{Bog}_{\bold{j}_*\,;\,i,i}(z)(R^{Bog}_{\bold{j}_*\,;\,i,i}(z))^{\frac{1}{2}}\Big]^{l_{i}}\| \leq \frac{1}{X_i}\,
\end{equation}
where $X_i$ is defined in  \emph{\underline{Lemma 3.6} of \cite{Pi1}} .
Starting from the definition
 \begin{equation}
[\check{\mathcal{G}}_{\bold{j}_{*}\,;\,i,i}(z)]_{\tau_h\,;\,\Delta n_{\bold{j}_{\bold{0}}}}:=\sum_{l_{i}=0}^{h-1}\{[\mathcal{W}_{\bold{j}_{*}\,;i,i-2}(z)\mathcal{W}^*_{\bold{j}_*\,;i-2,i}(z)]_{\Delta n_{\bold{j}_{\bold{0}}}}\,[\check{\mathcal{G}}_{\bold{j}_{*}\,;\,i-2,i-2}(z)]_{\tau_h\,;\,\Delta n_{\bold{j}_{\bold{0}}}}\}^{l_i}
\end{equation}
with $[\check{\mathcal{G}}_{\bold{j}_{*}\,;\,N-h-4, N-h-4}(z)]_{\tau_h\,;\,\Delta n_{\bold{j}_{\bold{0}}}}=1$,
the derivative with respect to $\Delta n_{\bold{j}_{\bold{0}}}$ yields the recursive relation
%We write
%\begin{equation}
%[\check{\mathcal{G}}_{\bold{j}_{*}\,;\,i,i}(z)]_{\Delta n_{\bold{j}_{\bold{0}}}}=\frac{1}{1-[\mathcal{W}_{\bold{j}_{*}\,;i,i-2}(z)\mathcal{W}^*_{\bold{j}_1\,;i-2,i}(z)]_{\Delta n_{\bold{j}_{\bold{0}}}}\,[\check{\mathcal{G}}_{\bold{j}_{*}\,;\,i-2,i-2}(z)]_{\Delta n_{\bold{j}_{\bold{0}}}}}
%\end{equation}
%Therefore,
\begin{eqnarray}
& &\frac{\partial [\check{\mathcal{G}}_{\bold{j}_{*}\,;\,i,i}(z)]_{\tau_h\,;\,\Delta n_{\bold{j}_{\bold{0}}}}}{\partial \Delta n_{\bold{j}_{\bold{0}}}}\label{der-G}\\
&=&
\Big\{\sum_{l_{i}=0}^{h-1}l_i\{[\mathcal{W}_{\bold{j}_{*}\,;i,i-2}(z)\mathcal{W}^*_{\bold{j}_*\,;i-2,i}(z)]_{\Delta n_{\bold{j}_{\bold{0}}}}\,[\check{\mathcal{G}}_{\bold{j}_{*}\,;\,i-2,i-2}(z)]_{\tau_h\,;\,\Delta n_{\bold{j}_{\bold{0}}}}\}^{l_i-1}\Big\}\times \\
%\Big(\frac{1}{1-[\mathcal{W}_{\bold{j}_{*}\,;i,i-2}(z)\mathcal{W}^*_{\bold{j}_1\,;i-2,i}(z)]_{\Delta n_{\bold{j}_{\bold{0}}}}\,[\check{\mathcal{G}}_{\bold{j}_{*}\,;\,i-2,i-2}(z)]_{\Delta n_{\bold{j}_{\bold{0}}}}}\Big)^2\times \\
& &\quad\quad \times \Big\{\frac{\partial [\mathcal{W}_{\bold{j}_{*}\,;i,i-2}(z)\mathcal{W}^*_{\bold{j}_*\,;i-2,i}(z)]_{\Delta n_{\bold{j}_{\bold{0}}}}}{\partial \Delta n_{\bold{j}_{\bold{0}}}}\, [\check{\mathcal{G}}_{\bold{j}_{*}\,;\,i-2,i-2}(z)]_{\tau_h\,;\,\Delta n_{\bold{j}_{\bold{0}}}} \nonumber\\
& &\quad\quad\quad\quad + [\mathcal{W}_{\bold{j}_{*}\,;i,i-2}(z)\mathcal{W}^*_{\bold{j}_*\,;i-2,i}(z)]_{\Delta n_{\bold{j}_{\bold{0}}}}\,\frac{\partial [\check{\mathcal{G}}_{\bold{j}_{*}\,;\,i-2,i-2}(z)]_{\tau_h\,;\,\Delta n_{\bold{j}_{\bold{0}}}}}{\partial \Delta n_{\bold{j}_{\bold{0}}}}\Big\}\,.
\end{eqnarray}
We recall $0\leq \Delta n_{\bold{j}_{\bold{0}}}\leq h$ and observe (see (\ref{def-G-n0-med})-(\ref{def-G-n0-fin})) that for $N-h-2\leq i \leq N-2$ (see the explicit computation in section 0.7 of \emph{supporting-file-Bose2.pdf})
\begin{equation}\label{W-rel}
[\mathcal{W}_{\bold{j}_{*}\,;i,i-2}(z)\mathcal{W}^*_{\bold{j}_*\,;i-2,i}(z)]_{\Delta n_{\bold{j}_{\bold{0}}}}\leq \mathcal{W}_{\bold{j}_{*}\,;i,i-2}(z+\frac{(h+4)\phi_{\bold{j}_*}}{N})\mathcal{W}^*_{\bold{j}_*\,;i-2,i}(z+\frac{(h+4)\phi_{\bold{j}_*}}{N})\,.
\end{equation}
The inequality in (\ref{W-rel}) together with the assumption on $z$ (see (\ref{range-z-G})) imply  that 
%for $N-h{\color{red}-2}\leq i\leq N-4$
\begin{eqnarray}\label{est-G}
& &[\check{\mathcal{G}}_{\bold{j}_{*}\,;\,i,i}(z)]_{\tau_h\,;\,\Delta n_{\bold{j}_{\bold{0}}}}\\
&\leq&\sum_{l_{i}=0}^{h-1}\{\mathcal{W}_{\bold{j}_{*}\,;i,i-2}(z+\frac{(h+4)\phi_{\bold{j}^*}}{N})\mathcal{W}^*_{\bold{j}_*\,;i-2,i}(z+\frac{(h+4)\phi_{\bold{j}^*}}{N})\,[\check{\mathcal{G}}_{\bold{j}_{*}\,;\,i-2,i-2}(z)]_{\tau_h\,;\,\Delta n_{\bold{j}_{\bold{0}}}}\}^{l_i}\,.\quad\quad \quad 
\end{eqnarray}
Since $[\check{\mathcal{G}}_{\bold{j}_{*}\,;\,N-h-4,N-h-4}(z)]_{\tau_h\,;\,\Delta n_{\bold{j}_{\bold{0}}}}\equiv [\check{\mathcal{G}}_{\bold{j}_{*}\,;\,N-h-4,N-h-4}(z+\frac{(h+4)\phi_{\bold{j}^*}}{N})]_{\tau_h\,;\,0}=1$, by induction the inequality  in (\ref{est-G}) implies 
\begin{eqnarray}
& &[\check{\mathcal{G}}_{\bold{j}_{*}\,;\,i,i}(z)]_{\tau_h\,;\,\Delta n_{\bold{j}_{\bold{0}}}}\\
&\leq&[\check{\mathcal{G}}_{\bold{j}_{*}\,;\,i,i}(z+\frac{(h+4)\phi_{\bold{j}^*}}{N})]_{\tau_h\,;\,0}\\
&\leq & \check{\mathcal{G}}_{\bold{j}_{*}\,;\,i,i}(z+\frac{(h+4)\phi_{\bold{j}^*}}{N})\,.\label{est-G-fin}
\end{eqnarray}
%From the positivity of $[\mathcal{W}_{\bold{j}_{*}\,;i,i-2}(z)\mathcal{W}^*_{\bold{j}_*\,;i-2,i}(z)]_{\tau_h\,;\,\Delta n_{\bold{j}_{\bold{0}}}}$ and $[\check{\mathcal{G}}_{\bold{j}_{*}\,;\,i-2,i-2}(z)]_{\tau_h\,;\,\Delta n_{\bold{j}_{\bold{0}}}}$, we deduce that
%\begin{equation}
%[\check{\mathcal{G}}_{\bold{j}_{*}\,;\,i-2,i-2}(z)]_{\tau_h\,;\,\Delta n_{\bold{j}_{\bold{0}}}}\leq \check{\mathcal{G}}_{\bold{j}_{*}\,;\,i-2,i-2}(z)
%\end{equation}
%and
Next, we make use of (\ref{W-rel}) and (\ref{est-G})-(\ref{est-G-fin}) to estimate
\begin{eqnarray}
%& &\frac{\partial [\check{\mathcal{G}}_{\bold{j}_{*}\,;\,i,i}(z)]_{\Delta n_{\bold{j}_{\bold{0}}}}}{\partial \Delta n_{\bold{j}_{\bold{0}}}}\\
& &
\sum_{l_{i}=0}^{h-1}l_i\{[\mathcal{W}_{\bold{j}_{*}\,;i,i-2}(z)\mathcal{W}^*_{\bold{j}_*\,;i-2,i}(z)]_{\Delta n_{\bold{j}_{\bold{0}}}}\,[\check{\mathcal{G}}_{\bold{j}_{*}\,;\,i-2,i-2}(z)]_{\tau_h\,;\,\Delta n_{\bold{j}_{\bold{0}}}}\}^{l_i-1}\\
&\leq &\Big[\frac{1}{1-[\mathcal{W}_{\bold{j}_{*}\,;i,i-2}(z)\mathcal{W}^*_{\bold{j}_*\,;i-2,i}(z)]_{\Delta n_{\bold{j}_{\bold{0}}}}\,[\check{\mathcal{G}}_{\bold{j}_{*}\,;\,i-2,i-2}(z)]_{\tau_h\,;\,\Delta n_{\bold{j}_{\bold{0}}}}}\Big]^2\\
&\leq&\Big[\frac{1}{1-\mathcal{W}_{\bold{j}_{*}\,;i,i-2}(z+\frac{(h+4)\phi_{\bold{j}^*}}{N})\mathcal{W}^*_{\bold{j}_*\,;i-2,i}(z+\frac{(h+4)\phi_{\bold{j}^*}}{N})\,\check{\mathcal{G}}_{\bold{j}_{*}\,;\,i-2,i-2}(z+\frac{(h+4)\phi_{\bold{j}^*}}{N})}\Big]^2\\
&= &\Big(\check{\mathcal{G}}_{\bold{j}_{*}\,;\,i,i}(z+\frac{(h+4)\phi_{\bold{j}^*}}{N})\Big)^2\,.\label{partial-sum-G}
%\Big[\frac{1}{1-[\mathcal{W}_{\bold{j}_{*}\,;i,i-2}(z)\mathcal{W}^*_{\bold{j}_1\,;i-2,i}(z)]_{\Delta n_{\bold{j}_{\bold{0}}}}\,[\check{\mathcal{G}}_{\bold{j}_{*}\,;\,i-2,i-2}(z)]_{\tau_h\,;\,\Delta n_{\bold{j}_{\bold{0}}}}}\Big]^2
\end{eqnarray}
Hence, we go back to (\ref{der-G}) and, thanks to (\ref{partial-sum-G}) we estimate
\begin{eqnarray}
& &\Big|\frac{\partial [\check{\mathcal{G}}_{\bold{j}_{*}\,;\,i,i}(z)]_{\Delta n_{\bold{j}_{\bold{0}}}}}{\partial \Delta n_{\bold{j}_{\bold{0}}}}\Big|\\
&\leq &\Big(\check{\mathcal{G}}_{\bold{j}_{*}\,;\,i,i}(z+\frac{(h+4)\phi_{\bold{j}^*}}{N})\Big)^2\times \\
& &\quad \times \Big\{\Big|\frac{\partial [\mathcal{W}_{\bold{j}_{*}\,;i,i-2}(z)\mathcal{W}^*_{\bold{j}_*\,;i-2,i}(z)]_{\Delta n_{\bold{j}_{\bold{0}}}}}{\partial \Delta n_{\bold{j}_{\bold{0}}}}\Big|\, [\check{\mathcal{G}}_{\bold{j}_{*}\,;\,i-2,i-2}(z)]_{\tau_h\,;\,\Delta n_{\bold{j}_{\bold{0}}}} \\
& &\quad\quad\quad\quad + [\mathcal{W}_{\bold{j}_{*}\,;i,i-2}(z)\mathcal{W}^*_{\bold{j}_*\,;i-2,i}(z)]_{\Delta n_{\bold{j}_{\bold{0}}}}\,\Big|\frac{\partial [\check{\mathcal{G}}_{\bold{j}_{*}\,;\,i-2,i-2}(z)]_{\tau_h\,;\,\Delta n_{\bold{j}_{\bold{0}}}}}{\partial \Delta n_{\bold{j}_{\bold{0}}}}\Big|\Big\}\,.
\end{eqnarray}
Furthermore, using (\ref{W-rel}) and the constraint on $z$ contained in (\ref{range-z-G}),  we can estimate
\begin{equation}\label{b-W-0}
 [\mathcal{W}_{\bold{j}_{*}\,;i,i-2}(z)\mathcal{W}^*_{\bold{j}_*\,;i-2,i}(z)]_{\Delta n_{\bold{j}_{\bold{0}}}}\leq  \mathcal{W}_{\bold{j}_{*}\,;i,i-2}(z+\frac{(h+4)\phi_{\bold{j}^*}}{N})\mathcal{W}^*_{\bold{j}_*\,;i-2,i}(z+\frac{(h+4)\phi_{\bold{j}^*}}{N})\leq \frac{1}{2}+\mathcal{O}( \sqrt{\epsilon_{\bold{j}_*}} )\,
\end{equation}
and, with the help of  (\ref{ineq-G}) and \emph{\underline{Lemma 3.6} of \cite{Pi1}} (see Section \ref{Feshbach}),
\begin{equation}\label{b-G}
\check{\mathcal{G}}_{\bold{j}_{*}\,;\,i,i}(z)\leq \check{\mathcal{G}}_{\bold{j}_{*}\,;\,i,i}(z+\frac{(h+4)\phi_{\bold{j}^*}}{N})\leq \frac{1}{X_i}\leq \frac{8}{3}+\mathcal{O}( \sqrt{\epsilon_{\bold{j}_*}} )\,.
\end{equation}
Next, from (\ref{b-W-0}) and the computation (see the definition in (\ref{def-G-n0-med})-(\ref{def-G-n0-fin}))
\begin{eqnarray}
& &\frac{\partial [\mathcal{W}_{\bold{j}_{*}\,;i,i-2}(z)\mathcal{W}^*_{\bold{j}_*\,;i-2,i}(z)]_{\Delta n_{\bold{j}_{\bold{0}}}}}{\partial \Delta n_{\bold{j}_{\bold{0}}}}\\
&= & \frac{-2i+2\Delta n_{\bold{j}_{\bold{0}}}+1}{N^2}\,\phi^2_{\bold{j}_{*}}\times \\
&  &\quad\times\,\frac{(N-i+2)^2}{4\Big[(\frac{i-\Delta n_{\bold{j}_{\bold{0}}}}{N}\phi_{\bold{j}_{*}}+(k_{\bold{j}_*})^2)(N-i)-z\Big]\Big[(\frac{i-2-\Delta n_{\bold{j}_{\bold{0}}}}{N}\phi_{\bold{j}_{*}}+(k_{\bold{j}_*})^2)(N-i+2)-z\Big]}\quad\quad\quad\\
& &+ \frac{(i-\Delta n_{\bold{j}_{\bold{0}}}-1)(i-\Delta n_{\bold{j}_{\bold{0}}})}{N^2}\,\phi^2_{\bold{j}_{*}}\times\\
&  &\quad\times\,\frac{\frac{\phi_{\bold{j}_{*}}(N-i)}{N}(N-i+2)^2}{4\Big[(\frac{i-\Delta n_{\bold{j}_{\bold{0}}}}{N}\phi_{\bold{j}_{*}}+(k_{\bold{j}_*})^2)(N-i)-z\Big]^2\Big[(\frac{i-2-\Delta n_{\bold{j}_{\bold{0}}}}{N}\phi_{\bold{j}_{*}}+(k_{\bold{j}_*})^2)(N-i+2)-z\Big]}\quad\quad\quad\\
& &+ \frac{(i-\Delta n_{\bold{j}_{\bold{0}}}-1)(i-\Delta n_{\bold{j}_{\bold{0}}})}{N^2}\,\phi^2_{\bold{j}_{*}}\times\\
&  &\quad\times\,\frac{\{\frac{\phi_{\bold{j}_{*}}(N-i+2)}{N}\}(N-i+2)^2}{4\Big[(\frac{i-\Delta n_{\bold{j}_{\bold{0}}}}{N}\phi_{\bold{j}_{*}}+(k_{\bold{j}_*})^2)(N-i)-z\Big]\Big[(\frac{i-2-\Delta n_{\bold{j}_{\bold{0}}}}{N}\phi_{\bold{j}_{*}}+(k_{\bold{j}_*})^2)(N-i+2)-z\Big]^2}\quad\quad\quad
\end{eqnarray}
we derive
\begin{equation}\label{b-W}
\Big|\frac{\partial [\mathcal{W}_{\bold{j}_{*}\,;i,i-2}(z)\mathcal{W}^*_{\bold{j}_*\,;i-2,i}(z)]_{\Delta n_{\bold{j}_{\bold{0}}}}}{\partial \Delta n_{\bold{j}_{\bold{0}}}}\Big|\leq \mathcal{O}(\frac{h}{N})\,.
\end{equation}

Therefore, for $z$ in the interval  (\ref{range-z-G}),  by using (\ref{partial-sum-G}), (\ref{b-G}), and (\ref{b-W}), we derive the bound
\begin{eqnarray}
& &\Big|\frac{\partial [\check{\mathcal{G}}_{\bold{j}_{*}\,;\,i,i}(z)]_{\Delta n_{\bold{j}_{\bold{0}}}}}{\partial \Delta n_{\bold{j}_{\bold{0}}}}\Big|\\
&\leq &\Big(\frac{8}{3}+\mathcal{O}( \sqrt{\epsilon_{\bold{j}_*}})\Big)^2\times \\
& &\quad \times \Big\{\mathcal{O}(\frac{h}{N}) \Big[\frac{8}{3}+\mathcal{O}( \sqrt{\epsilon_{\bold{j}_*}}\Big] + \Big[\frac{1}{2}+\mathcal{O}( \sqrt{\epsilon_{\bold{j}_*}})\Big]\,\Big|\frac{\partial [\check{\mathcal{G}}_{\bold{j}_{*}\,;\,i-2,i-2}(z)]_{\tau_h\,;\,\Delta n_{\bold{j}_{\bold{0}}}}}{\partial \Delta n_{\bold{j}_{\bold{0}}}}\Big|\Big\}\,.
\end{eqnarray}
Finally, for $N$ sufficiently large and $\sqrt{\epsilon_{\bold{j}_*}}$ sufficiently small, by induction we can conclude that 
\begin{equation}
\Big|\frac{\partial [\check{\mathcal{G}}_{\bold{j}_{*}\,;\,i,i}(z)]_{\tau_h\,;\,\Delta n_{\bold{j}_{\bold{0}}}}}{\partial \Delta n_{\bold{j}_{\bold{0}}}}\Big|\leq K\frac{h\cdot g^{i-N+h}}{\sqrt{N}}
\end{equation}
for some universal constants $K$ and  $g\in (3,4)$.
\qed
\\

\noindent
{\bf{Acknowledgements}}

I want to thank D.-A. Deckert, J. Fr\"ohlich, and P. Pickl for many stimulating discussions on the contents of this paper.  I am indebted to D.-A. Deckert for his contribution  to this project at its earliest stage and for helping with the implementation of crucial numerical simulations.

\end{document}

\
%%%%%%%%%%%%%%%%%%%%
$\quad\quad\quad\quad\quad\quad\quad\quad\quad\quad\quad\quad\quad \text{\underline{\bf{Supporting File}}}$
\\

\noindent
Notation: The formulae that are not  of the type (0...) are those in the manuscript (file BOSE2-2017.pdf). Theorems and corollaries are those in the manuscript (file BOSE2-2017.pdf) unless otherwise specified.
\subsection{Estimate of $\|\psi^{Bog}_{\bold{j}_1, \dots, \bold{j}_m}-(\psi^{Bog}_{\bold{j}_1, \dots, \bold{j}_m})'\|$ in (4.144)}
We recall that $z^{Bog}_{\bold{j}_1,\dots,\bold{j}_{m-1}}+(z^{(m)})'=:(z^{Bog}_{\bold{j}_1,\dots,\bold{j}_{m}})'$ and $z^{Bog}_{\bold{j}_1,\dots,\bold{j}_{m-1}}+z^{(m)}=:z^{Bog}_{\bold{j}_1,\dots,\bold{j}_{m}}$. We proceed as in step i) and ii) of Corollary 4.6, hence for some $\bar{j}< \frac{N}{2}$ to be fixed later  we write 
\begin{eqnarray}
(\psi^{Bog}_{\bold{j}_1, \dots, \bold{j}_{m}})'&= &\psi^{Bog}_{\bold{j}_1, \dots, \bold{j}_{m-1}}\\
& &+\Big[-\frac{1}{Q^{(N-2,N-1)}_{\bold{j}_m}\mathscr{K}^{Bog\,(N-4)}_{\bold{j}_1,\dots,\bold{j}_m}(z^{Bog}_{\bold{j}_1,\dots,\bold{j}_m})Q^{(N-2,N-1)}_{\bold{j}_m}}Q^{(N-2, N-1)}_{\bold{j}_m}W^*_{\bold{j}_m}\Big]\psi^{Bog}_{\bold{j}_1, \dots, \bold{j}_{m-1}}\label{gs-Hm-fin-bis-0}\\
%& &+\sum_{j=2}^{N/2}\prod^{2}_{r=j}\Big[-\frac{1}{Q_{\bold{j}_m}^{(N-2r, N-2r+1)}\mathscr{K}_{\bold{j}_1,\dots ,\bold{j}_m}^{Bog\,(N-2r-2)}(z^{Bog}_{\bold{j}_1,\dots ,\bold{j}_m})Q_{\bold{j}_m}^{(N-2r, N-2r+1)}}W^*_{\bold{j}_m;N-2r,N-2r+2}\Big]\psi^{Bog}_{\bold{j}_1, \dots, \bold{j}_{m-1}}\quad\quad\quad\label{gs-Hm-fin} \\
& &+\sum_{j=2}^{\bar{j}}\Big\{\prod^{2}_{r=j}\Big[-\frac{1}{Q_{\bold{j}_m}^{(N-2r, N-2r+1)}\mathscr{K}_{\bold{j}_1,\dots ,\bold{j}_m}^{Bog\,(N-2r-2)}((z^{Bog}_{\bold{j}_1,\dots ,\bold{j}_m})')Q_{\bold{j}_m}^{(N-2r, N-2r+1)}}W^*_{\bold{j}_m;N-2r,N-2r+2}\Big]\times\quad\quad\quad\quad\quad  \label{gs-Hm-fin-bis}\\
& &\quad\quad\quad\times \Big[-\frac{1}{Q^{(N-2,N-1)}_{\bold{j}_m}\mathscr{K}^{Bog\,(N-4)}_{\bold{j}_1,\dots,\bold{j}_m}((z^{Bog}_{\bold{j}_1,\dots,\bold{j}_m})')Q^{(N-2,N-1)}_{\bold{j}_m}}Q^{(N-2, N-1)}_{\bold{j}_m}W^*_{\bold{j}_m}\Big]\Big\}\psi^{Bog}_{\bold{j}_1, \dots, \bold{j}_{m-1}}\,\,\nonumber \\
&&+\mathcal{O}([\frac{1}{1+c\sqrt{\epsilon_{\bold{j}_m}}}]^{\bar{j}})+\mathcal{O}(\frac{1}{\Delta_{m-1} (\ln N)^{\frac{1}{4}}} )
\end{eqnarray}
 for some $c>0$. This is possible because, like in \emph{\underline{Corollary 4.6} of [Pi1]}, we can control the norm of the sum in (\ref{gs-Hm-fin-bis}) in terms of a geometric series. Indeed, by construction, for $w=(z^{Bog}_{\bold{j}_1,\dots,\bold{j}_{m}})'$,
\begin{eqnarray}
%& &\Big\|\Big\{\prod^{2}_{i=j}\Big[-\frac{1}{Q^{(N-2i)}_{\bold{j}_*}\mathscr{K}^{Bog\,(N-2i+1)}_{\bold{j}_*}(z_*)Q^{(N-2i)}}W^*_{\bold%{j}_*\,;\,N-2i,N-2i+2}\Big]\Big\}\frac{1}{Q^{(N-2)}_{\bold{j}_*}\mathscr{K}^{Bog\,(N-3)}_{\bold{j}_*}(z_*)Q^{(N-2)}_{\bold{j}_*}}Q^%{(N-2)}_{\bold{j}_*}W_{\bold{j}_*}^*\eta\Big\|  \nonumber\\
& &\Big\|\Big\{\prod^{2}_{l=j}\Big[-\frac{1}{Q^{(N-2l, N-2l+1)}_{\bold{j}_m}\mathscr{K}^{Bog\,(N-2l-2)}_{\bold{j}_1,\dots,\bold{j}_m}(w)Q^{(N-2l,N-2l+1)}_{\bold{j}_m}}W^*_{\bold{j}_m\,;\,N-2l,N-2l+2}\Big]\Big\}\times \\
& &\quad\quad\quad\quad\quad\quad\times \frac{1}{Q^{(N-2,N-1)}_{\bold{j}_m}\mathscr{K}^{Bog\,(N-4)}_{\bold{j}_1,\dots,\bold{j}_m}(w)Q^{(N-2,N-1)}_{\bold{j}_m}}Q^{(N-2, N-1)}_{\bold{j}_m}W^*_{\bold{j}_m}\psi^{Bog}_{\bold{j}_1, \dots, \bold{j}_{m-1}} \Big\| \nonumber\\
&\leq&\Big\|[R^{Bog}_{\bold{j}_1,\dots,\bold{j}_m\,;\,N-2j,N-2j}(w)]^{\frac{1}{2}}\Big\|\times\\
& &\quad \times \Big\{\prod^{2}_{l=j}\,\Big\|\check{\Gamma}^{Bog\,}_{\bold{j}_1,\dots,\bold{j}_m\,;\,N-2l,N-2l}(w)\Big\|\,\Big\|[R^{Bog}_{\bold{j}_1,\dots,\bold{j}_m\,;\,N-2l,N-2l}(w)]^{\frac{1}{2}}W^*_{\bold{j}_m\,;\,N-2l,N-2l+2}[R^{Bog}_{\bold{j}_*\,;\,N-2l+2,N-2l+2}(w)]^{\frac{1}{2}}\Big\|\Big\}\times \quad\quad\quad \nonumber\\
& &\quad\quad \times\,\Big\|\check{\Gamma}^{Bog\,}_{\bold{j}_1,\dots,\bold{j}_m\,;\,N-2,N-2}(w)\Big\|\,\Big\|[R^{Bog}_{\bold{j}_1,\dots,\bold{j}_m\,;\,N-2,N-2}(w)]^{\frac{1}{2}}Q^{(N-2,N-1)}_{\bold{j}_m}W^*_{\bold{j}_*}\psi^{Bog}_{\bold{j}_1, \dots, \bold{j}_{m-1}} \Big\|\,. \nonumber
\end{eqnarray}
where

\begin{equation}
\check{\Gamma}^{Bog\,}_{\bold{j}_1,\dots,\bold{j}_m\,;\,i,i}(w):=\sum_{l_i=0}^{\infty}\Big[(R^{Bog}_{\bold{j}_1,\dots,\bold{j}_m\,;\,i,i}(w))^{\frac{1}{2}}\Gamma^{Bog\,}_{\bold{j}_1,\dots,\bold{j}_m\,;\,i,i}(w)(R^{Bog}_{\bold{j}_1,\dots,\bold{j}_m\,;\,i,i}(w))^{\frac{1}{2}}\Big]^{l_i}\,,
\end{equation}
and  the estimates below hold true:
\begin{eqnarray}\label{key-estimate}
&  &\Big\|[R^{Bog}_{\bold{j}_1,\dots,\bold{j}_m\,;\,N-2l,N-2l}(w)]^{\frac{1}{2}}W^*_{\bold{j}_m\,;\,N-2l,N-2l+2}[R^{Bog}_{\bold{j}_1,\dots,\bold{j}_m\,;\,N-2l+2,N-2l+2}(w)]^{\frac{1}{2}}\Big\|\\
%&=&\Big\|[R^{Bog}_{\bold{j}_*\,;\,N-2l,N-2l}(z_*)]^{\frac{1}{2}}\Big\|\,\Big\|W_{\bold{j}_*\,;\,N-2l+2,N-2l}[R^{Bog}_{\bold{j}_*\,;\,N-2l,N-2l}(z_*)]W^*_{\bold{j}_*\,;\,N-2l,N-2l+2}\Big\|^{\frac{1}{2}}\\
&\leq&\frac{1}{2\Big[1+a_{\epsilon_{\bold{j}_m}}-\frac{2b_{\epsilon_{\bold{j}_m}}}{2l-1}-\frac{1-c_{\epsilon_{\bold{j}_m}}}{(2l-1)^2}]^{\frac{1}{2}}}
%\sqrt{(\ref{def-deltabog})}
\end{eqnarray}
%for $z\leq \xi E^{Bog}$
%\[
%???x_{2j}\geq\frac{1}{2}\Big[1+b_{\epsilon}-\frac{1}{N-2j+1-\xi}\Big]\,;???
%\]
\begin{equation}\label{control-2}
\|\check{\Gamma}^{Bog\,}_{\bold{j}_1,\dots,\bold{j}_m\,;\,N-2l,N-2l}(w)\|\leq  \frac{2}{\Big[1+\sqrt{\eta a_{\epsilon_{\bold{j}_m}}}-\frac{b_{\epsilon_{\bold{j}_m}}/ \sqrt{\eta a_{\epsilon_{\bold{j}_m}}}}{2l-\epsilon^{\Theta}_{\bold{j}_m}}\Big]}\,,
\end{equation}
and, 
\begin{equation}\label{control-3}
\Big\|[R^{Bog}_{\bold{j}_1,\dots,\bold{j}_m\,;\,N-2j,N-2j}(w)]^{\frac{1}{2}}\Big\|\Big\|[R^{Bog}_{\bold{j}_1,\dots,\bold{j}_m\,;\,N-2,N-2}(w)]^{\frac{1}{2}}Q^{(N-2,N-1)}_{\bold{j}_m}W^*_{\bold{j}_m}\psi^{Bog}_{\bold{j}_1, \dots, \bold{j}_{m-1}} \Big\|\leq \mathcal{O}(1)\,
\end{equation}
where $\eta=1-\sqrt{\epsilon_{\bold{j}_m}}$,  $a_{\epsilon_{\bold{j}_{m}}}, b_{\epsilon_{\bold{j}_{m}}}, c_{\epsilon_{\bold{j}_{m}}}$ are those defined in Corollary 5.1, and $\Theta$ is defined in Theorem 4.1. 
Therefore, we can bound $\|(\ref{gs-Hm-fin-bis})\|$ with a multiple of the series
\begin{eqnarray}\label{convergent-series}
& &\sum_{j=2}^{\infty}c_j:=\sum_{j=2}^{\infty}\Big\{\prod^{2}_{l=j}\frac{1}{\Big[1+\sqrt{\eta a_{\epsilon_{\bold{j}_m}}}-\frac{b_{\epsilon_{\bold{j}_m}}/ \sqrt{\eta a_{\epsilon_{\bold{j}_m}}}}{2l-\epsilon^{\Theta}_{\bold{j}_m}}\Big]\Big[1+a_{\epsilon_{\bold{j}_m}}-\frac{2b_{\epsilon_{\bold{j}_m}}}{2l-1}-\frac{1-c_{\epsilon_{\bold{j}_m}}}{(2l-1)^2}]^{\frac{1}{2}}}\Big\}
\end{eqnarray}
which is convergent because for $j$ sufficiently large
\begin{equation}\label{ratioc}
\frac{c_{j}}{c_{j-1}}=\frac{1}{\Big[1+\sqrt{\eta a_{\epsilon_{\bold{j}_m}}}-\frac{b_{\epsilon_{\bold{j}_m}}/ \sqrt{\eta a_{\epsilon_{\bold{j}_m}}}}{2j-\epsilon^{\Theta}_{\bold{j}_m}}\Big]\Big[1+a_{\epsilon_{\bold{j}_m}}-\frac{2b_{\epsilon_{\bold{j}_m}}}{2j-1}-\frac{1-c_{\epsilon_{\bold{j}_m}}}{(2j-1)^2}]^{\frac{1}{2}}}<\frac{1}{1+c\sqrt{\epsilon_{\bold{j}_m}}}<1
\end{equation}
for some $c>0$.

Now, starting from the very left in (\ref{gs-Hm-fin-bis}) we replace each operator of the type
\begin{eqnarray}
& &\frac{1}{Q^{(N-2r,N-2r+1)}_{\bold{j}_m}\mathscr{K}^{Bog\,(N-2r-2)}_{\bold{j}_1,\dots,\bold{j}_{m-1}}((z^{Bog}_{\bold{j}_1,\dots,\bold{j}_m})')Q^{(N-2r,N-2r+1)}_{\bold{j}_*}}\\
&=&\sum_{l_{N-2r}=0}^{\infty}R^{Bog}_{\bold{j}_1,\dots,\bold{j}_{m-1}\,;\,N-2r,N-2r}((z^{Bog}_{\bold{j}_1,\dots,\bold{j}_m})')\Big[\Gamma^{Bog\,}_{\bold{j}_1,\dots,\bold{j}_{m-1}\,;\,N-2r,N-2r}((z^{Bog}_{\bold{j}_1,\dots,\bold{j}_m})')\,R^{Bog}_{\bold{j}_1,\dots,\bold{j}_{m-1}\,;\,N-2r,N-2r}((z^{Bog}_{\bold{j}_1,\dots,\bold{j}_m})')\Big]^{l_{N-2r}}\quad\quad\quad \label{non-trunc}
\end{eqnarray}
with
\begin{eqnarray}
%& &\frac{1}{Q^{(N-2r,N-2r+1)}_{\bold{j}_m}\mathscr{K}^{Bog\,(N-2r-2)}_{\bold{j}_1,\dots,\bold{j}_{m-1}}(w)Q^{(N-2r,N-2r+1)}_{\bold{j}_*}}\\
&&\sum_{l_{N-2r}=0}^{h-1}R^{Bog}_{\bold{j}_1,\dots,\bold{j}_{m-1}\,;\,N-2r,N-2r}((z^{Bog}_{\bold{j}_1,\dots,\bold{j}_m})')\Big[\sum_{l=N-2r-h,\,l\, even}^{N-2r-2}[\Gamma^{Bog\,}_{\bold{j}_1,\dots,\bold{j}_m;\,N-2r,N-2r}((z^{Bog}_{\bold{j}_1,\dots,\bold{j}_m})')]_{(l,h_-,\,l+2,h_-,\, \dots \,, N-2r-2,h_-)}\times \quad\quad\quad \label{trunc} \\
& &\quad\quad\quad\quad\quad\quad\quad\quad\quad\quad  \times\,R^{Bog}_{\bold{j}_1,\dots,\bold{j}_{m-1}\,;\,N-2r,N-2r}((z^{Bog}_{\bold{j}_1,\dots,\bold{j}_m})')\Big]^{l_{N-2r}}\quad\quad \nonumber
\end{eqnarray}
where $h$ (even) is determined later. We estimate the difference between (\ref{non-trunc}) and (\ref{trunc}).
To this end, we define (for $2\leq r \leq j$)
\begin{equation}
\mathcal{T}_{N-2r,N-2r}:=(R^{Bog}_{\bold{j}_1,\dots,\bold{j}_{m-1}\,;\,N-2r,N-2r}((z^{Bog}_{\bold{j}_1,\dots,\bold{j}_m})'))^{\frac{1}{2}}\,\Gamma^{Bog\,}_{\bold{j}_1,\dots,\bold{j}_{m-1}\,;\,N-2r,N-2r}((z^{Bog}_{\bold{j}_1,\dots,\bold{j}_m})')\,(R^{Bog}_{\bold{j}_1,\dots,\bold{j}_{m-1}\,;\,N-2r,N-2r}((z^{Bog}_{\bold{j}_1,\dots,\bold{j}_m})'))^{\frac{1}{2}}\end{equation}
and
\begin{eqnarray}
& &\mathcal{S}_{N-2r,N-2r}:=(R^{Bog}_{\bold{j}_1,\dots,\bold{j}_{m-1}\,;\,N-2r,N-2r}((z^{Bog}_{\bold{j}_1,\dots,\bold{j}_m})'))^{\frac{1}{2}}\,\Big[\sum_{l=N-2r-h,\,l\, even}^{N-2r-2}[\Gamma^{Bog\,}_{\bold{j}_1,\dots,\bold{j}_m;\,N-2r,N-2r}((z^{Bog}_{\bold{j}_1,\dots,\bold{j}_m})')]_{(l,h_-,\,l+2,h_-,\, \dots \,, N-2r-2,h_-)}\Big]\times\quad\quad\quad\quad\\
& &\quad\quad\quad \times\,(R^{Bog}_{\bold{j}_1,\dots,\bold{j}_{m-1}\,;\,N-2r,N-2r}((z^{Bog}_{\bold{j}_1,\dots,\bold{j}_m})'))^{\frac{1}{2}}\nonumber
\end{eqnarray}
For $(2j\geq )2l\geq 4$, the estimates in (\ref{key-estimate}) and (\ref{control-2}) imply  $\|(\ref{key-estimate})\|^2\leq \frac{4}{15}+o(1)$ and $\|(\ref{control-2})\|\leq \frac{8}{3}+o(1)$. Then, using the relation in (4.52)-(4.54) (in Remark 4.3),  and invoking the argument of Remark 4.3,  we can estimate $\|\mathcal{T}_{N-2r,N-2r}\|\,,\,\|\mathcal{S}_{N-2r,N-2r}\|\leq \frac{4}{5}$. Furthermore, the estimates in (4.32), (4.33) (in Remark 4.2),  combined with the argument  in Remarks 4.3, imply
\begin{equation}
\|\mathcal{T}_{N-2r,N-2r}-\mathcal{S}_{N-2r,N-2r}\|\leq \mathcal{O}((\frac{1}{1+c\sqrt{\epsilon_{\bold{j}_m}}})^{h})
\end{equation}
for some $c>0$.
Then, we can write

%& &\frac{1}{Q^{(N-2r,N-2r+1)}_{\bold{j}_m}\mathscr{K}^{Bog\,(N-2r-2)}_{\bold{j}_1,\dots,\bold{j}_{m-1}}(w)Q^{(N-2r,N-2r+1)}_{\bold{j}_*}}\\
 \begin{eqnarray}
& &\Big\|\sum_{l_{N-2r}=0}^{\infty}\Big\{(R^{Bog}_{\bold{j}_1,\dots,\bold{j}_{m-1}\,;\,N-2r,N-2r}((z^{Bog}_{\bold{j}_1,\dots,\bold{j}_m})'))^{\frac{1}{2}}\Gamma^{Bog\,}_{\bold{j}_1,\dots,\bold{j}_{m-1}\,;\,N-2r,N-2r}((z^{Bog}_{\bold{j}_1,\dots,\bold{j}_m})')\,(R^{Bog}_{\bold{j}_1,\dots,\bold{j}_{m-1}\,;\,N-2r,N-2r}((z^{Bog}_{\bold{j}_1,\dots,\bold{j}_m})'))^{\frac{1}{2}}\Big\}^{l_{N-2r}}\quad\quad \nonumber\\
&&\quad -\sum_{l_{N-2r}=0}^{h-1}\Big\{(R^{Bog}_{\bold{j}_1,\dots,\bold{j}_{m-1}\,;\,N-2r,N-2r}((z^{Bog}_{\bold{j}_1,\dots,\bold{j}_m})'))^{\frac{1}{2}}\Big[\sum_{l=N-2r-h,\,l\, even}^{N-2r-2}[\Gamma^{Bog\,}_{\bold{j}_1,\dots,\bold{j}_m;\,N-2,N-2}((z^{Bog}_{\bold{j}_1,\dots,\bold{j}_m})')]_{(l,h_-,\,l+2,h_-,\, \dots \,, N-2r-2,h_-)}\Big]\times \quad\quad\quad\nonumber \\
& &\quad\quad\quad\quad\quad\quad\quad\quad\quad\quad  \times\,(R^{Bog}_{\bold{j}_1,\dots,\bold{j}_{m-1}\,;\,N-2r,N-2r}((z^{Bog}_{\bold{j}_1,\dots,\bold{j}_m})'))^{\frac{1}{2}}\Big\}^{l_{N-2r}}\,\Big\|\quad\quad \nonumber\\
&\leq&\|\sum_{l_{N-2r}=h}^{\infty}\Big\{R^{Bog}_{\bold{j}_1,\dots,\bold{j}_{m-1}\,;\,N-2r,N-2r}((z^{Bog}_{\bold{j}_1,\dots,\bold{j}_m})')\Big[\Gamma^{Bog\,}_{\bold{j}_1,\dots,\bold{j}_{m-1}\,;\,N-2r,N-2r}((z^{Bog}_{\bold{j}_1,\dots,\bold{j}_m})')\Big]\,(R^{Bog}_{\bold{j}_1,\dots,\bold{j}_{m-1}\,;\,N-2r,N-2r}((z^{Bog}_{\bold{j}_1,\dots,\bold{j}_m})'))^{\frac{1}{2}}\Big\}^{l_{N-2r}}\| \nonumber \\
& &+\|\sum_{l_{N-2}=1}^{h-1}\sum_{j=0}^{l_{N-2}-1}\mathcal{T}_{N-2r,N-2r}^j(\mathcal{T}_{N-2r,N-2r}-\mathcal{S}_{N-2r,N-2r})\mathcal{S}_{N-2r,N-2r}^{l_{N-2}-j-1}\|\nonumber\\
&\leq &\|\mathcal{T}_{N-2r,N-2r}-\mathcal{S}_{N-2r,N-2r}\|\sum_{l_{N-2}=1}^{\infty}l_{N-2}(\frac{4}{5})^{l_{N-2}-1}+\mathcal{O}((\frac{4}{5})^h)\nonumber\\
&\leq &\mathcal{O}((\frac{1}{1+c\sqrt{\epsilon_{\bold{j}_*}}})^{h})+\mathcal{O}((\frac{4}{5})^h)\nonumber
%&&+\sum_{l_{N-2r}=0}^{h-1}R^{Bog}_{\bold{j}_1,\dots,\bold{j}_{m-1}\,;\,N-2r,N-2r}((z^{Bog}_{\bold{j}_1,\dots,\bold{j}_m})')\Big\{\sum_{l=N-2r-h,\,l\, even}^{N-2r-2}\Big[\Gamma^{Bog\,}_{\bold{j}_1,\dots,\bold{j}_{m-1}\,;\,N-2r,N-2r}((z^{Bog}_{\bold{j}_1,\dots,\bold{j}_m})')-\\
%& &\quad\quad\quad\quad \Gamma^{Bog\,}_{\bold{j}_1,\dots,\bold{j}_m;\,N-2,N-2}((z^{Bog}_{\bold{j}_1,\dots,\bold{j}_m})')]_{(l,h_-,\,l+2,h_-,\, \dots \,, N-2r-2,h_-)}\Big]\,R^{Bog}_{\bold{j}_1,\dots,\bold{j}_{m-1}\,;\,N-2r,N-2r}((z^{Bog}_{\bold{j}_1,\dots,\bold{j}_m})')\Big\}^{l_{N-2r}}\quad\quad
\end{eqnarray}
We implement the same operations on $\psi^{Bog}_{\bold{j}_1,\dots,\bold{j}_{m}}$ defined like $(\psi^{Bog}_{\bold{j}_1,\dots,\bold{j}_{m}})'$ but with $z^{Bog}_{\bold{j}_1,\dots,\bold{j}_{m}}$ instead of $(z^{Bog}_{\bold{j}_1,\dots,\bold{j}_{m}})'$. The approximated expressions obtained for the two vectors coincide up to the replacement of $z^{Bog}_{\bold{j}_1,\dots,\bold{j}_{m}}$ with $(z^{Bog}_{\bold{j}_1,\dots,\bold{j}_{m}})'$.  In order to estimate the total error term,  it is enough to estimate the number of resolvents, bounded by $\mathcal{O}((2h)^{\frac{h+2}{2}})$, contained in each $$\sum_{l=N-2r-h,\,l\, even}^{N-2r-2}[\Gamma^{Bog\,}_{\bold{j}_1,\dots,\bold{j}_m;\,N-2,N-2}((z^{Bog}_{\bold{j}_1,\dots,\bold{j}_m})')]_{(l,h_-,\,l+2,h_-,\, \dots \,, N-2r-2,h_-)}$$ (see section 0.2 below),  and recall that $(z^{Bog}_{\bold{j}_1,\dots,\bold{j}_{m}})'-z^{Bog}_{\bold{j}_1,\dots,\bold{j}_{m}}=\mathcal{O}(\frac{1}{(\ln N)^{\frac{1}{4}}})$. Finally, thanks to the control of the sum in (\ref{gs-Hm-fin-bis}) in terms of a convergent geometric series, 
%provided by (\ref{key-estimate}), (\ref{control-2}), and (\ref{control-3}), 
we can conclude that, for $\epsilon_{\bold{j}_m}$ small but strictly positive,  the norm of the sum of all remainder terms can be estimated
\begin{equation} \label{error}
\mathcal{O}\Big(h(2h)^{\frac{h+2}{2}}\frac{1}{(\ln N)^{\frac{1}{4}}}\Big)+\mathcal{O}((\frac{1}{1+c\sqrt{\epsilon_{\bold{j}_*}}})^{h})
\end{equation}
By choosing $N$ large enough and $\bar{j}=h=[\ln(\ln N)]^{\frac{1}{2}}$ the error term in (\ref{error}) is bounded by $\mathcal{O}(\frac{1}{[\ln(\ln N)]^{\frac{1}{2}}})$.
\subsection{Estimate of the number of RW-blocks in $[\Gamma^{Bog\,}_{\bold{j}_1,\dots,\bold{j}_m\,;\,N-2,N-2}(w)]^{(r)}_{\mathcal{\tau}_h}$}
Using the definition (with $h\geq 2$ and even) in (5.27), we can write
\begin{eqnarray}\label{tau-exp}
& &[\Gamma^{Bog\,}_{\bold{j}_1,\dots,\bold{j}_m\,;\,N-h-2,N-h-2}(w)]_{\mathcal{\tau}_h}\\
&=&W_{\bold{j}_m} R^{Bog}_{\bold{j}_1,\dots,\bold{j}_m\,;\,N-h-4,N-h-4}(w)\,\sum_{l_{N-h-4}=0}^{h-1}\Big\{[\Gamma^{Bog\,}_{\bold{j}_1,\dots,\bold{j}_m\,;\,N-h-4,N-h-4}(w)]_{\mathcal{\tau}_h}\,R^{Bog}_{\bold{j}_1,\dots,\bold{j}_m\,;\,N-h-4,N-h-4}(w)\Big\}^{l_{N-h-4}} W^*_{\bold{j}_m}\quad\quad\quad \label{summands}
\end{eqnarray}
with
\begin{equation}\label{tau-exp-bis}
[\Gamma^{Bog\,}_{\bold{j}_1,\dots,\bold{j}_m\,;\,N-4-h,N-4-h}(w)]_{\mathcal{\tau}_h}
=  W_{\bold{j}_m}\,R^{Bog}_{\bold{j}_1,\dots,\bold{j}_m\,;\, N-4-h, N-4-h}(w)W_{\bold{j}_m}^*\,.\quad\quad
\end{equation}
Each summand in (\ref{summands}) is a monomial containing at most $$1+2(h-1)<2h$$ RW-blocks. Now, assume that for $j$ even and $h\geq j\geq 2$, $[\Gamma^{Bog\,}_{\bold{j}_1,\dots,\bold{j}_m\,;\,N-h-4+j,N-h-4+j}(w)]_{\mathcal{\tau}_h}$ is a sum of monomials of the form $W_{\bold{j}_m}$ times a product of  RW-blocks where the number of RW-blocks is $(2h)^{\frac{j}{2}}$ at most. Since, by the definition in (5.27), 
\begin{eqnarray}
& &[\Gamma^{Bog\,}_{\bold{j}_1,\dots,\bold{j}_m\,;\,N-h-4+(j+2),N-h-4+(j+2)}(w)]_{\mathcal{\tau}_h}\\
&=&W_{\bold{j}_m} R^{Bog}_{\bold{j}_1,\dots,\bold{j}_m\,;\,N-h-4+j,N-h-4+j}(w)\,\sum_{l=0}^{h-1}\Big\{[\Gamma^{Bog\,}_{\bold{j}_1,\dots,\bold{j}_m\,;\,N-h-4+j,N-h-4+j}(w)]_{\mathcal{\tau}_h}\,R^{Bog}_{\bold{j}_1,\dots,\bold{j}_m\,;\,N-h-4+j,N-h-4+J}(w)\Big\}^{l} W^*_{\bold{j}_m}\,,\quad\quad \label{summands-2}
\end{eqnarray}
we derive that each summand in (\ref{summands-2}) is a monomial containing at most $$(h-1)[(2h)^{\frac{j}{2}}+1]+1<(2h)^{\frac{j+2}{2}}\,$$
RW-blocks.
\subsection{Proof by induction to show the identity (5.136)=(5.137) in {\bf{Step a)}} of Lemma 5.3}
We recall the definition
\begin{eqnarray}\label{tau-exp}
& &[\Gamma^{Bog\,}_{\bold{j}_1,\dots,\bold{j}_m\,;\,i,i}(w)]_{\mathcal{\tau}_h}\\
&:=&W_{\bold{j}_m} R^{Bog}_{\bold{j}_1,\dots,\bold{j}_m\,;\,i-2,i-2}(w)\,\sum_{l_{i-2}=0}^{h-1}\Big\{[\Gamma^{Bog\,}_{\bold{j}_1,\dots,\bold{j}_m\,;\,i-2,i-2}(w)]_{\mathcal{\tau}_h}\,R^{Bog}_{\bold{j}_1,\dots,\bold{j}_m\,;\,i-2,i-2}(w)\Big\}^{l_{i-2}} W^*_{\bold{j}_m}\nonumber
\end{eqnarray}
for $N-2\geq i \geq N-h$, and
\begin{equation}\label{tau-exp-bis}
[\Gamma^{Bog\,}_{\bold{j}_1,\dots,\bold{j}_m\,;\,N-2-h,N-2-h}(w)]_{\mathcal{\tau}_h}
:=  W_{\bold{j}_m}\,R^{Bog}_{\bold{j}_1,\dots,\bold{j}_m\,;\, N-4-h, N-4-h}(w)W_{\bold{j}_m}^*\,.\quad\quad
\end{equation}

\noindent
By the normal ordering of the operators $a_{\bold{j}_m}$ and $a^*_{\bold{j}_m}$, it is evident that
\begin{eqnarray}\label{comp}
& &\tilde{Q}^{(N-2-h)}_{\bold{j}_m}(R^{Bog}_{\bold{j}_1,\dots,\bold{j}_m\,;\, N-2-h, N-2-h}(w))^{\frac{1}{2}}[\Gamma^{Bog\,}_{\bold{j}_1,\dots,\bold{j}_m\,;\,N-2-h,N-2-h}(w)]_{\mathcal{\tau}_h}(R^{Bog}_{\bold{j}_1,\dots,\bold{j}_m\,;\, N-2-h, N-2-h}(w))^{\frac{1}{2}}\tilde{Q}^{(N-2-h)}_{\bold{j}_m}\quad\quad\quad\\
&=&\tilde{Q}^{(N-2-h)}_{\bold{j}_m} (R^{Bog}_{\bold{j}_1,\dots,\bold{j}_m\,;\, N-2-h, N-2-h}(w))^{\frac{1}{2}} W_{\bold{j}_m}\,(R^{Bog}_{\bold{j}_1,\dots,\bold{j}_m\,;\, N-4-h, N-4-h}(w))^{\frac{1}{2}}\times \\
& &\quad\quad \times (R^{Bog}_{\bold{j}_1,\dots,\bold{j}_m\,;\, N-4-h, N-4-h}(w))^{\frac{1}{2}}W_{\bold{j}_m}^*(R^{Bog}_{\bold{j}_1,\dots,\bold{j}_m\,;\, N-2-h, N-2-h}(w))^{\frac{1}{2}}\tilde{Q}^{(N-2-h)}_{\bold{j}_m}\nonumber\\
&=&\tilde{Q}^{(N-2-h)}_{\bold{j}_m}\Big[(R^{Bog}_{\bold{j}_1,\dots,\bold{j}_m\,;\, N-2-h, N-2-h}(w))^{\frac{1}{2}} W_{\bold{j}_m}\,(R^{Bog}_{\bold{j}_1,\dots,\bold{j}_m\,;\, N-4-h, N-4-h}(w))^{\frac{1}{2}}\Big]_1\,\times\\
& &\quad\quad \times \Big[(R^{Bog}_{\bold{j}_1,\dots,\bold{j}_m\,;\, N-4-h, N-4-h}(w))^{\frac{1}{2}}W_{\bold{j}_m}^*(R^{Bog}_{\bold{j}_1,\dots,\bold{j}_m\,;\, N-2-h, N-2-h}(w))^{\frac{1}{2}}\Big]_1\tilde{Q}^{(N-2-h)}_{\bold{j}_m} \nonumber
\end{eqnarray}
where $\tilde{Q}^{(i)}_{\bold{j}_m}$  projects onto the subspace of vectors with exactly $\frac{N-i}{2}$ particles in the modes $\bold{j}_{m}$ and $-\bold{j_m}$, respectively.
Next, for $N-4\geq i-2\geq N-2-h$ and even, we assume
\begin{eqnarray}
& &\tilde{Q}^{(i-2)}_{\bold{j}_m}(R^{Bog}_{\bold{j}_1,\dots,\bold{j}_m\,;\, i-2, i-2}(w))^{\frac{1}{2}}[\Gamma^{Bog\,}_{\bold{j}_1,\dots,\bold{j}_m\,;\,i-2,i-2}(w)]_{\mathcal{\tau}_h}(R^{Bog}_{\bold{j}_1,\dots,\bold{j}_m\,;\, i-2, i-2}(w))^{\frac{1}{2}}\tilde{Q}^{(i-2)}_{\bold{j}_m}\\
&=&\Big[(R^{Bog}_{\bold{j}_1,\dots,\bold{j}_m\,;\, i-2, i-2}(w))^{\frac{1}{2}}[\Gamma^{Bog\,}_{\bold{j}_1,\dots,\bold{j}_m\,;\,i-2,i-2}(w)]_{\mathcal{\tau}_h}(R^{Bog}_{\bold{j}_1,\dots,\bold{j}_m\,;\, i-2, i-2}(w))^{\frac{1}{2}}\Big]_1\tilde{Q}^{(i-2)}_{\bold{j}_m}
\end{eqnarray}
and use the relation in (\ref{tau-exp}). We get
\begin{eqnarray}
& &\tilde{Q}^{(i)}_{\bold{j}_m}(R^{Bog}_{\bold{j}_1,\dots,\bold{j}_m\,;\, i, i}(w))^{\frac{1}{2}}[\Gamma^{Bog\,}_{\bold{j}_1,\dots,\bold{j}_m\,;\,i,i}(w)]_{\mathcal{\tau}_h}(R^{Bog}_{\bold{j}_1,\dots,\bold{j}_m\,;\, i, i}(w))^{\frac{1}{2}}\tilde{Q}^{(i)}_{\bold{j}_m}\\
&=&\tilde{Q}^{(i)}_{\bold{j}_m}(R^{Bog}_{\bold{j}_1,\dots,\bold{j}_m\,;\, i, i}(w))^{\frac{1}{2}}W_{\bold{j}_m} (R^{Bog}_{\bold{j}_1,\dots,\bold{j}_m\,;\, i-2, i-2}(w))^{\frac{1}{2}}\times \\
& &\quad \times \,\sum_{l_{i-2}=0}^{h-1}\Big\{(R^{Bog}_{\bold{j}_1,\dots,\bold{j}_m\,;\, i-2, i-2}(w))^{\frac{1}{2}}[\Gamma^{Bog\,}_{\bold{j}_1,\dots,\bold{j}_m\,;\,i-2,i-2}(w)]_{\mathcal{\tau}_h}\,(R^{Bog}_{\bold{j}_1,\dots,\bold{j}_m\,;\, i-2, i-2}(w))^{\frac{1}{2}}\Big\}^{l_{i-2}} \times\\
& &\quad\quad\quad\times (R^{Bog}_{\bold{j}_1,\dots,\bold{j}_m\,;\, i-2, i-2}(w))^{\frac{1}{2}}W^*_{\bold{j}_m}(R^{Bog}_{\bold{j}_1,\dots,\bold{j}_m\,;\, i, i}(w))^{\frac{1}{2}}\tilde{Q}^{(i)}_{\bold{j}_m}\nonumber\\
&=&\tilde{Q}^{(i)}_{\bold{j}_m}(R^{Bog}_{\bold{j}_1,\dots,\bold{j}_m\,;\, i, i}(w))^{\frac{1}{2}}W_{\bold{j}_m} (R^{Bog}_{\bold{j}_1,\dots,\bold{j}_m\,;\, i-2, i-2}(w))^{\frac{1}{2}}\times \\
& &\quad \times \,\sum_{l_{i-2}=0}^{h-1}\Big\{\tilde{Q}^{(i-2)}_{\bold{j}_m}(R^{Bog}_{\bold{j}_1,\dots,\bold{j}_m\,;\, i-2, i-2}(w))^{\frac{1}{2}}[\Gamma^{Bog\,}_{\bold{j}_1,\dots,\bold{j}_m\,;\,i-2,i-2}(w)]_{\mathcal{\tau}_h}\,(R^{Bog}_{\bold{j}_1,\dots,\bold{j}_m\,;\, i-2, i-2}(w))^{\frac{1}{2}}\tilde{Q}^{(i-2)}_{\bold{j}_m}\Big\}^{l_{i-2}} \times\quad\quad \\
& &\quad\quad\quad\times (R^{Bog}_{\bold{j}_1,\dots,\bold{j}_m\,;\, i-2, i-2}(w))^{\frac{1}{2}}W^*_{\bold{j}_m}(R^{Bog}_{\bold{j}_1,\dots,\bold{j}_m\,;\, i, i}(w))^{\frac{1}{2}}\tilde{Q}^{(i)}_{\bold{j}_m}\nonumber\\
&=&\tilde{Q}^{(i)}_{\bold{j}_m}(R^{Bog}_{\bold{j}_1,\dots,\bold{j}_m\,;\, i, i}(w))^{\frac{1}{2}}W_{\bold{j}_m} (R^{Bog}_{\bold{j}_1,\dots,\bold{j}_m\,;\, i-2, i-2}(w))^{\frac{1}{2}}\times \label{nord-1}\\
& &\quad \times \,\sum_{l_{i-2}=0}^{h-1}\Big\{\Big[(R^{Bog}_{\bold{j}_1,\dots,\bold{j}_m\,;\, i-2, i-2}(w))^{\frac{1}{2}}[\Gamma^{Bog\,}_{\bold{j}_1,\dots,\bold{j}_m\,;\,i-2,i-2}(w)]_{\mathcal{\tau}_h}(R^{Bog}_{\bold{j}_1,\dots,\bold{j}_m\,;\, i-2, i-2}(w))^{\frac{1}{2}}\Big]_1\tilde{Q}^{(i-2)}_{\bold{j}_m}\Big\}^{l_{i-2}} \times\quad\quad \label{sud}\\
& &\quad\quad\quad\times (R^{Bog}_{\bold{j}_1,\dots,\bold{j}_m\,;\, i-2, i-2}(w))^{\frac{1}{2}}W^*_{\bold{j}_m}(R^{Bog}_{\bold{j}_1,\dots,\bold{j}_m\,;\, i, i}(w))^{\frac{1}{2}}\tilde{Q}^{(i)}_{\bold{j}_m}\,.\label{nord-2}
\end{eqnarray}
The normal ordering of the variables $a_{\bold{j}_m}$ and $a^*_{\bold{j}_m}$ contained in (\ref{nord-1}) and (\ref{nord-2}) can be implemented as if the operator in (\ref{sud}) were absent,  because (due to the inductive hypothesis) in (\ref{sud}) the operators $a_{\bold{j}_m}$ and $a^*_{\bold{j}_m}$ can be replaced with c-numbers. Thus, eventually we get 
\begin{eqnarray}
& &\tilde{Q}^{(i)}_{\bold{j}_m}(R^{Bog}_{\bold{j}_1,\dots,\bold{j}_m\,;\, i, i}(w))^{\frac{1}{2}}[\Gamma^{Bog\,}_{\bold{j}_1,\dots,\bold{j}_m\,;\,i,i}(w)]_{\mathcal{\tau}_h}(R^{Bog}_{\bold{j}_1,\dots,\bold{j}_m\,;\, i, i}(w))^{\frac{1}{2}}\tilde{Q}^{(i)}_{\bold{j}_m}\\
&=&\Big[(R^{Bog}_{\bold{j}_1,\dots,\bold{j}_m\,;\, i, i}(w))^{\frac{1}{2}}[\Gamma^{Bog\,}_{\bold{j}_1,\dots,\bold{j}_m\,;\,i,i}(w)]_{\mathcal{\tau}_h}(R^{Bog}_{\bold{j}_1,\dots,\bold{j}_m\,;\, i, i}(w))^{\frac{1}{2}}\Big]_1\tilde{Q}^{(i)}_{\bold{j}_m}\,.
\end{eqnarray}
Using the previous result  for $i=N-2$,  with the same arguments we get that
\begin{eqnarray}\label{1-express}
& &\overline{\mathscr{P}_{\psi^{Bog}_{\bold{j}_1,\dots,\bold{j}_{m-1}}}}W_{\bold{j}_m}\,(R^{Bog}_{\bold{j}_1,\dots,\bold{j}_m\,;\,N-2,N-2}(w))^{\frac{1}{2}}\times\\
& &\times \sum_{l_{N-2}=0}^{h-1}\Big[(R^{Bog}_{\bold{j}_1,\dots,\bold{j}_m\,;\,N-2,N-2}(w))^{\frac{1}{2}}[\Gamma^{Bog}_{\bold{j}_1,\dots,\bold{j}_m\,;\,N-2,N-2}(w)]_{\tau_h}\,(R^{Bog}_{\bold{j}_1,\dots,\bold{j}_m\,;\,N-2,N-2}(w))^{\frac{1}{2}}\Big]^{l_{N-2}}\times\\
&&\quad \times (R^{Bog}_{\bold{j}_1,\dots,\bold{j}_m\,;\,N-2,N-2}(w))^{\frac{1}{2}}W^*_{\bold{j}_m}\,\overline{\mathscr{P}_{\psi^{Bog}_{\bold{j}_1,\dots,\bold{j}_{m-1}}}}\\
&= &\overline{\mathscr{P}_{\psi^{Bog}_{\bold{j}_1,\dots,\bold{j}_{m-1}}}}W_{\bold{j}_m}\,(R^{Bog}_{\bold{j}_1,\dots,\bold{j}_m\,;\,N-2,N-2}(w))^{\frac{1}{2}}\times\\
& &\times \sum_{l_{N-2}=0}^{h-1}\Big\{(R^{Bog}_{\bold{j}_1,\dots,\bold{j}_m\,;\,N-2,N-2}(w))^{\frac{1}{2}}[\Gamma^{Bog}_{\bold{j}_1,\dots,\bold{j}_m\,;\,N-2,N-2}(w)]_{\tau_h}\,(R^{Bog}_{\bold{j}_1,\dots,\bold{j}_m\,;\,N-2,N-2}(w))^{\frac{1}{2}}\tilde{Q}^{(N-2)}_{\bold{j}_m}\Big\}^{l_{N-2}}\times\\
& &\quad \times (R^{Bog}_{\bold{j}_1,\dots,\bold{j}_m\,;\,N-2,N-2}(w))^{\frac{1}{2}}W^*_{\bold{j}_m}\,\overline{\mathscr{P}_{\psi^{Bog}_{\bold{j}_1,\dots,\bold{j}_{m-1}}}}\\
&=&\overline{\mathscr{P}_{\psi^{Bog}_{\bold{j}_1,\dots,\bold{j}_{m-1}}}}\,W_{\bold{j}_m}\,(R^{Bog}_{\bold{j}_1,\dots,\bold{j}_m\,;\,N-2,N-2}(w))^{\frac{1}{2}}\times\\
& &\quad\times \sum_{l_{N-2}=0}^{h-1}\Big\{\Big[(R^{Bog}_{\bold{j}_1,\dots,\bold{j}_m\,;\,N-2,N-2}(w))^{\frac{1}{2}}[\Gamma^{Bog}_{\bold{j}_1,\dots,\bold{j}_m\,;\,N-2,N-2}(w)]_{\tau_h}\,(R^{Bog}_{\bold{j}_1,\dots,\bold{j}_m\,;\,N-2,N-2}(w))^{\frac{1}{2}}\Big]_1\Big\}^{l_{N-2}}\tilde{Q}^{(N-2)}_{\bold{j}_m}\times\\
& &\quad \times (R^{Bog}_{\bold{j}_1,\dots,\bold{j}_m\,;\,N-2,N-2}(w))^{\frac{1}{2}}W^*_{\bold{j}_m}\,\overline{\mathscr{P}_{\psi^{Bog}_{\bold{j}_1,\dots,\bold{j}_{m-1}}}}\\
&=&\overline{\mathscr{P}_{\psi^{Bog}_{\bold{j}_1,\dots,\bold{j}_{m-1}}}}\,\Big[W_{\bold{j}_m}\,(R^{Bog}_{\bold{j}_1,\dots,\bold{j}_m\,;\,N-2,N-2}(w))^{\frac{1}{2}}\Big]_1\times\\
& & \quad\times \sum_{l_{N-2}=0}^{h-1}\Big\{\Big[(R^{Bog}_{\bold{j}_1,\dots,\bold{j}_m\,;\,N-2,N-2}(w))^{\frac{1}{2}}[\Gamma^{Bog}_{\bold{j}_1,\dots,\bold{j}_m\,;\,N-2,N-2}(w)]_{\tau_h}\,(R^{Bog}_{\bold{j}_1,\dots,\bold{j}_m\,;\,N-2,N-2}(w))^{\frac{1}{2}}\Big]_1\Big\}^{l_{N-2}}\times\\
& &\quad \times \Big[(R^{Bog}_{\bold{j}_1,\dots,\bold{j}_m\,;\,N-2,N-2}(w))^{\frac{1}{2}}W^*_{\bold{j}_m}\Big]_1\,\overline{\mathscr{P}_{\psi^{Bog}_{\bold{j}_1,\dots,\bold{j}_{m-1}}}}\\
&= &\Big[\overline{\mathscr{P}_{\psi^{Bog}_{\bold{j}_1,\dots,\bold{j}_{m-1}}}}W_{\bold{j}_m}\,R^{Bog}_{\bold{j}_1,\dots,\bold{j}_m\,;\,N-2,N-2}(w)\sum_{l_{N-2}=0}^{h-1}\Big[[\Gamma^{Bog}_{\bold{j}_1,\dots,\bold{j}_m\,;\,N-2,N-2}(w)]_{\tau_h}\,R^{Bog}_{\bold{j}_1,\dots,\bold{j}_m\,;\,\,N-2,N-2}(w)\Big]^{l_{N-2}}W^*_{\bold{j}_m}\,\overline{\mathscr{P}_{\psi^{Bog}_{\bold{j}_1,\dots,\bold{j}_{m-1}}}}\Big]_1\,.\nonumber
\end{eqnarray}
%%%%%%%%%%%%%%%%%%%%%%%%%%%%%%%%%
\subsection{Proof by induction of the identity in (5.161)-(5.163)}

 We use the shorthand notation
\begin{equation}
Q:=\Big[(R^{Bog}_{i,i})^{\frac{1}{2}}[\Gamma^{Bog\,}_{i,i}]_{\mathcal{\tau}_h}\,(R^{Bog}_{i,i})^{\frac{1}{2}}\Big]_1
\end{equation}
so that the definition in (5.156) reads
\begin{eqnarray}
& &\mathcal{D}_i\Big(Q^{n} \Big) \label{def-D_i}:=\sum_{j=0}^{n-1}\Big\{\mathcal{L}_i\Big(Q\Big)\Big\}^j\mathcal{D}_i\Big(Q\Big)\,Q^{n-j-1}\,.\label{def}
\end{eqnarray}
For $n=1$ the identity in (5.161)-(5.163) corresponds to
\begin{equation}
Q=\mathcal{L}_i(Q)+\mathcal{D}_i(Q)
\end{equation}
and follows from the definitions in (5.151)-(5.153). Now, assuming that the property holds  for $n$, i.e.,
\begin{equation}
Q^n=\Big[\mathcal{L}_i(Q)\Big]^n+\mathcal{D}_i(Q^n)\,,
\end{equation}
we show that it is also true for $n+1$.
Using the inductive hypothesis and the definition in (\ref{def}), we can write
\begin{eqnarray}
Q^{n+1}
&=&Q^{n}Q\\
%& &\Big\{\Big[(R^{Bog}_{i,i})^{\frac{1}{2}}[\Gamma^{Bog\,}_{i,i}]_{\mathcal{\tau}_h}\,(R^{Bog}_{i,i})^{\frac{1}{2}}\Big]_1\Big\}^{n+1} \\
%& &\Big\{\Big[(R^{Bog}_{i,i})^{\frac{1}{2}}[\Gamma^{Bog\,}_{i,i}]_{\mathcal{\tau}_h}\,(R^{Bog}_{i,i})^{\frac{1}{2}}\Big]_1\Big\}^{n} \Big\{\Big[(R^{Bog}_{i,i})^{\frac{1}{2}}[\Gamma^{Bog\,}_{i,i}]_{\mathcal{\tau}_h}\,(R^{Bog}_{i,i})^{\frac{1}{2}}\Big]_1\Big\}\\
&=&\Big\{\Big[\mathcal{L}_i\Big(Q\Big)\Big]^{n}  +\mathcal{D}_i\Big(Q^{n}\Big)\Big\}Q\\
&=&\Big[\mathcal{L}_i\Big(Q\Big)\Big]^{n}Q+\mathcal{D}_i\Big(Q^{n}\Big)Q\\
&=&\Big[\mathcal{L}_i\Big(Q\Big)\Big]^{n}\Big\{\mathcal{L}_i\Big(Q\Big)+\mathcal{D}_i\Big(Q\Big)\Big\}+\mathcal{D}_i\Big(Q^{n}\Big)Q\\
&=&\Big[\mathcal{L}_i\Big(Q\Big)\Big]^{n+1}+\Big[\mathcal{L}_i\Big(Q\Big)\Big]^{n}\mathcal{D}_i\Big(Q\Big)+\mathcal{D}_i\Big(Q^{n}\Big)Q\\
&=&\Big[\mathcal{L}_i\Big(Q\Big)\Big]^{n+1}+\Big[\mathcal{L}_i\Big(Q\Big)\Big]^{n}\mathcal{D}_i\Big(Q\Big)+\sum_{j=0}^{n-1}\Big\{\mathcal{L}_i\Big(Q\Big)\Big\}^j\mathcal{D}_i\Big(Q\Big)\,Q^{n-j}\\
&=&\Big[\mathcal{L}_i\Big(Q\Big)\Big]^{n+1}+\sum_{j=0}^{n}\Big\{\mathcal{L}_i\Big(Q\Big)\Big\}^j\mathcal{D}_i\Big(Q\Big)\,Q^{n-j}\\
&=&\Big[\mathcal{L}_i\Big(Q\Big)\Big]^{n+1}+\mathcal{D}_i\Big(Q^{n+1} \Big)\,.
\end{eqnarray}
\subsection{Inequality in (5.194)-(5.195)}
We recall that
\begin{eqnarray}
& &\|[(R^{Bog}_{\bold{j}_1,\dots,\bold{j}_m\,;\,i,i}(w))^{\frac{1}{2}}W_{\bold{j}_m\,;\,i,i-2}(R^{Bog}_{\bold{j}_1,\dots,\bold{j}_m\,;\,i-2,i-2}(w))^{\frac{1}{2}}]_{\overline{\mathscr{P}}}\| \label{norm}\\
&\leq &\|\Big(\frac{1}{\overline{\mathscr{P}_{\psi^{Bog}_{\bold{j}_1,\dots,\bold{j}_{m-1}}}}\Big[\Delta_{m-1}+(\frac{a^*_{\bold{0}}a_{\bold{0}}}{N}\phi_{\bold{j}_{m}}+k_{\bold{j}_{m}}^2)(N-i)-z\Big]\overline{\mathscr{P}_{\psi^{Bog}_{\bold{j}_1,\dots,\bold{j}_{m-1}}}}}\Big)^{\frac{1}{2}}\times\\
& &\quad\times \overline{\mathscr{P}_{\psi^{Bog}_{\bold{j}_1,\dots,\bold{j}_{m-1}}}}\,\phi_{\bold{j}_{m}}\frac{(a^*_{\bold{0}}a_{\bold{0}}-1)}{2N}(N-i+2)^{\frac{1}{2}}\,(N-i+2)^{\frac{1}{2}}\overline{\mathscr{P}_{\psi^{Bog}_{\bold{j}_1,\dots,\bold{j}_{m-1}}}}\times\,\nonumber\\
& &\quad \times \Big(\frac{1}{\overline{\mathscr{P}_{\psi^{Bog}_{\bold{j}_1,\dots,\bold{j}_{m-1}}}}\Big[\Delta_{m-1}+(\frac{a^*_{\bold{0}}a_{\bold{0}}}{N}\phi_{\bold{j}_{m}}+k_{\bold{j}_{m}}^2)(N-i+2)-z\Big]\overline{\mathscr{P}_{\psi^{Bog}_{\bold{j}_1,\dots,\bold{j}_{m-1}}}}}\Big)^{\frac{1}{2}}\overline{\mathscr{P}_{\psi^{Bog}_{\bold{j}_1,\dots,\bold{j}_{m-1}}}}\|\nonumber\,
\end{eqnarray}
where $N-2\geq i \geq N-2-h$, with $h=\lfloor (\ln N)^{\frac{1}{2}} \rfloor$ and even, 
$w=z^{Bog}_{\bold{j}_1,\dots,\bold{j}_{m-1}}+z$, and
\begin{equation}\label{interval}
-4\phi_{\bold{j}_m}\leq z< z_{m}+\gamma \Delta_{m-1}\,\Big(<  E^{Bog}_{\bold{j}_m}+ \frac{1}{2}\sqrt{\epsilon_{\bold{j}_m}}\phi_{\bold{j}_m}\sqrt{\epsilon_{\bold{j}_m}^2+2\epsilon_{\bold{j}_m}}\,<0\Big)\,. \,
\end{equation}
We call $r$ the spectral variable associated with $(0\leq) \overline{\mathscr{P}_{\psi^{Bog}_{\bold{j}_1,\dots,\bold{j}_{m-1}}}}\frac{a^*_{\bold{0}}a_{\bold{0}}}{N} \overline{\mathscr{P}_{\psi^{Bog}_{\bold{j}_1,\dots,\bold{j}_{m-1}}}}\leq 1$ and we study
\begin{eqnarray}
& &\sup_{1\geq r\geq 0} \Big|\Big(\frac{1}{\Delta_{m-1}+(r\phi_{\bold{j}_{m}}+k_{\bold{j}_{m}}^2)(N-i)-z}\Big)^{\frac{1}{2}}\times\\
& &\quad\quad \quad\times \phi_{\bold{j}_{m}}\frac{(r-\frac{1}{N})}{2}(N-i+2)\Big(\frac{1}{\Delta_{m-1}+(r\phi_{\bold{j}_{m}}+k_{\bold{j}_{m}}^2)(N-i+2)-z}\Big)^{\frac{1}{2}}\Big|\,\,.
\end{eqnarray}
 If $N$ is sufficiently large,  
\begin{eqnarray}
& &\sup_{\frac{1}{N}\geq r\geq 0} \Big|\Big(\frac{1}{\Delta_{m-1}+(r\phi_{\bold{j}_{m}}+k_{\bold{j}_{m}}^2)(N-i)-z}\Big)^{\frac{1}{2}}\times\\
& &\quad\quad \quad\times \phi_{\bold{j}_{m}}\frac{(r-\frac{1}{N})}{2}(N-i+2)\Big(\frac{1}{\Delta_{m-1}+(r\phi_{\bold{j}_{m}}+k_{\bold{j}_{m}}^2)(N-i+2)-z}\Big)^{\frac{1}{2}}\Big|\,
\end{eqnarray}
is trivially bounded by the R-H-S of (5.195).
For $\frac{1}{N}<r\leq 1$ we observe that 
\begin{eqnarray}
& &\sup_{1\geq r> \frac{1}{N}}\Big(\frac{1}{\Delta_{m-1}+(r\phi_{\bold{j}_{m}}+k_{\bold{j}_{m}}^2)(N-i)-z}\Big)^{\frac{1}{2}}\times\\
& &\quad\quad \quad\times \phi_{\bold{j}_{m}}\frac{(r-\frac{1}{N})}{2}(N-i+2)\Big(\frac{1}{\Delta_{m-1}+(r\phi_{\bold{j}_{m}}+k_{\bold{j}_{m}}^2)(N-i+2)-z}\Big)^{\frac{1}{2}}\,\\
&= &\sup_{1\geq r> \frac{1}{N}}\Big(\frac{1}{\frac{\Delta_{m-1}}{(r-\frac{1}{N})}+(\frac{r}{(r-\frac{1}{N})}\phi_{\bold{j}_{m}}+\frac{k_{\bold{j}_{m}}^2}{(r-\frac{1}{N})})(N-i)-\frac{z}{(r-\frac{1}{N})}}\Big)^{\frac{1}{2}}\times\\
& &\quad\quad \quad\times \phi_{\bold{j}_{m}}\frac{1}{2}(N-i+2)\Big(\frac{1}{\frac{\Delta_{m-1}}{(r-\frac{1}{N})}+(\frac{r}{(r-\frac{1}{N})}\phi_{\bold{j}_{m}}+\frac{k_{\bold{j}_{m}}^2}{(r-\frac{1}{N})})(N-i+2)-\frac{z}{(r-\frac{1}{N})}}\Big)^{\frac{1}{2}}\\
&\leq &\frac{(N-i+2)^{\frac{1}{2}}\,(N-i+2)^{\frac{1}{2}}}{2\Big[\Delta_{m-1}+(\phi_{\bold{j}_{m}}+k_{\bold{j}_{m}}^2)(N-i+2)-z\Big]^{\frac{1}{2}} }\,\frac{\phi_{\bold{j}_{m}}}{\Big[\Delta_{m-1}+(\phi_{\bold{j}_{m}}+k_{\bold{j}_{m}}^2)(N-i)-z\Big]^{\frac{1}{2}} }\,.
\end{eqnarray}

\subsection{Inequality in (5.200)}
The inequality to be proven is 
\begin{eqnarray}
& &\frac{(N-i+2)^{2}}{4\Big[\Delta_{m-1}+(\phi_{\bold{j}_{m}}+k_{\bold{j}_{m}}^2)(N-i+2)-z\Big] }\,\frac{\phi^2_{\bold{j}_{m}}}{\Big[\Delta_{m-1}+(\phi_{\bold{j}_{m}}+k_{\bold{j}_{m}}^2)(N-i)-z\Big]}\quad\quad\quad \label{Delta-W}\\
&\leq &\mathcal{W}_{\bold{j}_{m}\,;\,i,i-2}(z-\Delta_{m-1}+\frac{U_{\bold{j}_m}}{\sqrt{N}})\mathcal{W}^*_{\bold{j}_{m}\,;\,i-2,i}(z-\Delta_{m-1}+\frac{U_{\bold{j}_m}}{\sqrt{N}})\nonumber
\end{eqnarray}
for \begin{equation}
-4\phi_{\bold{j}_m}\leq z< z_{m}+\gamma \Delta_{m-1}\,\Big(<  E^{Bog}_{\bold{j}_m}+ \frac{1}{2}\sqrt{\epsilon_{\bold{j}_m}}\phi_{\bold{j}_m}\sqrt{\epsilon_{\bold{j}_m}^2+2\epsilon_{\bold{j}_m}}<0\Big) \,
\end{equation}
and for $N-2\geq i \geq N-2-\mathcal{O}((\ln N)^{\frac{1}{2}})$ with $N$ sufficiently large.

We recall that
\begin{eqnarray}
& &\mathcal{W}_{\bold{j}_{m}\,;\,i,i-2}(z-\Delta_{m-1}+\frac{U_{\bold{j}_m}}{\sqrt{N}})\mathcal{W}^*_{\bold{j}_{m}\,;\,i-2,i}(z-\Delta_{m-1}+\frac{U_{\bold{j}_m}}{\sqrt{N}})\\
&=&\frac{(i-1)i}{N^2}\frac{(N-i+2)^{2}}{4\Big[\Big(\frac{(i-2)}{N}\phi_{\bold{j}_{m}}+k_{\bold{j}_{m}}^2\Big)(N-i+2)-z+\Delta_{m-1}-\frac{U_{\bold{j}_{m}}}{\sqrt{N}}\Big] }\times\\
& &\quad\quad\quad \times\,\frac{\phi^2_{\bold{j}_{m}}}{\Big[\Big(\frac{i}{N}\phi_{\bold{j}_{m}}+k_{\bold{j}_{m}}^2\Big)(N-i)-z+\Delta_{m-1}-\frac{U_{\bold{j}_{m}}}{\sqrt{N}}\Big]}\\
&=&\frac{(N-i+2)^{2}}{4\Big[\Big(\frac{(i-2)}{i-1}\phi_{\bold{j}_{m}}+\frac{N}{i-1}k_{\bold{j}_{m}}^2\Big)(N-i+2)+\frac{N}{i-1}[-z+\Delta_{m-1}-\frac{U_{\bold{j}_{m}}}{\sqrt{N}}]\Big] }\times\\
& &\quad\quad\quad \times\,\frac{\phi^2_{\bold{j}_{m}}}{\Big[\Big(\phi_{\bold{j}_{m}}+\frac{N}{i}k_{\bold{j}_{m}}^2\Big)(N-i)+\frac{N}{i}[-z+\Delta_{m-1}-\frac{U_{\bold{j}_{m}}}{\sqrt{N}}]\Big]}\,.
\end{eqnarray}
Hence, it is enough to observe that for $N-2\geq i \geq N-2-\mathcal{O}((\ln N)^{\frac{1}{2}})$ and for \begin{equation}\label{interval}
-4\phi_{\bold{j}_m}\leq z< z_{m}+\gamma \Delta_{m-1}\,\Big(<  E^{Bog}_{\bold{j}_m}+ \frac{1}{2}\sqrt{\epsilon_{\bold{j}_m}}\phi_{\bold{j}_m}\sqrt{\epsilon_{\bold{j}_m}^2+2\epsilon_{\bold{j}_m}}<0\Big) \,
\end{equation}
we have
 \begin{eqnarray}
& &\Big(\frac{(i-2)}{i-1}\phi_{\bold{j}_{m}}+\frac{N}{i-1}k_{\bold{j}_{m}}^2\Big)(N-i+2)+\frac{N}{i-1}[-z+\Delta_{m-1}-\frac{U_{\bold{j}_{m}}}{\sqrt{N}}] \\
&= &\Delta_{m-1}+(\phi_{\bold{j}_{m}}+k_{\bold{j}_{m}}^2)(N-i+2)-z-\frac{U_{\bold{j}_{m}}}{\sqrt{N}}+\mathcal{O}(\frac{\ln N}{N})
\end{eqnarray}
and
\begin{eqnarray}
& &\Big(\phi_{\bold{j}_{m}}+\frac{N}{i}k_{\bold{j}_{m}}^2\Big)(N-i)+\frac{N}{i}[-z+\Delta_{m-1}-\frac{U_{\bold{j}_{m}}}{\sqrt{N}}] \\
&=&\Delta_{m-1}+(\phi_{\bold{j}_{m}}+k_{\bold{j}_{m}}^2)(N-i)-z-\frac{U_{\bold{j}_{m}}}{\sqrt{N}}+\mathcal{O}(\frac{\ln N}{N})
\end{eqnarray}
with $U_{\bold{j}_{m}}>0$.
\subsection{Inequality in (5.270)}
We recall that by assumption $$i\geq N-h-2,\quad i-\Delta n_{\bold{j}_{\bold{0}}}-2\geq0, \quad \Delta n_{\bold{j}_{\bold{0}}}\geq 0\,.$$ Then, we estimate
\begin{eqnarray}
& &[\mathcal{W}_{\bold{j}_*\,;\,i,i-2}(z)\mathcal{W}^*_{\bold{j}_*\,;\,i-2,i}(z)]_{\Delta n_{\bold{j}_{\bold{0}}}}\label{def-G-n0-med}\\
&:= & \frac{(i-\Delta n_{\bold{j}_{\bold{0}}}-1)(i-\Delta n_{\bold{j}_{\bold{0}}})}{N^2}\,\phi^2_{\bold{j}_{*}}\\
&  &\times\,\frac{(N-i+2)^2}{4\Big[(\frac{i-\Delta n_{\bold{j}_{\bold{0}}}}{N}\phi_{\bold{j}_{*}}+(k_{\bold{j}_*})^2)(N-i)-z\Big]\Big[(\frac{i-2-\Delta n_{\bold{j}_{\bold{0}}}}{N}\phi_{\bold{j}_{*}}+(k_{\bold{j}_*})^2)(N-i+2)-z\Big]}\,.\quad\quad\quad\label{def-G-n0-fin}\\
&=& \frac{1}{N^2}\,\phi^2_{\bold{j}_{*}}\\
&  &\times\,\frac{(N-i+2)^2}{4\Big[(\frac{1}{N}\phi_{\bold{j}_{*}}+\frac{1}{(i-\Delta n_{\bold{j}_{\bold{0}}})}(k_{\bold{j}_*})^2)(N-i)-\frac{z}{(i-\Delta n_{\bold{j}_{\bold{0}}})}\Big]\Big[(\frac{1}{N}\phi_{\bold{j}_{*}}+\frac{(k_{\bold{j}_*})^2}{(i-\Delta n_{\bold{j}_{\bold{0}}}-1)})(N-i+2)-\frac{\phi_{\bold{j}_{*}}(N-i+2)}{N(i-\Delta n_{\bold{j}_{\bold{0}}}-1)}-\frac{z}{(i-\Delta n_{\bold{j}_{\bold{0}}}-1)}\Big]}\,\quad\quad\quad\\
&=&\frac{\phi^2_{\bold{j}_{*}}(N-i+2)^2}{4N^2\Big[(\frac{1}{N}\phi_{\bold{j}_{*}}+\frac{1}{(i-\Delta n_{\bold{j}_{\bold{0}}})}(k_{\bold{j}_*})^2)(N-i)-\frac{z}{(i-\Delta n_{\bold{j}_{\bold{0}}})}\Big]\Big[(\frac{1}{N}\phi_{\bold{j}_{*}}+\frac{(k_{\bold{j}_*})^2}{(i-\Delta n_{\bold{j}_{\bold{0}}}-1)})(N-i+2)-\frac{1}{(i-\Delta n_{\bold{j}_{\bold{0}}}-1)}\Big(\frac{\phi_{\bold{j}_{*}}(N-i+2)}{N}+z\Big)\Big]}\,.\quad\quad\quad\label{last}
\end{eqnarray}
Now we observe that for $i\geq N-h-2$ we have $\frac{\phi_{\bold{j}_{*}}(N-i+2)}{N}\leq \frac{\phi_{\bold{j}_{*}}(h+4)}{N}$. Therefore, if 
\begin{equation}
 z\leq E^{Bog}_{\bold{j}_*}+ \sqrt{\epsilon_{\bold{j}_*}}\phi_{\bold{j}_*}\sqrt{\epsilon_{\bold{j}_*}^2+2\epsilon_{\bold{j}_*}}-\frac{(h+4)\phi_{\bold{j}_*}}{N}\label{range-z-G}
\end{equation} and $\epsilon_{\bold{j}_*}$ is small, we have 
$\frac{\phi_{\bold{j}_{*}}(N-i+2)}{N}+z\leq \frac{\phi_{\bold{j}_{*}}(h+4)}{N}+z<0$ and we can bound 
\begin{eqnarray}
& &(\ref{last})\\
&\leq&\frac{1}{N^2}\,\phi^2_{\bold{j}_{*}}\\
&  &\times\,\frac{(N-i+2)^2}{4\Big[(\frac{1}{N}\phi_{\bold{j}_{*}}+\frac{1}{i}(k_{\bold{j}_*})^2)(N-i)-\frac{z}{i}\Big]\Big[(\frac{1}{N}\phi_{\bold{j}_{*}}+\frac{(k_{\bold{j}_*})^2}{(i-1)})(N-i+2)-\frac{1}{(i-1)}\Big(\frac{\phi_{\bold{j}_{*}}(N-i+2)}{N}+z\Big)\Big]}\\
&\leq&\frac{1}{N^2}\,\phi^2_{\bold{j}_{*}}\\
&  &\times\,\frac{(N-i+2)^2}{4\Big[(\frac{1}{N}\phi_{\bold{j}_{*}}+\frac{1}{i}(k_{\bold{j}_*})^2)(N-i)-\frac{1}{i}\Big(\frac{\phi_{\bold{j}_{*}}(h+4)}{N}+z\Big)\Big]\Big[(\frac{1}{N}\phi_{\bold{j}_{*}}+\frac{(k_{\bold{j}_*})^2}{(i-1)})(N-i+2)-\frac{1}{(i-1)}\Big(\frac{\phi_{\bold{j}_{*}}(h+4)}{N}+z\Big)\Big]}\nonumber \\
&=&\mathcal{W}_{\bold{j}_*\,;\,i,i-2}(z+\frac{\phi_{\bold{j}_{*}}(h+4)}{N})\mathcal{W}^*_{\bold{j}_*\,;\,i-2,i}(z+\frac{\phi_{\bold{j}_{*}}(h+4)}{N})
%&=&(\frac{(i-1)i}{N^2}+...)\,\phi^2_{\bold{j}_{*}}\\
%&  &\times\,\frac{(N-i+2)^2}{4\Big[(\frac{i}{N}\phi_{\bold{j}_{*}}+(k_{\bold{j}_*})^2)(N-i)-(z+\frac{(N-i)\Delta n_{\bold{j}_{\bold{0}}}\phi_{\bold{j}_{*}}}{N})\Big]\Big[(\frac{i-2-\Delta n_{\bold{j}_{\bold{0}}}}{N}\phi_{\bold{j}_{*}}+(k_{\bold{j}_*})^2)(N-i)+2(\frac{i-2-\Delta n_{\bold{j}_{\bold{0}}}}{N}\phi_{\bold{j}_{*}}+\epsilon_{\bold{j}_*})-z\Big]}\quad\quad\quad
\end{eqnarray}